\documentclass[]{jfm}
\usepackage[dvipsnames]{xcolor}

\usepackage{graphicx}
\usepackage{newtxtext}
\usepackage{newtxmath}
\usepackage{natbib}
\usepackage{hyperref}
\hypersetup{
	colorlinks = true,
	urlcolor   = blue,
	citecolor  = black,
}

\newcommand{\RomanNumeralCaps}[1]
\linenumbers

\usepackage{amsmath,accents}
\usepackage[labelfont={small}, font={normal}]{subfig} 
\usepackage{graphicx}
\usepackage{epstopdf}
\usepackage{amssymb}
\graphicspath{{./Figures/}}
\makeatletter
\renewcommand{\absfooterflag}{} 
\def\@oddfoot{}                 
\def\@evenfoot{}                
\makeatother

\title{Extensional rheology of dilute suspensions of spheres in polymeric liquids}

\author{Arjun Sharma\aff{1,2},
	\and Donald L. Koch\aff{3}
	\corresp{\email{dlk15@cornell.edu}}}

\affiliation{\aff{1}Sibley School of Mechanical and Aerospace Engineering, Cornell University, Ithaca, NY, 14853, USA
	\aff{2}Center for Computing Research, Sandia National Laboratories, Albuquerque, NM, 87185, USA
	\aff{3}Robert Frederick Smith School of Chemical and Biomolecular Engineering, Cornell University, Ithaca, NY, 14853, USA}

\begin{document}
	\maketitle
	
	\begin{abstract}
The extensional rheology of dilute suspensions of spheres in viscoelastic/polymeric liquids is studied computationally. At low polymer concentration $c$ and Deborah number $De$ (imposed extension rate times polymer relaxation time), a wake of highly stretched polymers forms downstream of the particles due to larger local velocity gradients than the imposed flow, indicated by $\Delta De_\text{local}>0$. This increases the suspension’s extensional viscosity with time and $De$ for $De < 0.5$. When $De$ exceeds 0.5, the coil-stretch transition value, the fully stretched polymers from the far-field collapse in regions with $\Delta De_\text{local} < 0$ (lower velocity gradient) around the particle's stagnation points, reducing suspension viscosity relative to the particle-free liquid. The interaction between local flow and polymers intensifies with increasing $c$. Highly stretched polymers impede local flow, reducing $\Delta De_\text{local}$, while $\Delta De_\text{local}$ increases in regions with collapsed polymers. Initially, increasing $c$ aligns $\Delta De_\text{local}$ and local polymer stretch with far-field values, diminishing particle-polymer interaction effects. However, beyond a certain $c$, a new mechanism emerges. At low $c$, fluid three particle radii upstream exhibits $\Delta De_\text{local} > 0$, stretching polymers beyond their undisturbed state. As $c$ increases, however, $\Delta De_\text{local}$ in this region becomes negative, collapsing polymers and resulting in increasingly negative stress from particle-polymer interactions at large $De$ and time. At high $c$, this negative interaction stress scales as $c^2$, surpassing the linear increase of particle-free polymer stress, making dilute sphere concentrations more effective at reducing the viscosity of viscoelastic liquids at larger $De$ and $c$.
	\end{abstract}
	
	
\section{Introduction}\label{sec:Intro}
Particle-filled polymeric or viscoelastic liquids are widely used in industrial applications such as hydraulic fracturing \citep{barbati2016complex}, fiber spinning, and extrusion molding \citep{breitenbach2002melt,huang2003review,nakajima1994advanced}. In hydraulic fracturing, polymeric liquids are chosen for their high viscosity, and spherical particles are added as proppants to keep fractures open. In extrusion molding and fiber spinning, solid particles enhance the strength of the final product. These processes often involve uniaxial extensional flow at critical stages, such as entering pores or passing through a die or spinneret. Understanding the extensional rheology of viscoelastic suspensions is crucial for optimizing the efficiency of such industrial processes, but remains less explored compared to shear rheology \citep{shaqfeh2019rheology}.

Previous theoretical and experimental studies have characterized the rheology of particle-free viscoelastic fluids, particularly the rapid increase in extensional viscosity due to the polymer's coil-stretch transition \citep{de1974coil}. The addition of particles can influence this mechanism, leading to distinct rheological behaviors in suspensions. Recently, \cite{jain2019extensional} examined short-term transient rheology within a narrow range of polymer relaxation times ($\lambda$) and concentrations ($c$), and \cite{SteadyStatePaper} looked at steady-state rheology across a broader range of $\lambda$ but small $c$. This study extends the understanding of transient rheology in dilute sphere suspensions over a wide range of $\lambda$ and $c$ through two computational approaches, identifying novel particle-polymer interaction mechanisms and qualitatively explaining the only existing experimental data-set on this subject \citep{SHERE_Report,soulages2010extensional,hall2009preliminary,SHERE2_Report,jaishankar2012shear}.

In a Newtonian fluid, in the absence of inertia, the particle stresslet \citep{batchelor1970stress} leads to additional stress or viscosity in the suspension \citep{einstein2005neue,batchelor1970stress,batchelor1972determination}. The stresslet represents the resistance of a rigid particle to deformation by the fluid and depends on the additional traction created on the particle surface by the fluid. In viscoelastic fluids, which consist of polymers dissolved in a Newtonian solvent, the stresslet also includes the effect of polymeric stress on the particle surface. The dissolved polymers are stretched differently in the presence of particles due to velocity perturbations. This accumulated effect of extra polymer stress throughout the fluid leads to additional stress in the suspension, termed particle-induced polymer stress (PIPS). The changing polymer configuration locally alters the solvent stress by modifying the local pressure and velocity. The sum of the stresslet due to polymer stress and the solvent stress modification is termed the interaction stresslet. Therefore, the two-way interaction between particles and polymers creates two additional stress mechanisms in the suspension: PIPS, where particles modify the polymer stretch or configuration, and the interaction stresslet, where polymers exert additional stress on the particle. The net particle-polymer interaction stress is the sum of PIPS and the interaction stresslet.

The stress measured in an experiment with a homogeneous suspension is an average over all possible particle configurations \citep{hinch1977averaged}. The stresslet, or the average of the extra stress inside the rigid particles, can be expressed in terms of a surface integral yielding the first moment of the fluid stress over the particle surface, as shown by \cite{batchelor1970stress}. This approach renders the knowledge of the constitutive relation within the particle unnecessary. In sufficiently dilute suspensions, particle-particle interactions are rare. Therefore, the flow and polymer configuration field in an unbounded fluid around an isolated particle fully determine the rheology of the suspension. While particle-induced polymer stress (PIPS) represents the extra polymeric stress in the fluid, previous attempts by \cite{greco2007rheology,housiadas2009rheology}; and \cite{jain2019extensional} to approximate PIPS through volume averaging have been inappropriate. As discussed by \cite{koch2016stress}, based on the far-field Newtonian velocity and strain rate disturbance scalings of $1/r^2$ and $1/r^3$, respectively, the additional polymeric stress in a steady linear flow around a sphere decays as $1/r^3$ at large distances $r$ from the particle. Consequently, volume averaging leads to logarithmic divergence, which causes slow or absent convergence in direct numerical simulations. \cite{koch2016stress} showed that this logarithmic divergence arises from the polymer stress linearized about the velocity and polymer stress in the fluid undisturbed by the particle. They also demonstrated that the ensemble average of this linearized stress is zero. Therefore, removing the linearized stress from the extra stress before approximating the ensemble averaging with volume averaging leads to convergent integrals. This approach is also numerically beneficial, as it accelerates convergence with domain size by removing the slower-decaying component of the stress from the volume integral. In a transient flow, the initial extra polymer stress is small, and equivalent PIPS is obtained with or without removing the linearized stress. However, as the extra polymer stress increases in both magnitude and spatial extent, the divergence in the linearized stress appears at a time proportional to the polymer relaxation time. 

In numerical or theoretical studies of viscoelastic fluids, polymers are typically modeled using a continuum equation for the polymer configuration tensor, which is proportional to the polymer stress \citep{bird2016polymer}. A simple way to capture polymer elasticity is through the dumbbell model, which represents a linear polymer molecule as a spring with two Brownian beads influenced by velocity gradients. The equations for the polymer configuration tensor, $\boldsymbol{\Lambda}$, are derived from the outer product of the dumbbell's end-to-end vector with itself and then averaged over all possible polymer orientations to obtain continuum-level governing equations. A non-linearly elastic spring is used to capture the finite extensibility of the polymer molecule, which is crucial in strong flows such as uniaxial extension. In simpler models like the Oldroyd-B model, the polymer undergoing extensional flow stretches indefinitely, and the polymer stress can become unbounded. The FENE (finite extensible non-linear elastic) models, particularly the FENE-P model \citep{peterlin1968non}, effectively capture the coil-stretch transition at small polymer concentrations, $c$ \citep{anna2008effect}, which is a key mechanism in this study. For higher $c$, the Giesekus model is more appropriate as it better describes the behavior of high-concentration {viscoelastic liquids} \citep{khan1987comparison}.

In the steady state of a liquid with dilute polymer concentration, $c$, particles and polymers interact differently before and after the coil-stretch transition of the undisturbed polymers \citep{SteadyStatePaper}, which occurs at $De = 0.5$. The Deborah number, $De$, is defined as the product of polymer relaxation time, $\lambda$, and imposed extension rate. When undisturbed polymers are in a coiled state ($De < 0.5$), velocity perturbations created by the particle cause extra stretching (beyond that caused by the imposed flow) around the extensional axis downstream of the particle, leading to a wake of highly stretched polymers. This results in additional stress due to particle-polymer interaction. When polymers are already in their stretched state, these regions of extra stretching do not significantly alter the polymer stretch near the particle compared to the far-field. However, low-stretching regions near the stagnation points on the particle surface, at the compressional (front) and extensional (rear) axes, cause the polymers to collapse nearly to their equilibrium or unstressed state. This creates regions of collapsed polymers around the particle surface. Therefore, beyond the coil-stretch transition of undisturbed polymers ($De > 0.5$), even at low $c$, the significant particle-polymer interaction stress can lead to a reduction in the suspension viscosity compared to the viscosity of the particle-free polymeric fluid.

Recent numerical results by \cite{jain2019extensional} have explored the extensional rheology of dilute particle-filled viscoelastic liquids. Our study enhances these results in several ways. Firstly, while they used volume averaging to obtain the suspension stress equivalent to particle-induced polymer stress (PIPS), we provide a quantitative correction using the more appropriate formulation by \cite{koch2016stress}. Secondly, we examine a broader range of parameters to characterize the suspension rheology. Although \cite{jain2019extensional} provided results for three different polymer constitutive models, their simulations were limited in terms of maximum time or Hencky strain (time non-dimensionalized with strain rate), $H$, and the range of $De$. Their simulations only extended up to the coil-stretch transition of undisturbed polymers, missing some profound features such as the negative particle-polymer interaction stress. We observe that significant particle-polymer interactions continue beyond the coil-stretch transition. In addition to considering larger $H$, we explore important qualitative and quantitative variations in suspension rheology by examining a wider range of $De$, $c$, maximum polymer extensibility ($L$) in {FENE-P, and polymer mobility ($1/\alpha$) in Giesekus liquids}. Lastly, we provide more detailed mechanistic explanations underlying the rheological observations.

To fully elucidate the rheology of suspensions, we implement two synergistic methods. Firstly, we adopt a semi-analytical approach, previously utilized by \cite{koch2016stress} and \cite{SteadyStatePaper}, which employs a regular perturbation expansion in $c$ and a generalized reciprocal theorem. This technique assesses the rheological behavior of particle suspensions in viscoelastic liquids with low $c$ values, across various $De$, maximum polymer extensibility ($L$), and $H$. Its computational efficiency is due to the straightforward numerical integration of ordinary differential equations obtained via the method of characteristics, allowing for exceptional spatial and temporal resolution and the capability to handle large computational domains. From this approach, we derive critical scaling relationships in $L$ and identify key mechanisms influencing suspension rheology across a wide range of $De$. Building on these insights, we then employ direct numerical simulations to investigate the rheology at higher $c$ values, where novel particle-polymer interaction mechanisms emerge. These simulations are facilitated by our in-house finite difference solver, introduced in \cite{NumericalMethodPaper}. This solver, formulated in prolate spheroidal coordinates, precisely models the particle surface. The chosen coordinate system naturally concentrates the grid near the particle surface in Euclidean space, achieving high resolution in areas with significant gradients of flow variables. The computational domain's outer boundary is defined by a far-field spherical surface where the extensional flow conditions are applied, simulating an unbounded fluid surrounding a particle.

The rest of the paper is organized as follows. Section \ref{sec:Formulation} discusses the equations governing fluid velocity and polymer stress around the particle, as well as the necessary rheological quantities. Since the additional stresses in the suspension arising from particle-polymer interaction, discussed in sections \ref{sec:DiluteRheology} and \ref{sec:ConcentratedRheology}, depend on the coil-stretch transition of the polymers far from the particles, the evolution of the undisturbed polymer stress is briefly reviewed in section \ref{sec:U0Rheology}. This discussion also helps compare the suspension rheology across different polymer constitutive models (FENE-P and Giesekus) used in our study. Section \ref{sec:DiluteRheology} characterizes the transient nature of the particle-polymer interaction on suspension rheology for dilute $c$ using a semi-analytical method. The effects of polymer concentration on suspension rheology using direct numerical simulations are illustrated in section \ref{sec:ConcentratedRheology}. We summarize the only available experimental results on the effect of particles on the extensional flow of polymeric fluids in section \ref{sec:ExperimentalResults} and relate these to our computational findings. The main conclusions are provided in section \ref{sec:Conclusions}. More details about the rheological quantities, interaction stresslet, and PIPS than mentioned in section \ref{sec:Formulation} are provided in appendix \ref{sec:Moredetails}, where the treatment of ensemble averaging by first removing the linearized stress is also discussed. Appendices \ref{sec:MethodSemiAnalytical} and \ref{sec:DNSMethodology} provide details about the semi-analytical method and the techniques and validation for the direct numerical simulations, respectively.

\section{Mathematical formulation and different stresses arising in a particle suspension in viscoelastic liquids}\label{sec:Formulation}
We consider the inertia-less and incompressible polymeric fluid flow in a dilute suspension of spheres. Thus, throughout the suspension, mass and momentum conservation imply a divergence-free velocity, $\mathbf{u}$, and stress, $\boldsymbol{\sigma}$, field, respectively,
\begin{equation}
	\nabla\cdot \mathbf{u}=0,\hspace{0.2in}\nabla\cdot \boldsymbol{\sigma}=0.\label{eq:MassMomentum}
\end{equation}
Since the fluid consists of polymers dissolved in a Newtonian solvent, the stress is the sum of a Newtonian solvent stress, $\boldsymbol{\tau}$, and a polymeric/viscoelastic stress, $\mathbf{\Pi}$,
\begin{equation}\label{eq:constitutive1}
	\boldsymbol{\sigma}=\boldsymbol{\tau}+c\mathbf{\Pi}=-p\boldsymbol{\delta}+2\boldsymbol{e}+c\mathbf{\Pi}.
\end{equation}
Here, $p$ is the hydrodynamic pressure, $\boldsymbol{e} = (\nabla\mathbf{u} + (\nabla\mathbf{u})^\text{T})/2$ is the strain rate tensor, and $\boldsymbol{\delta}$ is the identity tensor. According to the Huggins equation, for dilute polymer solutions, the volume or mass concentration of the polymers is linearly proportional to the ratio of polymer to solvent viscosity \citep{rubinstein2003polymer}. In non-dilute solutions, while remaining a monotonic function, the viscosity ratio becomes a quadratic function of the volume concentration. In this work, to maintain simplicity, we refer to the ratio of polymer to solvent viscosity as polymer concentration, $c$. We consider the effects of suspension under uniaxial extensional flow, such that the boundary conditions on equations \eqref{eq:MassMomentum} are
\begin{equation}
	\mathbf{u}=0,\text{ on particle surface, and, }\mathbf{u}=\langle\mathbf{u}\rangle \text{ as } |\mathbf{r}|\rightarrow\infty,\label{eq:BCs}
\end{equation} 
where {the imposed velocity is averaged over an ensemble of all particle configurations with this average  indicated by the angular brackets},
\begin{equation}
	\langle\mathbf{u}\rangle=\mathbf{r}\cdot\langle\boldsymbol{e}\rangle,\hspace{0.2in} \langle{e_{ij}}\rangle =\delta_{i1}\delta_{j1}-\frac{1}{2}(\delta_{i2}\delta_{j2}+\delta_{i3}\delta_{j3}). \label{eq:UndisturbedStrainRate}
\end{equation}
The radius of the spheres, the inverse of the imposed extensional rate, $\dot{\epsilon}$, and the solvent viscosity are used as the length, time, and viscosity scales to non-dimensionalize the governing equations.

We model the polymeric stress with either FENE-P or Giesekus constitutive equations, which consider the polymer to be a dumbbell as mentioned in section \ref{sec:Intro}. In a uniaxial extensional flow, the FENE-P model is appropriate for capturing the polymer stresses in liquids with dilute polymer concentrations \citep{anna2008effect}, whereas the Giesekus model is useful for {concentrated solutions} \citep{khan1987comparison}. The polymer's stress is a function of its conformation, $\boldsymbol{\Lambda}=\langle \mathbf{q}\mathbf{q}\rangle_\text{polymer orientation}$, defined as the average over orientations of the outer product of the relative vector of the ends of the dumbbell, $\mathbf{q}$, with itself. The stress-conformation relation for each model is,
\begin{equation}
	\boldsymbol{\Pi}=\begin{cases}
		\frac{1}{De}(f\boldsymbol{\Lambda}-b\boldsymbol{\delta}), \hspace{0.2in} f=\frac{L^2}{L^2-\text{tr}(\boldsymbol{\Lambda})}, \hspace{0.2in} b=\frac{L^2}{L^2-\text{tr}(\boldsymbol{\delta})},&\text{FENE-P},\\
		\frac{1}{De}(\boldsymbol{\Lambda}-\boldsymbol{\delta}),&\text{Giesekus}.
	\end{cases}\label{eq:constitutive2}
\end{equation}
The underlying velocity field influences $\boldsymbol{\Lambda}$ according to the following equation,
\begin{equation}
	\frac{\partial \boldsymbol{\Lambda}}{\partial t}+\mathbf{u}\cdot \nabla \boldsymbol{\Lambda}=\nabla \mathbf{u}^\text{T}\cdot\boldsymbol{\Lambda}+\boldsymbol{\Lambda}\cdot\nabla\mathbf{u}+\begin{cases}
		\frac{1}{De}(b\boldsymbol{\delta}-f\boldsymbol{\Lambda}),&\text{FENE-P}\\
		\frac{1}{De}((\boldsymbol{\delta}-\boldsymbol{\Lambda})-\alpha(\boldsymbol{\delta}-\boldsymbol{\Lambda})\cdot(\boldsymbol{\delta}-\boldsymbol{\Lambda})),&\text{Giesekus}
	\end{cases}.\label{eq:Configuration}
\end{equation}
In the FENE-P model, $L$ is the maximum extensibility of the polymer normalized with its radius of gyration. In the Giesekus model, $0\le\alpha\le1$ is the mobility parameter that models the ability of the entangled polymers to restrict the motion and stress of one another. These dumbbell models capture the polymer convection by the underlying velocity field (second term on the left), its stretching and rotation by the underlying velocity gradient (first two terms on the right), and its relaxation to the equilibrium state, $\boldsymbol{\Lambda}_\text{equil.}=\boldsymbol{\delta}$ (last term on the right, modeled differently in FENE-P and Giesekus).

In the dumbbell models, the polymer stretch, $\mathcal{S}$, or the 2-norm of the relative vector of the dumbbell's ends is 
\begin{equation}
	\mathcal{S}= \langle||\mathbf{q}||_2\rangle_\text{polymer orientation}=\sqrt{\text{tr}(\boldsymbol{\Lambda})}.
\end{equation}
A non-dimensional group, the Deborah number ($De$), appears in these equations and is the ratio of polymer relaxation time, $\lambda$, to the flow characteristic time, $1/\dot{\epsilon}$,
\begin{equation}
	De=\lambda \dot{\epsilon}.\label{eq:DeEqn}
\end{equation}
Neglecting fluid inertia leads to quasi-steady velocity and pressure fields that respond instantaneously to the polymer stress in equation~\eqref{eq:MassMomentum}. This requires the momentum diffusion time through the fluid (with kinematic viscosity $ \nu $) in the region surrounding a particle of radius $ a $, given by $ a^2/\nu $, to be small compared to $ \lambda $; i.e., the elasticity number (the ratio of Deborah to Reynolds number), $El = De/Re = \lambda \nu / a^2 \gg 1$, must be large—a common limit in polymeric fluids.

While equation \eqref{eq:constitutive1} captures the fluid stress, in a suspension of particles additional stress components arise whose physical origins are worth noting.
The stress in a suspension of particles is the average stress over the ensemble of all possible particle configurations. In a particular instance of particle configuration, the stress at any point within the suspension is the sum of the stress given by equation \eqref{eq:constitutive1} and the extra stress within the particle phase (which is zero in the fluid phase). In the rheology of an incompressible suspension, the deviatoric or traceless part of the suspension stress is most relevant, as the trace can be absorbed into a modified pressure. The ensemble-averaged deviatoric suspension stress in a particle suspension within a viscoelastic liquid is,
\begin{equation}\label{eq:SuspensionStress}
	\langle\hat{\boldsymbol{\sigma}}\rangle=\langle{\boldsymbol{\sigma}}\rangle-\frac{1}{3}\boldsymbol{\delta}\text{tr}(\langle\boldsymbol{\sigma}\rangle)=2\langle\boldsymbol{e}\rangle+c\langle \hat{\boldsymbol{\Pi}}\rangle+n\hat{\text{\textbf{S}}}(\boldsymbol{\sigma}),
\end{equation}
or,
\begin{equation}
	\langle\hat{\boldsymbol{\sigma}}\rangle=(2+5\phi)\langle\boldsymbol{e}\rangle+c\hat{\boldsymbol{\Pi}}^U+{c\phi}(\hat{\boldsymbol{\Pi}}^{PP}+\hat{\text{\textbf{S}}}^\text{PP}) \label{eq:ConstitutiveRheology}.
\end{equation}
Since $\hat{\boldsymbol{\sigma}}$ represents the stress field at any point within the suspension, its ensemble average incorporates the effects from both the particle and fluid phases. Here, $n$ is the number of particles per unit volume and $\phi=nV_p/(V_f+nV_p)$ is the particle volume fraction where the volume of each particle is $V_p$ and the total suspension volume is the sum of particle ($nV_p$) and fluid volume ($V_f$). The term $2\langle\boldsymbol{e}\rangle$ represents the imposed rate of strain, $c\langle \hat{\boldsymbol{\Pi}}\rangle=c(\hat{\boldsymbol{\Pi}}^U+\phi\hat{\boldsymbol{\Pi}}^{PP})$ is the ensemble average of the polymeric stress in the fluid, and $n\hat{\text{\textbf{S}}}(\boldsymbol{\sigma})=5\phi\langle\boldsymbol{e}\rangle+c\phi\hat{\text{\textbf{S}}}^\text{PP}$ arises due to the extra stress in the particle phase and is termed as the particle stresslet. The third term in equation \eqref{eq:ConstitutiveRheology} is the polymer stress in the absence of the particles, and the second, fourth, and fifth terms in equation \eqref{eq:ConstitutiveRheology} represent the effect of particles. The second term, $5\phi\langle\boldsymbol{e}\rangle$, is the Newtonian stresslet in a suspension of spheres (first calculated by \cite{einstein2005neue}). The last two terms, ${c\phi}(\hat{\boldsymbol{\Pi}}^{PP}+\hat{\text{\textbf{S}}}^\text{PP})$, represent the particle-polymer interaction stress in the suspension. The presence of particles leads to a deviation in the fluid velocity that alters the polymer stress within the fluid, and the extra stress hence created manifests as the particle-induced polymer stress or PIPS, ${c\phi}\hat{\boldsymbol{\Pi}}^{PP}$. The presence of polymers creates an additional stress in the particle phase which is represented by the interaction stresslet, ${c\phi}\hat{\text{\textbf{S}}}^\text{PP}$.

Due to the symmetry of the imposed uniaxial extensional flow around a sphere, we may express,
\begin{equation}
	\langle\hat{\boldsymbol{\sigma}}\rangle=2\mu_\text{ext}\langle\boldsymbol{e}\rangle, \text{with, }\mu_\text{ext}= 1+2.5\phi+0.5 c \hat{\Pi}^U\Big[1+{\phi}\frac{\hat{\Pi}^{PP}+\hat{\text{S}}^{PP}}{\hat{\Pi}^U}\Big],\label{eq:ExtVisc}
\end{equation}
being the extensional viscosity of the suspension. It is the sum of the extensional viscosity of the polymeric fluid, $\mu_\text{fluid}$, and the net effect of the addition of particles to the suspension stress, $\mu_\text{part}$,
\begin{equation}
	\mu_\text{fluid}=1+0.5c\hat{\Pi}^U,\hspace{0.2in}\mu_\text{part}=\phi(2.5+0.5c(\hat{\Pi}^{PP}+\hat{\text{S}}^{PP})).\label{eq:PartVisc}
\end{equation}
Further details on the constituents of particle-polymer interaction stress, ${c\phi}(\hat{\boldsymbol{\Pi}}^{PP}+\hat{\text{\textbf{S}}}^\text{PP})={c\phi}(\hat{\Pi}^{PP}+\hat{\text{S}}^{PP})\langle\boldsymbol{e}\rangle$ are provided in appendix \ref{sec:Moredetails}. {We will consider the extensional component of the stress when illustrating the polymer stress for both the undisturbed fluid, $\hat{\boldsymbol{\Pi}}^U$, in figures \ref{fig:UndisturbedStress} and \ref{fig:UndisturbedStresGiesekusFENEPs} in section \ref{sec:U0Rheology} and that due to the effect of particle polymer interaction $\hat{\Pi}^{PP}+\hat{\text{S}}^{PP}$ in figures \ref{fig:TotalInteractionStressUnNorm} to \ref{fig:TotalPIPSandStresslet}, and, \ref{fig:VarywithcDep4} to \ref{fig:L50De2_} in sections \ref{sec:DiluteRheology} and \ref{sec:ConcentratedRheology}.}

\section{Undisturbed polymer stress and the coil-stretch transition phenomenon}\label{sec:U0Rheology} 
In extensional rheology {of a startup flow with constant strain rate}, time is non-dimensionalized with the imposed extension rate and is termed as Hencky strain, $H$. In a uniaxial extensional flow, the transient behavior of the undisturbed polymer stress, $c\hat{\Pi}^U\langle\boldsymbol{e}\rangle$, is significant and qualitatively impacts the particle-polymer interaction stress. As we will later see, the polymer behavior in various regions around the particle relative to the undisturbed polymers is crucial for understanding the particle-polymer interaction stress. The undisturbed polymer stress varies linearly with $c$ for the entire range of relevant parameters, so we consider this stress normalized with $c$, i.e., $\hat{\Pi}^U\langle\boldsymbol{e}\rangle$, for the FENE-P and Giesekus liquids in this section.

Since the undisturbed stress is spatially homogeneous for imposed linear flows, such as the uniaxial extensional flow considered here, the governing equations for this stress are ordinary differential equations (ODEs). Specifically, the convective term in equation \eqref{eq:Configuration} is absent for the undisturbed polymer configuration equations. We solve the corresponding ODEs using MATLAB's ode15s for the small $c$ results in section \ref{sec:DiluteRheology} and a variable time step second-order Runge-Kutta method for the results at an arbitrary $c$ in section \ref{sec:ConcentratedRheology}.

\subsection{FENE-P}\label{sec:FENEU0}
\begin{figure}
	\centering
	\subfloat{\includegraphics[width=0.33\textwidth]{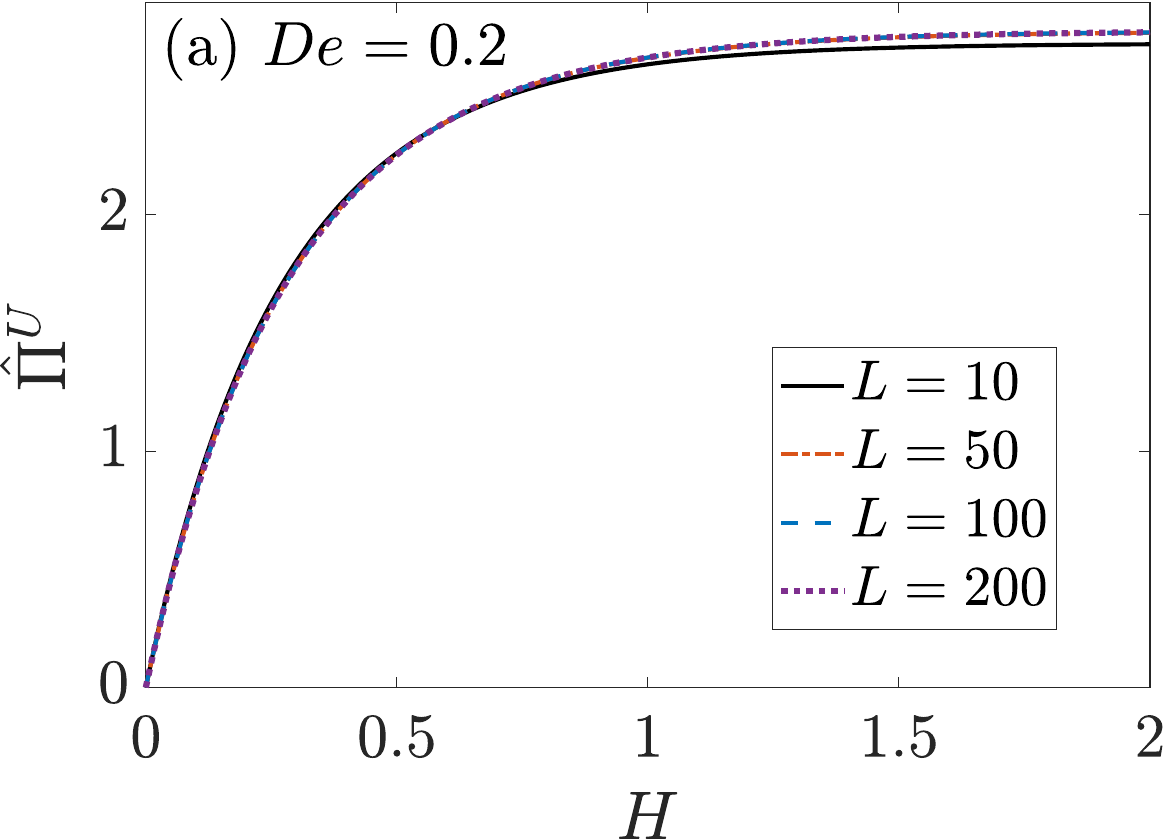}\label{fig:U0Dep2}}
	\subfloat{\includegraphics[width=0.33\textwidth]{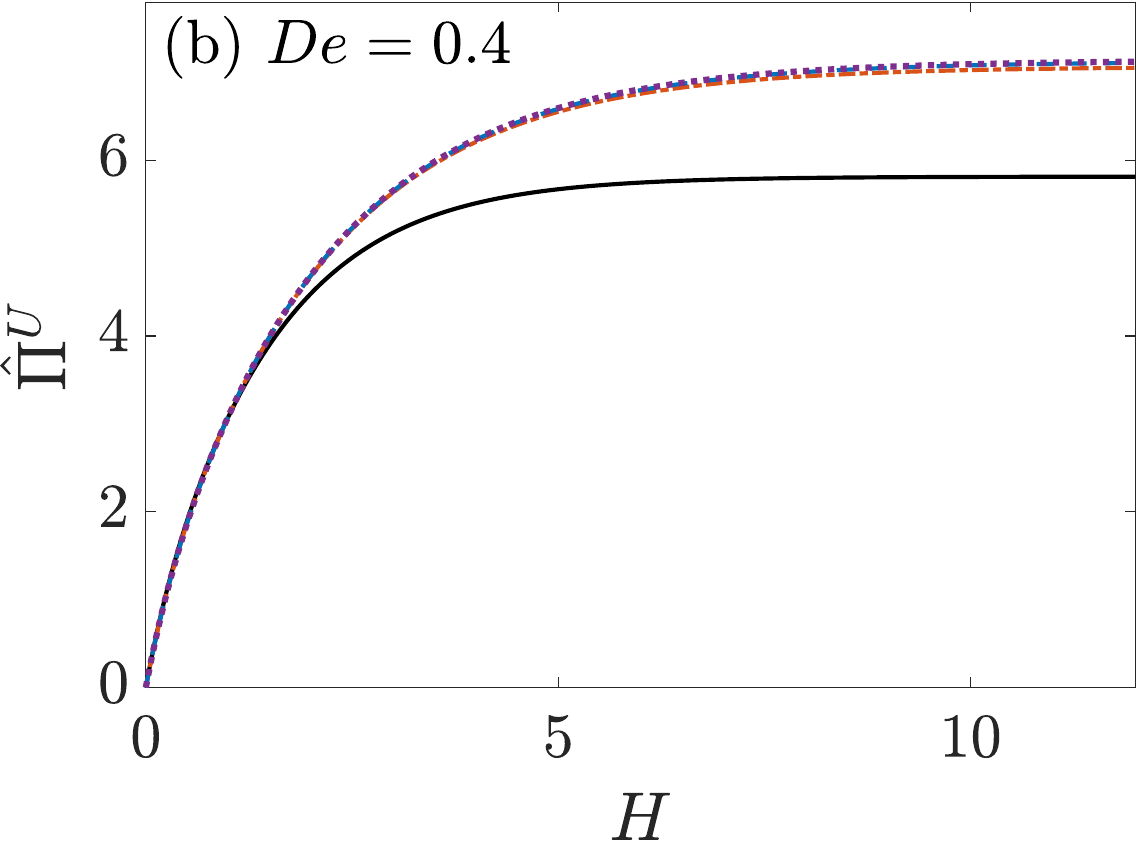}\label{fig:U0Dep4}}\\ 
	\subfloat{\includegraphics[width=0.33\textwidth]{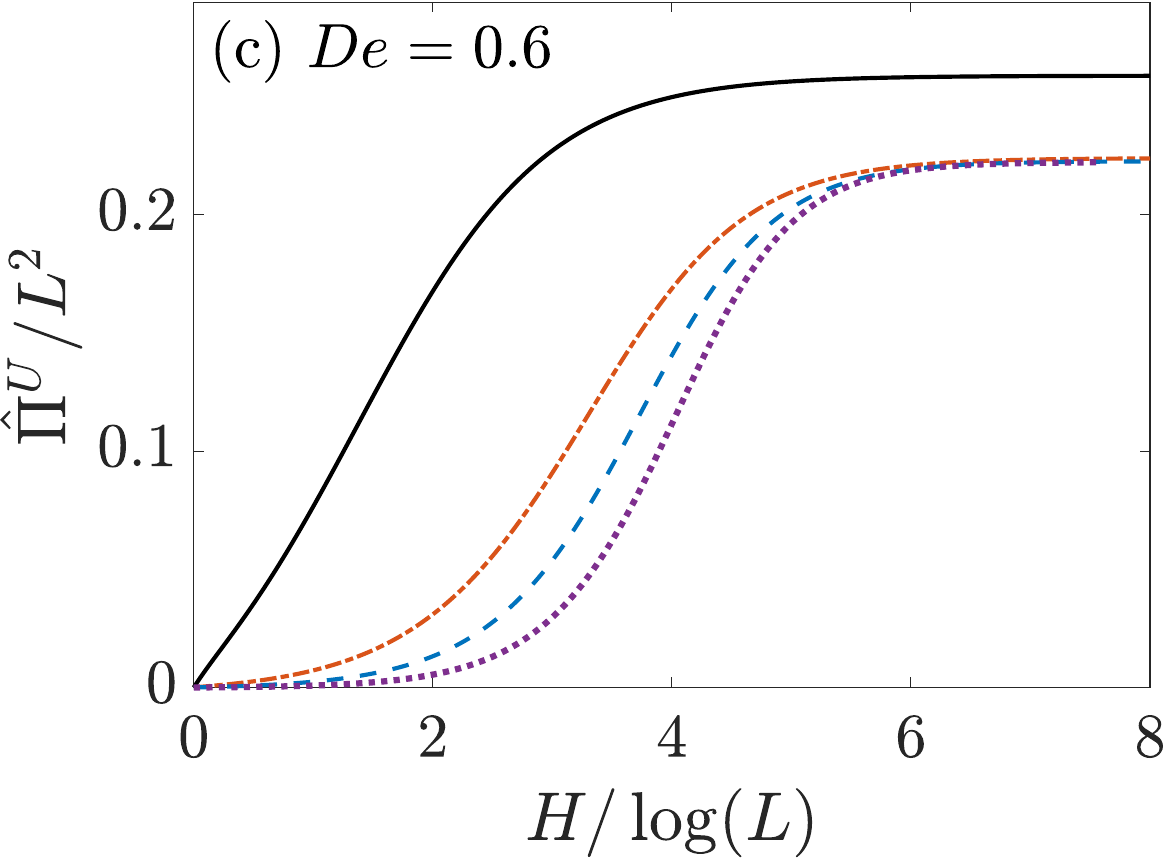}\label{fig:U0Dep6}}
	\subfloat{\includegraphics[width=0.33\textwidth]{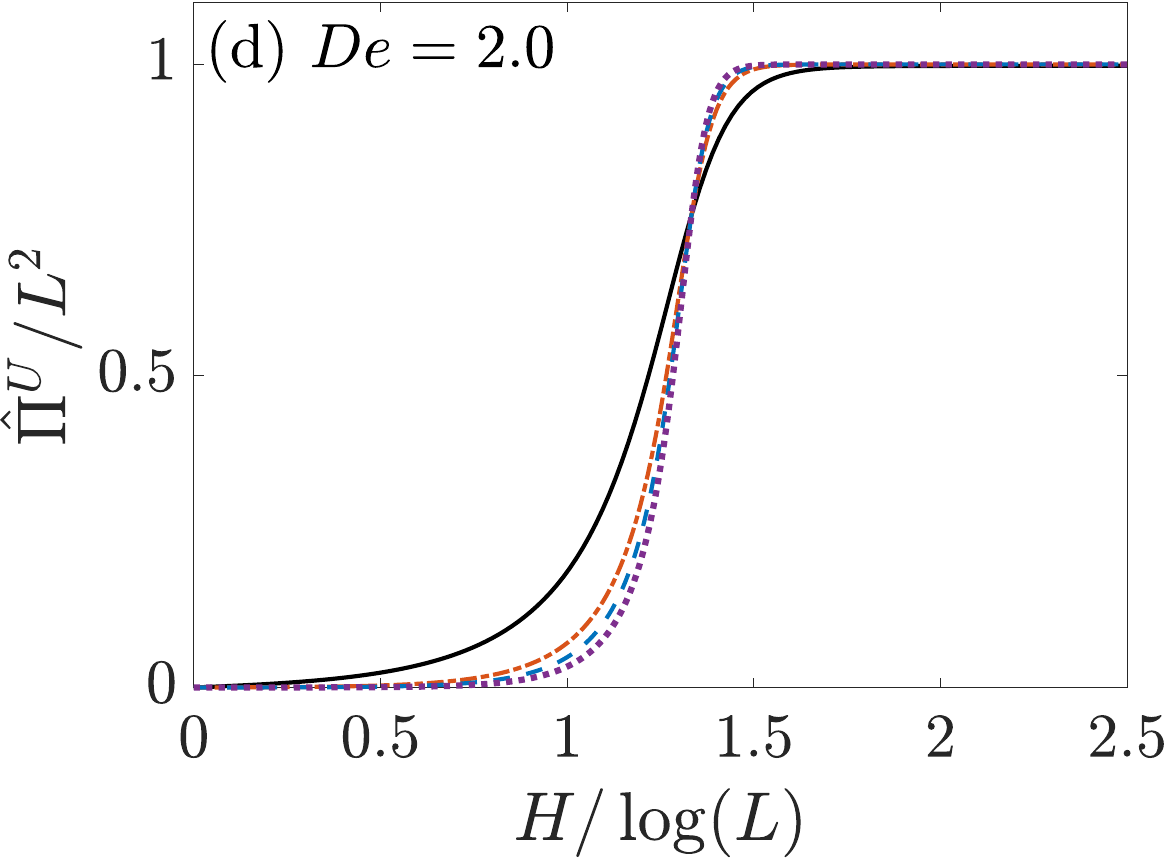}\label{fig:U0De2}}
	\subfloat{\includegraphics[width=0.33\textwidth]{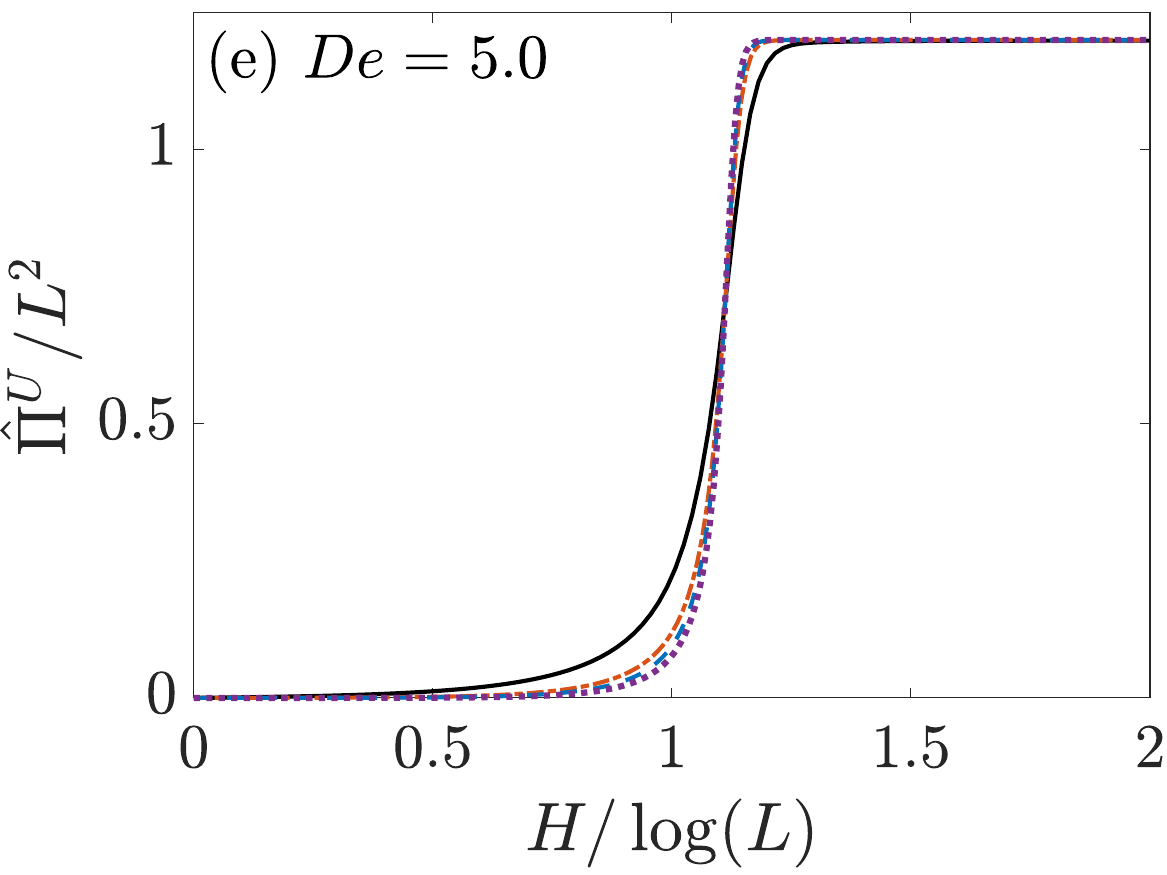}\label{fig:U0De5}}
	\caption {Evolution of undisturbed (no particles) polymer stress, $\hat{\Pi}^U$ with Hencky strain, $H$ for: (a) $De=0.2$, (b) $De=0.4$, (c) $De=0.6$, (d) $De=2.0$, and (e) $De=5.0$ for different $L$. For the latter three cases $\hat{\Pi}^U$ is normalized with $L^2$ and $H$ with $\log(L)$. All cases share the same legend for $L$ shown in (a). \label{fig:UndisturbedStress}}
\end{figure}

The evolution of undisturbed polymer stress in a FENE-P polymeric liquid is shown in figure \ref{fig:UndisturbedStress} for different Deborah numbers, $De$, and maximum polymer extensibility, $L$. Initially, the polymers are in equilibrium (zero stress) and are subsequently stretched due to the imposed extension, causing $\hat{\Pi}^U$ to increase monotonically with $H$. For $De<0.5$, the polymer stretch, $\sqrt{\text{tr}(\boldsymbol{\Lambda}^{U})}$, is much smaller than $L$, resulting in almost identical curves for different $L\ge 50$ at $De=0.2$ and 0.4 in figure \ref{fig:UndisturbedStress}. Additionally, because $\sqrt{\text{tr}(\boldsymbol{\Lambda}^{U})}\ll L$ in the $De<0.5$ regime for all $H$, increasing $De$ causes the steady-state to be reached later, as larger $De$ implies a larger final polymer stretch.  

For $De\ge0.5$, as shown in figure \ref{fig:UndisturbedStress}, we observe the well-known coil-stretch transition \citep{de1974coil,perkins1997single}—the polymer stress increases much more rapidly with $H$ beyond a certain point as the coiled polymer chains uncoil and approach full extension. Consequently, the polymer stress (proportional to the polymer configuration, which is the outer product of the polymer end-to-end vector) increases as $L^2$. Therefore, the $L^2$-normalized graphs for sufficiently large $L$ at $De=0.6$, 2.0, and 5.0 are identical in the steady-state. The polymers stretch towards $L$ at a rate inversely proportional to the logarithm of $L$. The transient behavior of different curves is similar for a normalized Hencky strain, $H/\log(L)$, for very large $De$, as shown here for 2.0 and 5.0 at $L\ge50$. In this regime of large $De$, polymers stretch faster towards the steady-state as $De$ increases, since the polymers have a maximum limit, $L$, that can be approached more quickly with a larger extension rate.

The effects of limited polymer extensibility are observed close to the coil-stretch transition point of $De=0.5$. For $De=0.4$, with $\sqrt{\text{tr}(\boldsymbol{\Lambda}^{U})}\ll L$, the collapsed polymers stretch much less for $L=10$ than larger $L$, resulting in a lower steady-state $\hat{\Pi}^U$. For $De=0.6$, polymers undergo the coil-stretch transition, and $\text{tr}(\boldsymbol{\Lambda}^{U})\sim L^2$, but the stretch of larger $L$ polymers is still not as close to their respective $L$ as it is for $L=10$. When $De$ is larger ($De=$2.0 and 5.0 in figure \ref{fig:UndisturbedStress}), the coil-stretch transition is more rapid. Due to the large extension rate, even the larger $L$ polymers get extended very close to their respective $L$, and the (normalized) steady state at all $L$ is similar. 

\subsection{Giesekus}\label{sec:GiesekusU0}
\begin{figure}
	\centering
	\subfloat{\includegraphics[width=0.33\textwidth]{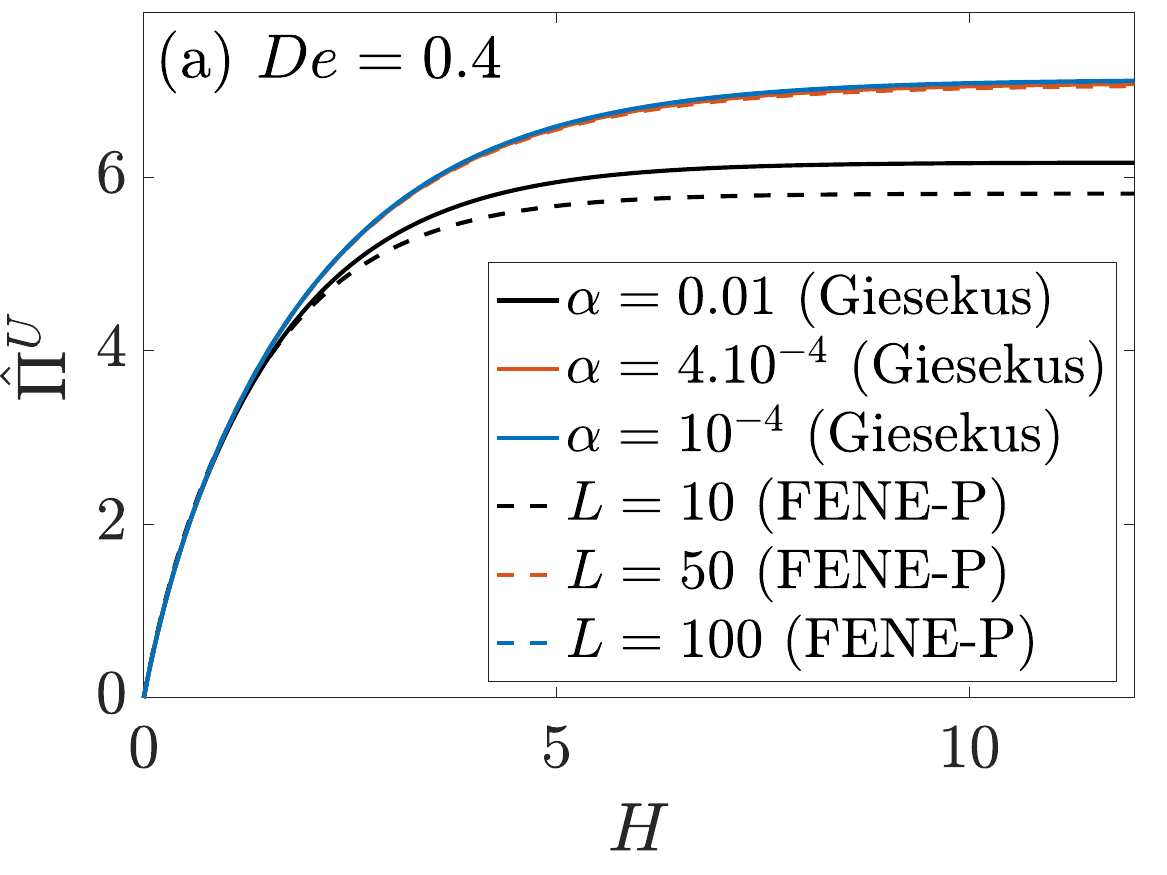}\label{fig:U0Dep4GieskFENEP}}\hfill
	\subfloat{\includegraphics[width=0.33\textwidth]{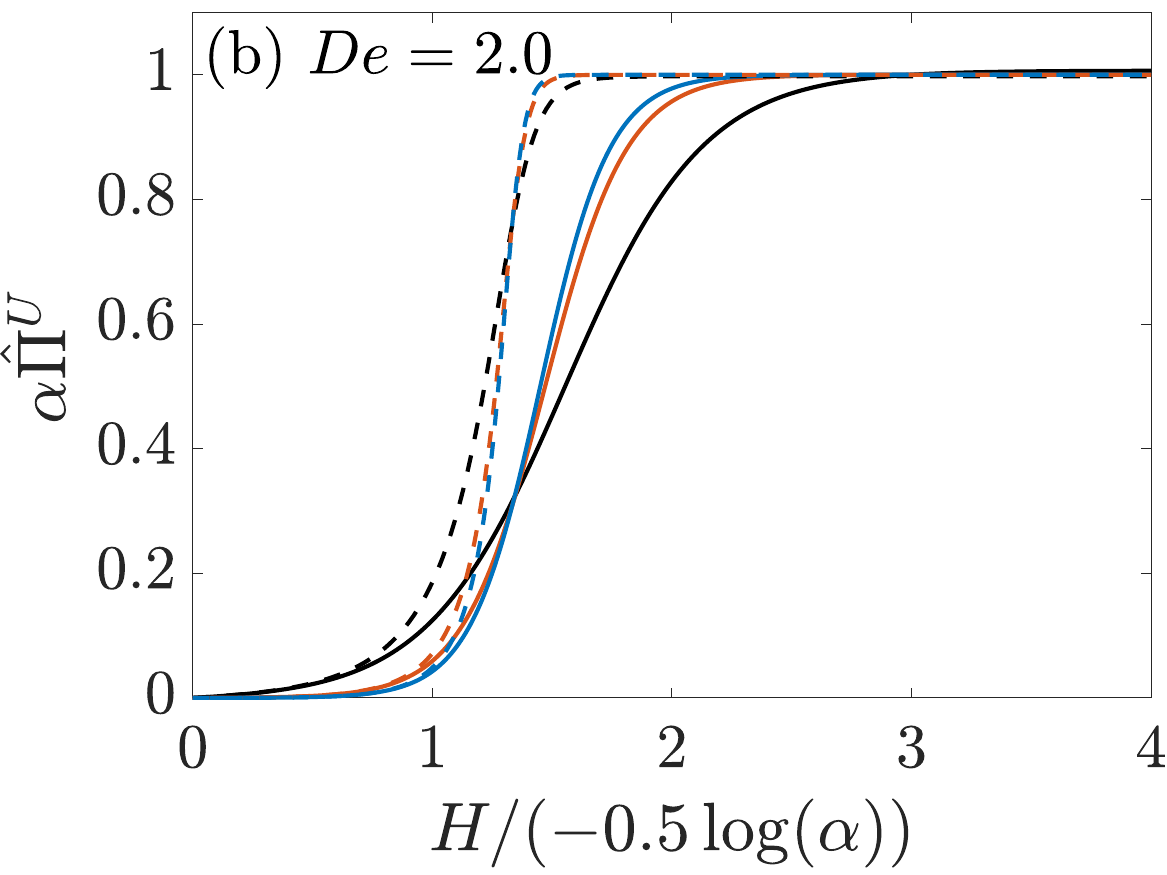}\label{fig:U0De2GieskFENEP}}\hfill
	\subfloat{\includegraphics[width=0.33\textwidth]{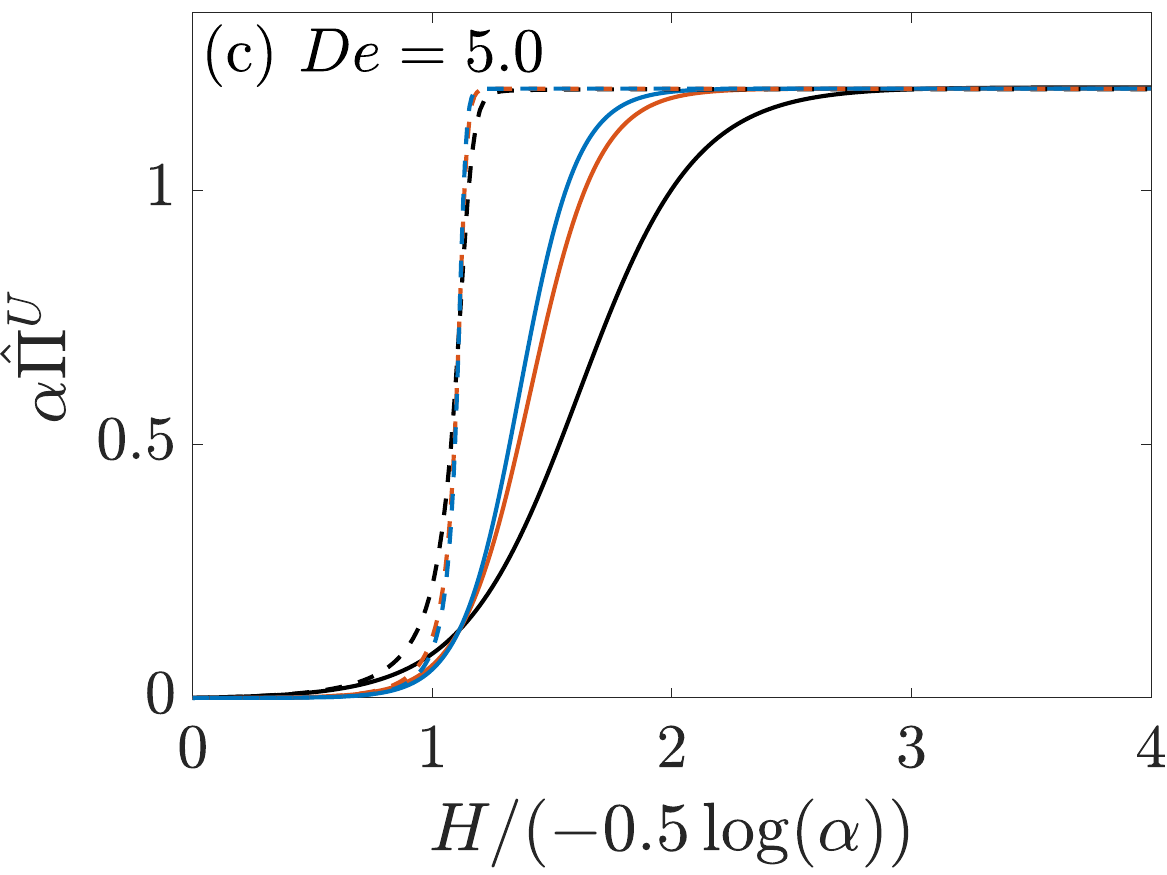}\label{fig:U0De5GieskFENEP}}
	\caption {Evolution of undisturbed (no particles) polymer stress, $\hat{\Pi}^U$ with Hencky strain, $H$ for: (a) $De=0.4$, (b) $De=2.0$, and (c) $De=5.0$ for different $\alpha$ in the Giesekus model (solid lines) and equivalent $L=\alpha^{-0.5}$ in the FENE-P model (dashed lines). For (b) and (c) $\hat{\Pi}^U$ is normalized with $\alpha^{-1}$ and $H$ with $-0.5 \log(\alpha)$. All cases share the same legend for $\alpha$ or $L$ shown in (a). \label{fig:UndisturbedStresGiesekusFENEPs}}
\end{figure}

The evolution of the undisturbed polymer stress for different levels of polymer entanglement (proportional to the mobility parameter $\alpha$) predicted by the Giesekus model is shown as solid lines for three different $De=0.4$, 2.0, and 5.0 in figure \ref{fig:UndisturbedStresGiesekusFENEPs}. While the FENE-P model imposes a maximum polymer extensibility explicitly, the Giesekus model implicitly restricts the polymer's maximum extension through the mobility parameter $\alpha$, which increases for more entangled polymers. By analyzing the relaxation term in equation \eqref{eq:Configuration}, we can define an effective maximum extensibility, $L_\text{Giesekus}=\alpha^{-0.5}$. This equivalence between the FENE-P and Giesekus models was previously known. \cite{housiadas2013skin} obtained it through the expression for the steady-state extensional viscosity at large $De$. Therefore, in figure \ref{fig:UndisturbedStresGiesekusFENEPs}, we also show the curves obtained from the FENE-P model as dashed lines with $L=\alpha^{-0.5}$ for the same values of $\alpha$ as those considered in the solid lines representing the Giesekus model. 
The qualitative features of the undisturbed stress in Giesekus fluids are similar to those in FENE-P fluids discussed in section \ref{sec:FENEU0} above. The stress is independent of $\alpha$ (or $L_\text{Giesekus}$) for $De<0.5$ for small enough $\alpha$ (large enough $L_\text{Giesekus}$) throughout the transient and scales as $\alpha^{-1}$ or $L_\text{Giesekus}^2$ above $De=0.5$ in the steady state. In the large $De$ regime, the polymers stretch towards $L$ at a rate inversely proportional to the logarithm of $\alpha$. The steady-state polymer stress of a Giesekus fluid for small enough $\alpha$ (large enough $L$) in the small $De$ regime and in the large $De$ regime for all $\alpha$ considered is the same as that for a FENE-P fluid with $L=\alpha^{-0.5}$. However, in the large $De$ regime, it takes longer for the polymers in a Giesekus fluid to reach the steady state relative to those in a FENE-P fluid. This is likely because, in addition to uncoiling, the Giesekus polymers also have to disentangle from one another.

\section{Suspension rheology of dilute polymeric solutions (FENE-P liquids)}\label{sec:DiluteRheology}
At very low polymer concentrations ($c$), the polymer configuration is primarily influenced by the velocity fluctuations generated by the particles in a Newtonian fluid. This means that the Newtonian velocity field is sufficient to describe the polymer behavior. This simplification allows for more efficient calculations than those using computational fluid dynamics (CFD) techniques, which involve discretization of the equations governing fluid mass and momentum along with the polymer constitutive equations. Here, the extensional viscosity of the suspension, denoted as $\mu_\text{ext}$ in equation \eqref{eq:ExtVisc}, is obtained using a combination of regular perturbation in $c$, a generalized reciprocal theorem, and a method of characteristics.

The leading-order polymer constitutive equation is obtained by solving the evolution of polymers along the streamlines (characteristics) of the Newtonian flow. The leading-order effect of polymers on the fluid velocity and pressure field, which modifies the particle stresslet, is evaluated using a generalized reciprocal theorem. This approach circumvents the need for a numerical solution of the $\mathcal{O}(c)$ mass and momentum equations. The use of the method of characteristics makes this method similar to Lagrangian particle methods (LPM) for tracer transport \citep{bosler2017lagrangian}, in contrast to CFD methods used more commonly in Eulerian reference frame. However, unlike the use of LPM in a general flow scenario, here the fluid velocity is steady, making the implementation more straightforward. This semi-analytical methodology is similar to that considered by \cite{koch2016stress} and \cite{SteadyStatePaper}. Further details and the required changes due to the transient nature of polymer evolution in a uniaxial extensional flow are discussed in appendix \ref{sec:MethodSemiAnalytical}.

The results of these calculations, spanning a wide range of Deborah numbers ($De$) and maximum polymer extensibility ($L$), are presented in section \ref{sec:ResultsSemiAnalytical} before the underlying mechanisms are elucidated in section \ref{sec:PolymerStretch}. 

\subsection{Rheological observations}\label{sec:ResultsSemiAnalytical}
The extensional viscosity of the suspension and its contributions from the particle-polymer stress discussed in this section are formally $\mathcal{O}(c)$ and are calculated from equations \eqref{eq:ExtVisc1} to \eqref{eq:StressetSplit1}. Since we only consider $\mathcal{O}(c)$ effects, for brevity of notation, we do not include $0$ in the superscript of various quantities appearing in these equations while discussing the following results.

\subsubsection{Net effect of particle-polymer interaction on the extensional viscosity of the suspension}\label{sec:RheologyNet} 
\begin{figure}
\centering
\subfloat{\includegraphics[width=0.33\textwidth]{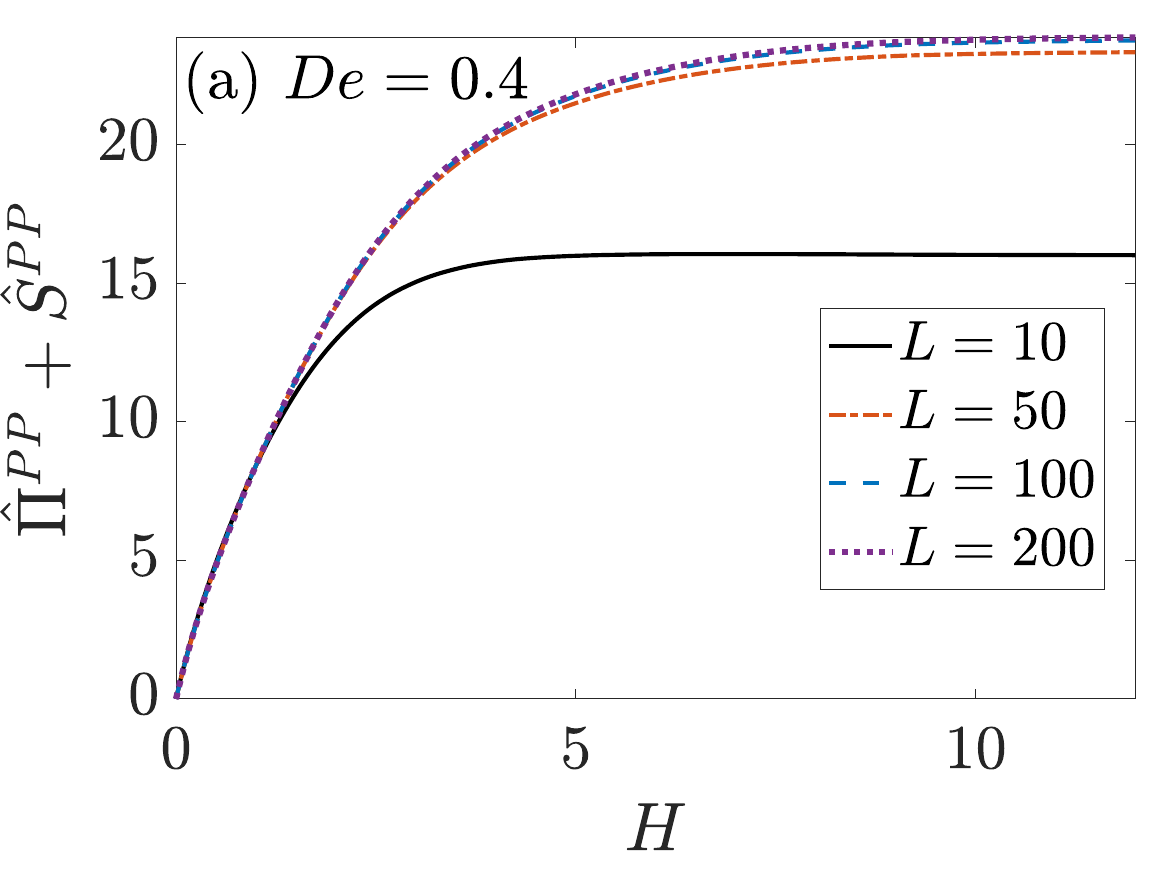}\label{fig:TotalInteractUnNormDep4}}\hfill
\subfloat{\includegraphics[width=0.33\textwidth]{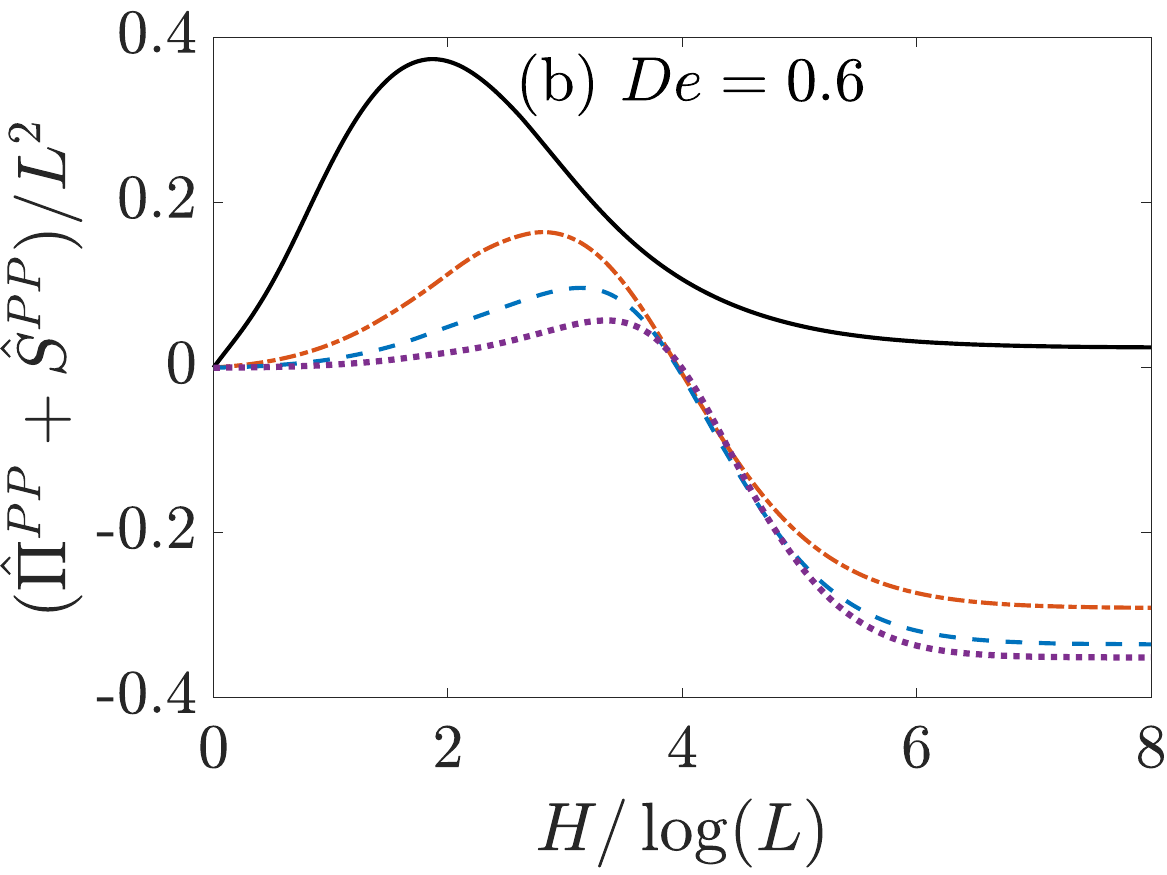}\label{fig:TotalInteractUnNormDep6}}\hfill
\subfloat{\includegraphics[width=0.33\textwidth]{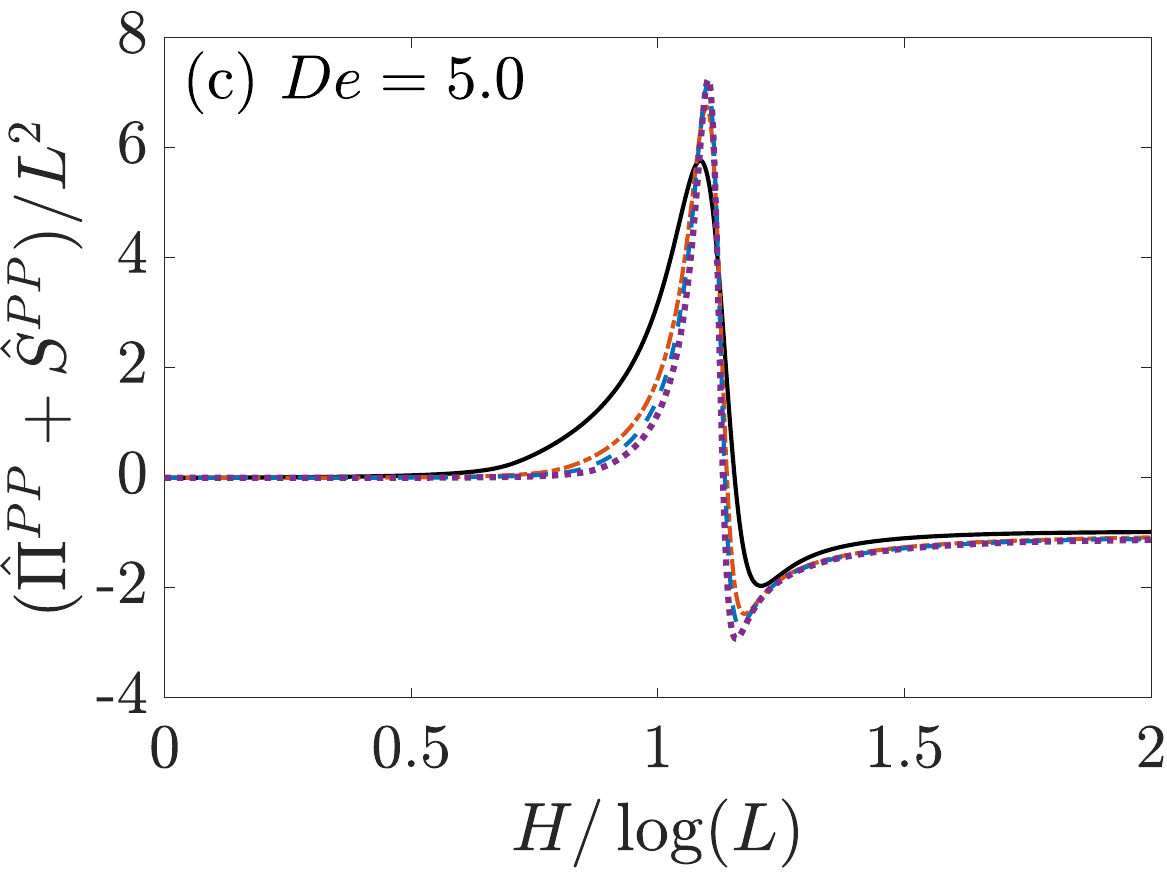}\label{fig:TotalInteractUnNormDe5}}
\caption {Evolution of total particle-polymer interaction stress, $(\hat{\Pi}^{PP}+\hat{\text{S}}^{PP})$ with Hencky strain, $H$ for: (a) $De=0.4$, (b) $De=0.6$, and (c) $De=5.0$. For the latter two cases $H$ is normalized with $\log(L)$ and stress with $L^2$. All cases share the same legend for $L$ shown in (a). \label{fig:TotalInteractionStressUnNorm}}
\end{figure}

\begin{figure}
\centering
\subfloat{\includegraphics[width=0.33\textwidth]{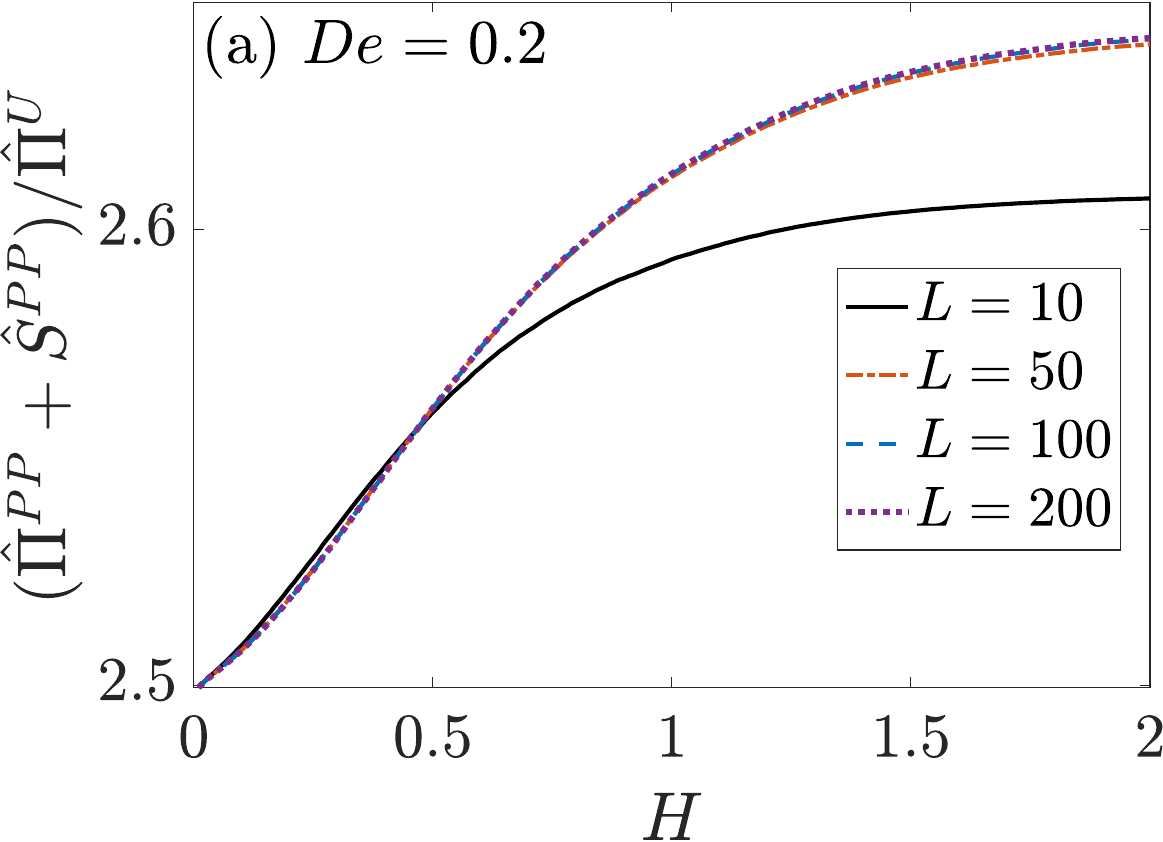}\label{fig:TotalInteractDep2}}
\subfloat{\includegraphics[width=0.33\textwidth]{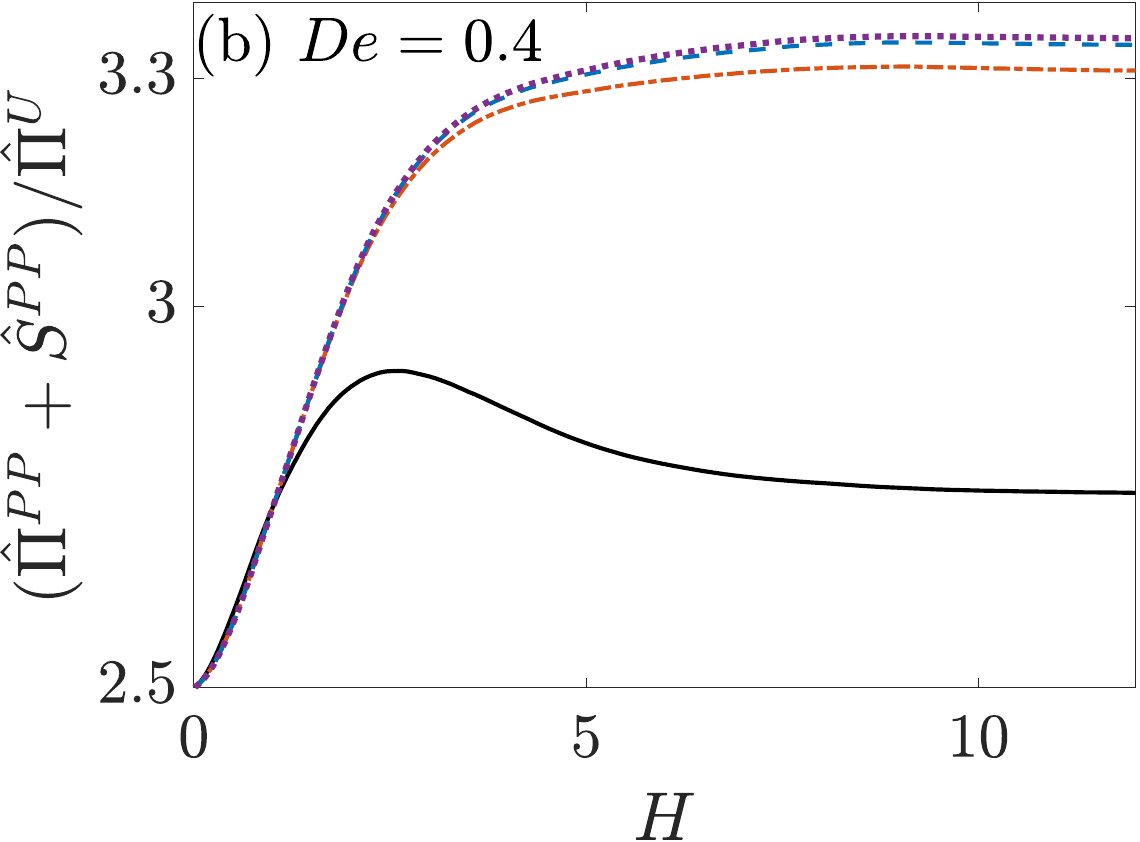}\label{fig:TotalInteractDep4}}\\ 
\subfloat{\includegraphics[width=0.33\textwidth]{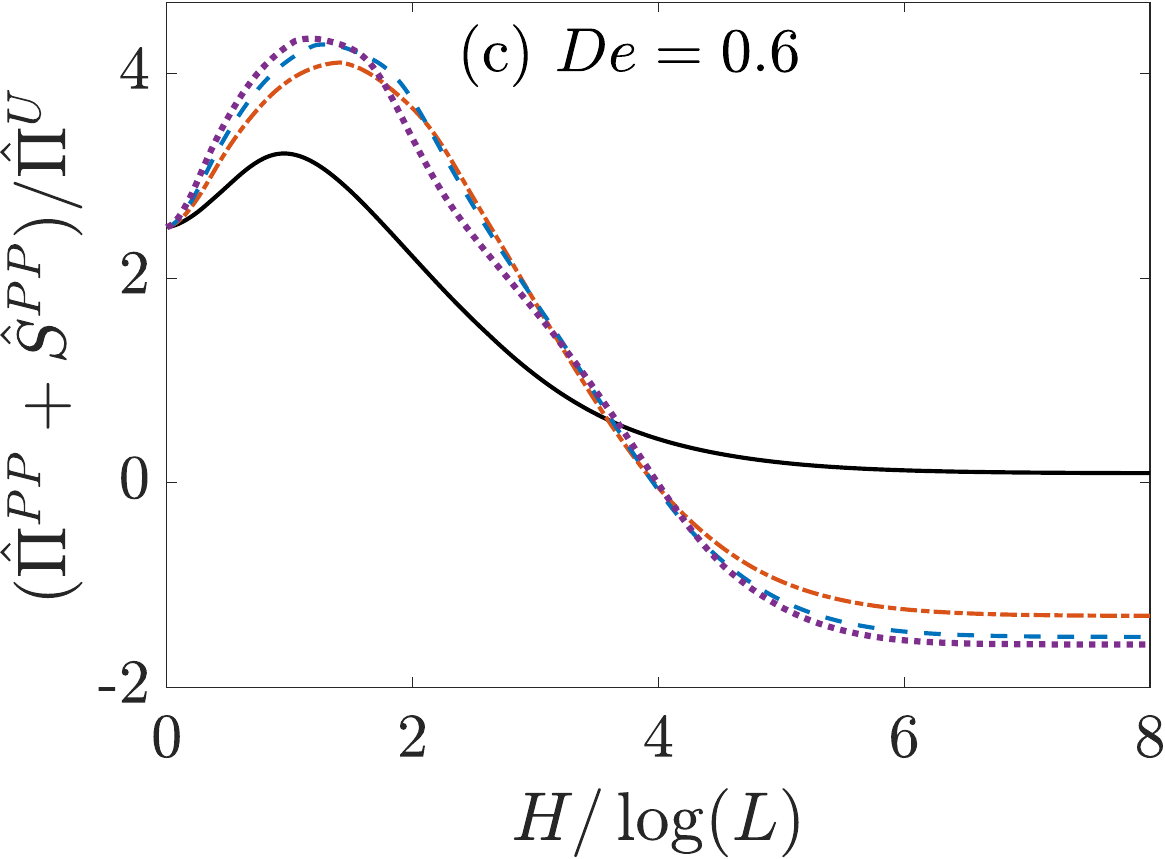}\label{fig:TotalInteractDep6}}
\subfloat{\includegraphics[width=0.33\textwidth]{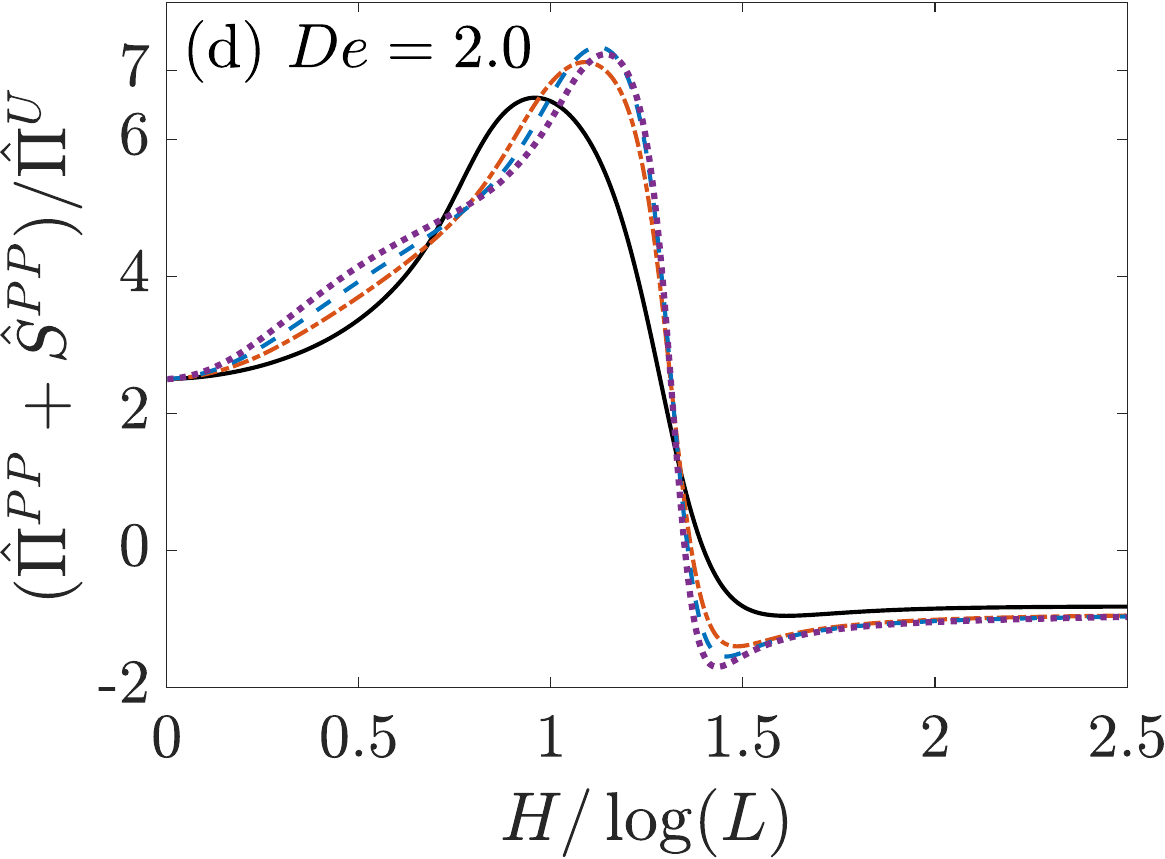}\label{fig:TotalInteractDe2}}
\subfloat{\includegraphics[width=0.33\textwidth]{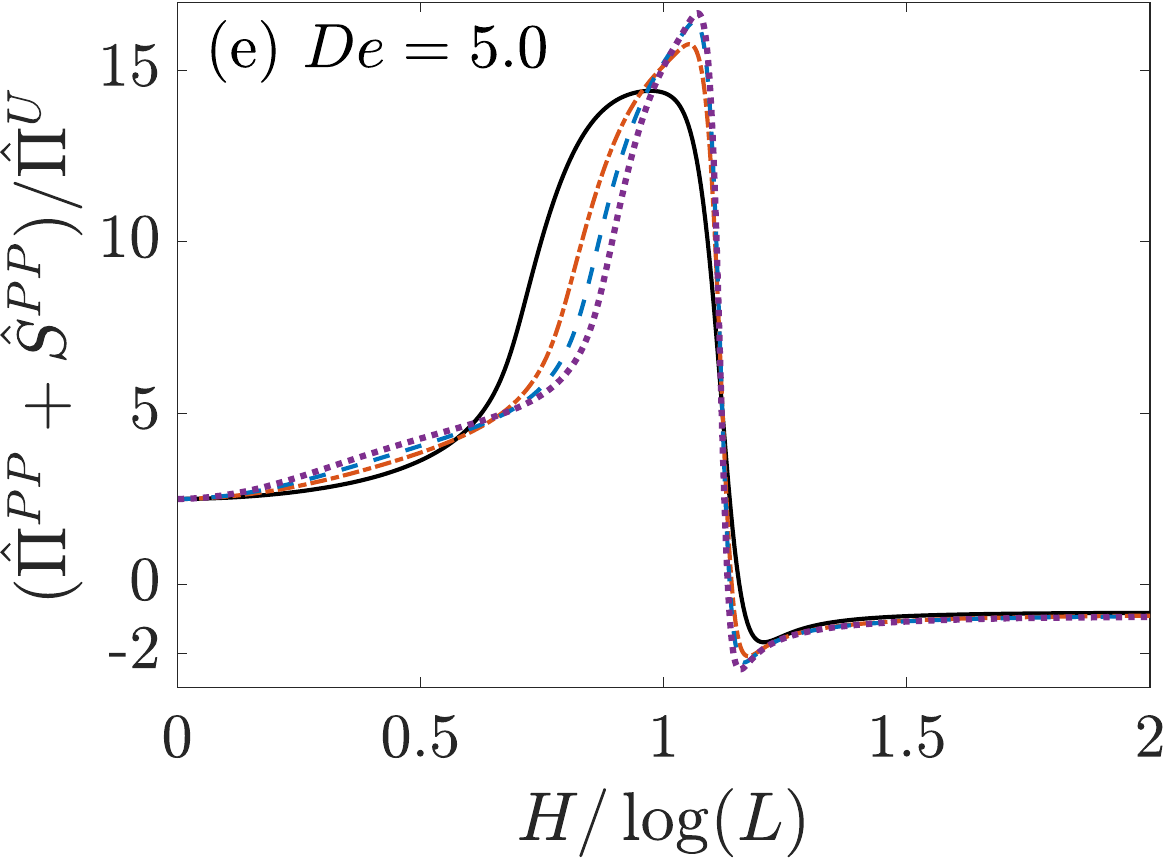}\label{fig:TotalInteractDe5}}
\caption {Evolution of total interaction stress normalized with the undisturbed (no particles) polymer stress, $(\hat{\Pi}^{PP} + \hat{\text{S}}^{PP}) / \hat{\Pi}^U$, with Hencky strain, $H$, for: (a) $De = 0.2$, (b) $De = 0.4$, (c) $De = 0.6$, (d) $De = 2.0$, and (e) $De = 5.0$ for different $L$. For the latter three cases, $H$ is normalized with $\log(L)$. All cases share the same legend for $L$ shown in (a). \label{fig:TotalInteractionStress}}
\end{figure}

Figure \ref{fig:TotalInteractionStressUnNorm} illustrates the evolution of the net particle-polymer interaction stress, $\hat{\Pi}^{PP} + \hat{\text{S}}^{PP}$, for different Deborah numbers ($De = 0.4$, 0.6, and 5.0) and maximum polymer extensibilities ($L = 10$, 50, 100, and 200) with respect to the Hencky strain, $H$. In the case of $De = 0.6$ and 5.0, $H$ is normalized with $\log(L)$, and the stress is normalized with $L^2$. To focus specifically on the interaction stress, figure \ref{fig:TotalInteractionStress} presents the evolution of $(\hat{\Pi}^{PP} + \hat{\text{S}}^{PP}) / \hat{\Pi}^U$, i.e., the particle-polymer interaction stress normalized with $\hat{\Pi}^U$ from figure \ref{fig:UndisturbedStress}. This allows for a clearer understanding of the relative contribution of the interaction stress compared to the undisturbed stress. The figure includes data for $De = 0.2, 0.4, 0.6, 2.0$, and 5.0, and for each Deborah number, it shows the results for $L = 10$, 50, 100, and 200. At $H = 0$, $(\hat{\Pi}^{PP} + \hat{\text{S}}^{PP}) / \hat{\Pi}^U = 2.5$, consistent with the small time estimates obtained by equation \eqref{eq:SmallTime3}. For $De = 0.2$ and 0.4, the stress at large $L \geq 50$ is independent of $L$ for all values of $H$, while for $De = 0.6$, 2.0, and 5.0, the stress scales with $L^2$ for all values of $H / \log(L)$. The attainment of steady state occurs more slowly for larger Deborah numbers below $De = 0.5$, but faster for $De > 0.5$. These scaling relationships and trends with respect to $L$ and $H$ observed in the particle-polymer interaction stress are similar to those observed in the undisturbed polymer stress shown in figure \ref{fig:UndisturbedStress} and discussed in section \ref{sec:U0Rheology}.

For $De=0.2$ we observe a monotonic increase in the normalized interaction stress at each $L$ in figure \ref{fig:TotalInteractDep2}. This indicates that particle-polymer interaction contributes to an increase in the extensional viscosity, $\mu_\text{ext}$, of the suspension with the Hencky strain until reaching a steady state. This positive effect arises from the net contribution of particles, $\mu_\text{part}$ (equation \eqref{eq:PartVisc}). At the lowest $L=10$ shown, the normalized interaction stress behaves similarly to the higher $L$ cases but with a lower magnitude. Similar to the undisturbed stress, $\hat{\Pi}^U$, in figure \ref{fig:UndisturbedStress}, the normalized interaction stress, $(\hat{\Pi}^{PP}+\hat{\text{S}}^{PP})/\hat{\Pi}^U$ in figure \ref{fig:TotalInteractionStress} is larger for $De=0.4$ than $De=0.2$. From this observation and additional data for $De = 0.1$ and 0.3 (not shown), we conclude that in the $De < 0.5$ regime, increasing $De$ enhances the positive contribution of $\mu_\text{part}$ to $\mu_\text{ext}$ at all $H$. For $De = 0.4$, the interaction of particles with $L = 10$ polymers is qualitatively different from that with higher $L$ polymers, as shown by the transient peak for $L = 10$ in figure \ref{fig:TotalInteractDep4}.

The effect of particle-polymer interaction becomes more profound and interesting for $De$ beyond the undisturbed coil-stretch transition value of 0.5. In this regime, the evolution of $(\hat{\Pi}^{PP} + \hat{\text{S}}^{PP}) / \hat{\Pi}^U$ with $H$ is non-monotonic, as shown in the $De = 0.6$, 2.0, and 5.0 plots in figure \ref{fig:TotalInteractionStress}. Initially, at smaller $H$, it increases rapidly to large positive values and then sharply falls to negative values. Before settling into a negative steady state for all $L$ at $De = 2.0$ and 5.0, there is a valley in $(\hat{\Pi}^{PP} + \hat{\text{S}}^{PP}) / \hat{\Pi}^U$ (and also in $\hat{\Pi}^{PP} + \hat{\text{S}}^{PP}$ as depicted in figure \ref{fig:TotalInteractUnNormDe5}). The magnitude of this valley increases with $De$. The sharp drop to negative values occurs around the same $H$ as when undisturbed polymers undergo the coil-stretch transition (figure \ref{fig:UndisturbedStress}).

We recently considered the steady-state extensional rheology in \cite{SteadyStatePaper}. In that study, we found that the interaction stress $\hat{\Pi}^{PP} + \hat{\text{S}}^{PP}$ increases with $De$ in the $De < 0.5$ regime and is independent of $L$ for $L \ge 50$. However, for a value of $De$ in the range $0.55 \lessapprox De \lessapprox 0.65$, depending on $L$, $\hat{\Pi}^{PP} + \hat{\text{S}}^{PP}$ changes from positive to negative and scales as $L^2$ for $L \ge 50$ (for larger $De$, the $L^2$ scaling extends to a lower $L = 10$). Following this negative value, further increases in $De$ make $\hat{\Pi}^{PP} + \hat{\text{S}}^{PP}$ more negative. In the current study, the same conclusions about the steady-state effect of the particle-polymer interaction as \cite{SteadyStatePaper} are obtained from the largest $H$ values shown here.

From figure \ref{fig:TotalInteractUnNormDep6}, we find the maximum and steady-state values of interaction stress $\hat{\Pi}^{PP} + \hat{\text{S}}^{PP}$ to be $0.1L^2$ and $-0.33L^2$ for $De = 0.6$ and $L = 100$. The maximum, minimum, and steady-state values for $De = 5.0$ are $7.1L^2$, $-2.7L^2$, and $-1.1L^2$, as shown in figure \ref{fig:TotalInteractUnNormDe5}. Therefore, in the initial part of the transient, the maximum net particle extensional viscosity is $\mu_\text{part} = (2.5 + 1000c)\phi$ and $(2.5 + 71000c)\phi$ for $De = 0.6$ and 5.0, respectively. Considering a very small value of $c = 0.001$, expected to be well within the range of validity of our low $c$ methodology, this implies a maximum particle-polymer interaction led increase in $\mu_\text{part}$ of 0.4 and 28.4 times compared to the Newtonian $\mu_\text{part} = 2.5\phi$. From similar calculations at the largest $H$ considered here, $\mu_\text{part} = (2.5 - 3300c)\phi$ and $(2.5 - 11200c)\phi$ for $De = 0.6$ and 5.0, respectively. Even with a low $c$ of 0.001, the net particle effect of $\mu_\text{part}$ is negative. This is a surprising and interesting result as it indicates that adding particles reduces the extensional viscosity of the suspension relative to the fluid viscosity, i.e., $\mu_\text{ext} < \mu_\text{fluid}$. For large $De$ values such as 2.0 and 5.0 shown here, due to the negative peak, the reduction in $\mu_\text{ext}$ is even larger in the transient phase just after its large initial increase. While the low $c$ methodology implicitly assumes the polymer stress to remain smaller than the Newtonian stress, in the above-mentioned estimates, this assumption is violated. However, direct numerical simulations, which make no assumption on the smallness of $c$, in the forthcoming section \ref{sec:ConcentratedRheology} will demonstrate that the broad observations from this low $c$ methodology are valid even $c$ larger than 0.001. Thus, the effect of particle-polymer interaction on the extensional rheology of a dilute suspension of spheres in a polymeric liquid is expected to be profound.

\subsubsection{Decomposition of particle-polymer interaction stress}\label{sec:RheologySplit}
\begin{figure}
	\centering
	\subfloat{\includegraphics[width=0.33\textwidth]{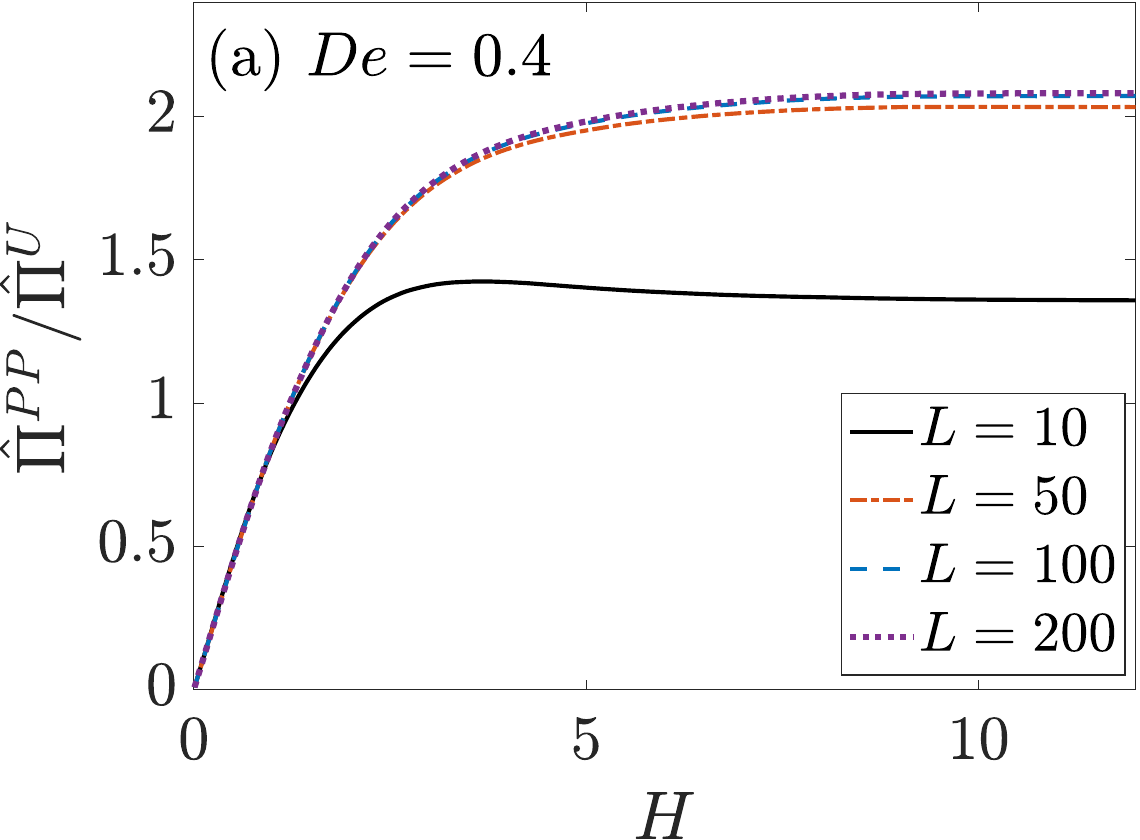}\label{fig:PIPSTotalDep4}}\hfill
	\subfloat{\includegraphics[width=0.33\textwidth]{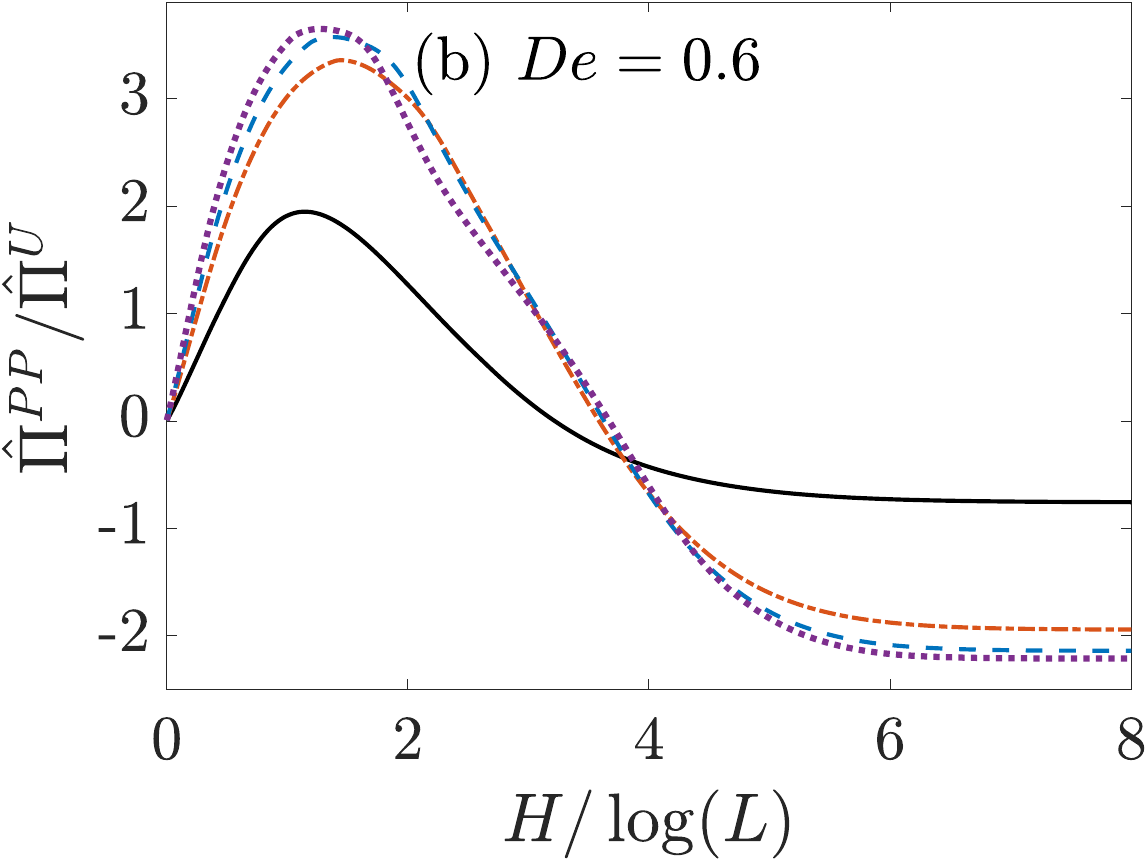}\label{fig:PIPSTotalDep6}}\hfill
	\subfloat{\includegraphics[width=0.33\textwidth]{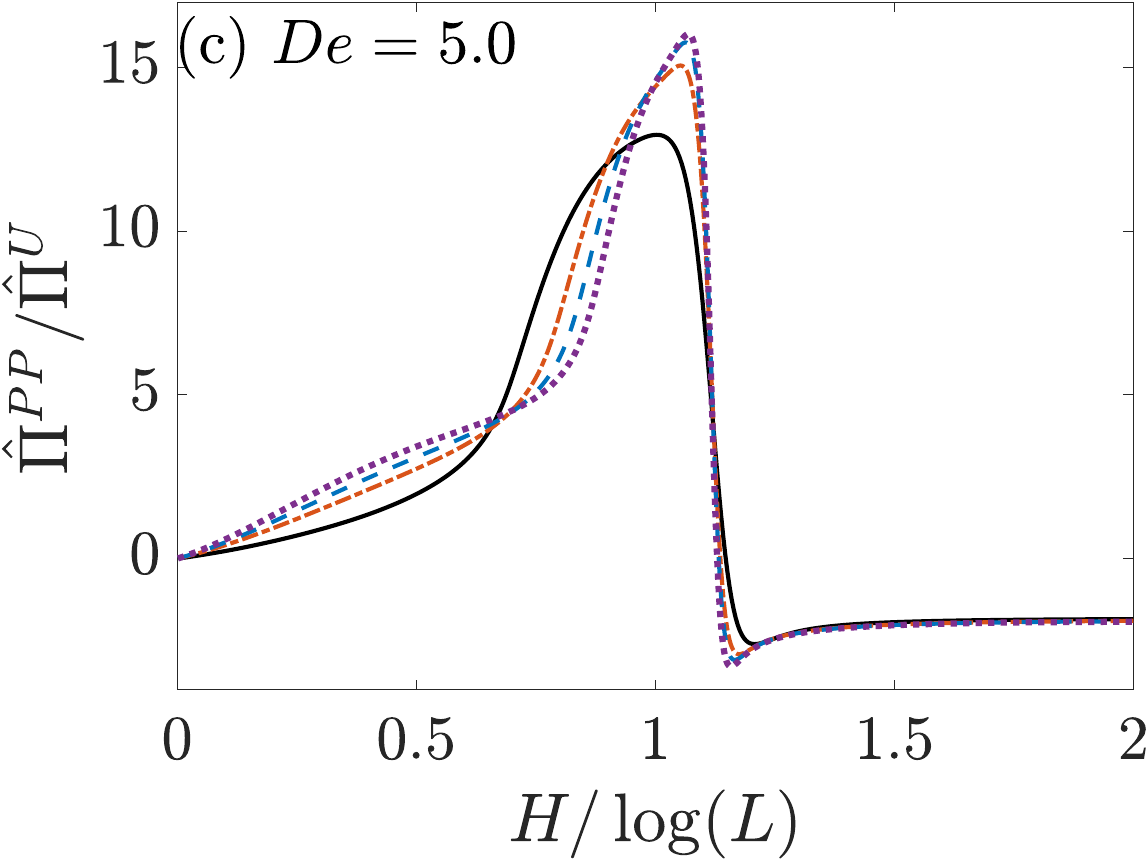}\label{fig:PIPSTotalDe5}}\\
	\subfloat{\includegraphics[width=0.33\textwidth]{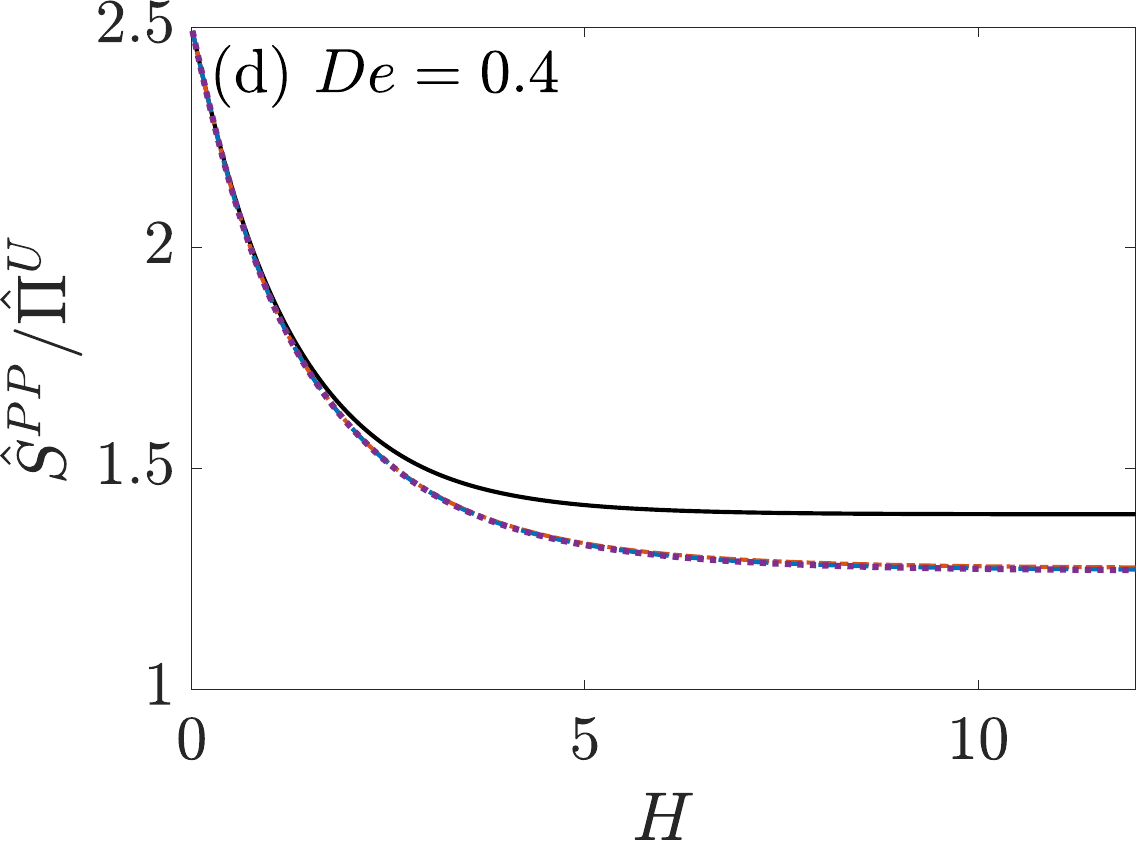}\label{fig:StressletTotalDep4}}\hfill
	\subfloat{\includegraphics[width=0.33\textwidth]{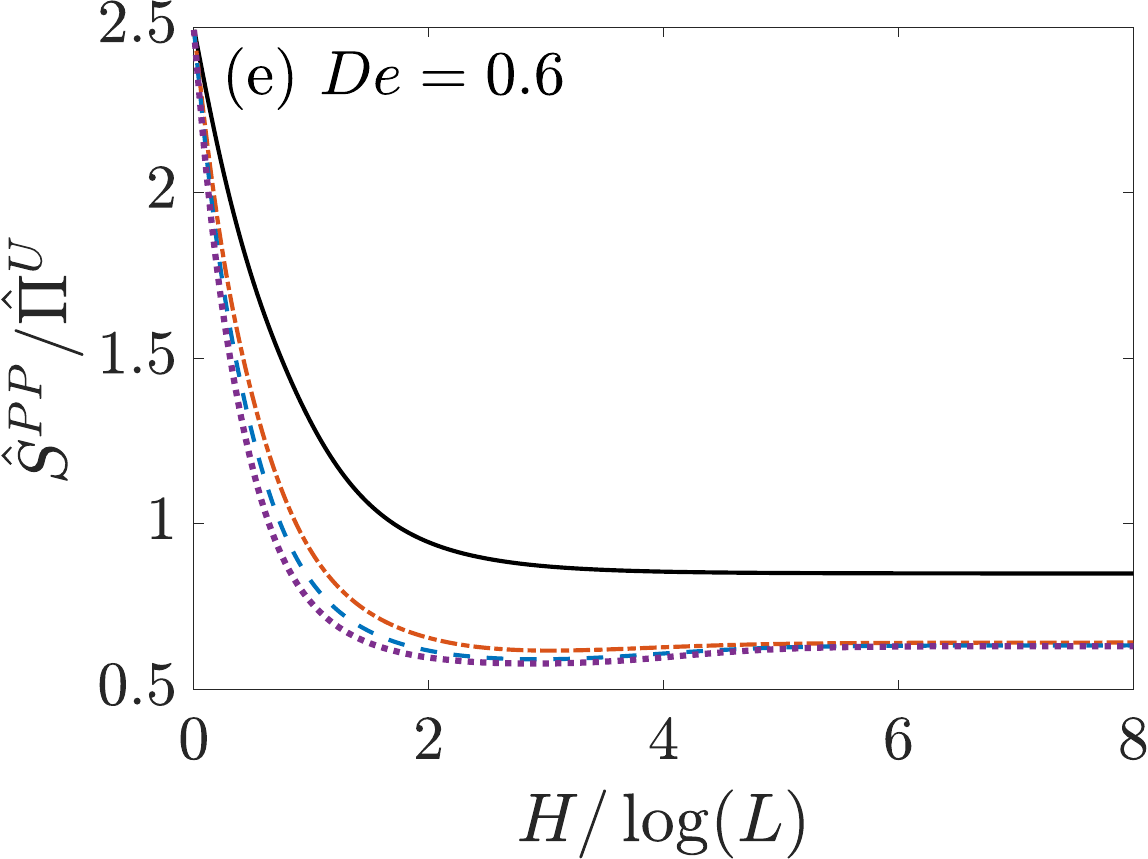}\label{fig:StressletTotalDep6}}\hfill
	\subfloat{\includegraphics[width=0.33\textwidth]{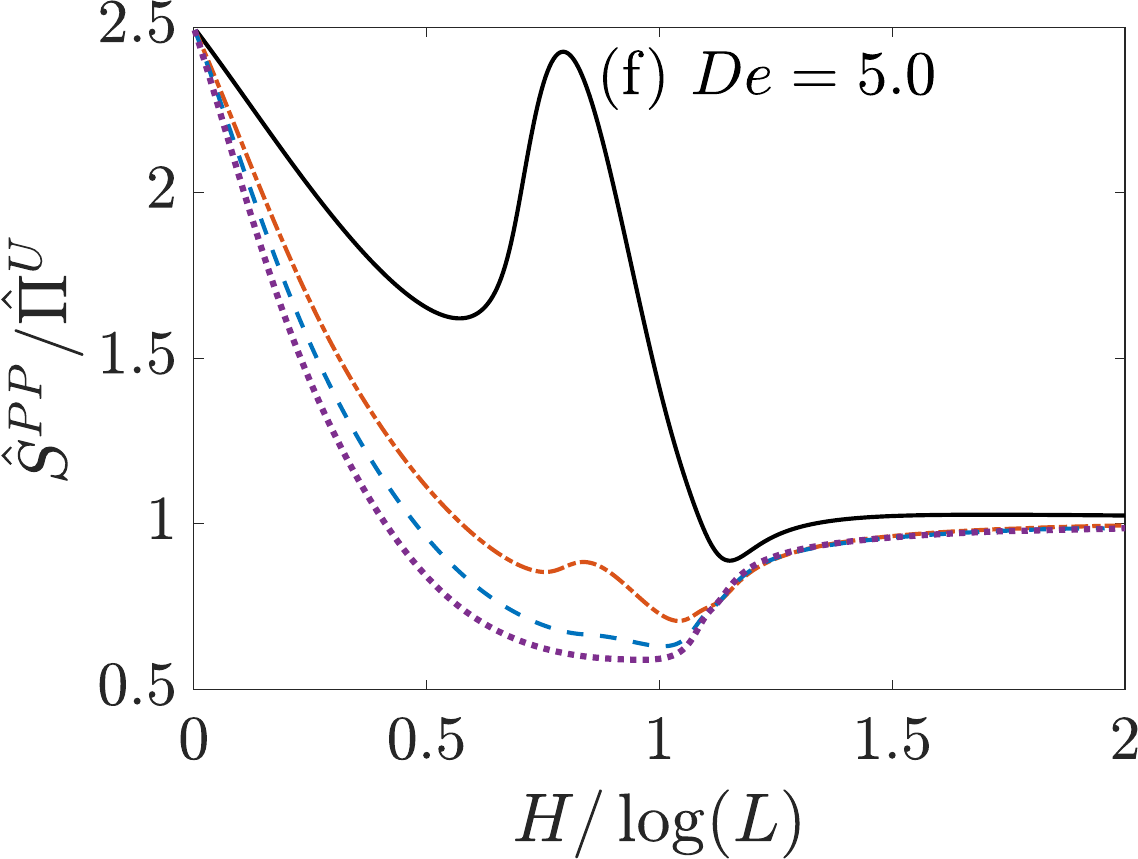}\label{fig:StressletTotalDe5}}
	\caption {Evolution of the $\hat{\Pi}^U$ normalized particle-induced polymer stress (PIPS, $\hat{\Pi}^{PP} / \hat{\Pi}^U$) and interaction stresslet ($\hat{\text{S}}^{PP} / \hat{\Pi}^U$) with Hencky strain, $H$, for different $De$ and $L$. The $De$ for each plot is marked, and different $L$ values are shown in the legend in (a), which is shared by all figures. \label{fig:TotalPIPSandStresslet}}
\end{figure}

To obtain further insight into the behavior of the interaction stress, its primary constituents, particle-induced polymer stress (PIPS, $\hat{\Pi}^{PP}$) and the interaction stresslet ($\hat{S}^{PP}$), normalized with $\hat{\Pi}^U$, are shown in figure \ref{fig:TotalPIPSandStresslet} for $De = 0.4$, 0.6, and 5.0 with $L = 10$, 50, 100, and 200 for each $De$. Figures \ref{fig:PIPSTotalDep4} to \ref{fig:PIPSTotalDe5} display PIPS, while figures \ref{fig:StressletTotalDep4} to \ref{fig:StressletTotalDe5} show the interaction stresslet. We do not show $De = 0.2$ and 2.0 considered earlier, but the figures for these $De$ values are similar to $De = 0.4$ and 5.0, respectively.

The qualitative features of the total interaction stress are almost fully captured by PIPS. This is observed by the similarity in figures \ref{fig:TotalInteractDep4} \& \ref{fig:PIPSTotalDep4}, \ref{fig:TotalInteractDep6} \& \ref{fig:PIPSTotalDep6}, and \ref{fig:TotalInteractDe5} \& \ref{fig:PIPSTotalDe5}, which compare the total interaction stress with PIPS for $De = 0.4$, 0.6, and 5.0, respectively. However, the contribution of $\hat{S}^{PP}$ is quantitatively important during the transient. It is always positive and decreases with $H$ for all $L$ and $De$, starting from 2.5 at $H = 0$. Beyond the undisturbed coil-stretch transition point of $De = 0.5$, the decrease in $\hat{S}^{PP} / \hat{\Pi}^U$ with $H$ is not monotonic, as observed from the $De = 0.6$ and 5.0 plots in figures \ref{fig:StressletTotalDep6} and \ref{fig:StressletTotalDe5}. For large $L$ ($\ge 50$) and $De = 2.0$ and 5.0, $\hat{S}^{PP} / \hat{\Pi}^U$ decreases to a value below the steady state during the transient. In these cases, the steady-state $\hat{S}^{PP} / \hat{\Pi}^U = 1$ (see figure \ref{fig:StressletTotalDe5} for example). With limited polymer extensibility, $L = 10$ for $De = 2.0$ (not shown) and $L = 10$ and 50 for $De = 5.0$, there is a brief period during the transient where $\hat{S}^{PP} / \hat{\Pi}^U$ increases before decreasing again. Both types of intermediate non-monotonic transient behavior of $\hat{S}^{PP} / \hat{\Pi}^U$ occur around the values of $H$ when the undisturbed polymers (figure \ref{fig:UndisturbedStress}) are undergoing coil-stretch transition. This is also the period when $\hat{\Pi}^{PP} / \hat{\Pi}^U$ undergoes a rapid decrease to negative values after its rapid initial increase, as previously noted for $(\hat{\Pi}^{PP} + \hat{\text{S}}^{PP}) / \hat{\Pi}^U$. Hence, at small $De$, both PIPS and the interaction stresslet contribute in the same way towards the positive $\mu_\text{part}$ for all $L$ and $H$. At large $De$, when $\mu_\text{part} < 0$ at larger $H$, the effect of just PIPS is even more negative, and the positive interaction stresslet reduces PIPS' influence.

According to the alternative decomposition of the interaction stresslet, $\hat{\text{S}}^{PP}$, into the undisturbed, $\hat{\text{S}}^{U} = \hat{\Pi}^U$, and the volumetric, $\hat{\text{S}}^{\text{Vol}}$ stresslet introduced in appendix \ref{sec:Stresslet}, the value of $\hat{\text{S}}^{PP} / \hat{\Pi}^U$ in figures \ref{fig:StressletTotalDep4} to \ref{fig:StressletTotalDe5} is simply $1 + \hat{\text{S}}^{\text{Vol}} / \hat{\Pi}^U$. Therefore, the variation in normalized interaction stresslet with $H$ is fully explained by the variation of $\hat{\text{S}}^{\text{Vol}} / \hat{\Pi}^U$, and the alternative stresslet decomposition into the undisturbed and volumetric stresslet is a clear and direct way to understand the mechanism at play here. The volumetric stresslet (defined in equation \eqref{eq:NewStressletDecomp}) is an integral of a quantity proportional to the difference in the polymer stress relative to the undisturbed value. As the Hencky strain accumulates, the distribution of the polymer stress disturbance, and not the total polymer stress, throughout the fluid volume plays a role in the variation of the interaction stresslet normalized with undisturbed stress (equation \eqref{eq:NewStressletDecomp}). From figure \ref{fig:StressletTotalDe5}, we notice that at large $De$ and large $H$, $\hat{\text{S}}^{PP} / \hat{\Pi}^U \rightarrow 1$, implying $\hat{\text{S}}^{\text{Vol}} / \hat{\Pi}^U \rightarrow 0$. Therefore, for polymers with larger relaxation time or at large extension rates, the polymer stress disturbance in the fluid volume does not affect the interaction stresslet. In the next section, we will observe that this is because, in a large volume around the sphere, polymers stretch similarly to the undisturbed ones.

\subsection{Deciphering the interaction mechanism through polymer stretch around the particle}\label{sec:PolymerStretch}
In this section, we gain insight into the changes in the extensional viscosity of the suspension, $\mu_\text{ext}$, through the net effect of the particle, $\mu_\text{part}$, discussed above, by observing the influence of the particle on the polymer configuration. As mentioned at the beginning of section \ref{sec:DiluteRheology}, only the Newtonian velocity field affects the leading-order polymer configuration. Therefore, to understand why the particle changes the polymer configuration in any particular way, we must delve into the kinematics of the Newtonian velocity field around a sphere in uniaxial extensional flow. {We take a brief detour to reintroduce a useful scalar diagnostic field for this purpose in section \ref{sec:KDM} before illustrating the particle-polymer interaction mechanism in section \ref{sec:InteractionMech}.}

\subsubsection{{Kinematic diagnostic measures}}\label{sec:KDM}
\begin{figure}
\centering	
\subfloat{\includegraphics[width=0.4\textwidth]{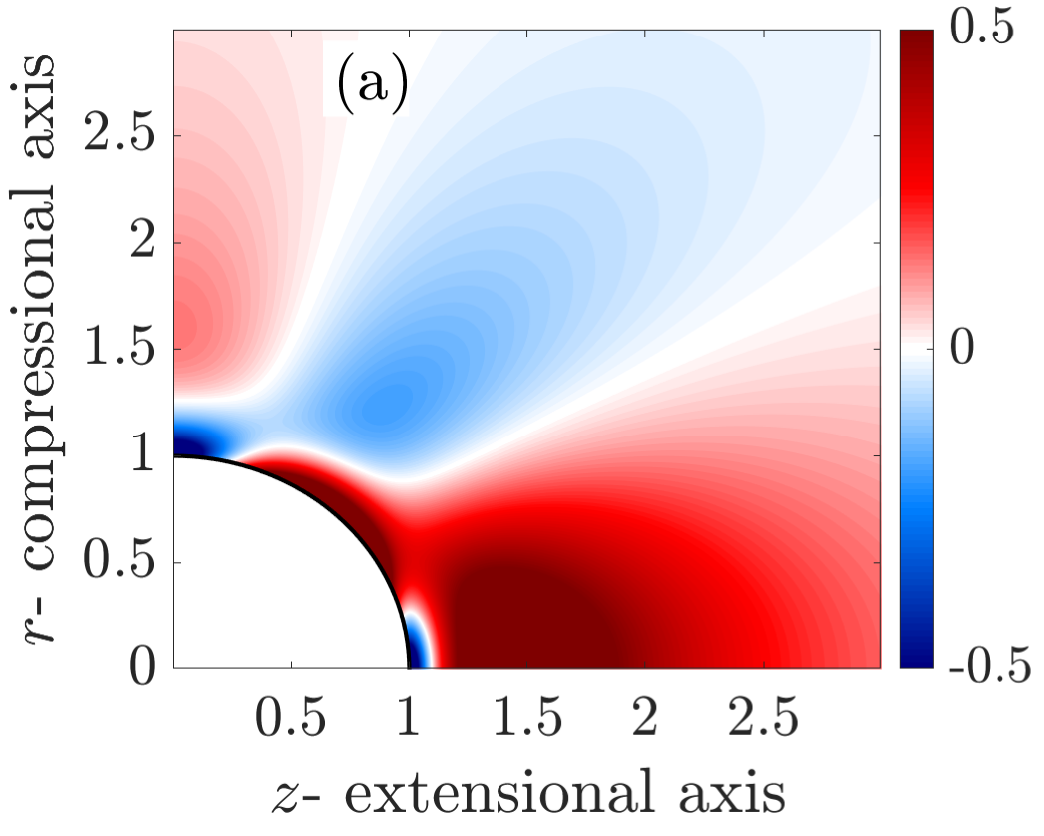}\label{fig:Local_De.pdf}}\hspace{0.2in}
\subfloat{\includegraphics[width=0.35\textwidth]{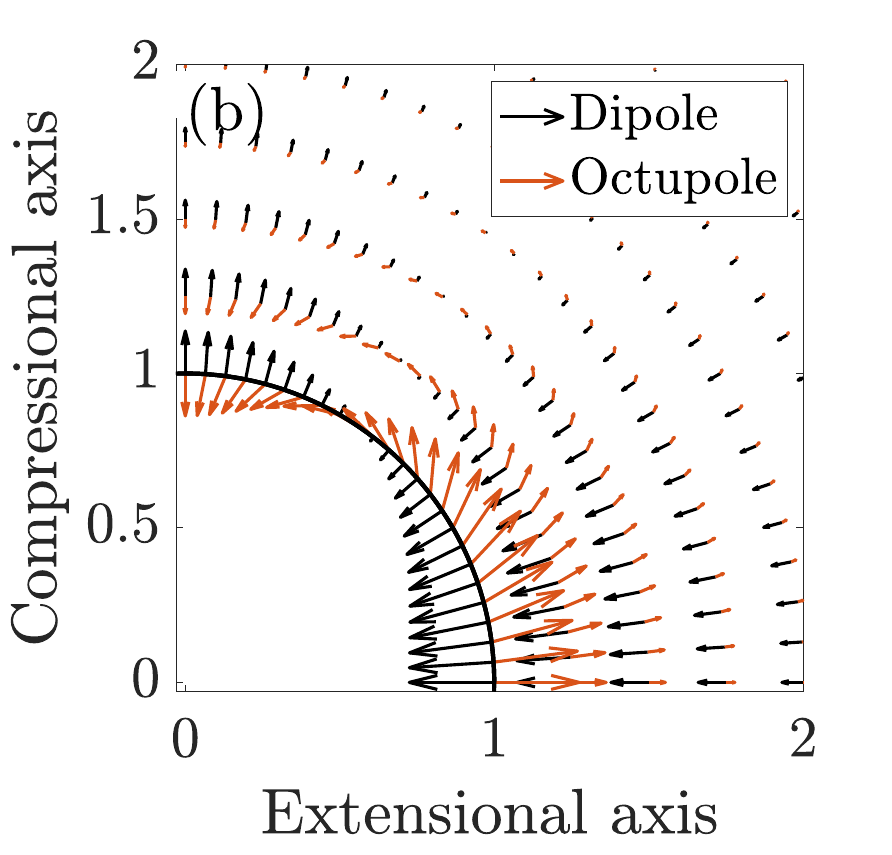}\label{fig:MultipoleDisturbances}}
\caption {(a) Fractional change in the local Deborah number field, $\Delta De_\text{local}$, due to a sphere in an imposed extensional flow used to locally diagnose the kinematics of the velocity field. (b) Multipole disturbances created by the sphere in a Newtonian fluid.}
\end{figure}

In \cite{SteadyStatePaper}, we found the scalar diagnostic field termed as the fractional change in the local Deborah number field, or $\Delta De_\text{local}$,
\begin{equation}
\Delta De_\text{local}=\sqrt{\frac{\boldsymbol{e}:\boldsymbol{e}}{\langle\boldsymbol{e}\rangle:\langle\boldsymbol{e}\rangle}}-1,\label{eq:DeLocal}
\end{equation}
to be helpful in explaining the steady-state changes in polymer configuration by the particle. Here, $\boldsymbol{e} = 0.5(\nabla \mathbf{u} + (\nabla \mathbf{u})^T)$ is the local rate of strain field, and $\langle \boldsymbol{e} \rangle$ is the undisturbed or imposed rate of strain. $\Delta De_\text{local}$ captures the local changes in stretching by the velocity gradient field compared to the far-field stretching. A positive $\Delta De_\text{local}$ implies more stretching locally than in the far field, while $\Delta De_\text{local} < 0$ implies less local stretching. In a uniaxial extensional flow, $\langle \boldsymbol{e} \rangle : \langle \boldsymbol{e} \rangle = 1.5$. The $\Delta De_\text{local}$ field is shown in figure \ref{fig:Local_De.pdf}.
From figure \ref{fig:Local_De.pdf}, we observe a region of largest (an intense red) local stretching compared to the far-field around the extensional axis of the flow. {Additionally, there is a stretching region (red region) near the particle surface at about $45^\circ$ from the extensional axis. A red region centered around the compressional axis represents extra local stretching, as the streamlines on either side of the axis separate. There are two notable regions of intense reduction in local stretching relative to the far-field (blue regions) around the front stagnation circle and rear stagnation point. The no slip boundary condition on the particle surface implies that the velocity gradient goes to zero at these points.  There is a less intense region of reduction in local stretching away from the particle surface at $45^\circ$ from the axes.}

The Newtonian velocity field around a sphere comprises dipole ($1/r^2$) and octupole ($1/r^4$) disturbances relative to the undisturbed velocity field. These disturbances contribute to the features of $\Delta De_\text{local}$, with the velocity vectors for both shown in figure \ref{fig:MultipoleDisturbances}. It can be observed that the black arrows (representing the dipole velocity) increase in length along their direction, towards the particle, around the extensional axis. Consequently, the front of a fluid element convecting with only this velocity field will experience a larger dipole velocity magnitude than the back, resulting in stretching. In this region, the octupole velocity (orange arrows) decreases in magnitude along their direction, reducing the stretch of a fluid element. The net effect is less stretching from the velocity gradients very close to the particle surface due to the octupole. Further away where the octupole disturbances have rapidly decayed, the net effect is due to the dipole.

{Around the compressional axis, the streamlines of the overall velocity field separate, causing a stretch in the direction of the extensional axis ($z$-axis) (this appears as the red region around the compressional axis in figure \ref{fig:Local_De.pdf}). However, the overall velocity vectors reduce in magnitude as they approach the sphere, causing lower stretching in the direction of the compressional axis ($r$-axis) (this appears as the blue region around the compressional axis near the particle surface in figure \ref{fig:Local_De.pdf}). From figure \ref{fig:MultipoleDisturbances}, we can observe that both these effects—increased stretching away from and reduced stretching near the particle—are enhanced due to the dipole velocity disturbance, while the octupole velocity disturbance causes an opposite effect. Around the compressional axis, both close to and away from the particle, the dipole effect dominates. The far-field stretching region does not play a dominant role in the upcoming discussion to elucidate the interaction mechanisms in the low $c$ regime of the current section \ref{sec:DiluteRheology}. However, this stretching along the compressional axis far upstream of the particle, created by the dipole disturbance of the Newtonian velocity field, will explain in section \ref{sec:ConcentratedRheology} the activation of another particle-polymer interaction mechanism at larger $c$. }

The $\Delta De_\text{local}$ field shown in figure \ref{fig:Local_De.pdf} will be used in conjunction with a field describing the fractional change in the polymer stretch created by the particle, defined as $\Delta\mathcal{S}$,
\begin{equation}
	\Delta\mathcal{S} = \sqrt{\text{tr}(\boldsymbol{\Lambda})} / \sqrt{\text{tr}(\boldsymbol{\Lambda}^U)} - 1, \label{eq:DeltaS}
\end{equation}
to understand how the polymer stretch field changes in the presence of a particle. Here, $\sqrt{\text{tr}(\boldsymbol{\Lambda})}$ and $\sqrt{\text{tr}(\boldsymbol{\Lambda}^U)}$ are the local and undisturbed polymer stretches, respectively. A positive $\Delta\mathcal{S}$ indicates an increase in the local polymer stretch, while a negative $\Delta\mathcal{S}$ implies a reduced local polymer stretch or polymer collapse. We performed a similar analysis in \cite{SteadyStatePaper} but used the change in the polymer stretch field, $\sqrt{\text{tr}(\boldsymbol{\Lambda})} - \sqrt{\text{tr}(\boldsymbol{\Lambda}^U)}$. In the rest of this section, whenever comparative statements are made, they are relative to the current undisturbed polymers or the stretching of the imposed velocity field in the far-field. Also, we remind the reader that the Hencky strain and time are the same according to our non-dimensionalization and hence can be used interchangeably.

\subsubsection{{Interaction mechanism}}\label{sec:InteractionMech}
In this section, we use $De=0.2$ and $0.4$ to illustrate the particle-polymer interaction mechanism in the $De<0.5$ regime, and $De=0.6$ and $5.0$ for the large $De$ regime. In \cite{SteadyStatePaper}, we found that the dominant contribution of $\Delta\mathcal{S}$ along the extensional axis in the far field at steady state arises from changes in the extensional component of the configuration tensor, $\Lambda_{zz}'$. Specifically, $\Lambda_{zz}'\sim z^{-\alpha}$, with $\alpha=1/De-2$ for $0.2<De<0.5$, $\alpha=4De-2$ for $0.5<De<1.25$, and $\alpha=-3$ otherwise. Therefore, the length scale at which the particle influences the steady-state polymer stretch is largest in the intermediate range of $0.2<De<1.25$. We will observe the signature of this steady-state region of influence of the particles on polymer stretch in the discussion of transient results below.
	
In the small $De$ regime, a wake of extra polymer stretch downstream of the particle around the extensional axis, due to $\Delta De_\text{local}>0$, leads to a positive particle-polymer interaction stress at all times. In contrast, in the large $De$ regime, a region of polymer collapse around the particle is observed beyond a $De$ and $L$ dependent time. This region originates from the $\Delta De_\text{local}<0$ region around the stagnation point on the particle at the compressional axis and leads to a negative particle-polymer interaction stress. The memory effects alter the interaction behavior with $De$ within each regime, as described below.

\begin{figure}
	\centering
	\subfloat{\includegraphics[width=0.33\textwidth]{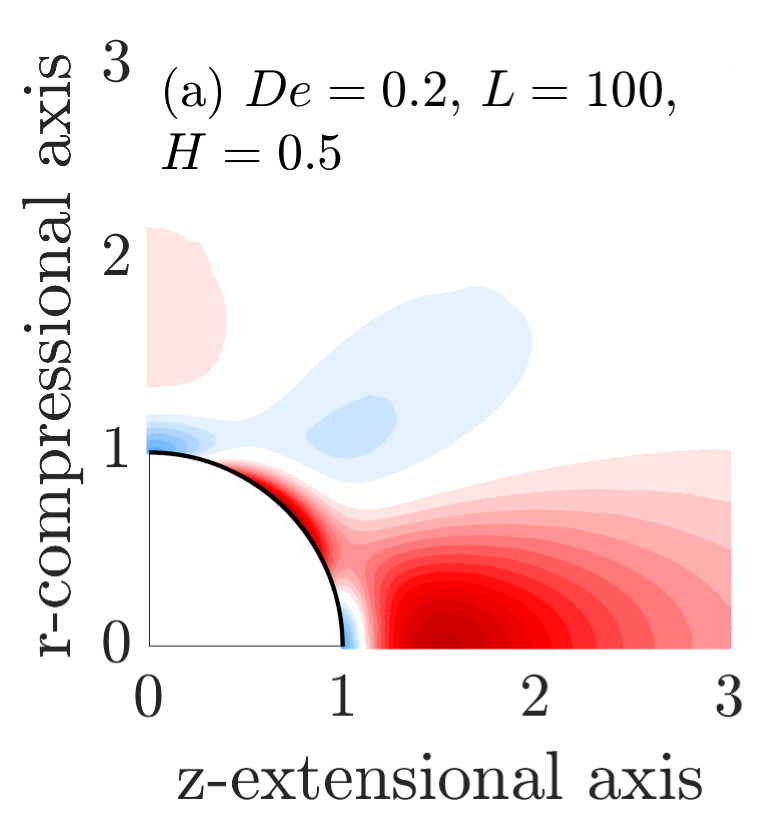}\label{fig:DelSNormHenckyp5Dep2L100}}\hspace{0.2in}
	\subfloat{\includegraphics[width=0.33\textwidth]{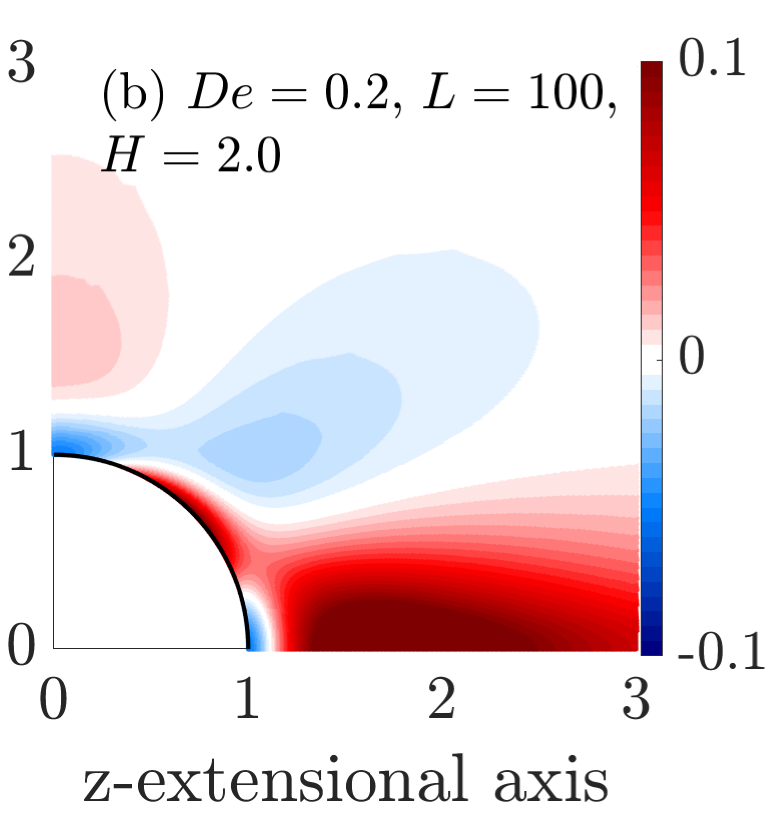}\label{fig:DelSNormHencky2Dep2L100}}
	\caption {$\Delta\mathcal{S}$ for $De = 0.2$ and $L = 100$ for Hencky strain, $H$: (a) 0.5 and (b) 2.0. Compared to the currently undisturbed polymers, in the red regions the polymers are stretched more, and in the blue regions they are collapsed.   \label{fig:DelSNormDept2L100}}
\end{figure}

A small $De = 0.2$ is a case where the polymers react mostly to the local flow. Figure \ref{fig:DelSNormDept2L100} shows $\Delta\mathcal{S}$ for $De = 0.2$ and $L = 100$ at two different $H = 0.5$ and 2.0 (close to steady-state). At $H = 0$, the polymers are in equilibrium everywhere; as $H$ increases, the polymers further from the particle start to get stretched by the imposed extensional flow. At a small but finite $H = 0.5$ for $De = 0.2$, the memory effects of polymer relaxation are limited such that the polymers respond only to the local velocity gradients. At this instant, there is an almost perfect positive correlation between the $\Delta\mathcal{S}$ field in figure \ref{fig:DelSNormHenckyp5Dep2L100} and the $\Delta De_\text{local}$ field in figure \ref{fig:Local_De.pdf}. This indicates a good local kinematic diagnostic capability of the latter. Upon further increasing $H$ to 2.0, the polymers are stretched further in the undisturbed regions and even more so in the regions of larger local stretch. This shows itself as more intense red regions in figure \ref{fig:DelSNormHencky2Dep2L100} compared to figure \ref{fig:DelSNormHenckyp5Dep2L100}. A more intense blue region indicates that locally in the blue or $\Delta De_\text{local} < 0$ regions, the polymers are even more collapsed compared to the currently undisturbed polymers at $H = 2.0$ than they were at $H = 0.5$. The weak memory effects due to a small but nonzero $De = 0.2$ can be observed in figure \ref{fig:DelSNormHencky2Dep2L100}, where the red or highly stretched polymer region around the extensional axis is elongated compared to the red or $\Delta De_\text{local} > 0$ region in figure \ref{fig:Local_De.pdf} around the extensional axis. This occurs because once the polymers get stretched during $H < 2.0$, they need some time or downstream distance along their convecting streamlines to relax from this stretched state. The $\Delta\mathcal{S}$ shown in figure \ref{fig:DelSNormHencky2Dep2L100} has already reached the steady state for $De = 0.2$ and $L = 100$.

\begin{figure}
	\centering
	\subfloat{\includegraphics[width=0.33\textwidth]{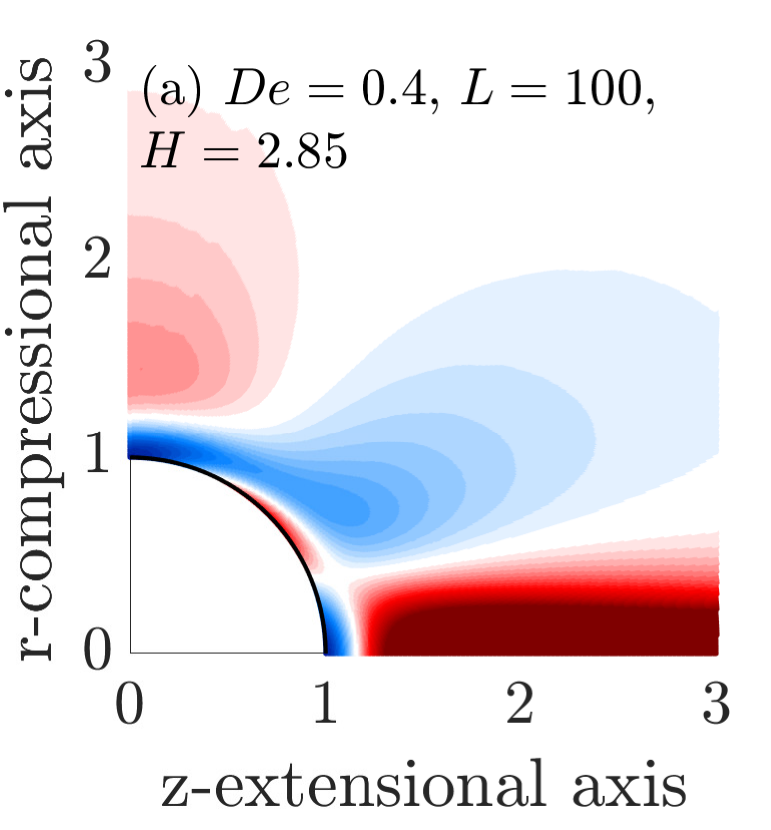}\label{fig:DelSNormHenckyp2p85Dep4L100}}\hspace{0.2in}
	\subfloat{\includegraphics[width=0.33\textwidth]{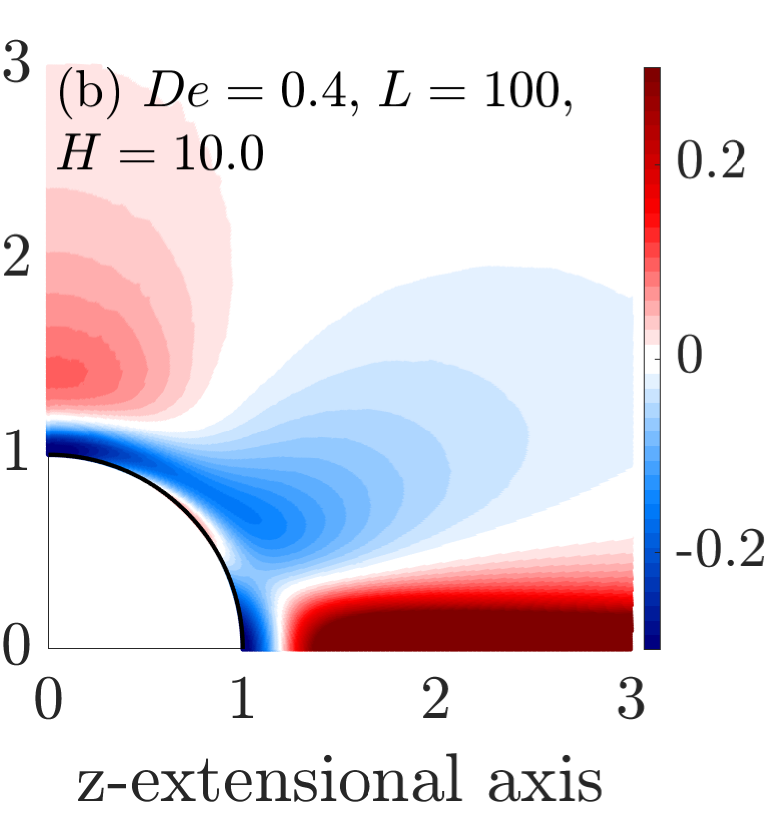}\label{fig:DelSNormHenckyp10Dep4L100}}
	\caption {$\Delta\mathcal{S}$ for $De = 0.4$ and $L = 100$ for Hencky strain, $H$: (a) 2.85 and (b) 10.0. Compared to the currently undisturbed polymers, in the red regions the polymers are stretched more, and in the blue regions they are collapsed. \label{fig:DelSNormDept4L100}}
\end{figure}

The polymer stretch behavior is qualitatively similar for $De = 0.4$ and $L = 100$ in figure \ref{fig:DelSNormDept4L100} as that discussed above for $De = 0.2$ and $L = 100$, but with a few differences. At $H$ smaller than shown in figure \ref{fig:DelSNormDept4L100}, $\Delta\mathcal{S}$ for $De = 0.4$ is similar to the $\Delta\mathcal{S}$ plots for the $De = 0.2$ case shown in figure \ref{fig:DelSNormDept2L100}. Initially, the polymers always respond to the local velocity as they are starting from equilibrium or a completely un-stretched configuration. However, due to the larger extension rate, the intensification from the un-stretched configuration at $H = 0$ to $0.5$ and then through $2.0$ shown by figure \ref{fig:DelSNormDept2L100} for $De = 0.2$ is more rapid for $De = 0.4$. This continues to larger strains, represented by $H = 2.85$ in this case of $De = 0.4$, where the intensity of the highly stretched polymers around the extensional axis increases even more. Due to a larger polymer relaxation time, once the polymers are stretched by $\Delta De_\text{local} > 0$ regions around the extensional axis, they take a longer time to relax from their highly stretched state. This is particularly evident around the extensional axis as the flow speed in that region is large and increases along the streamline. Therefore, the region of stretched polymers forms an elongated wake that extends to a larger downstream distance from the particle than for $De = 0.2$ discussed earlier. The memory effects are not particularly pronounced in the $\Delta\mathcal{S} > 0$ region around the compressional axis near the front stagnation point because the flow speed in that region is low and decreases along the streamlines. Hence, the polymers have enough time in this low-speed flow to relax and respond only to the local stretching. {However, upon increasing $H$ from 2.85 to 10, the intensity of the region of extra polymer stretch around the compressional axis increases as observed by comparing figures \ref{fig:DelSNormHenckyp2p85Dep4L100} and \ref{fig:DelSNormHenckyp10Dep4L100}.} The highly stretched region around the extensional axis starts at roughly the same location along the axis for $De = 0.2$ and $De = 0.4$ at all $H$ because just before this region, there is a stagnation point, and the flow speed is very low. Therefore, the start of extensional stretching is not much affected by polymer relaxation time. For this $De = 0.4$, there are significant regions of polymer collapse ($\Delta\mathcal{S} < 0$) around the front and rear stagnation points and on the particle surface around $45^\circ$ from the axes. These coincide with the regions where $\Delta De_\text{local} < 0$. They are more larger intense at $De = 0.4$ than $De = 0.2$ because the undisturbed polymer stretch is larger in the former.

The features of $\Delta\mathcal{S}$ just observed for $De=0.2$ and $0.4$ explain the rheological results of sections \ref{sec:RheologyNet} and \ref{sec:RheologySplit}. Since the intensity of the highly stretched polymers around the extensional axis is more than the collapsed polymers anywhere at each $H$, a positive PIPS or $\hat{\Pi}^{PP}$, is obtained for $De=0.2$. An intense wake of highly stretched polymers explains the positive PIPS in figure \ref{fig:PIPSTotalDep4} for $De=0.4$. Here PIPS initially increases rapidly with $H$ because the wake intensifies from when it first appears as a highly stretched region around the extensional axis with a shape similar to plots of figure \ref{fig:DelSNormDept2L100}. While still remaining very intense, this wake loses some of its thickness from $H=2.85$ to 10.0 (the red region around the z axis in \ref{fig:DelSNormHenckyp10Dep4L100} is slightly thinner than figure \ref{fig:DelSNormHenckyp2p85Dep4L100}). Yet PIPS increases monotonically between $H=2.85$ and 10.0 (figure \ref{fig:PIPSTotalDep4}), because the highly stretched region around the compressional axis intensifies with $H$. Dominance of stretching regions over the collapsed regions at all $H$ explains positive $\hat{\text{S}}^\text{Vol}/\hat{\Pi}^U=\hat{\text{S}}^{PP}/\hat{\Pi}^U-1$ from figure \ref{fig:StressletTotalDep4}. 

\begin{figure}
	\centering
	\subfloat{\includegraphics[width=0.33\textwidth]{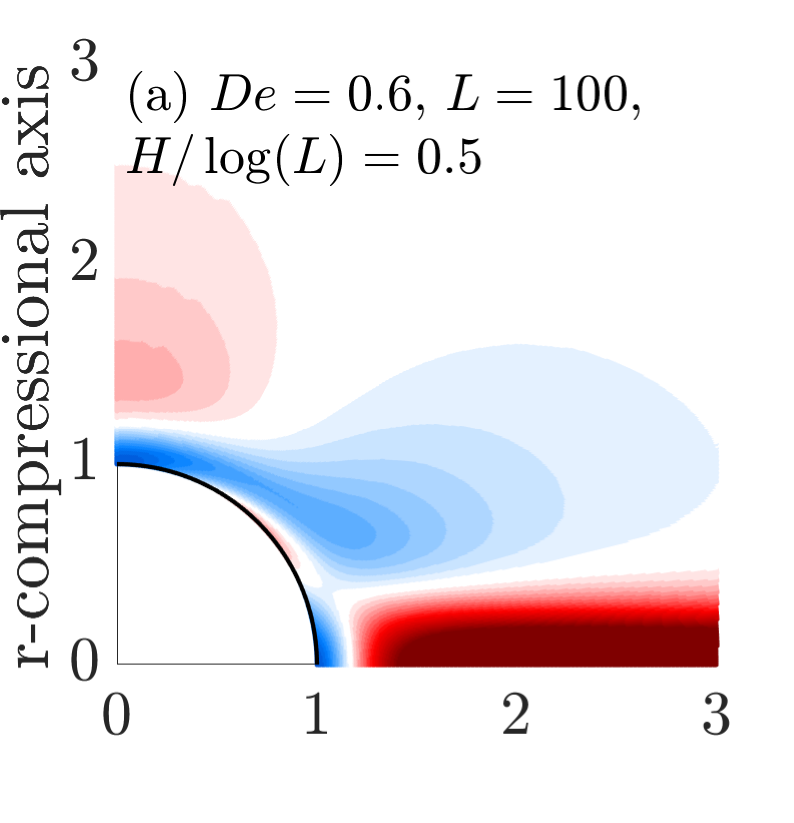}\label{fig:DelSNormHenckyp5Dep6L100}}
	\subfloat{\includegraphics[width=0.33\textwidth]{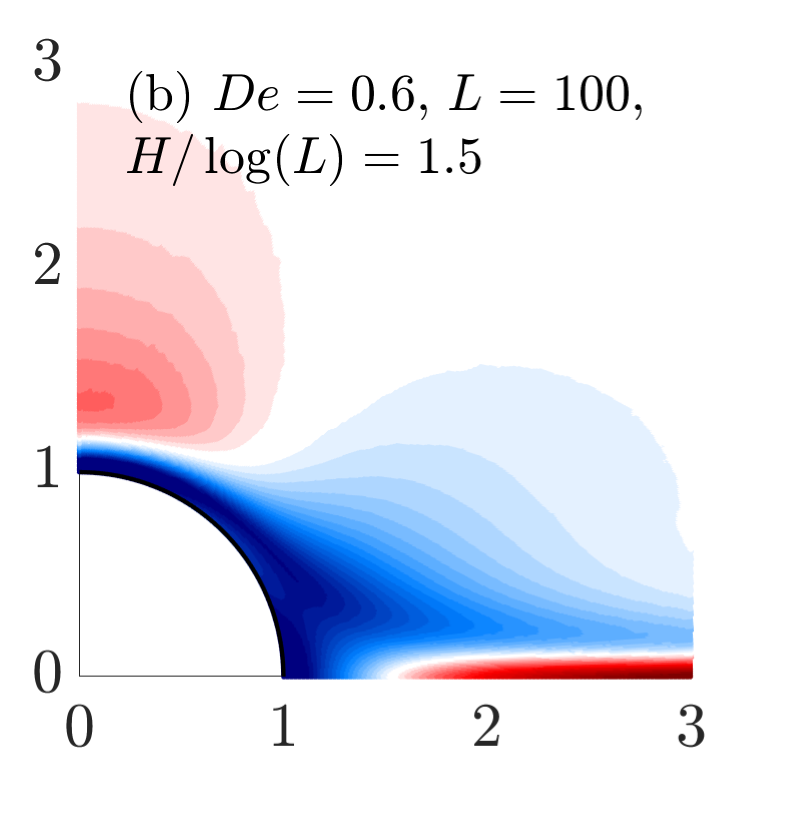}\label{fig:DelSNormHencky1p5Dep6L100}}
	\subfloat{\includegraphics[width=0.33\textwidth]{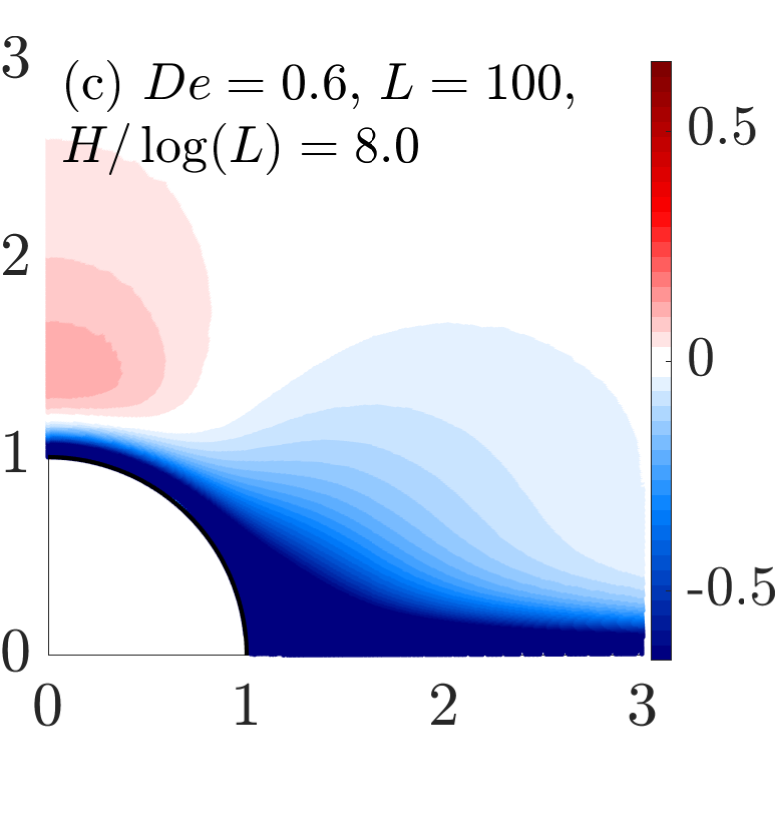}\label{fig:DelSNormHencky8Dep6L100}}\\
	\subfloat{\includegraphics[width=0.33\textwidth]{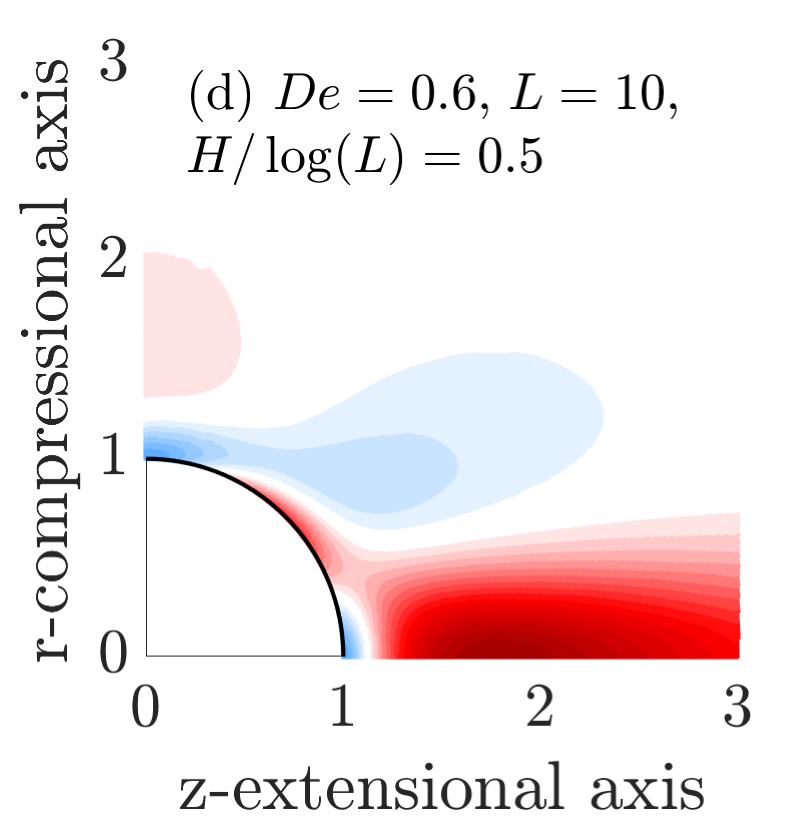}\label{fig:DelSNormHenckyp5Dep6L10}}
	\subfloat{\includegraphics[width=0.33\textwidth]{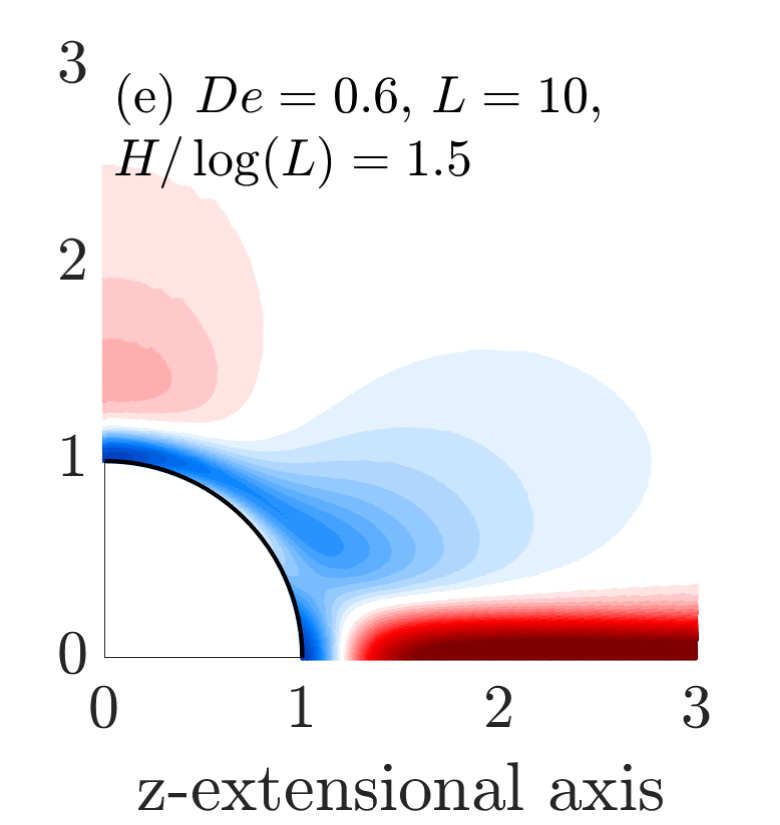}\label{fig:DelSNormHencky1p5Dep6L10}}
	\subfloat{\includegraphics[width=0.33\textwidth]{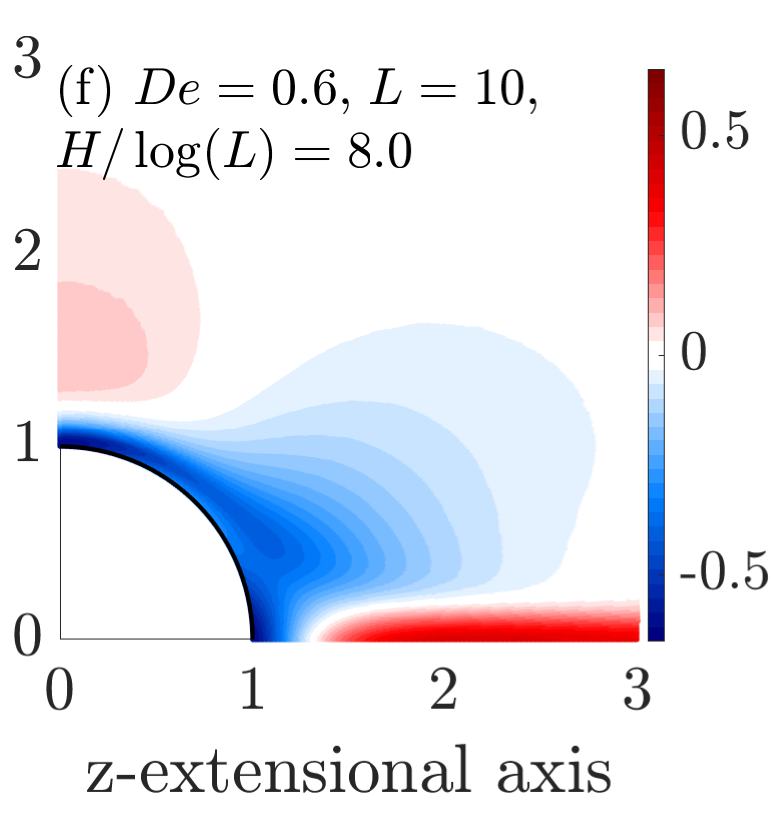}\label{fig:DelSNormHencky8Dep6L10}}
	\caption {$\Delta\mathcal{S}$ for $De = 0.6$ at $L = 100$ and 10 for $\log(L)$ normalized Hencky strain, $H / \log(L) = 0.5$, 1.5, and 8.0. $L$ and $H$ are indicated on each plot. Compared to the currently undisturbed polymers, in the red regions the polymers are stretched more, and in the blue regions they are collapsed. \label{fig:DelSNormDept6L10L100}}
\end{figure}

In figure \ref{fig:DelSNormDept6L10L100}, we show $\Delta\mathcal{S}$ for $De=0.6$ at $L=10$ and 100 and three different $H/\log(L)=0.5$, 1.5 and 8.0. Figures \ref{fig:DelSNormHenckyp5Dep6L100} to \ref{fig:DelSNormHencky8Dep6L100} are for $L=100$ and figures \ref{fig:DelSNormHenckyp5Dep6L10} to \ref{fig:DelSNormHencky8Dep6L10} are for $L=10$. The steady state is represented by $H/\log(L)=8.0$. First consider $L=100$. Up to $H/\log(L)$ similar mechanisms are at play those we described above for $De=0.4$ and $L=100$ up to its steady state. This also explains the similar initial variation of PIPS, $\hat{\Pi}^{PP}/\hat{\Pi}^U$ in the case of $De=0.6$ and $L=100$ as the variation of PIPS up to the steady state of $De=0.4$ and $L=100$ (compare figures \ref{fig:PIPSTotalDep4} and \ref{fig:PIPSTotalDep6}). For the $De=0.6$ case the effects are more pronounced. The wake of stretched polymers around the extensional axis is more intense (compare figures \ref{fig:DelSNormHenckyp10Dep4L100} and \ref{fig:DelSNormHenckyp5Dep6L100}-- both their wakes appear similar but the color scale for the latter is over a larger range) that leads to larger positive increase in the PIPS (compare figures \ref{fig:PIPSTotalDep4} and initial part of \ref{fig:PIPSTotalDep6}). But, unlike $De=0.4$, this is not the steady-state for $De=0.6$.

For $De=0.6$, the undisturbed polymers keep getting more stretched with $H$ and undergo coil-stretch transition where they stretch close to their maximum extensibility $L$. Hence, the polymers around the particle surface undergo an intense relative collapse compared to the highly stretched, undisturbed polymers. This is observed by a much larger and more intense blue region in figure \ref{fig:DelSNormHencky1p5Dep6L100} at $H/\log(L)=1.5$ as compared to $H/\log(L)=0.5$ in figure \ref{fig:DelSNormHenckyp5Dep6L100} or at the steady-state $H=8.0$ for De=0.4 in figure \ref{fig:DelSNormHenckyp10Dep4L100}. {The collapsed polymer region extends further downstream from the surface due to memory effects as the polymers need time and downstream distance to recover from their collapsed state.} While initially, the collapse arises because the far-field polymers undergo coil-stretch transition before the polymers near the particle surface have time to respond. The collapse persists into the steady-state because when the highly stretched polymers from the far-field arrive at the front stagnation point, the low stretching region represented by blue or $\Delta De_\text{local}<0$ around this location shown in figure \ref{fig:Local_De.pdf} makes them undergo a stretch-to-coil transition. They cannot fully recover from this collapsed state until a large distance downstream of the particle for $De=0.6$ and L=100. The region of collapsed polymers ultimately engulfs the wake of highly stretched polymers around the extensional axis as shown in figure \ref{fig:DelSNormDept6L10L100} for $H/\log(L)=8.0$. A more detailed discussion about the steady-state collapse and other features of $\Delta\mathcal{S}$ in the steady-state for a wide range of $De$ and $L$ is made in \cite{SteadyStatePaper}.

%

The beginning of the collapse marks the end of the increasing PIPS in figure \ref{fig:PIPSTotalDep6}. Beyond this point, PIPS starts to decrease, and due to a large region of collapsed polymers around the particle surface in the steady state represented by figure \ref{fig:DelSNormHencky8Dep6L100}, it becomes negative at large $H / \log(L)$. The volumetric stresslet $\hat{\text{S}}^\text{Vol} / \hat{\Pi}^U = \hat{\text{S}}^{PP} / \hat{\Pi}^U - 1$ from figure \ref{fig:StressletTotalDep6} is also negative because of this collapse. These large negative values make the total interaction stress ($\hat{\Pi}^{PP} + \hat{\text{S}}^\text{Vol}$) negative. Hence, the net negative impact of adding particles discussed in section \ref{sec:RheologyNet} arises from this large region of collapsed polymers for $De = 0.6$ and $L = 100$.

At $L = 10$, the initial intensification of the wake of highly stretched polymers is weaker, and in the steady state or at large $H / \log(L)$, the region of collapse is smaller and less intense than in the $L = 100$ case. This can be observed by comparing figures \ref{fig:DelSNormHenckyp5Dep6L100} to \ref{fig:DelSNormHencky8Dep6L100} with the corresponding $H / \log(L)$ from figures \ref{fig:DelSNormHenckyp5Dep6L10} to \ref{fig:DelSNormHencky8Dep6L10}. A less intense wake in the initial phase for $L = 10$ implies a smaller increase in $\hat{\Pi}^{PP} / \hat{\Pi}^U$ shown in figure \ref{fig:PIPSTotalDep6}. A smaller and less intense region of collapsed polymers implies a smaller and less rapid decrease of $\hat{\Pi}^{PP} / \hat{\Pi}^U$ in figure \ref{fig:PIPSTotalDep6} and a less negative $\hat{\text{S}}^\text{Vol} / \hat{\Pi}^U = \hat{\text{S}}^{PP} / \hat{\Pi}^U - 1$ in figure \ref{fig:StressletTotalDep6}. Therefore, the rheological changes shown in figure \ref{fig:TotalInteractDep6} for the total interaction stress at $L = 10$ are less profound than at $L = 100$ for all $H / \log(L)$.

\begin{figure}
	\centering
	\subfloat{\includegraphics[width=0.33\textwidth]{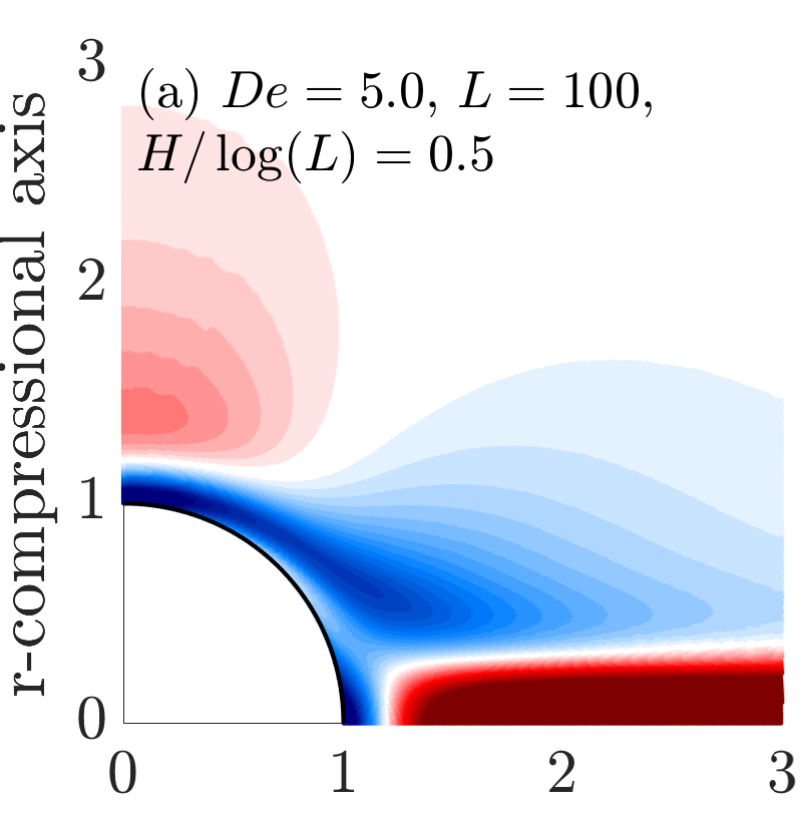}\label{fig:DelSNormHenckyp5De5L100}}
	\subfloat{\includegraphics[width=0.33\textwidth]{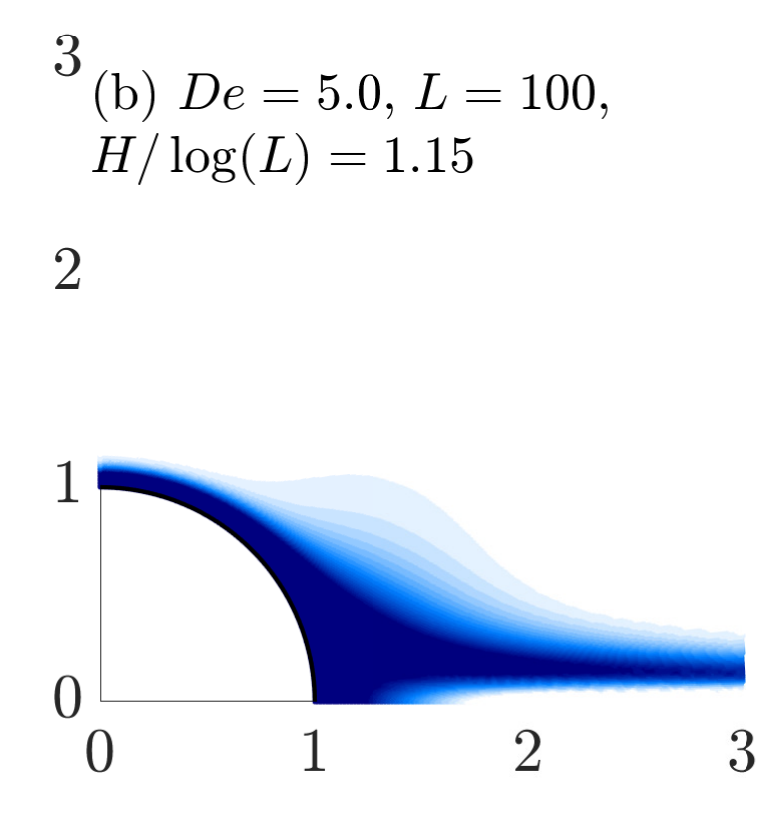}\label{fig:DelSNormHencky1p15De5L100}}
	\subfloat{\includegraphics[width=0.33\textwidth]{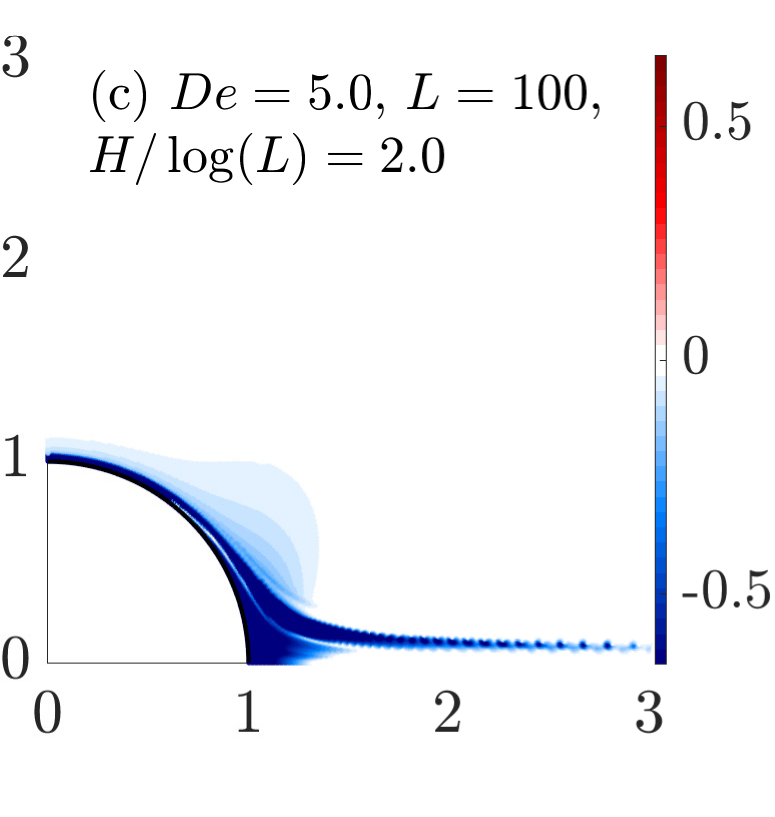}\label{fig:DelSNormHencky2De5L100}}\\
	\subfloat{\includegraphics[width=0.33\textwidth]{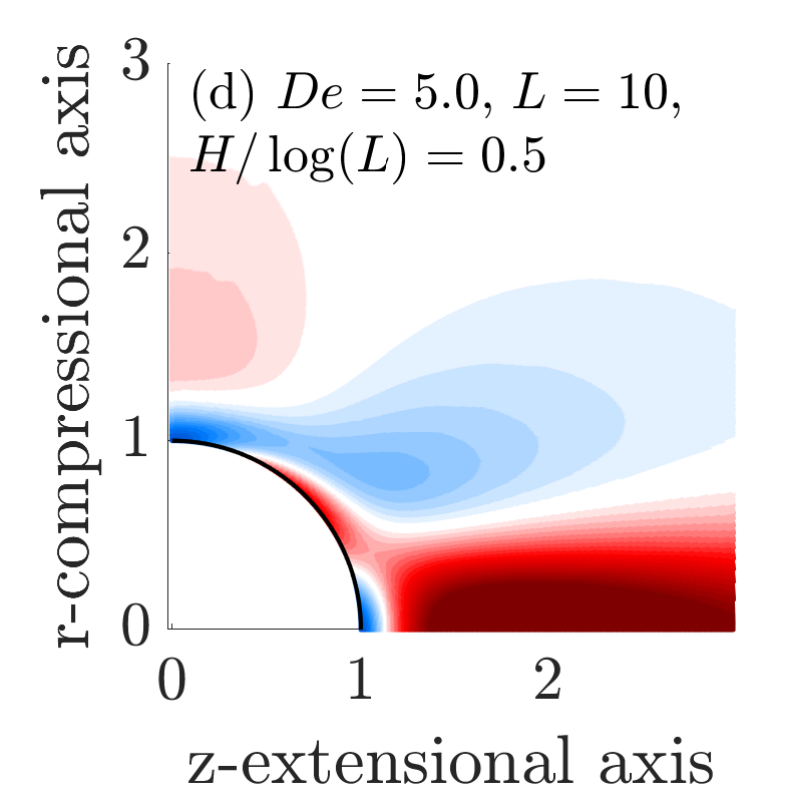}\label{fig:DelSNormHenckyp5De5L10}}
	\subfloat{\includegraphics[width=0.33\textwidth]{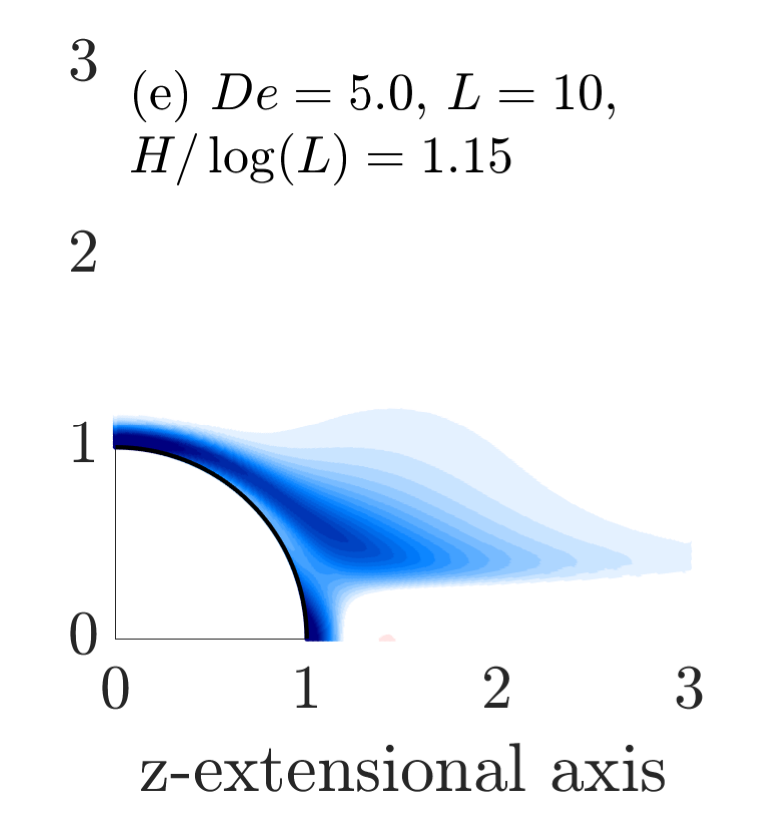}\label{fig:DelSNormHencky1p15De5L10}}
	\subfloat{\includegraphics[width=0.33\textwidth]{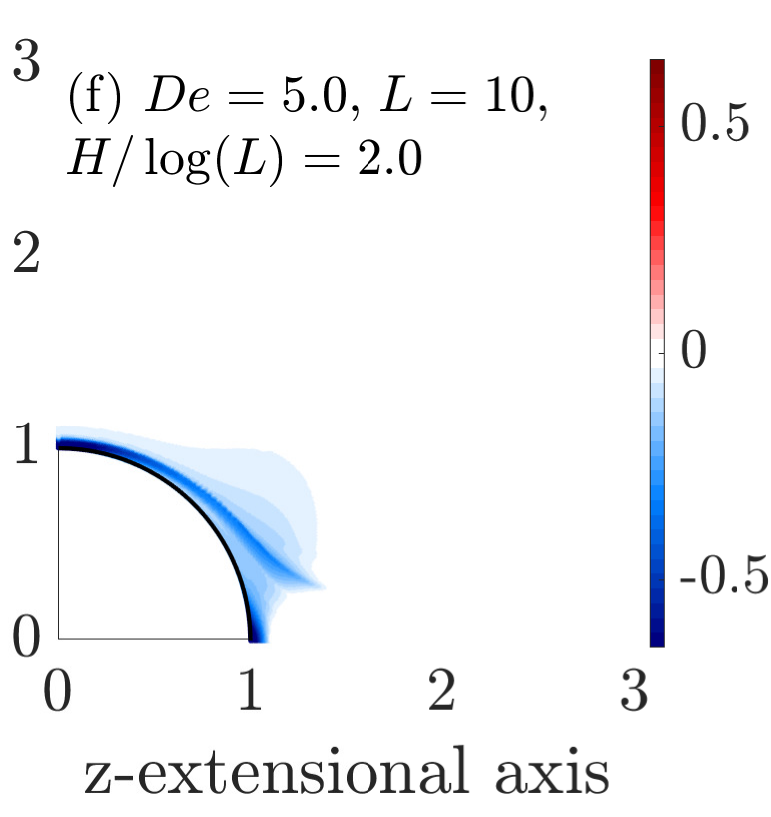}\label{fig:DelSNormHencky2De5L10}}
	\caption {$\Delta\mathcal{S}$ for $De = 5$ at $L = 100$ and 10 for $\log(L)$ normalized Hencky strain, $H / \log(L) = 0.5$, 1.15, and 2.0. $L$ and $H$ are indicated on each plot. Compared to the currently undisturbed polymers, in the red regions the polymers are stretched more, and in the blue regions they are collapsed. \label{fig:DelSNormDe5L10L100}}
\end{figure}

Finally, we explain the behavior of rheological changes observed in sections \ref{sec:RheologyNet} and \ref{sec:RheologySplit} at large $De$ through the $\Delta\mathcal{S}$ field of $De = 5.0$ shown in figure \ref{fig:DelSNormDe5L10L100}. Initially, a similar mechanism as that previously explained for the initial phase for $De = 0.6$ occurs. For both $L = 10$ and 100, as $H$ increases from 0, a wake of highly stretched polymers forms around the extensional axis, intensifying upon initially increasing $H$. Along with this, polymers around the particle surface start to collapse, as shown in figures \ref{fig:DelSNormHenckyp5De5L100} and \ref{fig:DelSNormHenckyp5De5L10} for $L = 100$ and 10, respectively. This is accompanied by a rapid increase in $\hat{\Pi}^{PP} / \hat{\Pi}^U$ up to $H / \log(L) \approx 1.0$, as shown in figure \ref{fig:PIPSTotalDe5}.

The undisturbed polymers at $De = 5.0$ undergo a local coil-stretch transition starting at $H / \log(L) \approx 1.0$, as shown in figure \ref{fig:U0De5}. At larger $De$, the undisturbed coil-stretch transition is much more rapid and leads to larger polymer stress, as observed by comparing $De = 0.6$, 2.0, and 5.0 plots in figure \ref{fig:UndisturbedStress}. Polymer stretch follows the same trend in $De$. In a short time frame from $H / \log(L) \approx 1.0$ to $H / \log(L) \approx 1.15$, the undisturbed polymer stretch increases rapidly to very large values (figure \ref{fig:U0De5}). {Hence, the polymers in the wake around the extensional axis that were highly stretched relative to the undisturbed polymers before the coil-stretch transition (figures \ref{fig:DelSNormHenckyp5De5L100} and \ref{fig:DelSNormHenckyp5De5L10}) are now less stretched (figures \ref{fig:DelSNormHencky1p15De5L100} and \ref{fig:DelSNormHencky1p15De5L10}) than the even more stretched undisturbed polymers after a rapid transition. This manifests as a rapid reduction in the intensity of the highly-stretched-polymer wake around the extensional axis and an increased intensity of the polymer collapse around the particle surface.}

A sharp reduction in $\hat{\Pi}^{PP} / \hat{\Pi}^U$ and $(\hat{\text{S}}^{PP} + \hat{\Pi}^{PP}) / \hat{\Pi}^U$ in figures \ref{fig:PIPSTotalDe5} and \ref{fig:TotalInteractDe5}, respectively, to negative values reflects these polymer stretch dynamics in the suspension rheology. The region of collapsed polymers has a large spatial extent due to the memory of the polymers that were previously collapsed in the regions close to the front stagnation line before traveling to the far-field along extensional axis. Upon further increasing $H / \log(L)$ to 2.0 in figures \ref{fig:DelSNormHencky2De5L100} and \ref{fig:DelSNormHencky2De5L10} for $L = 100$ and 10, respectively, the region of collapse shrinks towards the particle. This occurs because at large $De$, a large imposed extension rate allows a quicker recovery of polymers from collapse to attain stretch close to the undisturbed polymer stretch within small distances from the particle surface. This shrinking of the collapsed polymer region appears as a negative peak in $\hat{\Pi}^{PP} / \hat{\Pi}^U$ for $De = 2.0$ and 5.0, as shown for the latter in figure \ref{fig:PIPSTotalDe5}.

At large $H$ and $De$, as shown for $H / \log(L) = 2.0$ and $De = 5.0$ in figures \ref{fig:DelSNormHencky2De5L100} and \ref{fig:DelSNormHencky2De5L10} for $L = 100$ and 10, respectively, polymer collapse is only in a thin region close to the particle surface. The local polymer stretch is the same as the undisturbed stretch elsewhere in the fluid, and the collapse in the thin region is very intense, as $\hat{\Pi}^{PP} / \hat{\Pi}^U$ beyond the aforementioned negative peak is still negative and significant in figure \ref{fig:PIPSTotalDe5}. Therefore, the negative effect of the addition of particles for $De = 5.0$ and $L = 100$ mentioned at the beginning of section \ref{sec:RheologyNet} arises from this thin region of collapsed polymers.

Due to the local polymer stretch being the same as the undisturbed stretch everywhere except in a thin region in the steady state, $\hat{\text{S}}^\text{Vol} / \hat{\Pi}^U = \hat{\text{S}}^{PP} / \hat{\Pi}^U - 1$ is zero for $De = 5.0$ (figure \ref{fig:StressletTotalDe5}) and 2.0 (not shown) at large $H$. Before this steady state, the transient behavior of $\hat{\text{S}}^\text{Vol} / \hat{\Pi}^U$ is different for $L = 10$ and 100, as shown in figure \ref{fig:StressletTotalDe5}. For larger $L$, the higher intensity of the region of collapsed polymers around the particle surface compared to the wake of stretched polymers around the extensional axis (figure \ref{fig:DelSNormHenckyp5De5L100}) leads to a negative $\hat{\text{S}}^\text{Vol} / \hat{\Pi}^U$ in the transient. However, for smaller $L$, a more intense wake and less intense collapsed polymer region (figure \ref{fig:DelSNormHenckyp5De5L10}) lead to a sharp transient increase in $\hat{\text{S}}^\text{Vol} / \hat{\Pi}^U$ with $H$ before it decreases again.

{An analogy between the dynamics of polymer stretch shown in this section and that of a line of dye in a Newtonian fluid is offered in appendix \ref{sec:PolymerDye}.}

\section{Suspension rheology of concentrated polymeric {solutions}}\label{sec:ConcentratedRheology}
In the preceding section, we examined the effects of particle-polymer interactions in a liquid with low polymer concentration, $c$, where the influence of polymer configuration on the velocity and pressure fields is minimal. The polymer-induced velocity, $c\mathbf{u}^\text{PI}$ (defined as $\mathbf{u} - \mathbf{u}^\text{Stokes}$ in equation \eqref{eq:NonNewtonianFields}), was small enough that $| c\mathbf{u}^\text{PI} | \ll | \mathbf{u}^\text{Stokes} |$, allowing the Newtonian velocity field, $\mathbf{u}^\text{Stokes}$, to predominantly govern the evolution of the polymer configuration. However, as $c$ increases, $\mathcal{O}(| c\mathbf{u}^\text{PI} |) = \mathcal{O}(|\mathbf{u}^\text{Stokes}|)$, and the polymer configuration depends on the total fluid velocity, $c\mathbf{u}^\text{PI} + \mathbf{u}^\text{Stokes}$.

To address this complexity, we utilize direct numerical simulations to study the extensional rheology of dilute suspensions of spheres in concentrated polymeric fluids. These simulations are conducted with our in-house numerical solver. For a detailed methodology, we refer to our previous publication \cite{NumericalMethodPaper}, which includes validation of flow around spheres and spheroids in various viscoelastic flows. We summarize our numerical method in appendix \ref{sec:DNSMethodology}, including modifications for evaluating and validating the polymer-induced particle stress (PIPS).

Here, we first illustrate the rheology results in section \ref{sec:ResultsFinitec} before discussing the underlying mechanisms in sections \ref{sec:Mech1} and \ref{sec:Mech2}. The data from the simulations with $c = 10^{-5}$ (equivalent to that from the low $c$ methodology as shown in appendix \ref{sec:DNSMethodology}) represents the low $c$ results to be compared with numerical data at larger $c$. As mentioned earlier in sections \ref{sec:Intro} and \ref{sec:Formulation}, upon increasing $c$, the {Giesekus model} becomes more appropriate than the FENE-P model for polymer solutions. Thus, for the largest $c$ results shown below, only the Giesekus model is used. For intermediate values of $c$, we use both models, where comparison between the two provides a measure of the effect of polymer entanglement in {concentrated solutions in addition to their coiling in dilute solutions}. While making this comparison, similar to the undisturbed stress in section \ref{sec:U0Rheology}, we ensure that parameters are chosen such that $\alpha = L^{-2}$.

\subsection{Rheological observations}\label{sec:ResultsFinitec}
To understand the effect of polymer concentration on the particle-polymer interaction stress and extensional viscosity, $\mu_\text{ext}$ (defined in equation \eqref{eq:ExtVisc}), we briefly review the effects at low polymer concentration, $c$, from section \ref{sec:DiluteRheology} and our previous publication \citep{SteadyStatePaper}. Regardless of $c$, the undisturbed polymer extensional viscosity, $0.5c\hat{\Pi}^U$, varies linearly with $c$ at all $De$. In the low $c$ regime, the interaction stress contribution to the extensional viscosity, $0.5c\phi(\hat{\Pi}^{PP} + \hat{\text{S}}^{PP})$, is also linear in $c$. Thus, the particle-polymer interaction stress normalized with the undisturbed polymer stress, $(\hat{\Pi}^{PP} + \hat{\text{S}}^{PP}) / \hat{\Pi}^U$, is independent of $c$ at low $c$.

In section \ref{sec:DiluteRheology}, we found that for $De < 0.5$, $(\hat{\Pi}^{PP} + \hat{\text{S}}^{PP}) / \hat{\Pi}^U$ increases with Hencky strain, $H$, or time. This increase also occurs for $De > 0.5$ initially, but at a faster rate, leading to a larger increase in interaction stress. For $De < 0.5$, a steady state is reached with a positive interaction stress value. However, for $De > 0.5$, due to the coil-stretch transition of the undisturbed polymers, as $H$ increases, $(\hat{\Pi}^{PP} + \hat{\text{S}}^{PP}) / \hat{\Pi}^U$ decreases and becomes negative after the initial increase. At higher $De$, this reduction is more rapid and results in a more negative steady-state value, as shown by \cite{SteadyStatePaper}.

\begin{figure}
	\centering	
	\subfloat{\includegraphics[width=0.4\textwidth]{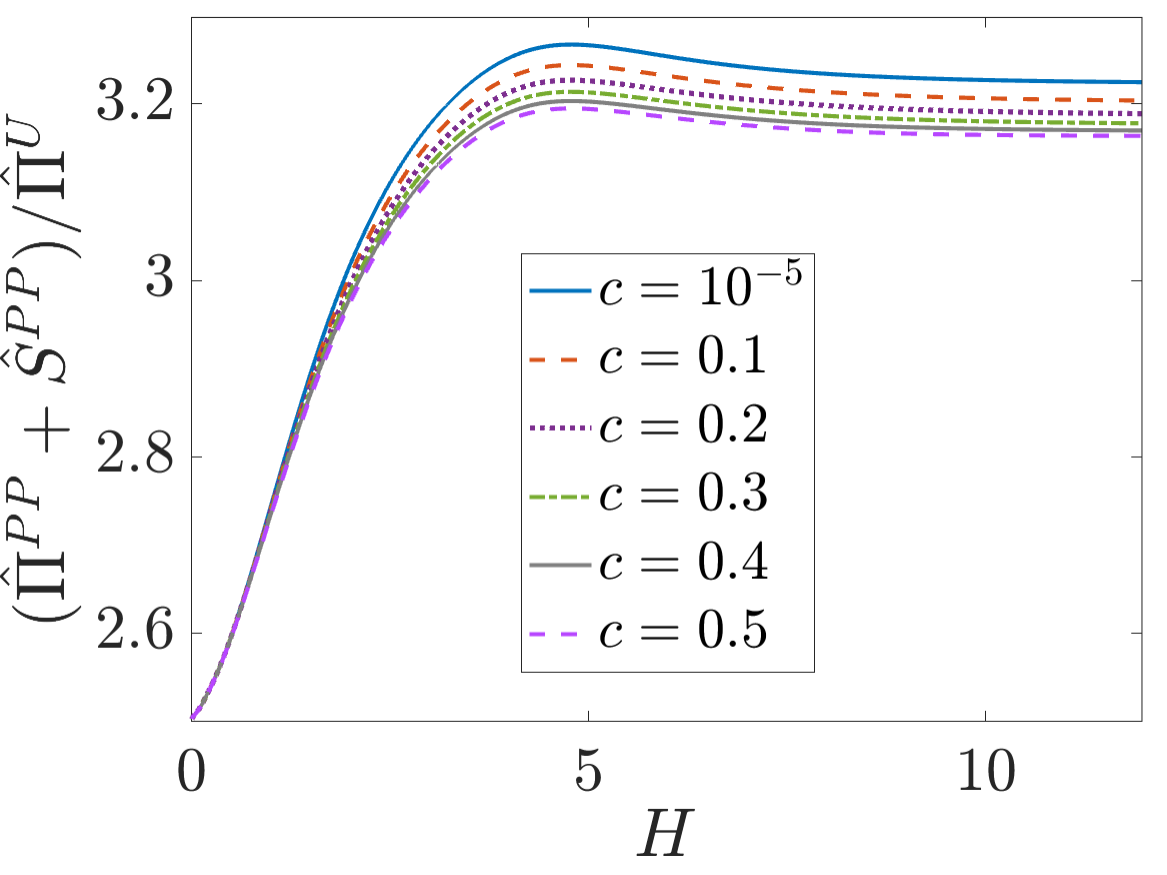}\label{fig:Dep4L50diffctimeplot}}\hspace{0.2in}
	\subfloat{\includegraphics[width=0.4\textwidth]{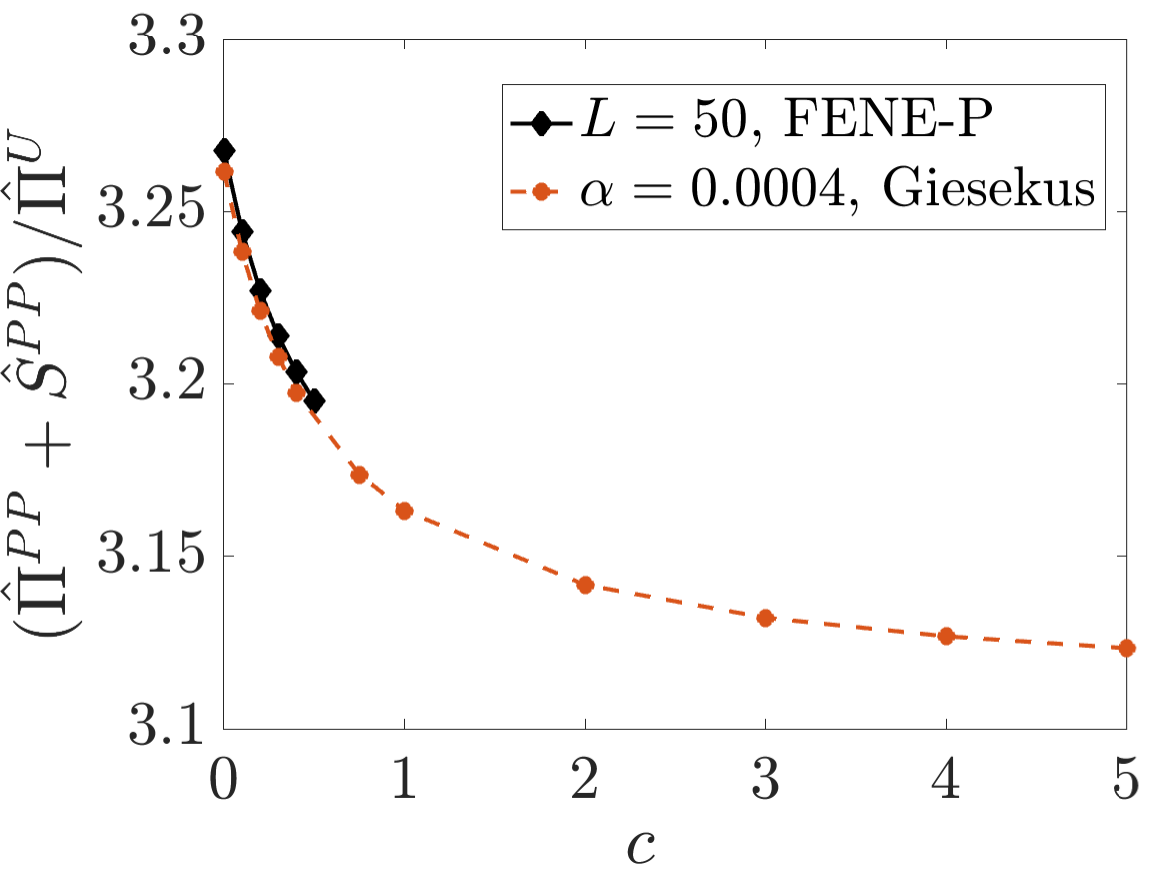}\label{fig:PlotvscDept4L10}}
	\caption{Effect of polymer concentration, $c$, on $(\hat{\Pi}^{PP} + \hat{\text{S}}^{PP}) / \hat{\Pi}^U$ : (a) Time or Hencky strain, $H$, evolution for Giesekus liquids with $\alpha = 0.01$ at $De = 0.4$, and (b) Comparison of the steady-state values for Giesekus fluids with $\alpha=0.0004$ with FENE-P liquids at $L = 50$.}
	\label{fig:VarywithcDep4}
\end{figure}

\begin{figure}
	\centering	
	\subfloat{\includegraphics[width=0.4\textwidth]{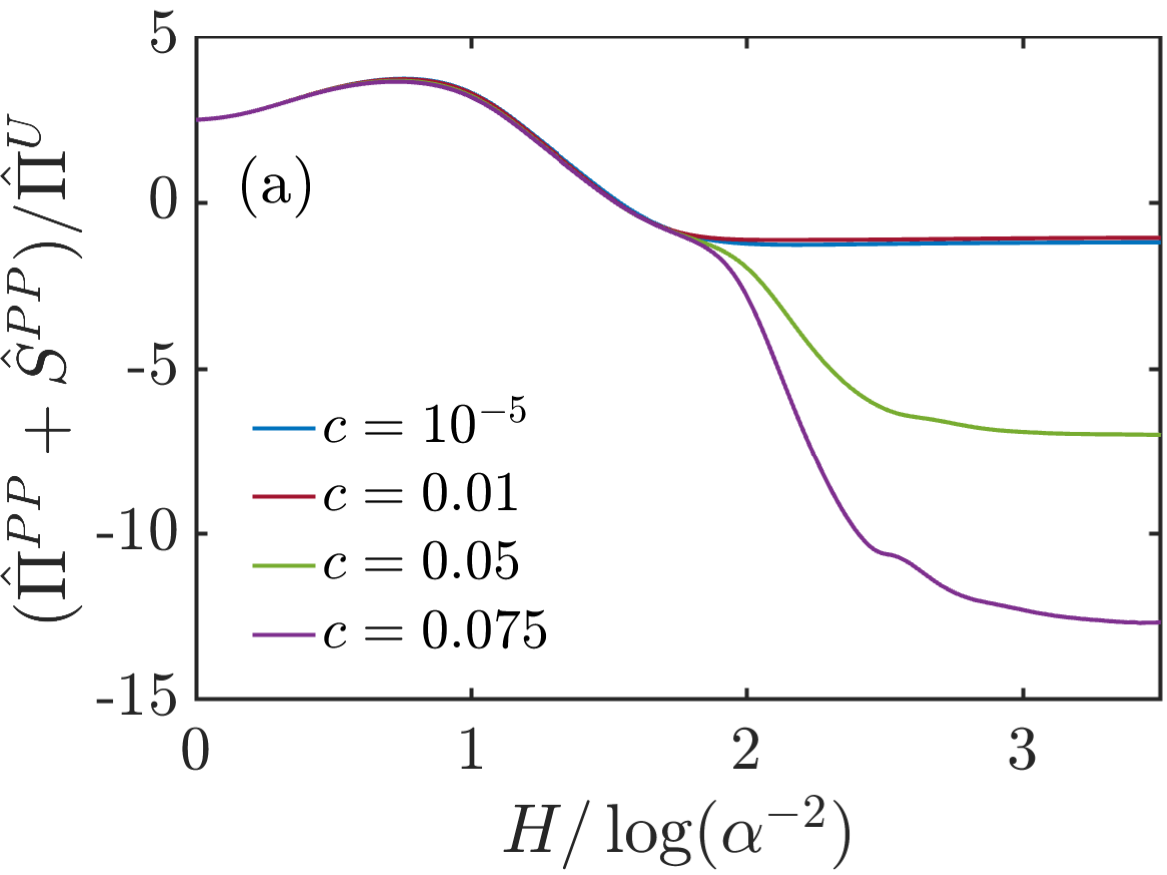}\label{fig:De2alfap0004diffctimeplot}}\hspace{0.2in}
	\subfloat{\includegraphics[width=0.4\textwidth]{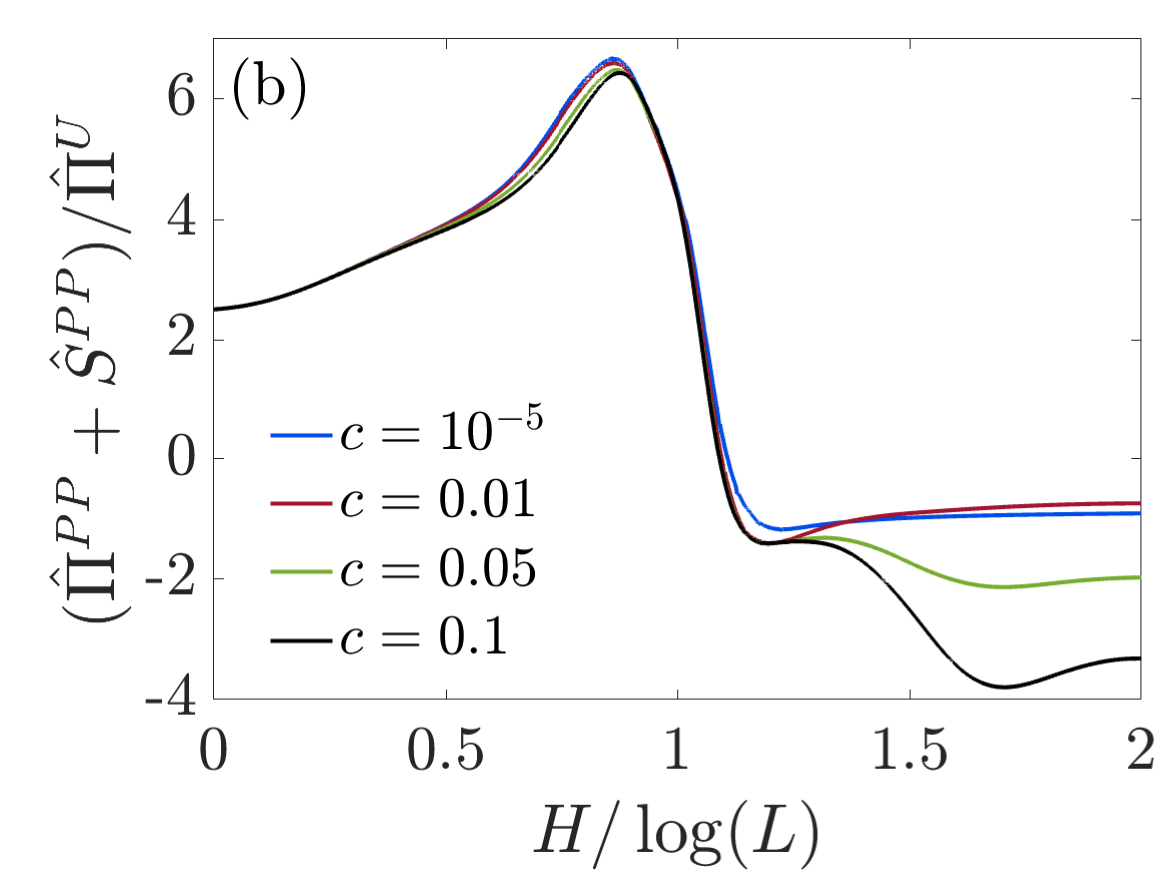}\label{fig:TotalInteractionStressL50De2variousc2}}
	\caption{Effect of polymer concentration, $c$, on the evolution of $(\hat{\Pi}^{PP} + \hat{\text{S}}^{PP}) / \hat{\Pi}^U$ at $De = 2.0$ for (a) Giesekus liquids with $\alpha = 0.0004$ (simulations running), and (b) FENE-P liquids with $L = 50$.}
	\label{fig:VarywithcDe2L50}
\end{figure}

\begin{figure}
	\centering	
	\subfloat{\includegraphics[width=0.4\textwidth]{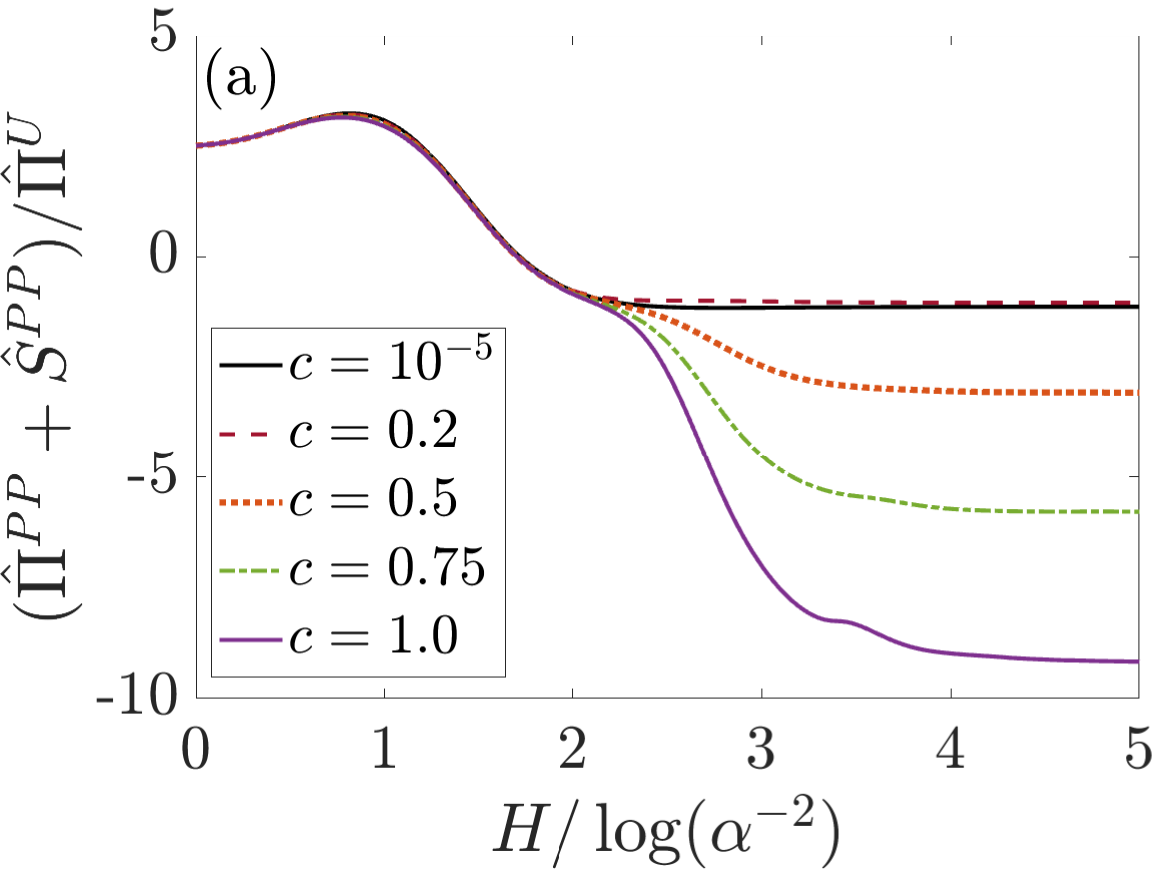}\label{fig:De5alfap01diffctimeplot}}\hspace{0.2in}
	\subfloat{\includegraphics[width=0.4\textwidth]{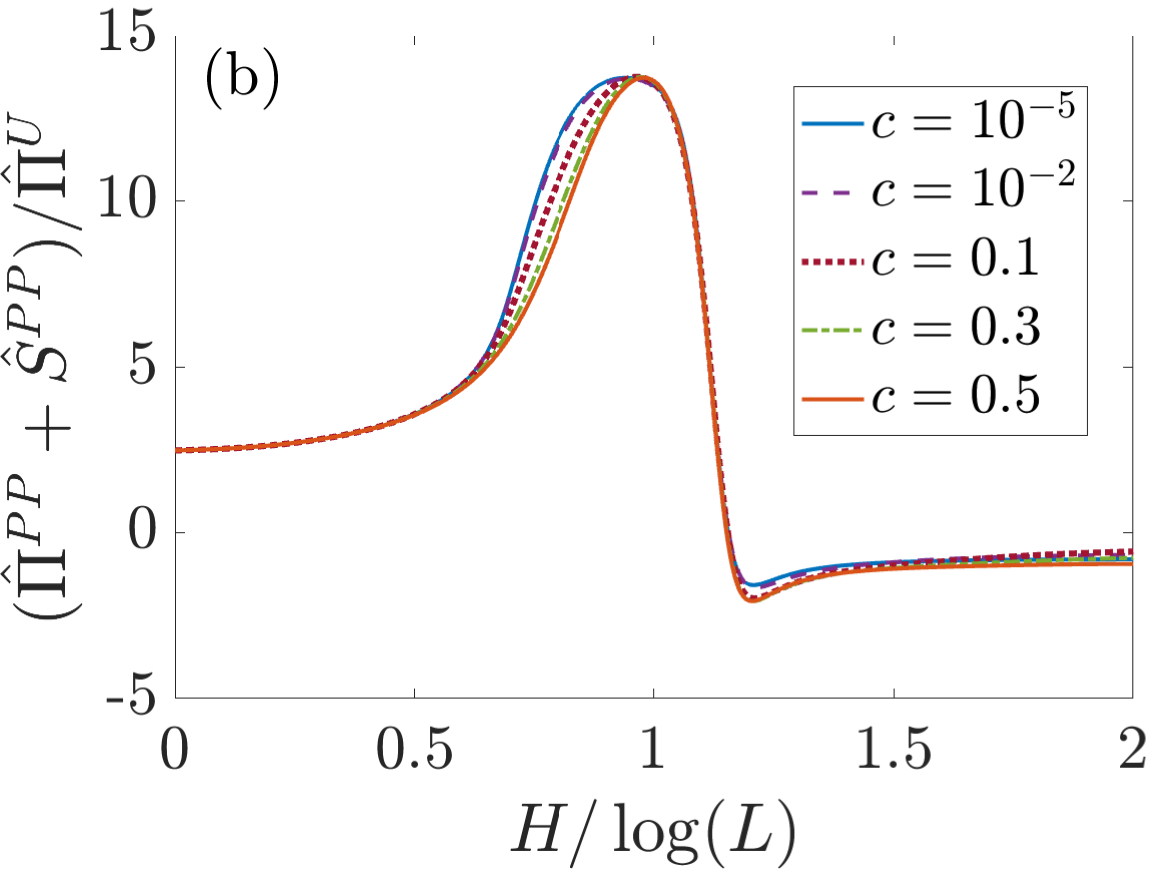}\label{fig:De5L10diffctimeplot}}
	\caption{Effect of polymer concentration, $c$, on the evolution of $(\hat{\Pi}^{PP} + \hat{\text{S}}^{PP}) / \hat{\Pi}^U$ at $De = 5.0$ for (a) Giesekus liquids with $\alpha = 0.01$, and (b) FENE-P liquids with $L = 10$.}
	\label{fig:VarywithcDe5L10}
\end{figure}

\begin{figure}
	\centering	
	\subfloat{\includegraphics[width=0.4\textwidth]{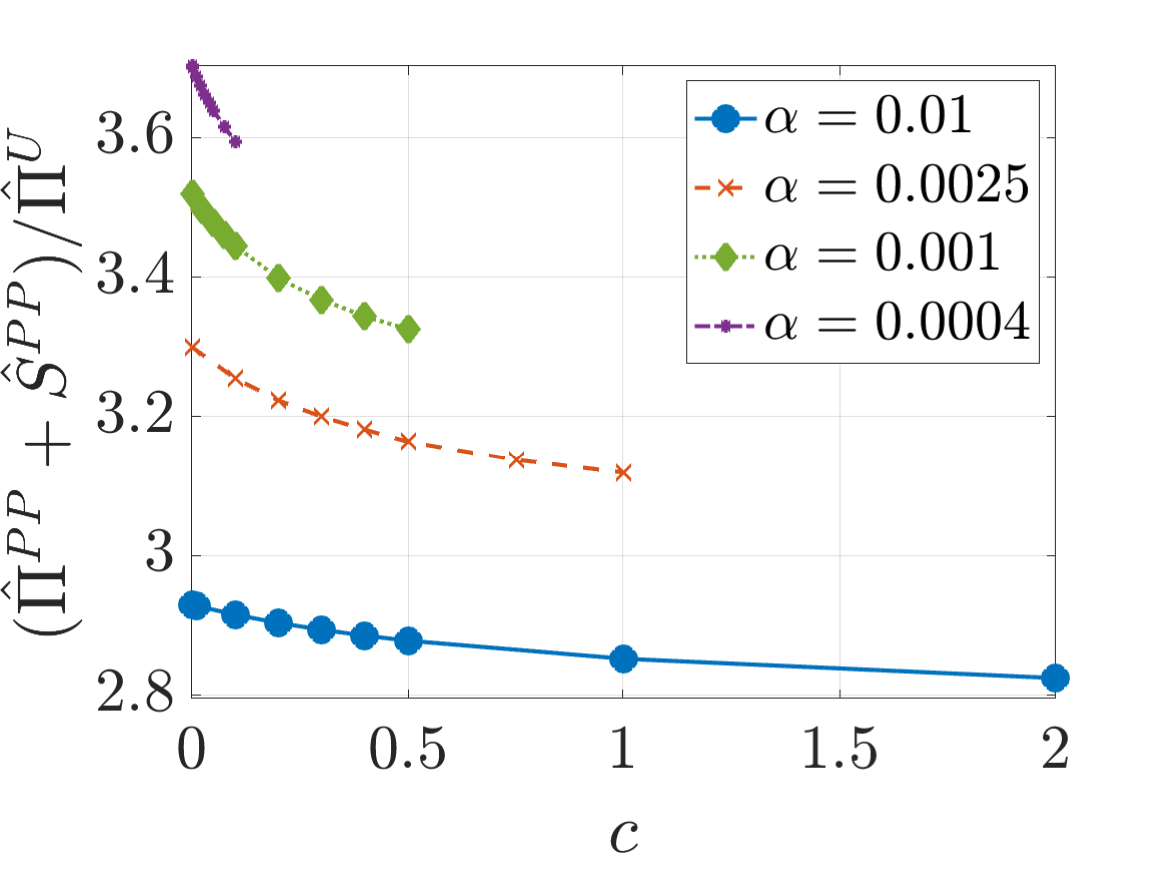}\label{fig:RhRheologyalfavaryDe1Max}}\hspace{0.2in}
	\subfloat{\includegraphics[width=0.4\textwidth]{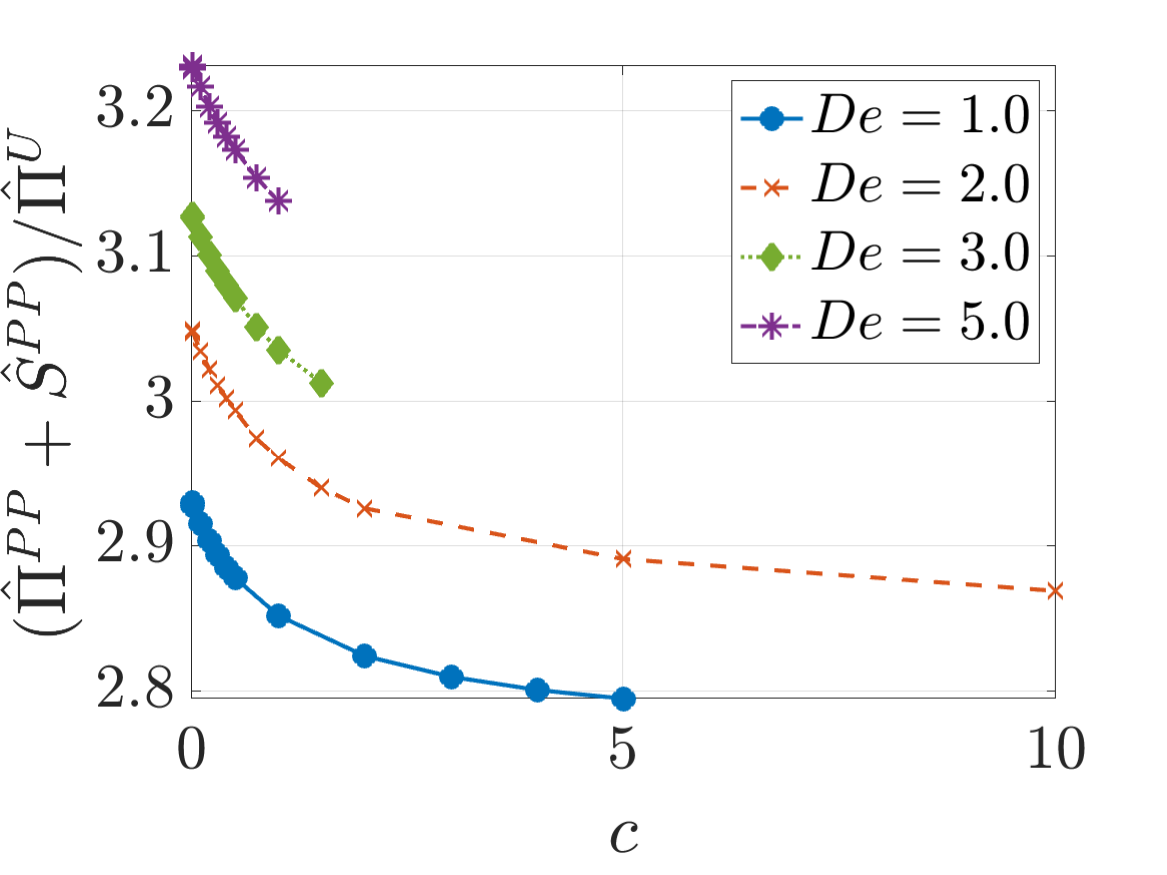}\label{fig:RhelogyvsDdiffDe_max}}
	\caption{Effect of polymer concentration, $c$, on the maximum $(\hat{\Pi}^{PP} + \hat{\text{S}}^{PP}) / \hat{\Pi}^U$ in a Giesekus fluid with (a) different $\alpha$ at $De = 1.0$ and (b) with $\alpha = 0.01$ at different $De$. \label{fig:DeandAlfaGiesekusMax}}
\end{figure}

\begin{figure}
	\centering	
	\subfloat{\includegraphics[width=0.4\textwidth]{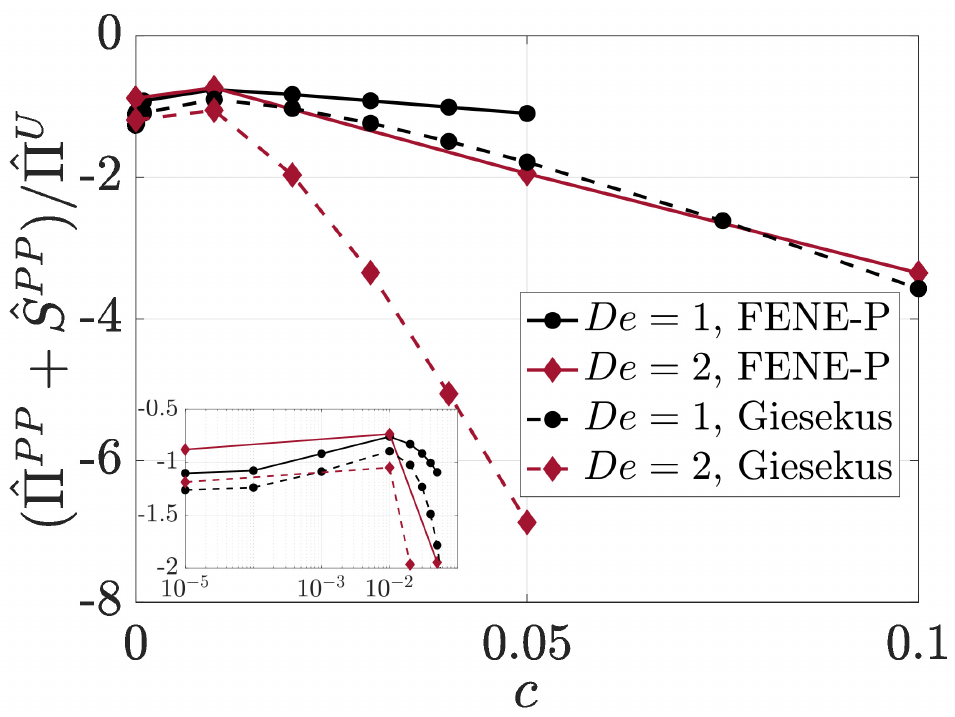}\label{fig:RheologyFENEPandGiesekusL50100Steady}}\hspace{0.2in}
	\subfloat{\includegraphics[width=0.4\textwidth]{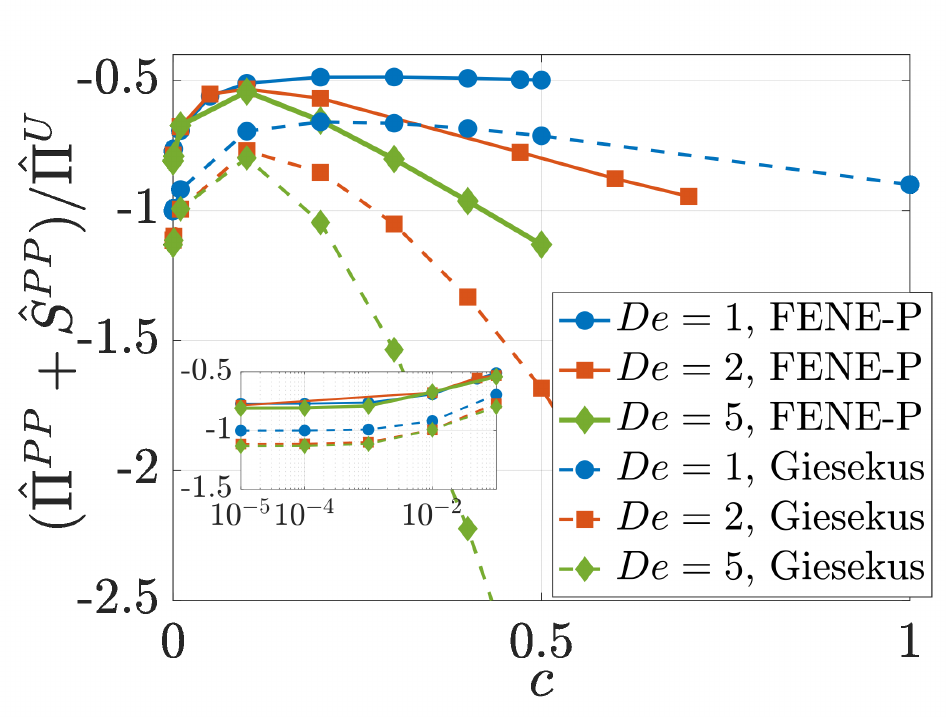}\label{fig:RheologyFENEPSteadL10y}}
	\caption{Effect of polymer concentration, $c$, on steady-state $(\hat{\Pi}^{PP} + \hat{\text{S}}^{PP}) / \hat{\Pi}^U$ for a FENE-P (solid lines) and Giesekus (dashed lines) fluid for different $De$ and equivalent $\alpha$ and $L$ such that $\alpha = L^{-2}$ for (a) $L = 50$, and (b) $L = 10$.}
	\label{fig:VaryDeGiesekusandFENE}
\end{figure}

\begin{figure}
	\centering	
	\subfloat{\includegraphics[width=0.4\textwidth]{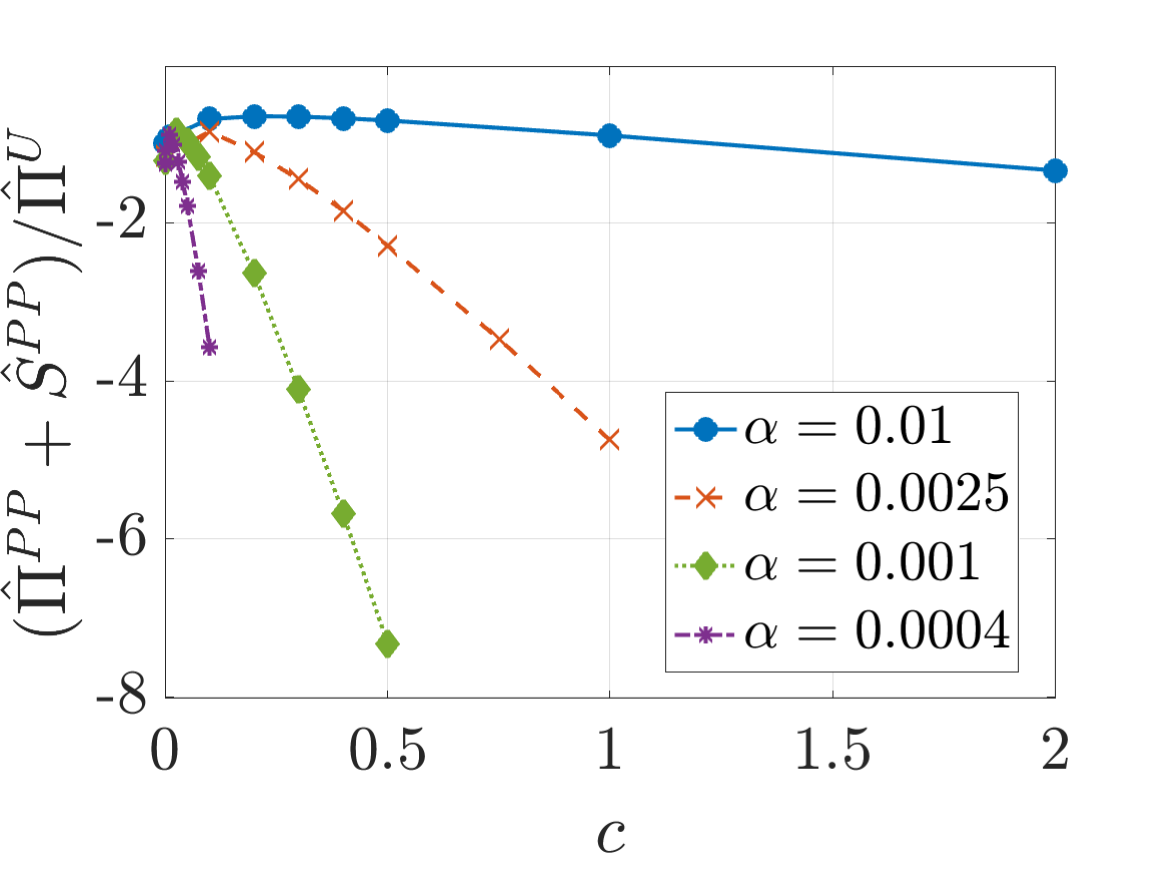}\label{fig:RheologyalfavaryDe1Steady}} \hspace{0.2in}
	\subfloat{\includegraphics[width=0.4\textwidth]{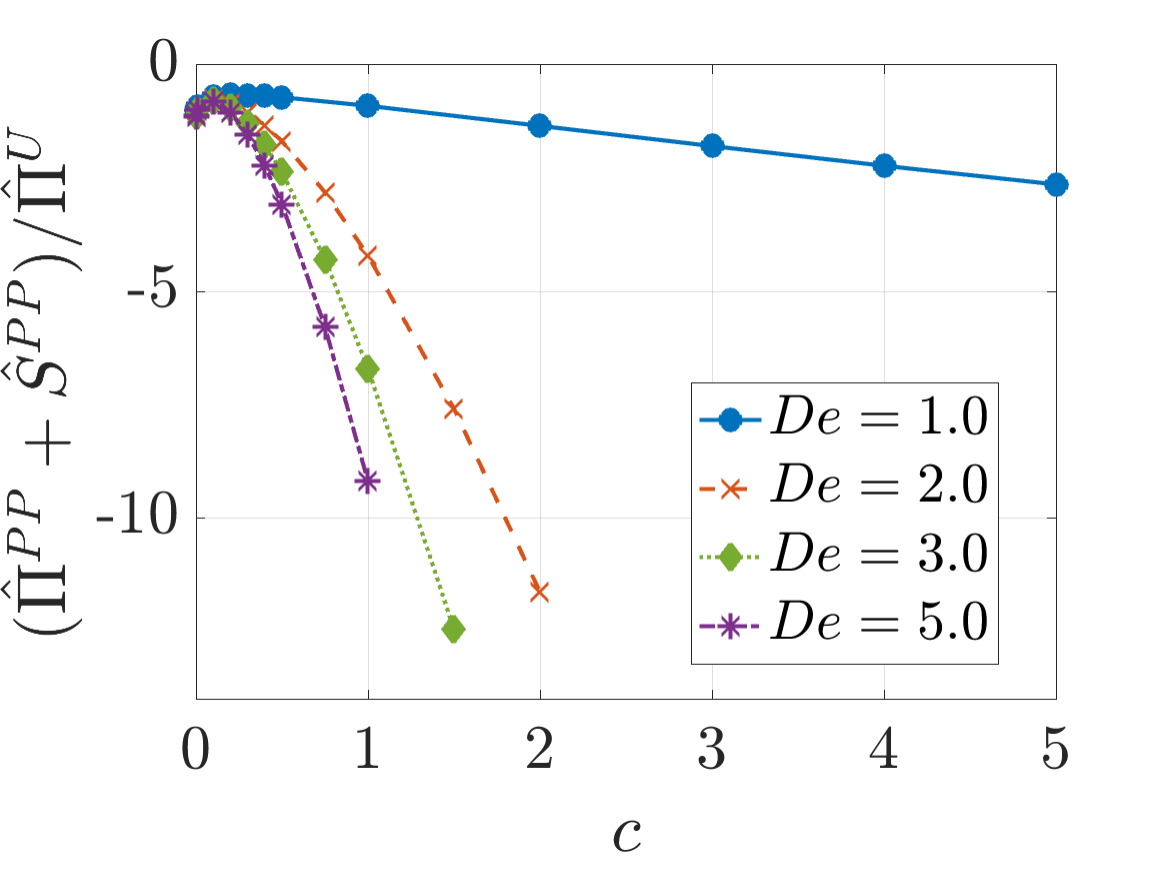}\label{fig:RhelogyvsDdiffDe_Steady}} 
	\caption{Effect of polymer concentration, $c$, on the steady state $(\hat{\Pi}^{PP} + \hat{\text{S}}^{PP}) / \hat{\Pi}^U$ in a Giesekus fluid with (a) different $\alpha$ at $De = 1.0$ and (b) with $\alpha = 0.01$ at different $De$. With $\alpha = 0.0004$ attempting to run $c = 0.15, 0.2$, and 0.25.}\label{fig:DeandAlfaGiesekusSteady}
\end{figure}

\begin{figure}
	\centering	
	\subfloat{\includegraphics[width=0.4\textwidth]{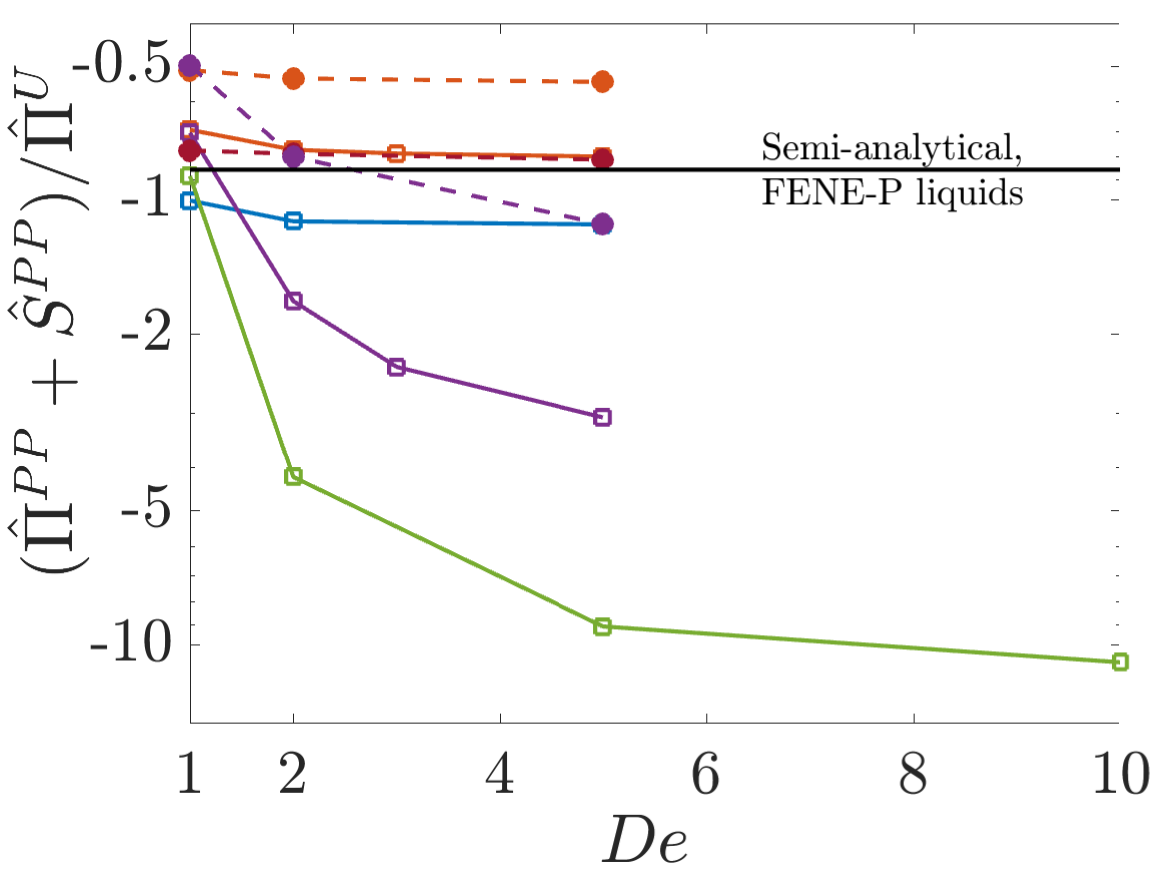}\label{fig:SteadyvsDe}}\hspace{0.2in}
	\subfloat{\includegraphics[width=0.4\textwidth]{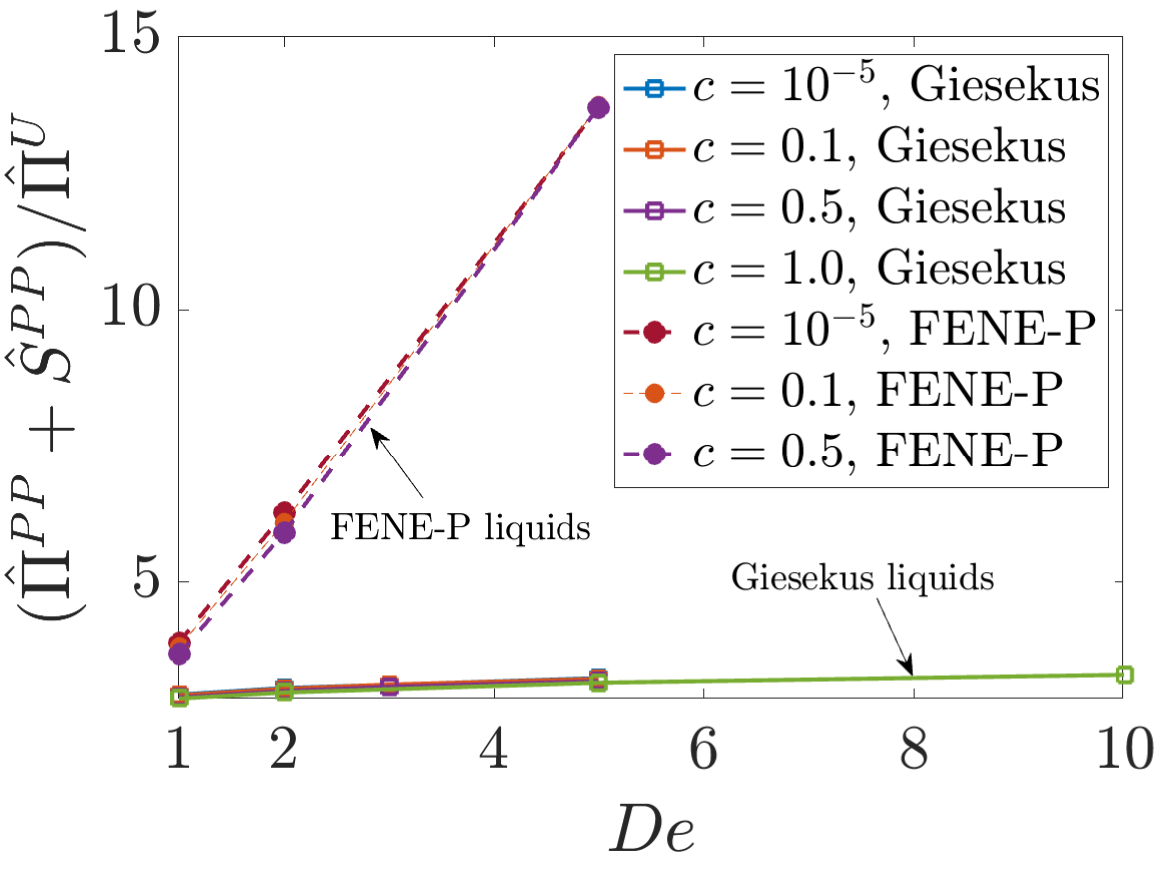}\label{fig:MaxvalsvsDe}}
	\caption{Effect of polymer concentration, $c$, on the variability of $(\hat{\Pi}^{PP} + \hat{\text{S}}^{PP}) / \hat{\Pi}^U$ with $De$ for Giesekus (solid lines) fluids with $\alpha = 0.1$ and FENE-P (dashed lines) with $L = 10$: (a) steady state, and (b) maximum. Both figures share the same legend, and the semi-analytical curve drawn corresponds to a value of -0.853 as estimated by \cite{SteadyStatePaper} for FENE-P liquids with small $c$.}
	\label{fig:VarywithDe}
\end{figure}

With this low $c$ rheology in mind, we first summarize the key findings at larger $c$ before discussing them in more detail using the observations from figures \ref{fig:VarywithcDep4} to \ref{fig:VarywithDe}. For $De > 0.5$, increasing $c$ significantly impacts the negative $(\hat{\Pi}^{PP} + \hat{\text{S}}^{PP}) / \hat{\Pi}^U$ found at larger Hencky strains, $H$. As $c$ surpasses a certain threshold dependent on $De$, and the Giesekus parameter $\alpha$, or the FENE-P parameter $L$, the interaction stress at larger $H$ becomes increasingly negative, decreasing at a rate of approximately $c^2$. This decrease is more pronounced for {Giesekus than for FENE-P liquids} and is amplified by higher $De$, larger $L$, and lower $\alpha$. These changes in $De$, $L$, and $\alpha$ also increase the undisturbed polymer stress (section \ref{sec:U0Rheology}). Consequently, at sufficient $H$, also the parameter regime where undisturbed polymer stress is larger, the interaction of a dilute concentration of spheres with the polymers counteracts the linear increase in undisturbed polymer stress with $c$. In other words, adding a small concentration of spheres leads to an even more effective reduction in suspension stress at larger $c$, $De$, $L$, $1 / \alpha$, and $H$, where the undisturbed fluid stress is expected to be higher.

Figure \ref{fig:Dep4L50diffctimeplot} shows the evolution of the normalized interaction stress for a Giesekus liquid with $\alpha = 0.01$ at $De = 0.4$ for different polymer concentrations, $c$. For each $c$, the evolution of $(\hat{\Pi}^{PP} + \hat{\text{S}}^{PP}) / \hat{\Pi}^U$ is qualitatively similar to the small $c$ result, showing an increase towards a positive steady state. However, the magnitude of the steady-state value decreases with increasing $c$. This effect is more clearly illustrated in figure \ref{fig:PlotvscDept4L10}, which shows the steady-state value of $(\hat{\Pi}^{PP} + \hat{\text{S}}^{PP}) / \hat{\Pi}^U$ for a Giesekus fluid with $\alpha = 0.0004$ (orange dashed curve). An identical effect of $c$ is also observed for a FENE-P liquid with $L = 50$ (solid black curve). From this and simulations (not shown) at other values of $De$, $L$, and $\alpha$ at different $c$, we find that, for $De < 0.5$, the interaction stress of a FENE-P liquid with a given $L$ is similar to that of a Giesekus liquid with $\alpha = L^{-2}$ when $L \ge 50$. This observation aligns with the undisturbed polymer stress behavior across these models (section \ref{sec:U0Rheology}, figure \ref{fig:U0Dep4GieskFENEP}).

For $De > 0.5$, the polymer concentration $c$ has a more profound influence on $(\hat{\Pi}^{PP} + \hat{\text{S}}^{PP}) / \hat{\Pi}^U$, particularly affecting the negative interaction stress values at large $H$. This is illustrated for $De = 2.0$ in figure \ref{fig:VarywithcDe2L50} for Giesekus and FENE-P liquids with $\alpha = 0.0004$ and $L = 50$, and for $De = 5.0$ in figure \ref{fig:VarywithcDe5L10} with $\alpha = 0.01$ and $L = 10$. The qualitative behavior of $(\hat{\Pi}^{PP} + \hat{\text{S}}^{PP}) / \hat{\Pi}^U$ evolution with $H$ is similar across different $c$ values: an initial increase to large positive values followed by a decrease to large negative values. Similar to the steady-state interaction stress for $De < 0.5$, the maximum value for $De > 0.5$ slightly reduces with increasing $c$.

By comparing the sub-figures in figures \ref{fig:VarywithcDe2L50} and \ref{fig:VarywithcDe5L10}, it is evident that this effect on maximum interaction stress is more pronounced for FENE-P than for Giesekus liquids. A clearer indication of this decrease with $c$ is shown for Giesekus liquids at $De = 1.0$ with different $\alpha$ values in figure \ref{fig:RhRheologyalfavaryDe1Max}, and for $\alpha = 0.01$ at different $De$ values in figure \ref{fig:RhelogyvsDdiffDe_max}.

Comparing figures \ref{fig:De2alfap0004diffctimeplot} and \ref{fig:TotalInteractionStressL50De2variousc2} reveals a larger magnitude of the negative $(\hat{\Pi}^{PP} + \hat{\text{S}}^{PP}) / \hat{\Pi}^U$ at larger $H$ for a Giesekus liquid at $De = 2.0$ and $\alpha = 0.0004$ than for a FENE-P liquid at the same $De$ and equivalent $L = 50$. Similarly, figures \ref{fig:De5alfap01diffctimeplot} and \ref{fig:De5L10diffctimeplot} at $De = 5.0$ show that increasing $c$ makes $(\hat{\Pi}^{PP} + \hat{\text{S}}^{PP}) / \hat{\Pi}^U$ at larger $H$ more negative for a Giesekus liquid with $\alpha = 0.01$ than for a FENE-P liquid with $L = 10$. A direct comparison of the effect of $c$ on steady-state $(\hat{\Pi}^{PP} + \hat{\text{S}}^{PP}) / \hat{\Pi}^U$ between the two polymer constitutive models at different $De$ and $L$ values is shown in figure \ref{fig:VaryDeGiesekusandFENE} (comparing solid with dashed lines). Therefore, the steady state $(\hat{\Pi}^{PP} + \hat{\text{S}}^{PP}) / \hat{\Pi}^U$ is more sensitive to $c$ for {Giesekus than FENE-P liquids}. In the former, the steady state is approached much slower, consistent with the slower evolution of undisturbed polymer stress in Giesekus fluids shown in figures \ref{fig:U0De2GieskFENEP} and \ref{fig:U0De5GieskFENEP} at $De = 2.0$ and 5.0 with $L = \alpha^{-0.5}$. However, the steady-state $(\hat{\Pi}^{PP} + \hat{\text{S}}^{PP}) / \hat{\Pi}^U$ is more negative for the Giesekus model, even though the steady-state undisturbed stress of the two models are identical.  As shown by the curves in figures \ref{fig:VaryDeGiesekusandFENE} and \ref{fig:DeandAlfaGiesekusSteady}, the steady-state normalized interaction stress first reduces in magnitude with $c$ but then becomes increasingly negative with $c$. This latter decrease with $c$ is almost linear for both FENE-P and Giesekus models.

Comparing the two models for polymer-induced drag reduction in wall-bounded flows, \cite{housiadas2013skin} showed that similar drag reduction is achieved for the Giesekus and FENE-P models when $L = \alpha^{-0.5}$. In contrast, under these conditions, we observe a significant difference in the steady-state interaction stress (and hence extensional viscosity) between the two models at large $De$. Even when the steady state is reached in the suspension, the individual polymers will continue to stretch and relax transiently as they traverse the steady-state velocity streamlines around the sphere and experience different local flows along their trajectories.  Whereas the steady-state stresses in the two models differ significantly at a large Deborah numbers such as $De=2$ and $5$ (figures \ref{fig:U0De2GieskFENEP} and \ref{fig:U0De5GieskFENEP}), they are similar to one another at $De=0.4$ (compare figure \ref{fig:PlotvscDept4L10} to figure \ref{fig:U0Dep4GieskFENEP}). This suggests that the differences  in the results of the Giesekus and FENE-P models at large $De$ are likely due to the differing coil-stretch transition behaviors of the individual polymers in these constitutive models (figures \ref{fig:U0De2GieskFENEP} and \ref{fig:U0De5GieskFENEP}).

The rate of decrease in steady-state $(\hat{\Pi}^{PP} + \hat{\text{S}}^{PP}) / \hat{\Pi}^U$ at large $c$ increases with $1 / \alpha$ for Giesekus liquids and $L$ for FENE-P liquids. The influence of $\alpha$ can be observed by comparing different $\alpha$ curves at $De = 1.0$ in figure \ref{fig:RheologyalfavaryDe1Steady} or by comparing the dashed $De = 1.0$ or 2.0 curves between figures \ref{fig:RheologyFENEPandGiesekusL50100Steady} and \ref{fig:RheologyFENEPSteadL10y}. Similarly, the effect of $L$ is observed by comparing different $L$ curves for $De = 1.0$ or 2.0 represented by solid lines in figures \ref{fig:RheologyFENEPandGiesekusL50100Steady} and \ref{fig:RheologyFENEPSteadL10y}. The Deborah number, $De$, also increases the rate of decrease in steady-state $(\hat{\Pi}^{PP} + \hat{\text{S}}^{PP}) / \hat{\Pi}^U$ at large $c$, as found by comparing different solid lines in figures \ref{fig:RheologyFENEPandGiesekusL50100Steady} for $L = 50$ and \ref{fig:RheologyFENEPSteadL10y} for $L = 10$, different dashed lines in figures \ref{fig:RheologyFENEPandGiesekusL50100Steady} for $\alpha = 0.0004$ and \ref{fig:RheologyFENEPSteadL10y} for $\alpha = 0.01$, or the different curves in figure \ref{fig:RhelogyvsDdiffDe_Steady} for a Giesekus fluid with $\alpha = 0.01$.


In our previous work \citep{SteadyStatePaper}, we found that at large $De$, the steady-state normalized interaction stress for low $c$ FENE-P liquids is $-0.853$, i.e., independent of $De$. (We have used slightly different normalizing factors across the two studies to define the interaction stress contribution to the suspension stress as found by comparing equation \eqref{eq:ExtVisc} with equation 93 of the previous work.) Figure \ref{fig:SteadyvsDe} shows that simulations of FENE-P liquid (dashed red line with filled circles) with $c = 10^{-5}$ asymptote to this value at large $De$. At a given $c$, the value that the steady-state normalized interaction stress approaches at large $De$ is different in Giesekus liquids. However, the stress is more sensitive to $De$ at a given $c$ in Giesekus liquids. At larger $c$, the steady-state $(\hat{\Pi}^{PP} + \hat{\text{S}}^{PP}) / \hat{\Pi}^U$ becomes more sensitive to $De$, and the value of $De$ at which it plateaus is proportional to $c$. As observed in figure \ref{fig:MaxvalsvsDe}, the maximum value of normalized interaction stress during its evolution increases with $De$. Here, we can observe that this maximum value is more sensitive for FENE-P than Giesekus liquid.

\begin{figure}
	\centering	
	\subfloat{\includegraphics[width=0.4\textwidth]{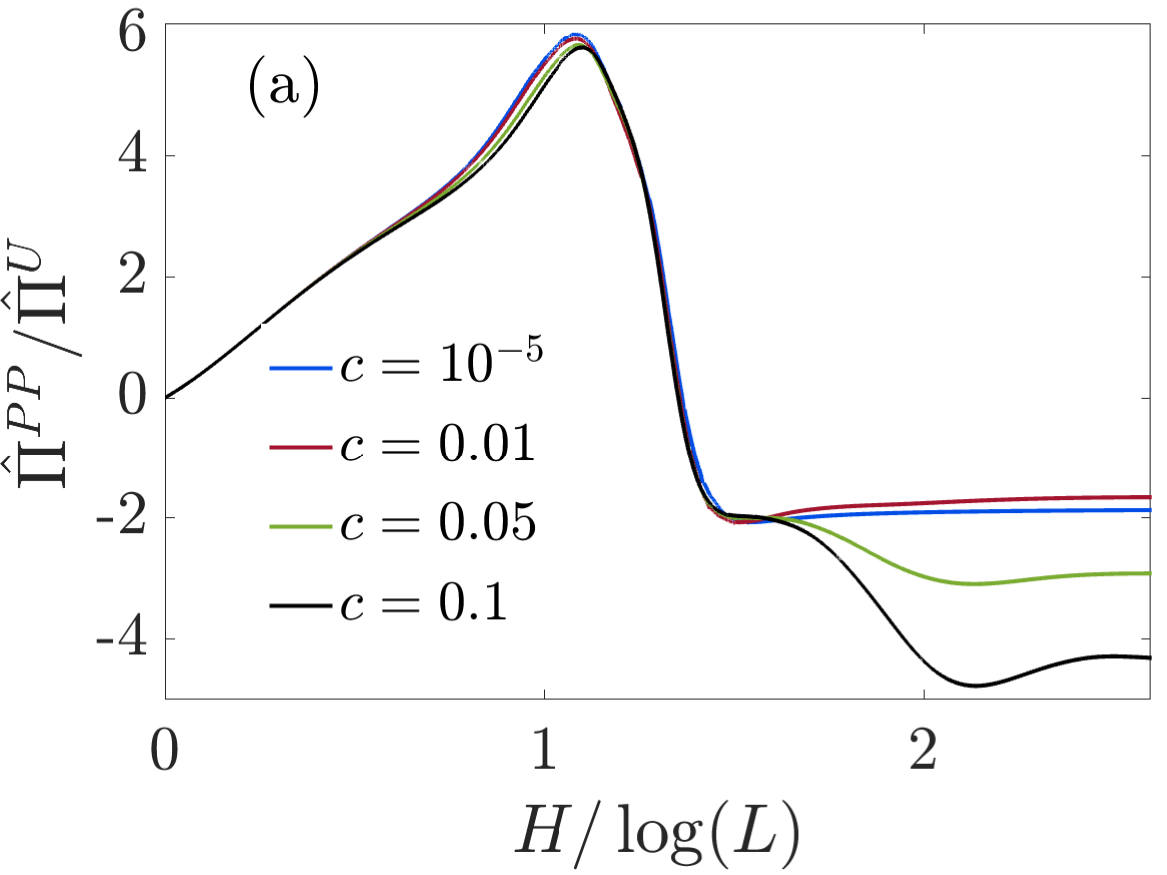}\label{fig:TotalPIPSL50De2variousc_}} \hspace{0.2in}
	\subfloat{\includegraphics[width=0.4\textwidth]{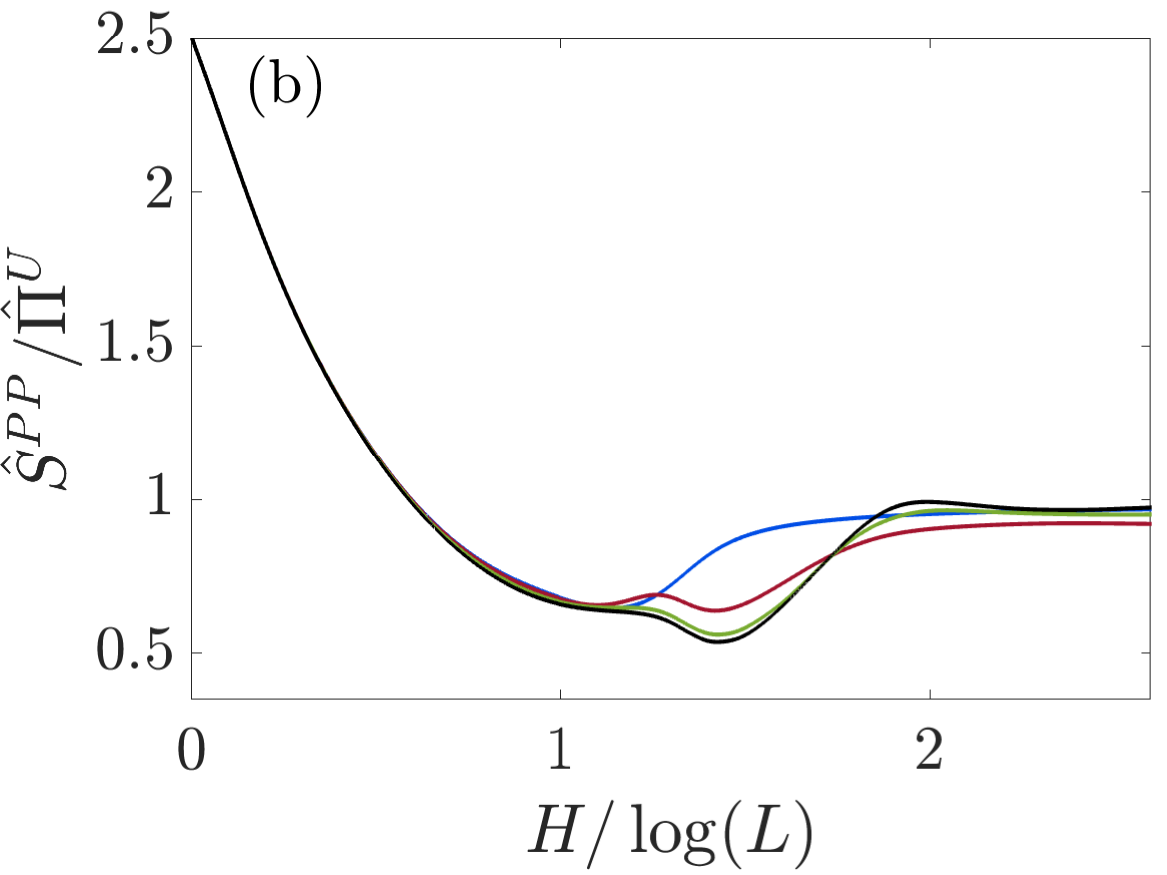}\label{fig:TotalInteractionStressletL50De2variousc_}}
	\caption{Effect of polymer concentration, $c$, on (a) $(\hat{\Pi}^{PP} + \hat{\text{S}}^{PP}) / \hat{\Pi}^U$, (b) $\hat{\Pi}^{PP} / \hat{\Pi}^U$, and (c) $\hat{\text{S}}^{PP} / \hat{\Pi}^U$ at $De = 2.0$ and $L = 50$. Both figures share the same legend.}
	\label{fig:L50De2_}
\end{figure}

Figure \ref{fig:L50De2_} shows the decomposition of $(\hat{\Pi}^{PP} + \hat{\text{S}}^{PP}) / \hat{\Pi}^U$ for different $c$ for a FENE-P liquid with $De = 2.0$ and $L = 50$ into its constituents. The total interaction stress $(\hat{\Pi}^{PP} + \hat{\text{S}}^{PP}) / \hat{\Pi}^U$ for this case was shown earlier in figure \ref{fig:TotalInteractionStressL50De2variousc2}. We can observe that while the stresslet is quantitatively important, the qualitative changes reported above are primarily driven by PIPS, $\hat{\Pi}^{PP}$. {The details of the qualitative contributions of the stresslet and PIPS to $(\hat{\Pi}^{PP} + \hat{\text{S}}^{PP}) / \hat{\Pi}^U$ at large $c$ are the same as those at small $c$ discussed in section \ref{sec:RheologySplit}.} In sections \ref{sec:Mech1} and \ref{sec:Mech2}, we will explore the mechanisms driving the rheological changes with $c$ illustrated in this section.

\subsection{{Mechanisms reducing the particle-polymer interaction stress magnitude at higher polymer concentrations}}\label{sec:Mech1}
{As shown in figure \ref{fig:VarywithcDep4} for $De=0.4$, the interaction stress for $De<0.5$ reduces at all times upon increasing $c$. For $De>0.5$, the peak interaction stress decreases with $c$, as shown in figure \ref{fig:DeandAlfaGiesekusMax}. While the aforementioned changes occur for all $c$, the magnitude of the steady-state interaction stress at $De>0.5$ decreases with $c$ but only up to a moderate $c$ valud that depends on $De$ and $L$ or $\alpha$, as shown in figures \ref{fig:VaryDeGiesekusandFENE} and \ref{fig:DeandAlfaGiesekusSteady}. In this section, we investigate the mechanism responsible for this reduction in the effectiveness of the particles to alter the suspension viscosity with $c$.}

                   

\begin{figure}
		\centering	
		\subfloat{\includegraphics[width=0.33\textwidth]{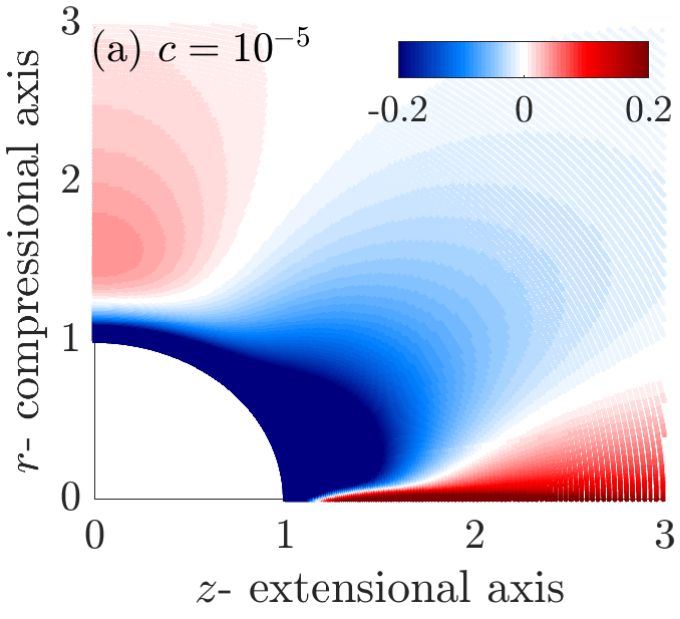}\label{fig:Giesc1em5De5L10}}
		\subfloat{\includegraphics[width=0.33\textwidth]{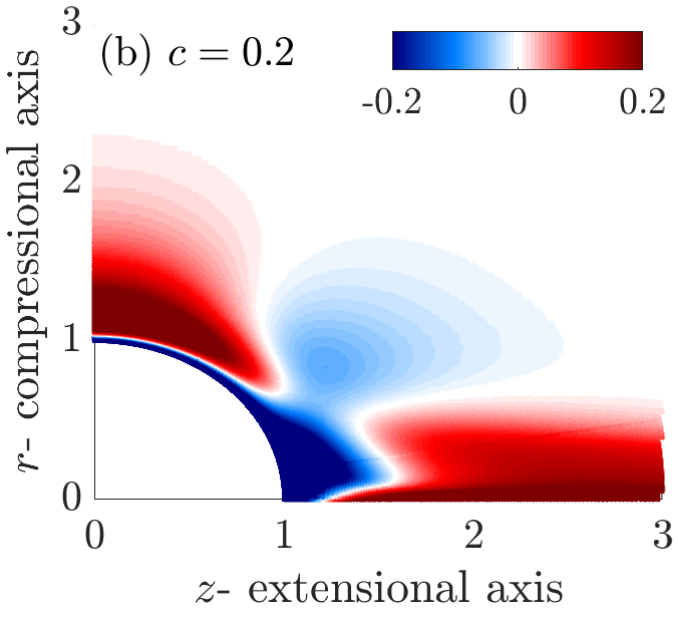}\label{fig:Giescp2De5L10}}
		\subfloat{\includegraphics[width=0.33\textwidth]{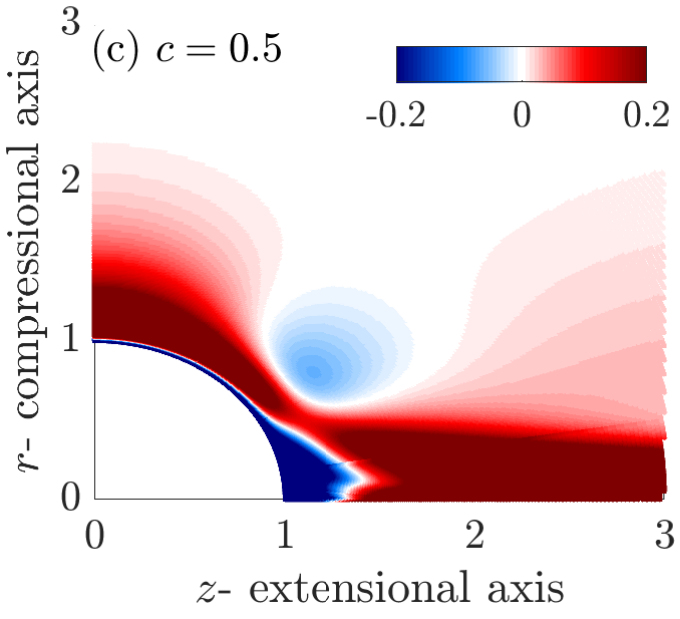}\label{fig:Giescp5De5L10}}
		\caption{Steady-state $\Delta \mathcal{S}$ at $De = 5.0$ for a Giesekus liquid with $\alpha = 0.01$ at different $c$ values labeled on the figure.}
		\label{fig:DeltaSDe5L10alfap01Zoom}
	\end{figure}

{Figure \ref{fig:DeltaSDe5L10alfap01Zoom} illustrates the steady-state extra polymer stretch $\Delta \mathcal{S}$ around a sphere for $De = 5.0$ for a Giesekus liquid with different $c$ values. Increasing $c$ leads to a reduction in the size of the polymer collapse region and an enhancement of the extra polymer stretch in the surrounding areas. These changes in $\Delta \mathcal{S}$ contribute to the observed decrease in the magnitude of the negative particle-polymer stress in figure \ref{fig:VaryDeGiesekusandFENE} as $c$ increases from very small values (insets show the numerically calculated values approaching an asymptote expected in the small $c$ limit). The $\Delta \mathcal{S}$ plots in figure \ref{fig:DeltaSDe5L10alfap01Zoom} correspond to the green dashed line in figure \ref{fig:RheologyFENEPSteadL10y}. The correlation between $\Delta \mathcal{S}$ and $(\hat{\Pi}^{PP} + \hat{\text{S}}^{PP}) / \hat{\Pi}^U$ remains consistent across other values (not shown) of $De$ and $\alpha$ considered in figures \ref{fig:VaryDeGiesekusandFENE} and \ref{fig:DeandAlfaGiesekusSteady}. This correlation is also similar for FENE-P liquids (not shown).}
	
\begin{figure}
		\centering	
		\subfloat{\includegraphics[width=0.33\textwidth]{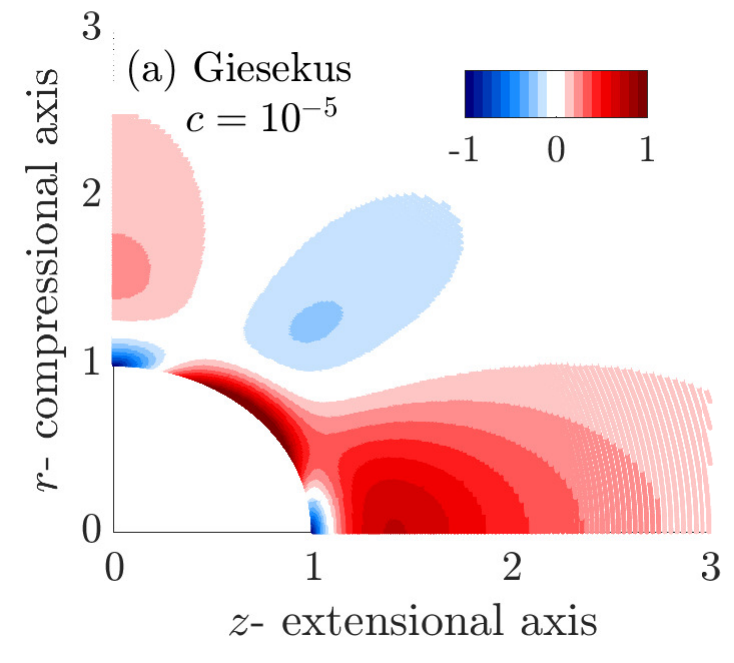}}
		\subfloat{\includegraphics[width=0.33\textwidth]{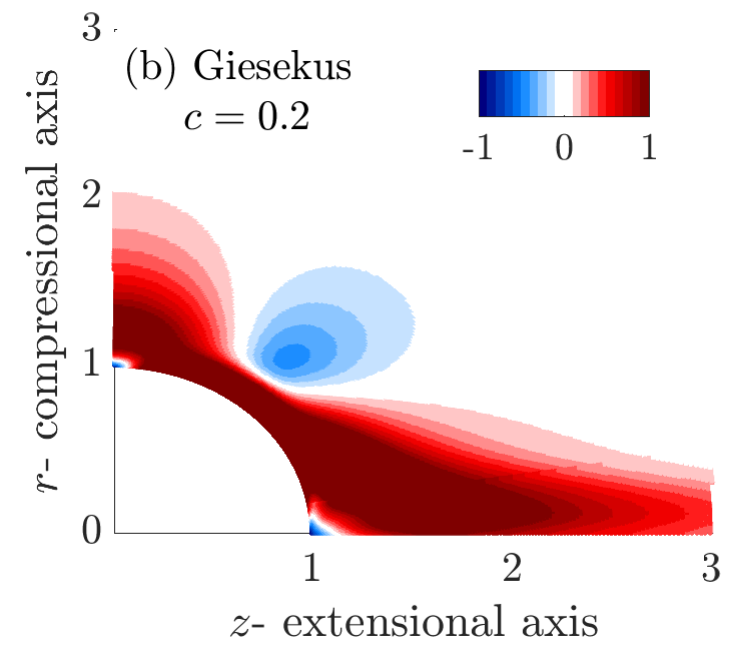}}
		\subfloat{\includegraphics[width=0.33\textwidth]{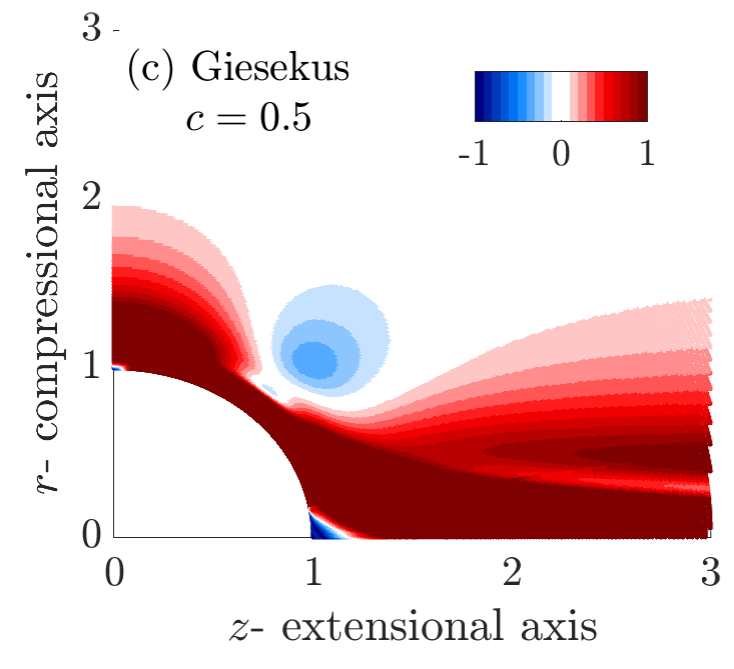}}
		\caption{Steady-state $\Delta De_\text{local}$ at $De = 5.0$ for Giesekus liquid with $\alpha = 0.01$ and different $c$ values.}
		\label{fig:DeltaDeDe5L10alfap01Zoom}
	\end{figure}

{The reduction in the magnitude of negative particle-polymer interaction stress at high $De$ and $H$ with increasing $c$ is due to the diminished intensity of the underlying polymer collapse region, influenced by the enhanced interaction between fluid velocity and polymer stress at larger $c$. Figure \ref{fig:DeltaDeDe5L10alfap01Zoom} illustrates the steady-state $\Delta De_\text{local}$ for three different values of $c$, corresponding to the steady-state $\Delta \mathcal{S}$ for Giesekus liquids shown in figure \ref{fig:DeltaSDe5L10alfap01Zoom}. }

Polymer collapse in the small $c$ regime occurs when polymers enter the low-speed, negative $\Delta De_\text{local}$ (blue) region around the stagnation point along the compressional axis, as discussed in section \ref{sec:PolymerStretch}. In this regime, the negative $\Delta De_\text{local}$ persists due to limited alteration of the flow field by the polymers. However, at larger $c$, as the polymers collapse before the steady state (not shown) in this region, the local fluid acts as one with a lower viscosity, allowing for a more rapid flow with enhanced local stretching capabilities. Consequently, in this region, increasing $c$ results in a rise in steady-state $\Delta De_\text{local}$ to positive values for sufficiently high $c$, as shown in the two right panels of figures \ref{fig:DeltaDeDe5L10alfap01Zoom}. This shift effectively reduces the extent of the polymer collapse region. Moreover, the changes in negative $\Delta De_\text{local}$ exhibit greater sensitivity to $c$ at larger $De$, $L$, and $1 / \alpha$ (not shown), leading to a more pronounced dependence of steady-state interaction stress on $c$, as illustrated by the figures discussed in section \ref{sec:ResultsFinitec}.
	
The positive steady-state particle-polymer interaction stress observed at small $De$, as well as the increase to large positive transient values before the undisturbed coil-stretch transition at high $De$, is attributed to the wake of extra polymer stretch along the extensional axis, as discussed in section \ref{sec:PolymerStretch}. When $c$ is increased, the intensity of this wake diminishes (not shown), paralleling the reduction in the intensity of the polymer collapse region in the steady state for high $De$ cases discussed above. This highly stretched wake results from the region of $\Delta De_\text{local} > 0$. Once the polymers start stretching, the local fluid behaves as if it has a higher viscosity than the surrounding fluid, inhibiting flow and reducing the local stretching capability. Thus, as $c$ increases, the steady-state $\Delta De_\text{local} > 0$ becomes less positive (not shown). This reduction limits the extent to which the polymers can be locally stretched, thereby explaining why increasing $c$ leads to a decrease in interaction stress in these scenarios, as illustrated in figures \ref{fig:VarywithcDep4} and \ref{fig:DeandAlfaGiesekusMax}.

\subsection{Mechanism for increasingly negative interaction stress with $c$ beyond moderate $c$}\label{sec:Mech2}
As described in section \ref{sec:ResultsFinitec} and illustrated in figures \ref{fig:VaryDeGiesekusandFENE} and \ref{fig:DeandAlfaGiesekusSteady}, increasing $c$ beyond a moderate value results in a greater magnitude of the negative particle-polymer interaction stress at large times for $De > 0.5$. This section investigates the physical mechanisms responsible for these changes as $c$ increases.

\begin{figure}
		\centering	
		\subfloat{\includegraphics[width=0.33\textwidth]{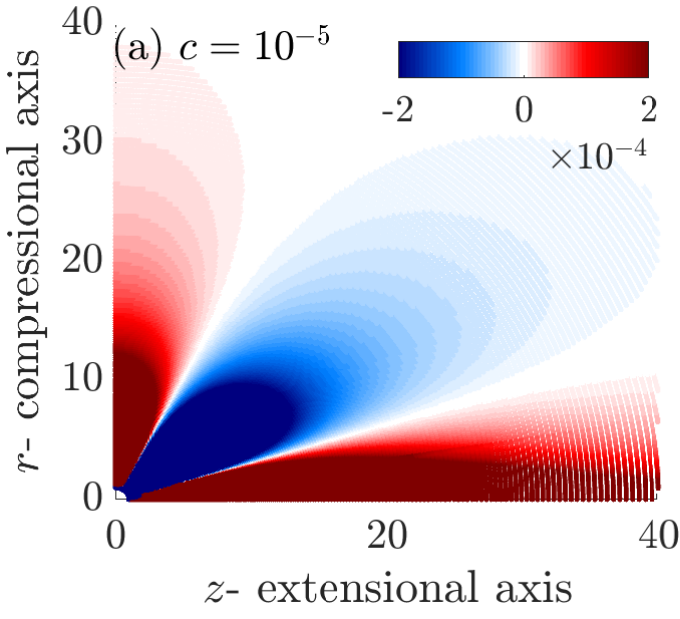}\label{fig:DeltaSDe5ap01c1em5}}
		\subfloat{\includegraphics[width=0.33\textwidth]{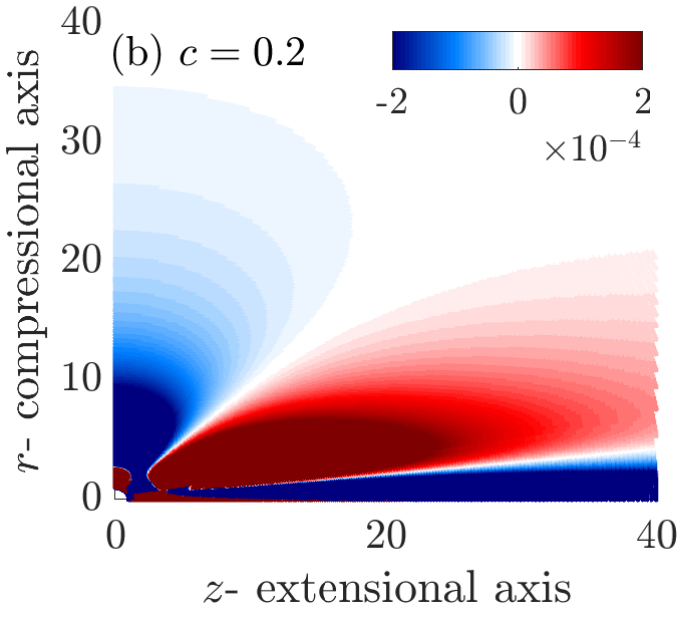}\label{fig:DDeltaSDe5ap01cp2}}
		\subfloat{\includegraphics[width=0.33\textwidth]{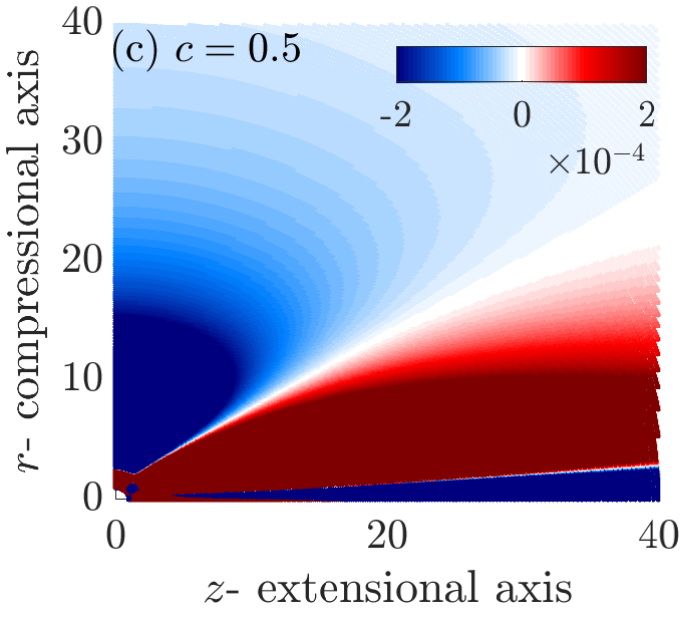}\label{fig:DeltaSDe5ap01cp5}}
		\caption{Steady-state $\Delta \mathcal{S}$ at $De = 5.0$ for Giesekus liquid with $\alpha = 0.01$ and different $c$ values in a larger region around the particle than figure \ref{fig:DeltaSDe5L10alfap01Zoom}.}
		\label{fig:DeltaSDe5L10alfap01}
	\end{figure}

The changes in $\Delta \mathcal{S}$ and $\Delta De_\text{local}$ near the particle surface, as described in section \ref{sec:Mech1}, persist as $c$ increases. If these were the dominant mechanisms, we would expect the magnitude of the steady-state negative interaction stress at high $De$ to continue decreasing with $c$, eventually becoming positive as the region of polymer collapse at $c = 10^{-5}$ transitions to one with extra polymer stretch at $c = 0.5$, as illustrated in figure \ref{fig:DeltaSDe5L10alfap01Zoom}. However, as shown in the figures in section \ref{sec:ResultsFinitec}, beyond a certain moderate $c$ (which depends on $De$ and $\alpha$ or $L$), further increases in $c$ lead to a progressively more negative value of $(\hat{\Pi}^{PP} + \hat{\text{S}}^{PP}) / \hat{\Pi}^U$. The underlying cause of these changes occurs around the compressional axis, further upstream of the particle than previously considered. These changes are depicted for suspensions in Giesekus liquids with $De = 5.0$ and $\alpha = 0.01$ at three different $c$ values in figure \ref{fig:DeltaSDe5L10alfap01}. These values bracket the transition (at $c=0.1$) along the dashed green line in figure \ref{fig:RheologyFENEPSteadL10y} from a reduction in magnitude of the interaction stress to an increase with $c$.

At small $c = 10^{-5}$, the region around the compressional (radial) axis is characterized by a radially decaying region of extra polymer stretch as shown in figure \ref{fig:DeltaSDe5ap01c1em5} for a Giesekus liquid with $De=5.0$ and $\alpha=0.01$. As $c$ increases, the portion of this region, more than 3-5 particle radii upstream along the compressional axis transforms into one of polymer collapse (figures \ref{fig:DDeltaSDe5ap01cp2} and \ref{fig:DeltaSDe5ap01cp5}). The intensity of this collapse increases with $c$ (compare figure \ref{fig:DeltaSDe5ap01cp5} with \ref{fig:DDeltaSDe5ap01cp2}). Although the change in polymer stretch due to this collapse is smaller than the increase in local stretch closer to the particle (figure \ref{fig:DeltaSDe5L10alfap01Zoom}), it affects a larger volume, outweighing the local effects.

Similar trends in $\Delta \mathcal{S}$ with $c$ at other $De$ and $L$ or $\alpha$ values (not shown) explain the change in slope of $(\hat{\Pi}^{PP} + \hat{\text{S}}^{PP}) / \hat{\Pi}^U$ vs. $c$ at moderate $c$ values in the various curves presented in figures \ref{fig:VaryDeGiesekusandFENE} and \ref{fig:DeandAlfaGiesekusSteady}. Compared to Giesekus liquids, FENE-P liquids are more resistant to changes in polymer stretch due to the presence of the particle (not shown) at all polymer concentrations. This leads to a larger slope of $(\hat{\Pi}^{PP} + \hat{\text{S}}^{PP}) / \hat{\Pi}^U$ vs. $c$ in Giesekus than in FENE-P liquids at an equivalent $\alpha = L^{-2}$ in figure \ref{fig:VaryDeGiesekusandFENE}.

\begin{figure}
	\centering	
	\subfloat{\includegraphics[width=0.33\textwidth]{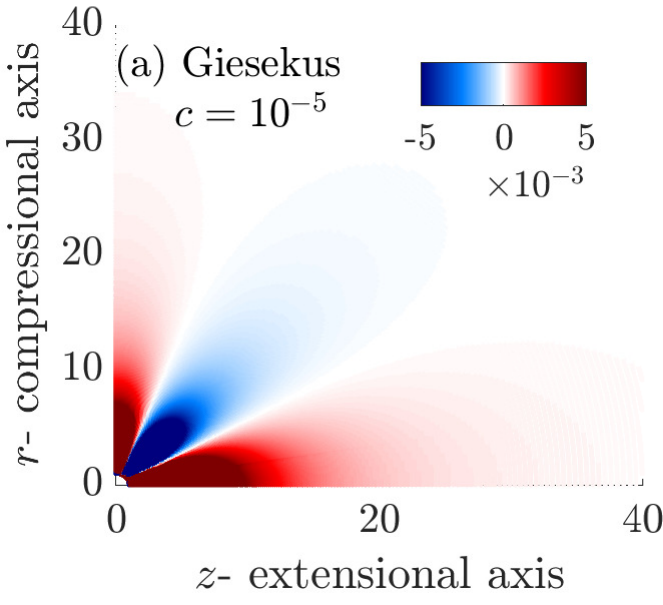}\label{fig:DeltaDeDe5ap01c1em5}}
	\subfloat{\includegraphics[width=0.33\textwidth]{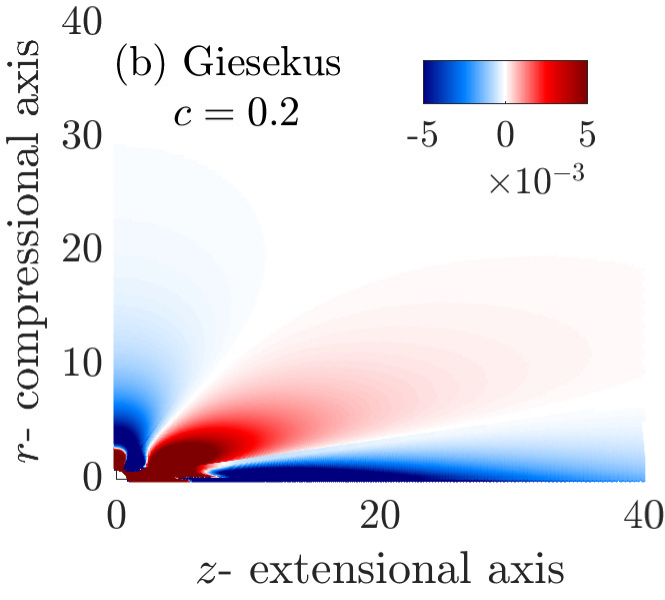}}
	\subfloat{\includegraphics[width=0.33\textwidth]{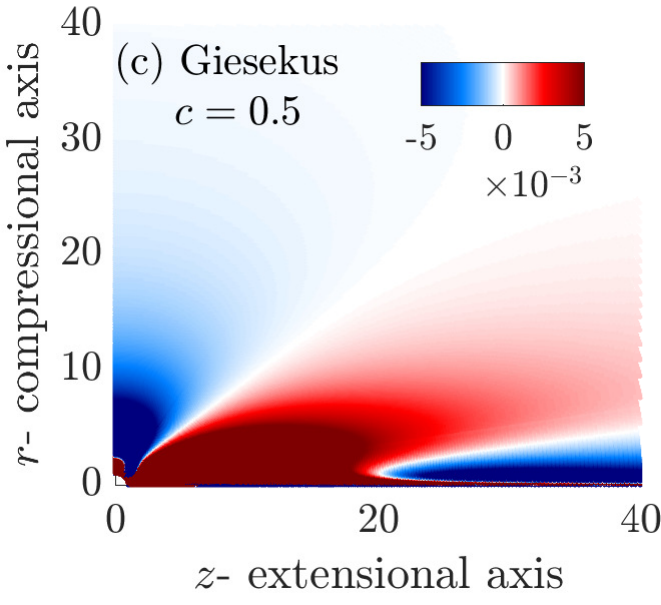}}
	\caption{Steady-state $\Delta De_\text{local}$ at $De = 5.0$ for Giesekus liquids with $\alpha = 0.01$ and different $c$ values in a larger region around the particle than figure \ref{fig:DeltaDeDe5L10alfap01Zoom}.}
	\label{fig:DeltaDeDe5L10alfap01}
\end{figure}

Figure \ref{fig:DeltaDeDe5L10alfap01} shows $\Delta De_\text{local}$ for the same parameters as the $\Delta \mathcal{S}$ plots from figures \ref{fig:DeltaSDe5ap01c1em5} to \ref{fig:DeltaSDe5ap01cp5}. Comparing plots across and within these figures reveals that the switch in the sign of the extra polymer stretch, $\Delta \mathcal{S}$, correlates with the switch in the sign of the extra velocity stretching, $\Delta De_\text{local}$, around the compressional axis. This correlation is consistent for FENE-P liquids and across other values of $De$ and $\alpha$ or $L$ discussed in section \ref{sec:ResultsFinitec}.

\begin{figure}
	\centering
	\subfloat{\includegraphics[width=0.33\textwidth]{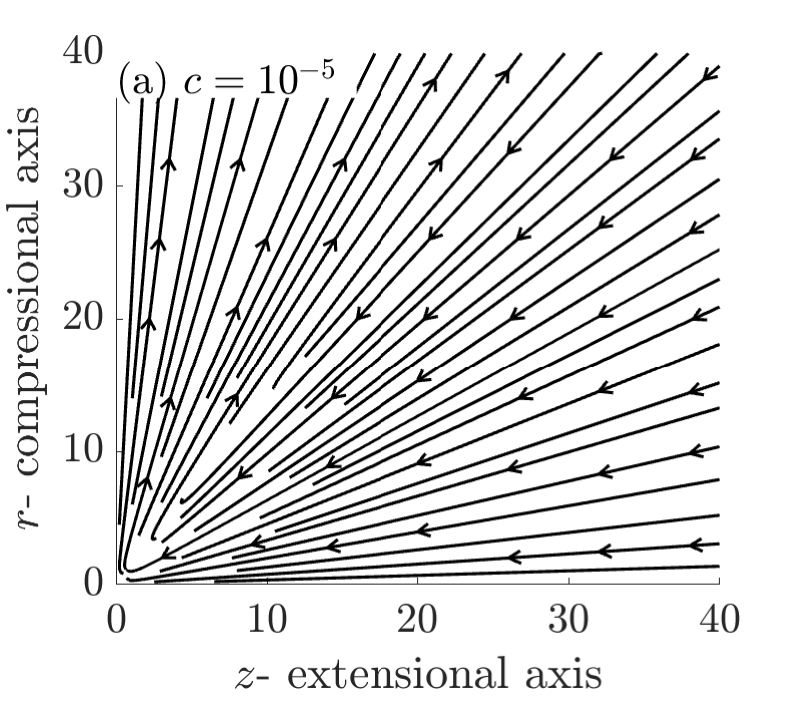}\label{fig:DistStreamlinesc1em5}}
	\subfloat{\includegraphics[width=0.33\textwidth]{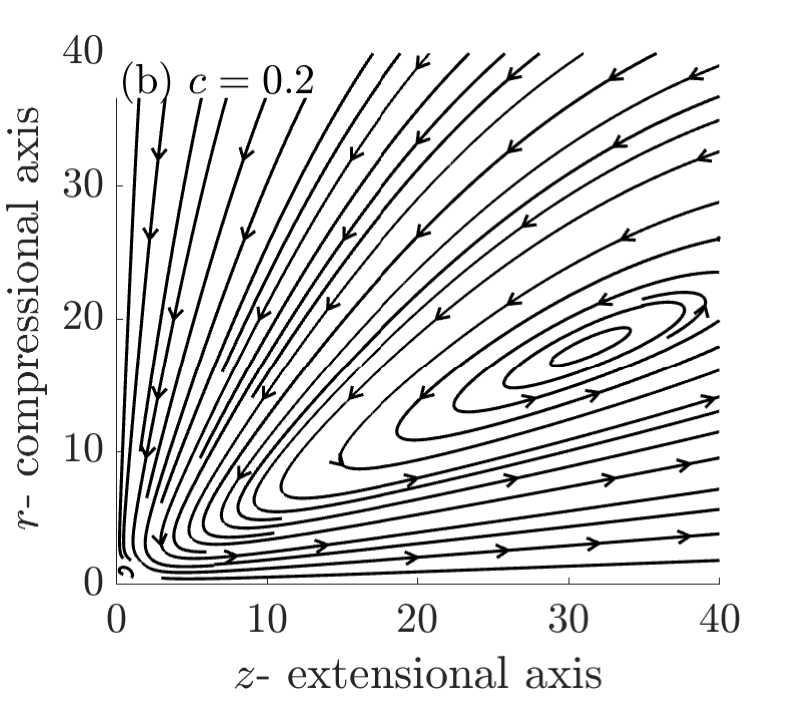}\label{fig:DistStreamlinescp2}}
	\subfloat{\includegraphics[width=0.33\textwidth]{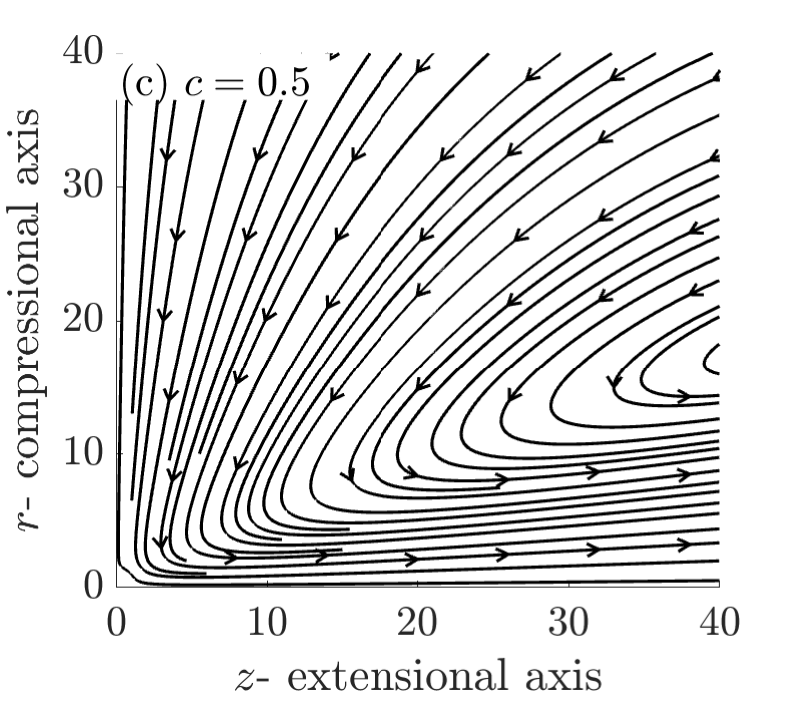}\label{fig:DistStreamlinescp5}}
	\caption{{Streamlines of the disturbance velocity field for $De = 5.0$ for Giesekus liquid with $\alpha = 0.01$ and different $c$.}}
	\label{fig:DisturbanceSteamlines}
\end{figure}

As discussed in section \ref{sec:KDM} and shown in figure \ref{fig:MultipoleDisturbances}, the extra stretching of the polymers at low $c$ in the region 3-5 particle radii upsteream along the compressional axis results from the separation of streamlines created by the dipole disturbance of the Newtonian velocity field. The streamlines of the net flow disturbance, in the steady state, created by the particle in a Giesekus fluid with $\alpha=0.01$ and different values of $c$ are shown in figure \ref{fig:DisturbanceSteamlines}. The leftmost panel, corresponding to $c=10^{-5}$, depicts primarily the Newtonian disturbance field and shows the separation of streamlines near the compressional axis. As the value of $c$ increases to 0.2 and 0.5, the disturbance velocity field shown in the middle and right panels indicates a change in the direction of disturbance velocity. This change in streamlines aligns with the transition in the region near the compressional axis from $\Delta De_\text{local}>0$ at $c=10^{-5}$ to $\Delta De_\text{local}<0$ at $c=0.2$ and 0.5, as discussed above using figure \ref{fig:DeltaDeDe5L10alfap01}. 

\begin{figure}
	\centering
	\includegraphics[width=0.5\linewidth]{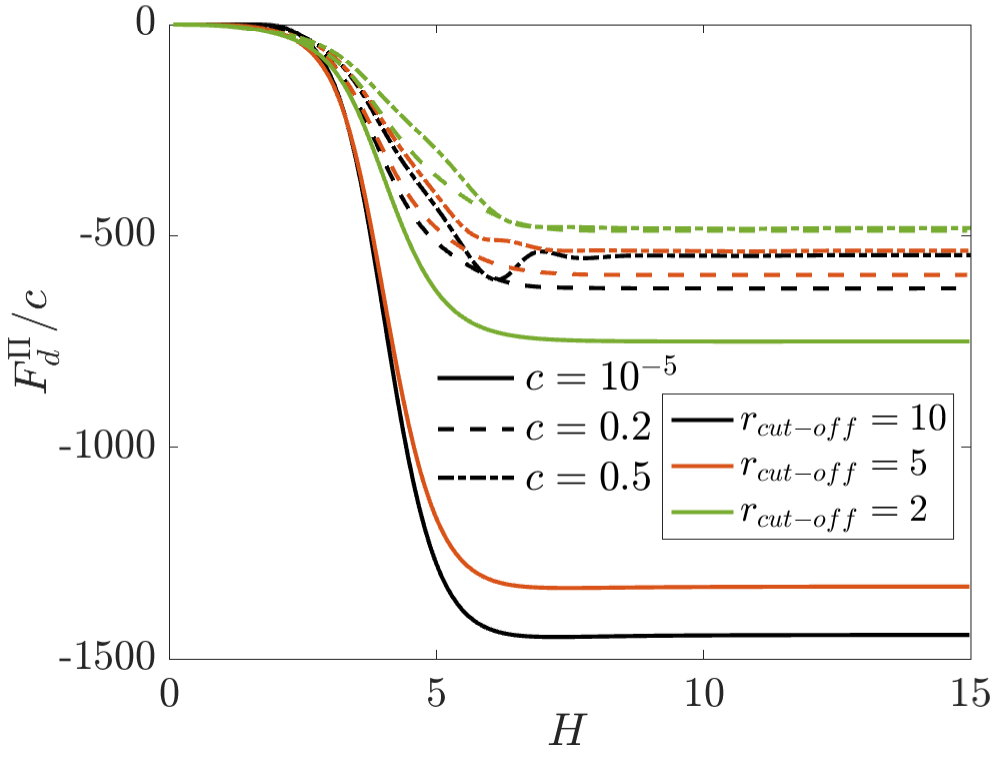}
	\caption{{Polymer-induced force dipole, $F_d^\Pi/c$, for a Giesekus liquid with $\alpha = 0.01$ and $De = 5.0$. Different colors represent various integrating radii cut-offs, $r_\text{cut-off}$, and different line styles represent various $c$.}}
	\label{fig:polymerdipoleforce}
\end{figure}

{The change in the disturbance streamlines further from the particle can be explained by the polymer force field induced near the particle surface. The reversal of streamline direction near the compressional axis arises from the reversal of the dipole disturbance relative to that in the Newtonian velocity field due to the polymer stress. In a linear flow, such as the uniaxial extensional flow considered here, the leading-order far-field disturbance in an unbounded fluid around a particle arises from a force dipole placed at the particle center. Based on a multipole expansion, far from the particle, the dipole disturbance is given by
	\begin{equation}
		\mathbf{u}_\text{dipole}=\boldsymbol{D}:\nabla \boldsymbol{G},
	\end{equation}
	where $\boldsymbol{D}$ is the force dipole tensor, and $\boldsymbol{G}$ is the Green's function of the Stokes equation. For a sphere in a Newtonian fluid, the dipole tensor is $20\pi/3 \langle\boldsymbol{e}\rangle$. In the presence of polymer force, $\mathbf{f}=\nabla\cdot\boldsymbol{\Pi}$, we can approximately estimate the net force dipole tensor as
	\begin{equation}
		\boldsymbol{D}=\frac{20\pi}{3}\langle\boldsymbol{e}\rangle+F_d^\Pi\langle\mathbf{e}\rangle=\frac{20\pi}{3}\langle\boldsymbol{e}\rangle-\int dV \frac{(\widehat{\mathbf{fr}+\mathbf{r f}})}{2},\label{eq:DipoleTensor}
	\end{equation}
	where $F_d^\Pi\langle\mathbf{e}\rangle$ is the force dipole tensor induced by the polymer force, $\mathbf{f}=\nabla\cdot\boldsymbol{\Pi}$. As shown in the above equation, $F_d^\Pi\langle\mathbf{e}\rangle$ is obtained via a volume integral involving $\mathbf{f}$ in the fluid volume. The time evolution of $F_d^\Pi$ (normalized with $c$) for different values of $c$ and the radial limit in the volume integration, $r_\text{cut-off}$, from 2 to 10 is shown in figure \ref{fig:polymerdipoleforce} for the flow around a sphere in a Giesekus liquid with $\alpha=0.01$ and $De=5$. Different colors and line styles represent different $r_\text{cut-off}$ and $c$, respectively. We observe that the integral converges with $r_\text{cut-off}$ for each $c$, indicating that near-field polymers constitute the majority of the polymer-induced dipole, $F_d^\Pi\langle\mathbf{e}\rangle$. As evident from the closeness of the curves for $c=0.2$ and $0.5$ for a given $r_\text{cut-off}$, we note that as $c$ increases, $F_d^\Pi/c$ collapses. This partially explains the $c^2$ scaling of the interaction stress discussed in section \ref{sec:ResultsFinitec}; as the dipole velocity disturbance scales as $c$, leading to the resulting polymer stress scaling as $c^2$.}

{The polymer-induced dipole, $F_d^\Pi$, is negative at all $H$ and $c$ shown in figure \ref{fig:polymerdipoleforce} and decreases monotonically with $H$. Therefore, if $c$ and $H$ are sufficiently large, the magnitude of $F_d^\Pi\langle\mathbf{e}\rangle$ exceeds that of the Newtonian dipole, ${20\pi}/{3}\langle\boldsymbol{e}\rangle$, resulting in a net dipole tensor, $\boldsymbol{D}$ (equation \eqref{eq:DipoleTensor}), that is negative. In other words, at sufficiently high $c$ and $H$, the net dipole disturbance created by the interaction of the particle and polymers opposes the Newtonian dipole disturbance.}

\begin{figure}
	\centering	
	\subfloat{\includegraphics[width=0.33\textwidth]{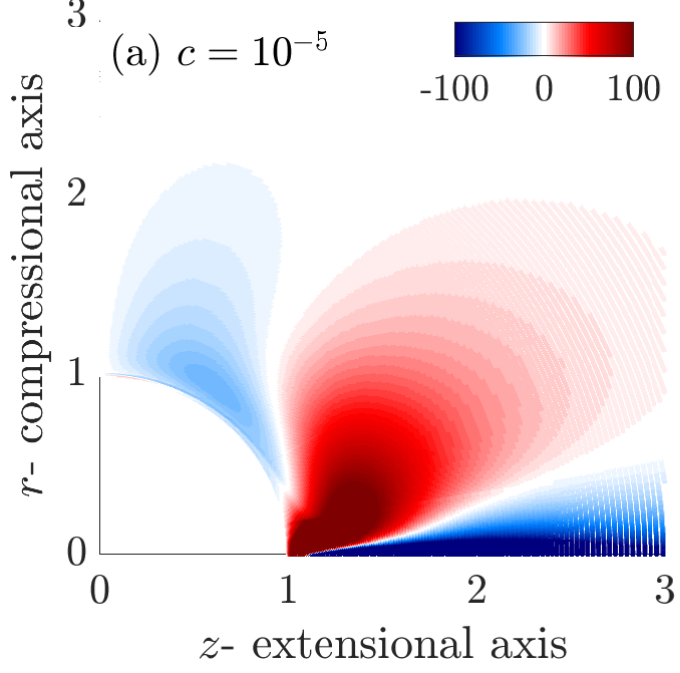}}
	\subfloat{\includegraphics[width=0.33\textwidth]{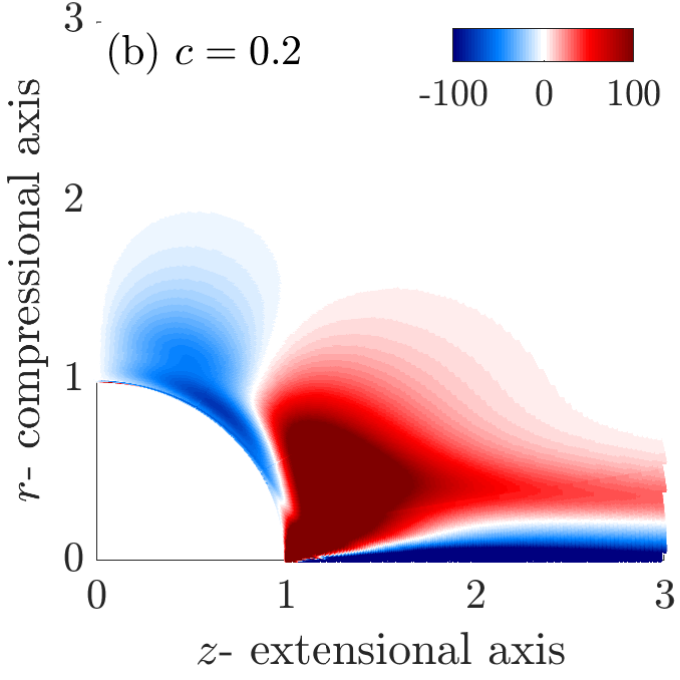}}
	\subfloat{\includegraphics[width=0.33\textwidth]{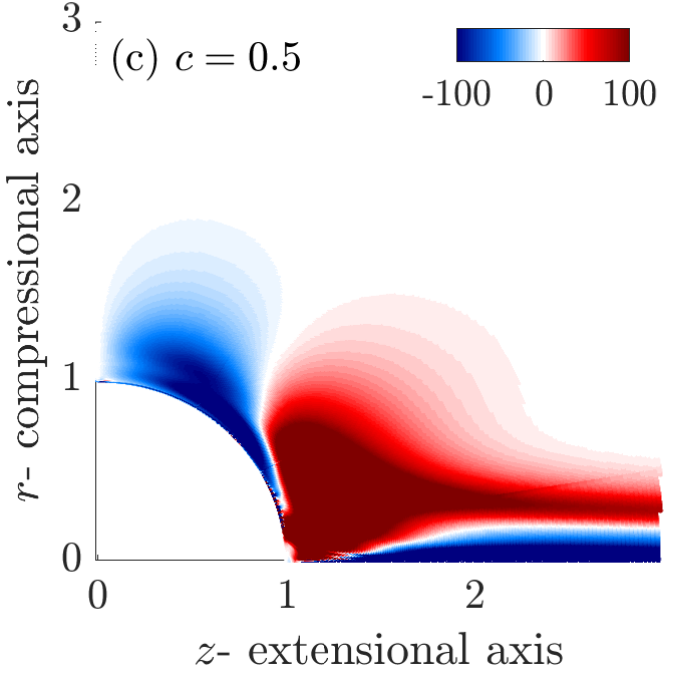}}
	\caption{{Steady-state extensional component of the polymeric force induced on the fluid flow, $\mathbf{f}/c=\nabla\cdot\boldsymbol{\Pi}/c$, for $De = 5.0$ for Giesekus liquid with $\alpha = 0.01$ at different $c$.}}
	\label{fig:Fext}
\end{figure}

{In section \ref{sec:Mech1}, we noted that as $c$ increases, the near-field polymers around the particle transition from a collapsed state to a more stretched state relative to the undisturbed polymers (figure \ref{fig:DeltaSDe5L10alfap01Zoom}). While the polymer stretch just off the particle changes with $c$, there remains a layer (which becomes thinner with increasing $c$) of collapsed polymers at the particle surface at all $c$ (figure \ref{fig:DeltaSDe5L10alfap01Zoom}). This thin layer of collapsed polymers ensures that the structure of the polymer stress gradients does not qualitatively change with $c$. This is depicted by the extensional component of the steady-state $\mathbf{f}=\nabla\cdot\boldsymbol{\Pi}$ for different $c$ in figure \ref{fig:Fext}. Therefore, $F_d^\Pi$, as estimated by the volume integral in the near-field region around the particle, does not change sign with $c$.}

{As discussed above, at large $H$ and $De$, increasing $c$ causes the polymer stress to alter the velocity field. At larger $c$, this field exhibits reduced stretching capability in a region approximately 3-5 particle radii upstream of the particle around the compressional axis. The highly stretched polymers arriving from the far field in this large upstream region, with lower velocity stretching capability, undergo a mild stretch-to-coil transition, resulting in an increasingly negative particle-polymer interaction stress with $c$, as discussed in section \ref{sec:ResultsFinitec}. In the next section, we analyze the structure of this wake of collapsed polymers and use it to explain the $c^2$ scaling of the interaction stress.}

\subsubsection{Structure and scaling of the wake of polymer collapse}
\begin{figure}
	\centering	
	\subfloat{\includegraphics[width=0.45\textwidth]{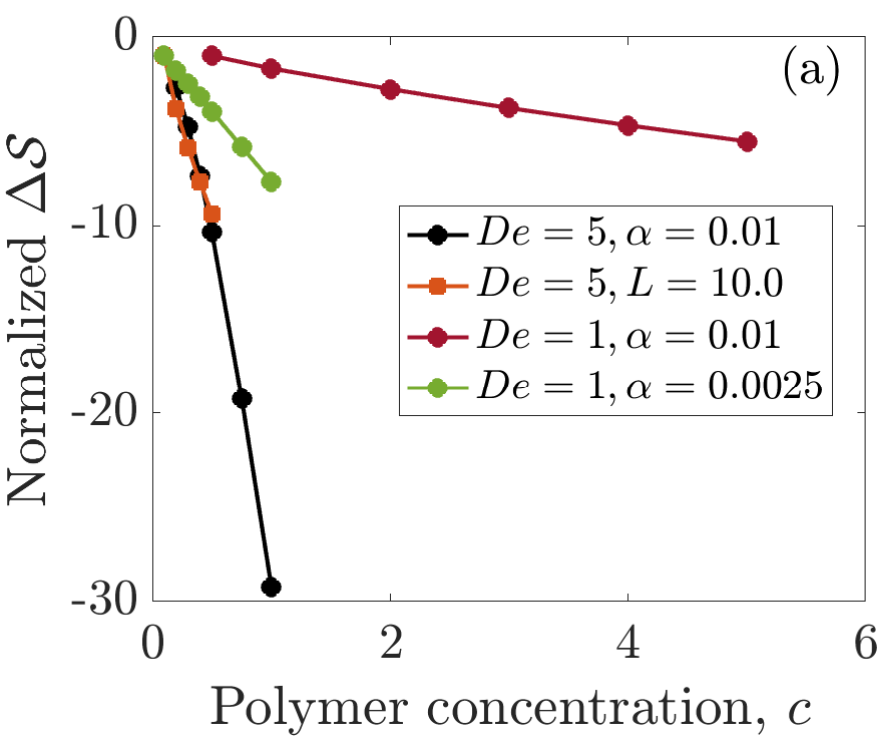}\label{fig:Stretchalongcompresionmanycases}}\hspace{0.1in}
	\subfloat{\includegraphics[width=0.45\textwidth]{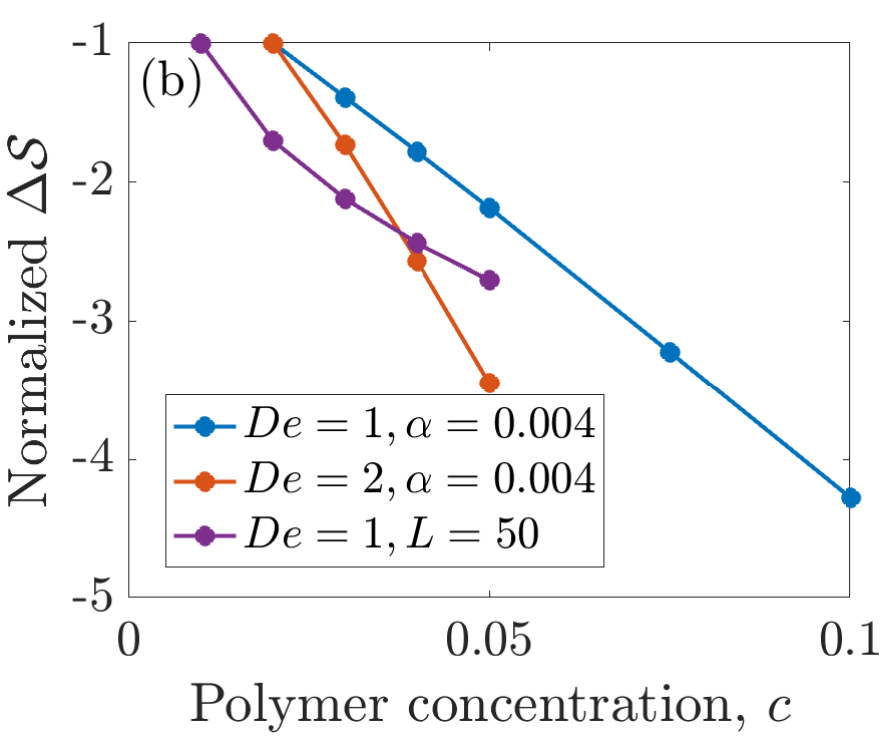}\label{fig:Stretchalongcompresionmanycases2}} 
	\caption{The variation of normalized $\Delta \mathcal{S}(z = 0, r = 20; c, De, \alpha)$, representing the polymer stretch on the compressional axis at a distance of 20 particle radii upstream of the particle, within the region of self-similar polymer collapse, for different $De$ and $\alpha$ in a Giesekus liquid. One curve for FENE-P liquid is also included in each figure. Normalization is applied to facilitate visualization, using the magnitude of $\Delta \mathcal{S}$ corresponding to the first point on each curve.}
	\label{fig:Slopesofpolymerstretch}
\end{figure}

The mechanism discussed above explains the change in slope of $(\hat{\Pi}^{PP} + \hat{\text{S}}^{PP}) / \hat{\Pi}^U$ vs. $c$ from positive to negative at a moderate $c$, which depends on $De$ and $L$ or $\alpha$. This transition occurs due to the formation of a wake of polymer collapse around the compressional axis, extending upstream by three particle radii. The linear negative slope observed from moderate to large $c$, or equivalently, the scaling of $(\hat{\Pi}^{PP} + \hat{\text{S}}^{PP})$ with $-c^2$, is attributed to the additional polymer stretch within the wake (as illustrated in figure \ref{fig:DeltaSDe5L10alfap01}). Figure \ref{fig:Slopesofpolymerstretch} shows the change in polymer stretch,$\Delta \mathcal{S}$, for a point on the compressional axis, located 20 particle radii upstream from the particle center. Here, we observe that $\Delta \mathcal{S}$ varies linearly with $c$ across a range of $\alpha$ and $De$ in Giesekus liquids. Notably, the magnitude of the slope of this variation increases with $De$ and $1/\alpha$, mirroring the variation of the slope of $(\hat{\Pi}^{PP} + \hat{\text{S}}^{PP}) / \hat{\Pi}^U$ shown earlier in figure \ref{fig:DeandAlfaGiesekusSteady}.

\begin{figure}
	\centering	
	\subfloat{\includegraphics[width=0.244\textwidth]{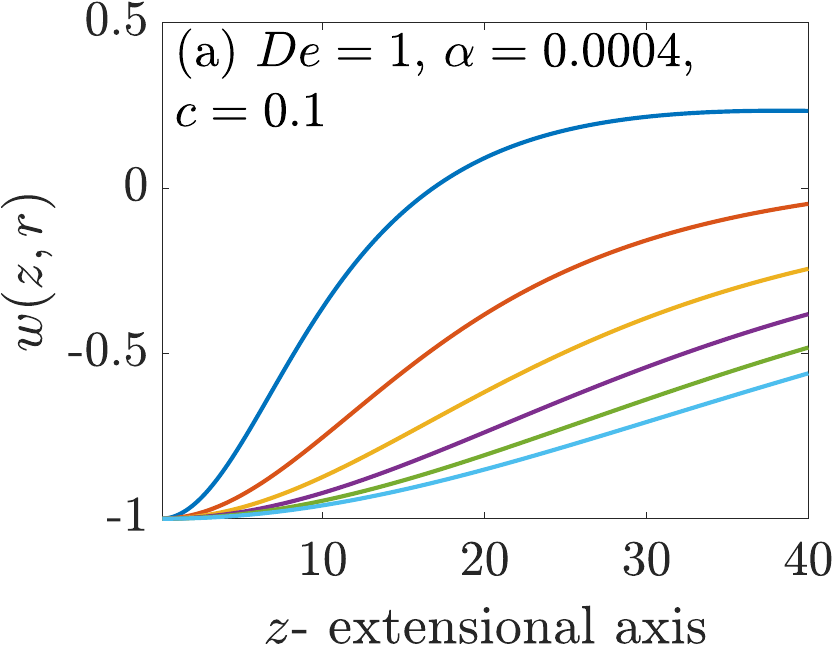}}\hfill
	\subfloat{\includegraphics[width=0.244\textwidth]{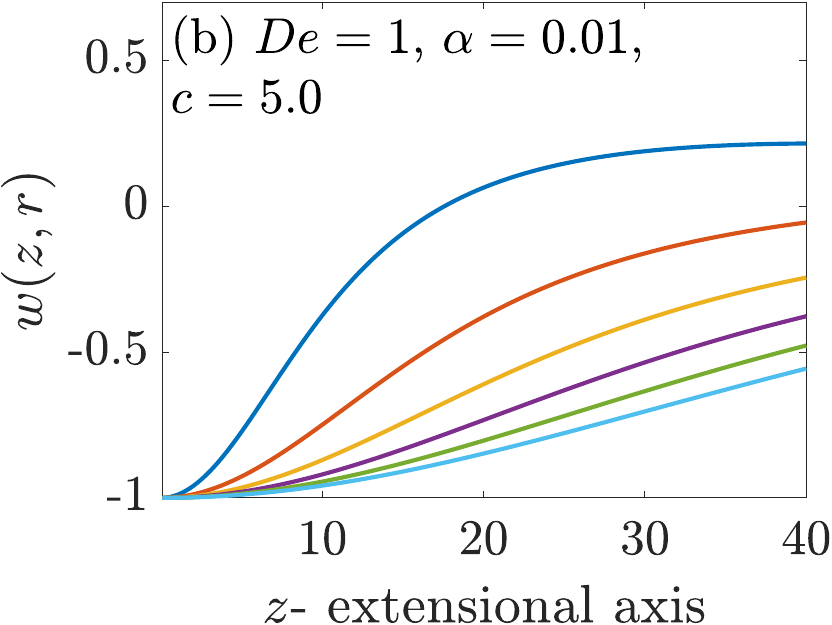}} \hfill
	\subfloat{\includegraphics[width=0.244\textwidth]{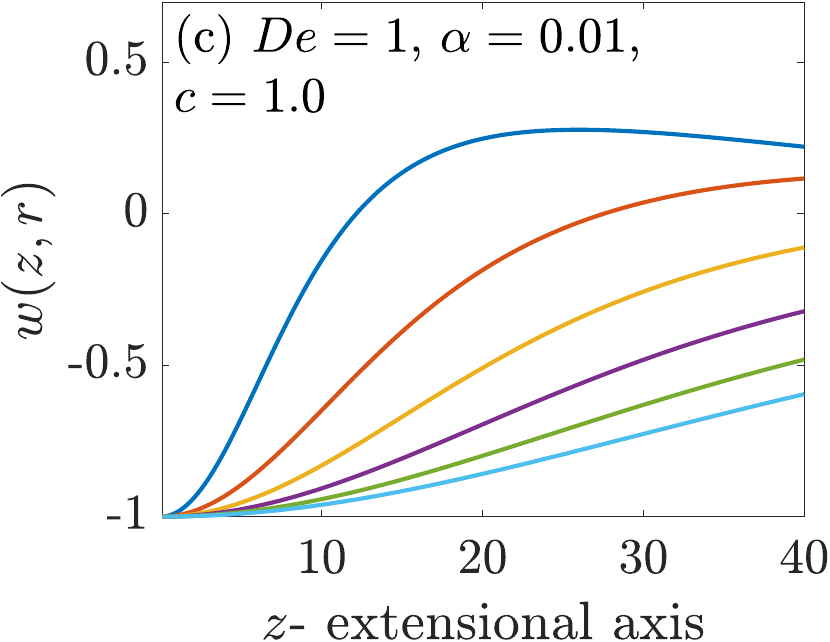}}\hfill
	\subfloat{\includegraphics[width=0.244\textwidth]{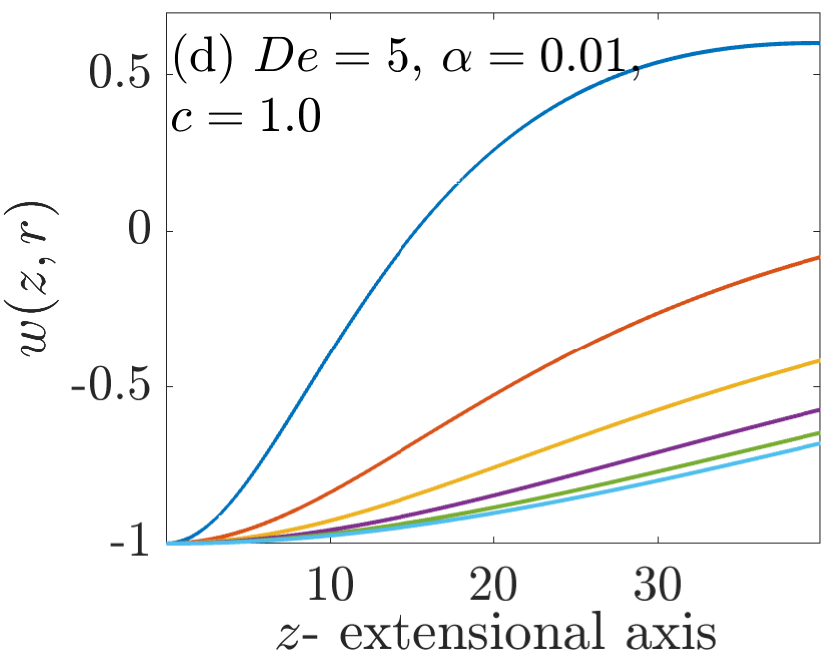}}\\
	\subfloat{\includegraphics[width=0.244\textwidth]{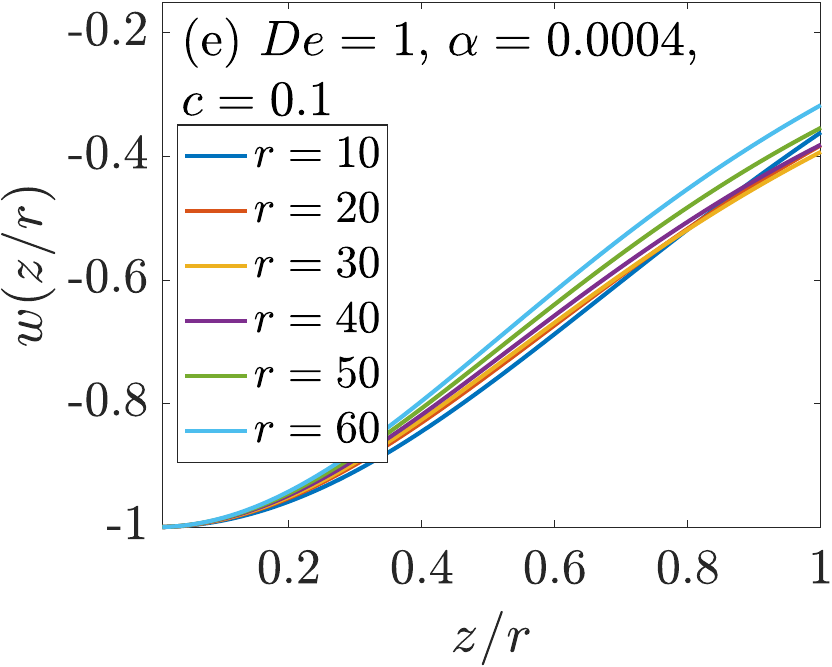}}\hfill
	\subfloat{\includegraphics[width=0.244\textwidth]{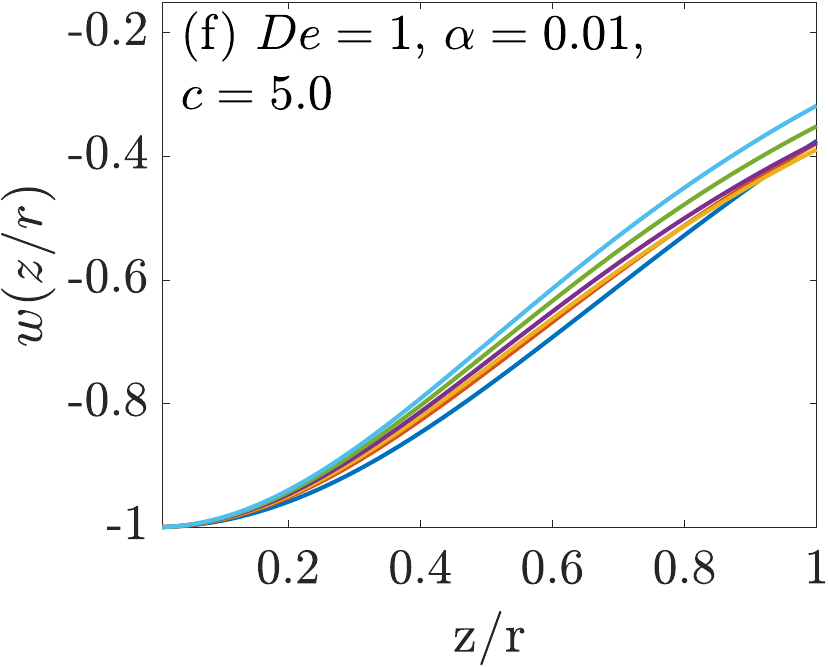}} \hfill
	\subfloat{\includegraphics[width=0.244\textwidth]{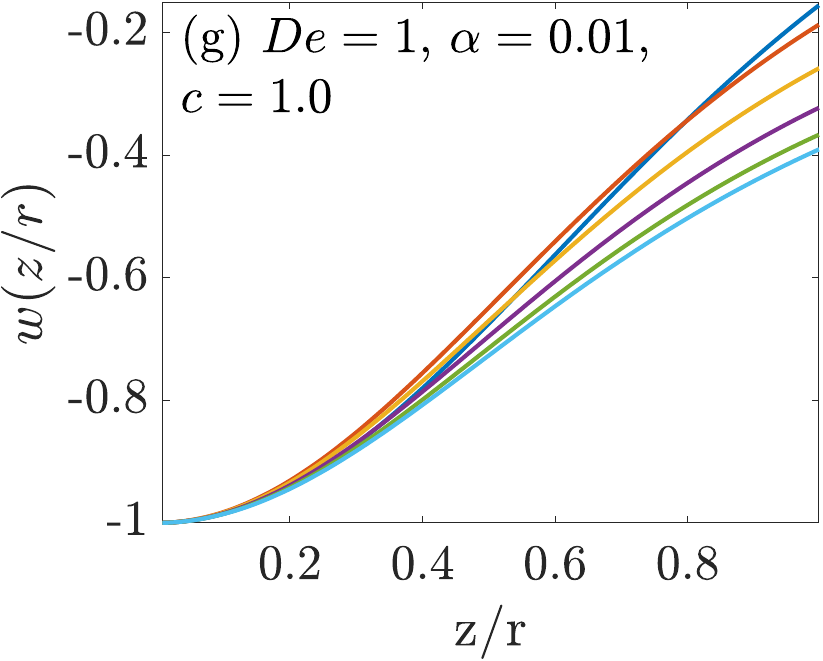}}\hfill
	\subfloat{\includegraphics[width=0.244\textwidth]{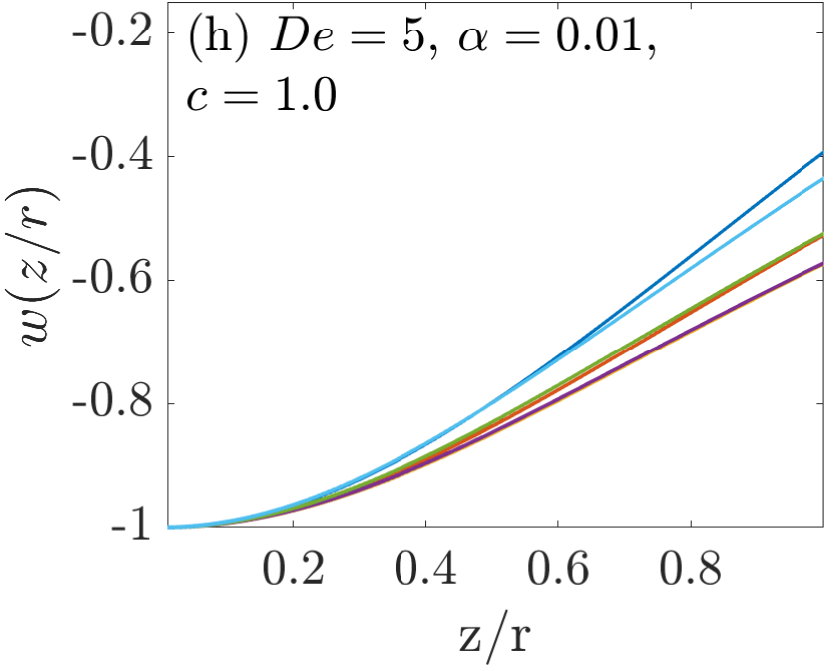}}
	\caption{The variation of $w(z, r; c, De, \alpha) = \frac{\Delta \mathcal{S}(z, r; c, De, \alpha)}{|\Delta \mathcal{S}(z = 0, r; c, De, \alpha)|}$ with distance along the extensional axis, $z$ (top panel), and the rescaled variable $z/r$ (bottom panel) for different values of $r$ (locations along the compressional direction). The specific parameter sets are as follows: (a), (e) $De = 1$, $\alpha = 0.0004$, $c = 0.1$; (b), (f) $De = 1$, $\alpha = 0.01$, $c = 5.0$; (c), (g) $De = 1$, $\alpha = 0.01$, $c = 1.0$; and (d), (h) $De = 5$, $\alpha = 0.01$, $c = 1.0$. All figures share the same legend displayed in figure (e).}
	\label{fig:SelfSimilarCollapse}
\end{figure}

Furthermore, this wake behaves in a self-similar manner. The top panel of figure \ref{fig:SelfSimilarCollapse} illustrates the function,
\begin{equation}
	w(z, r; c, De, \alpha) = \frac{\Delta \mathcal{S}(z, r; c, De, \alpha)}{|\Delta \mathcal{S}(z = 0, r; c, De, \alpha)|}, \hspace{0.1in} r > r_\text{wake-self-similar}\approx 5-10,
\end{equation}
which represents the variation of $\Delta \mathcal{S}(z, r)$ (normalized by the magnitude of $\Delta \mathcal{S}$ at the compressional axis, $z = 0$) along $z$, starting from different locations on the $r$ axis for various values of $De$, $\alpha$, and $c$. These curves correspond to the region of polymer collapse upstream of the particle along the compressional axis. In the bottom panel of figure \ref{fig:SelfSimilarCollapse}, the same $w(z, r; c, De, \alpha)$ variation as the top panel is presented, but with the horizontal axis rescaled to $z/r$. This rescaling allows the curves for different $r$ to collapse, with a slightly broader range of $z/r$ for larger values of $c$, $De$, and $1/\alpha$, demonstrating the self-similarity.

The linear scaling of the reduction polymer stretch with the polymer concentration, $c$ in the self-similar wake upstream of the particle around the compressional axis translates into a $c^2$ scaling of the reduction in polymer stress relative to the undisturbed value. Hence, beyond a moderate $c$ the particle-polymer interaction stress becomes increasingly negative at a rate proportional to $c^{2}$.

\section{Comparison with previous experiments}\label{sec:ExperimentalResults}
Experiments conducted during the SHERE (Shear History Extensional Rheology Experiment) \citep{SHERE_Report, soulages2010extensional, hall2009preliminary, SHERE2_Report, jaishankar2012shear} campaign aboard the International Space Station measured the extensional viscosity of a polymer solution with a dilute concentration of spheres using a filament stretching rheometer \citep{mckinley2002filament}. The rheometer initially consists of a liquid bridge between two circular plates. The plates are pulled apart at an exponential rate, generating homogeneous uniaxial extension at the center of the liquid bridge. In some experiments, a pre-shear is imparted before the extension by rotating one of the plates. The Weissenberg number, $Wi = \lambda \dot{\gamma}$, characterizes the pre-shear, where $\dot{\gamma}$ is the shear rate and $\lambda$ is the polymer relaxation time. A force transducer on the fixed plate and video recordings of the liquid bridge's mid-plane diameter enable the estimation of the extensional viscosity.

The experimental campaign comprises three distinct sets: SHERE, SHERE-R, and SHERE-II. The SHERE and SHERE-R fluids consist of 0.025 wt\% narrow polydispersity high molecular weight polystyrene (the polymer) in oligomeric styrene oil (a Newtonian solvent), with a polymer concentration of $c = 0.09$. In the SHERE-II experiments, a 3.5\% volume fraction of 6$\mu$m diameter poly(methyl methacrylate) microspheres is suspended in the same polymer solution used in SHERE and SHERE-R. Although the lowest $De$ in these experiments exceeds the highest reported from the simulations in sections \ref{sec:ResultsSemiAnalytical} and \ref{sec:ResultsFinitec}, a qualitative comparison remains valid. Comparing results from SHERE/SHERE-R and SHERE-II provides unique experimental evidence of the impact of particle-polymer interactions on extensional rheology. The kinematic viscosity of the SHERE fluid is $0.036\, \mathrm{m}^2/\mathrm{s}$, and the polymer relaxation time is approximately $3\,\mathrm{s}$, leading to an elasticity number $El = {De}/{Re} = {\lambda \nu}/{a^2} \sim \mathcal{O}(10^{10}).$ This indicates that inertial effects are negligible and the fluid velocity and pressure fields are quasi-steady, as assumed in this work and discussed in section~\ref{sec:Formulation}.

\begin{figure}
	\centering
	\subfloat{\includegraphics[width=0.33\textwidth]{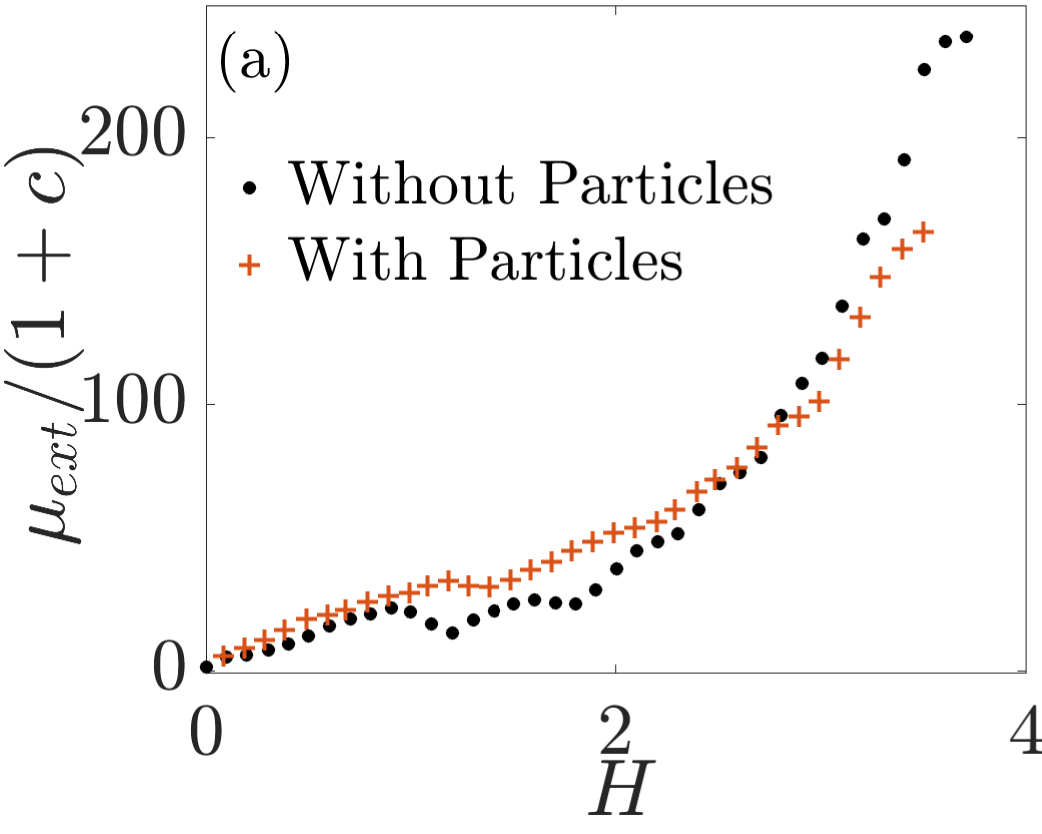}\label{fig:SHEREISS_De_15_Wi_0}}\hspace{0.1in}
		\subfloat{\includegraphics[width=0.33\textwidth]{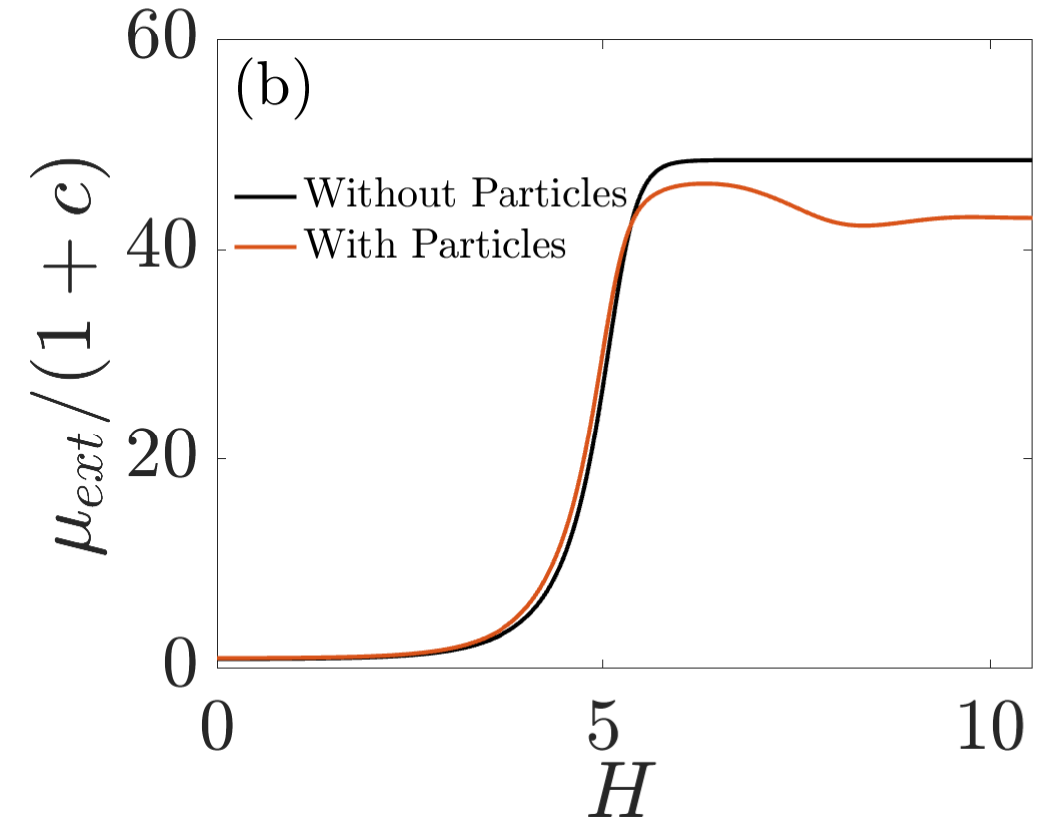}\label{fig:NumericalDe2L50FENEPphip035}}
	\caption {{Comparison of the effect of particles on the extensional viscosity of the suspension, $\mu_\text{ext}$ (equation \eqref{eq:ExtVisc}), for a dilute suspension of spheres with a volume fraction of 3.5\% in a viscoelastic liquid from: (a) SHERE/SHERE-R and SHERE-II experiments by \cite{SHERE_Report, soulages2010extensional, hall2009preliminary, SHERE2_Report, jaishankar2012shear} at $De = 15$ and $c=0.09$; and (b) our numerical simulations using the FENE-P model with $De=2.0$, $L=50$, and $c=0.1$.} \label{fig:ExperimentalvsSims}}
\end{figure}

\begin{figure}
	\centering
	\subfloat{\includegraphics[width=0.33\textwidth]{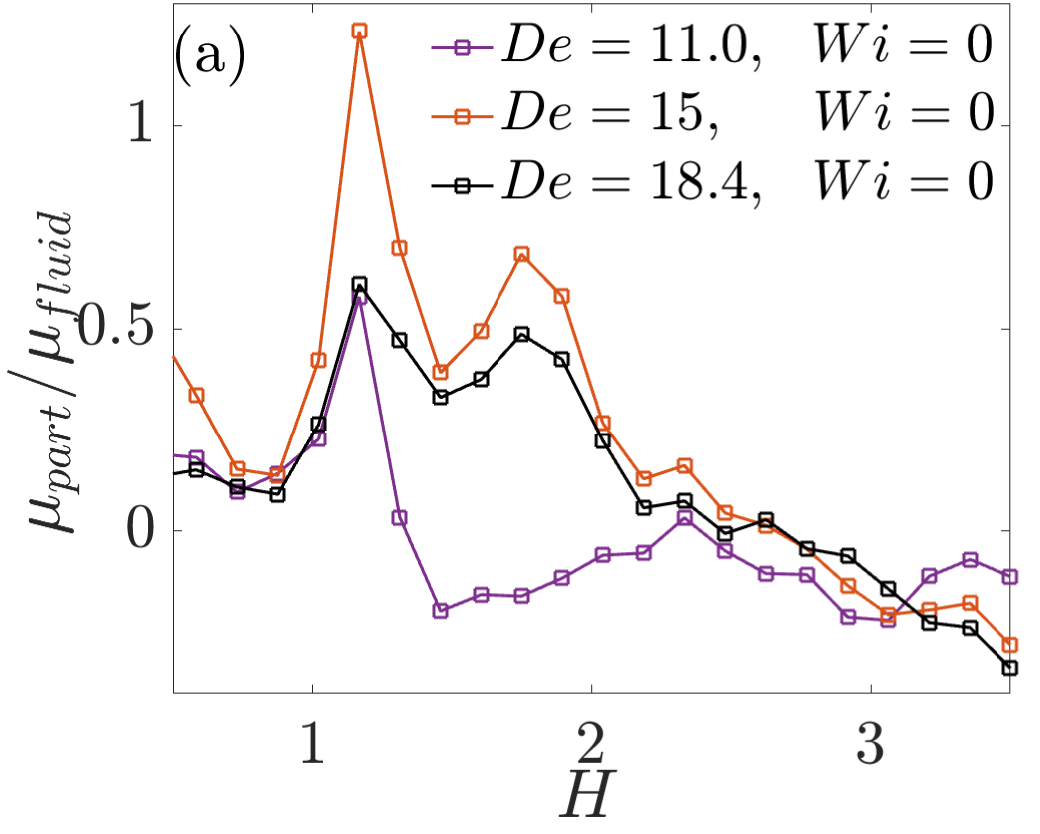}\label{fig:ParticleDifference}}\hspace{0.1in}
	\subfloat{\includegraphics[width=0.33\textwidth]{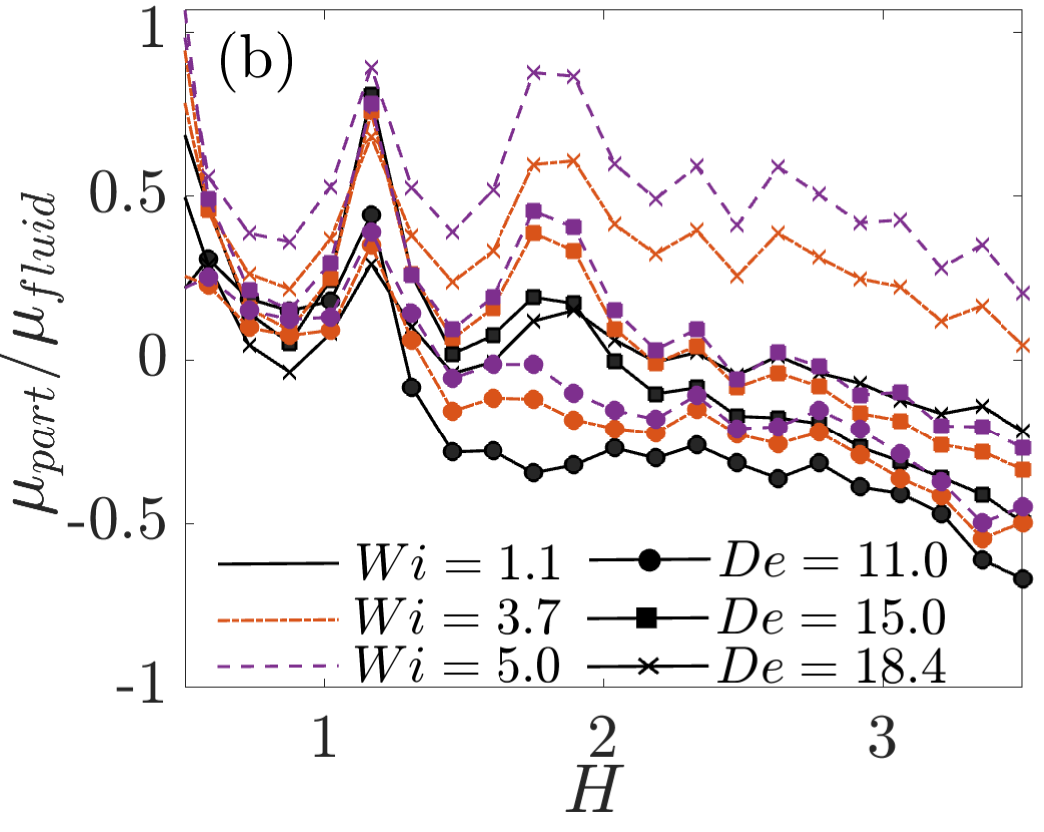}\label{fig:NonZeroWi}}
	\caption {{Relative effect of the particles, $\mu_\text{part} / \mu_\text{fluid}$ (equation \eqref{eq:PartVisc}), in SHERE experiments at (a) three different $De$ values with no pre-shear ($Wi = 0$); and (b) at three different $De$ values with varying $Wi$. In figure (b), different colors and line styles correspond to different $Wi$ values, and different $De$ values are indicated by different symbols.}\label{fig:ExperimentalResults}}
\end{figure}

{Figure \ref{fig:SHEREISS_De_15_Wi_0} shows the effect of particles at $De = 15$ and no pre-shear ($Wi = 0$) as $H$ evolves from the SHERE experiments. Initially, the particles increase the extensional viscosity ($\mu_\text{ext}$, defined in equation \eqref{eq:ExtVisc}), which then decreases as strain develops. This behavior qualitatively resembles the numerically obtained (from the method of \cite{NumericalMethodPaper}) evolution of particle-polymer interaction stress at $De > 0.5$, as illustrated in figure \ref{fig:NumericalDe2L50FENEPphip035} for a particle volume fraction $\phi=0.035$ in a FENE-P liquid with $De=2.0$ and $L=50$. This qualitative similarity, where particles reduce extensional viscosity at large $H$, was also observed in SHERE experiments at other $De$ values (as shown in figure \ref{fig:SHEREISS_De_15_Wi_0}) as well as the numerical results for small $c$ in section  \ref{sec:ResultsSemiAnalytical} and a general $c$ in section \ref{sec:ResultsFinitec}. The effect of particles at each Hencky strain is quantified by the relative change in viscosity, defined as $\mu_\text{part} / \mu_\text{fluid}$ (equation \eqref{eq:PartVisc}). For large $H$, this relative change decreases linearly with $H$ for $De = 11$, 15, and 18.4, as shown in figure \ref{fig:ParticleDifference}. This trend persists even with moderate pre-shear ($Wi \ne 0$), as seen in figure \ref{fig:NonZeroWi}.}

Due to noise within the SHERE data, a linear mixed-effects model is used to assess the statistical significance of the reduction in strain hardening induced by particles at $H \ge 2.25$. The mixed-effects model is defined as
\begin{equation}
	\frac{\mu_\text{part}}{\mu_\text{fluid}} = \beta_0 + \beta_H (H^k - 2.25) + \beta_{De} De^k + \beta_{Wi} Wi^k + \gamma_0^m + \gamma_H^m (H - 2.25) + \psi^k, \hfill H \ge 2.25.
\end{equation}
Here, $\beta_H$, $\beta_{De}$, and $\beta_{Wi}$ are the physical (fixed) parameters modeling the effects of $H$ above 2.25 across various $De$ and $Wi$, while $\psi^k$, $\gamma_H^m$, and $\gamma_0^m$ represent Gaussian-distributed random errors. Using $k \in [1, 120]$ points across $m \in [1, 12]$ curves in figures \ref{fig:ParticleDifference} and \ref{fig:NonZeroWi} leads to $\beta_0 = -0.84$, $\beta_H = -0.25$, $\beta_{De} = 0.056$, and $\beta_{Wi} = 0.034$ with 95\% confidence intervals of $[-1.3, -0.37]$, $[-0.29, -0.21]$, $[-0.03, 0.09]$, and $[-0.01, 0.08]$, respectively. The p-values associated with $\beta_H$, $\beta_{De}$, and $\beta_{Wi}$ are $6 \times 10^{-8}$, 0.002, and 0.1. Thus, $\beta_H = -0.25$ provides statistically significant evidence of reduced extensional viscosity upon particle addition at large $H > 2.25$. This effect is independent of pre-shear ($Wi$) and extension rate ($De$), as $\beta_{De}$ and $\beta_{Wi}$ are an order of magnitude smaller than $\beta_H$. Therefore, the SHERE experiments offer compelling evidence of negative particle-polymer interaction stress in a dilute particle suspension of viscoelastic liquid undergoing uniaxial extensional flow at high $De$ and $H$ through the mechanisms discussed in sections \ref{sec:PolymerStretch}, \ref{sec:Mech1}, and \ref{sec:Mech2}.

\section{Conclusions}\label{sec:Conclusions}
{
	In this paper, we investigate the impact of particle-polymer interactions on the extensional rheology of a dilute suspension of spheres in a polymeric solution using a combination of a semi-analytical method and direct numerical simulations. The particles and polymers engage in a two-way interaction that generates additional stress, referred to as particle-polymer interaction stress, in the suspension. This stress is in addition to the combined stress from a particle-free viscoelastic fluid and the Newtonian particle-induced suspension stress. The latter is $5\phi \langle\mathbf{e}\rangle$, where $\phi$ is the particle volume fraction and $\langle\mathbf{e}\rangle$ is the imposed rate of strain tensor. Particles affect the polymers by modifying the flow field and changing their configuration, while the polymers impact the particles by inducing additional elastic traction on the particle surface and altering the local flow field. As a result, a change in the particle-induced extensional viscosity, $\mu_\text{part}$ (equation \eqref{eq:PartVisc}), beyond the Newtonian value of $2.5\phi$, is observed. Assuming a dilute particle concentration, i.e., negligible particle-particle interactions, allows us to use ensemble averaging techniques and the flow field around an isolated particle in an unbounded fluid to characterize the suspension rheology.
}

{
	When the Deborah number ($De$), defined as the non-dimensional product of polymer relaxation time and the imposed extension rate, is smaller than 0.5, the interaction stress evolves from zero to a positive steady-state value. For $De > 0.5$, the suspension exhibits a transient behavior similar to that of lower $De$ suspensions, but with a more pronounced initial increase in interaction stress, which subsequently peaks and then falls to a negative steady state. The steady-state extensional viscosity of particle-free viscoelastic fluids ($\mu_\text{fluid}$, as defined in equation \eqref{eq:PartVisc}) increases with $De$ at large $De$. This increase results from significant polymer stretching due to the coil-stretch transition \citep{de1974coil}. However, in the $De > 0.5$ regime at large times, $\mu_\text{part}$ is negative, and thus the suspension viscosity is found to be lower than that of the particle-free fluid. The fluid viscosity, $\mu_\text{fluid}$, increases linearly with polymer concentration ($c$) and can become substantial in industrial applications. As $c$ increases, the intensity of particle-polymer interactions rises, and at large $De$ and times, leads to an increased magnitude of the negative interaction stress that scales as $c^2$ beyond a moderate polymer concentration. Thus, the particles lead to a greater reduction in viscoelastic stress at higher $c$ and $De$.
}

{
	At very low $c$, the velocity field in the fluid remains unaffected by the presence of polymers. Thus, using a regular perturbation in $c$ and a generalized reciprocal theorem, we obtain the suspension rheology up to $\mathcal{O}(c)$ by using just the analytically available Newtonian velocity field around the particle. These computations are highly efficient as the equations governing mass and momentum conservation in the fluid need not be solved. In this low $c$ regime, viscoelastic fluids can be characterized as polymer solutions within a Newtonian solvent, well described by the FENE-P constitutive equation \citep{anna2008effect}. Thus, we solve the FENE-P equations using the method of characteristics to determine the leading-order polymer configuration. This approach is similar to the calculations performed by \cite{koch2016stress} and \cite{SteadyStatePaper}; however, due to the extreme stretching of fluid elements along the streamlines around the extensional axis and the transient nature of polymer evolution, additional care is taken, as detailed in Appendix \ref{sec:MethodSemiAnalytical}.
}

The increase in steady-state interaction stress at $De < 0.5$ and during transients for larger $De$ arises from a wake of highly stretched polymers (relative to those undisturbed by the sphere) around the extensional axis. This wake intensifies until steady state for small $De$ and peaks for larger $De$. {The polymers are stretched in this wake because, near the extensional axis, the disturbances created by the sphere enhance the stretching capability of the underlying velocity field beyond that of the imposed flow.} In large $De$ cases, the peak interaction stress occurs just before the polymers far from the sphere transition from a coiled to a stretched state, a mechanism that leads to a significant increase in polymer stress. While the undisturbed polymers in the far field experience substantial stretching, the region near the stagnation point on the sphere's compressional axis experiences low stretching due to small velocity gradients and low flow speed. Consequently, the polymers close to the particle around the compressional axis undergo collapse, resulting in negative particle-polymer interaction stress at large $De$ and large Hencky strains, $H$. In extensional flow rheology, $H$ (product of extension rate and time) is used as the non-dimensional time scale and represents the strain accumulated in the sample being tested. As the polymers first collapse, they occupy a large volume around the particle. However, since the stretching capability of the velocity gradients is greater downstream, around the extensional axis, these collapsed polymers recover some of their stretch, leading to a reduction in the collapse region. This recovery is reflected in a negative peak of the interaction stress value before it achieves the steady state. These values are consistent with our previous work \citep{SteadyStatePaper}, where at large $De$, the interaction stress was estimated to be -0.853$\phi$ times the undisturbed polymer stress at steady state.

Increasing the polymer concentration ($c$) enhances the interaction between local polymer stretching and the velocity field, significantly altering the fluid velocity due to the stress induced by the underlying polymers.  {While the FENE-P model is suitable up to moderate $c$, the Giesekus model is more appropriate for high-concentration polymer solutions}. Therefore, we perform direct numerical simulations using our in-house numerical solver \citep{NumericalMethodPaper} to analyze the flow of FENE-P and Giesekus liquids around a sphere across a wide range of $De$, $c$, and the model parameters $L$ and $\alpha$. This method employs finite difference discretization of the mass and momentum conservation equations, as well as the polymer constitutive equations, and is briefly described and validated in the context of extensional rheology in appendix \ref{sec:DNSMethodology}.

At large $c$, the effect of $De$ on suspension rheology resembles that observed in the $\mathcal{O}(c)$ calculations described above. However, as $c$ increases from very small values, the magnitude of interaction stress decreases at all $De$ and $H$ because stretched polymers alter the underlying flow. In regions where polymers collapse, the local fluid stress is lower, allowing for swifter flow than the far-field if $c$ is sufficiently large. Consequently, as $c$ increases, velocity stretching capability diminishes in areas of highly stretched polymers while being enhanced in regions of collapse. This results in the polymer stretch approaching the undisturbed state in both the wake of highly stretched polymers around the extensional axis at low $De$ and up to the peak interaction stress at large $De$, as well as in the polymer collapse region around the particle at large $De$ and $H$.

In Newtonian flow, the region upstream of approximately three particle radii along the compressional axis exhibits additional velocity stretching capability, primarily driven by dipole disturbance. Although this region occupies a large volume, the difference in stretching capability relative to the imposed flow is minimal. Consequently, when $c$ is small, polymers in this region experience only a slight extra stretch compared to the undisturbed flow, a phenomenon observed at all $De$ and $H$. Even at large values of these parameters, the contribution to rheology from this region is overshadowed by that from the polymer collapse near the particle. As $c$ increases, the polymers begin to resist the Newtonian dipole-induced flow, ultimately transforming it into a region with lower velocity stretching capability. This transformation is attributed to the change in sign of the net force dipole caused by the polymer force field near the particle surface. Thus, with the dipole velocity disturbance acting in the opposite direction to the Newtonian disturbance at large $De$ and $H$, the polymers in this region collapse relative to those in the far field. Although upon increasing $c$, the intensity of polymer collapse near the particle surface diminishes and may even turn into extra stretch, as mentioned above, beyond a certain $c$ (dependent on $De$ and $L$ or $\alpha$), the particle-polymer interaction stress becomes increasingly negative with $c$. This $c^2$ scaling of the negative interaction stress arises because the gentle polymer collapse in this large region around the compressional axis scales as $c$. At different radial locations (compressional axis coordinate) within the wake of polymer collapse upstream of about three particle radii, the polymer stretch exhibits self-similarity along the extensional axis coordinate $z$ under the rescaling $z/r$.

Partial qualitative experimental validation of the numerically obtained negative particle-polymer interaction stress at large $De$ and $H$ is provided by previous experiments conducted by McKinley and co-authors on the International Space Station \citep{SHERE_Report, soulages2010extensional, hall2009preliminary, SHERE2_Report, jaishankar2012shear}. These experiments utilized a polymer solution at a single concentration of $c = 0.09$. Notably, our simulations indicate that the interaction stress varies much more rapidly with $c$ than the linear increase of stress observed in particle-free fluids. This finding highlights the need for a more extensive experimental campaign involving particle suspensions in both {FENE-P and Giesekus liquids} across a broader range of industrially relevant concentrations. Future experiments with larger $c$ could be conducted on Earth, where the fluid would be more viscous and the sagging of the liquid bridge in the filament stretching extensional rheometer due to gravity would be minimized. 

One aspect not addressed here is polymer depletion near non-interacting, impenetrable surfaces, as observed in previous experiments \citep{ausserre1985concentration, omari1989wall, omari1989hydrodynamic, ausserre1991hydrodynamic}. The thickness of these depletion layers increases with the polymer radius of gyration ($R_g$) and decreases with shear rate \citep{ausserre1991hydrodynamic} and $c$ \citep{omari1989wall}. Depletion around suspended particles depends qualitatively on the ratio $R_g/a$ as larger polymers wrap around the particle and smaller ones perceive the particle as an infinite wall  \citep{fuchs2002structure, lekkerkerker2024colloids}. For SHERE fluids with molecular weight $2.25 \times 10^6\,\mathrm{g/mol}$, $R_g \approx 80$\,nm \citep{jaishankar2012shear, terao2004line}. Thus, for particles of radius $a = 3\,\mu$m used in these experiments, $R_g/a \ll 1$—a regime in which polymer chains flatten and align parallel to the surface \citep{doxastakis2004polymer}. In this limit, the depletion layer is thin and negligible at low $De$ for particle sizes and polymeric liquids commonly used in rheology experiments. However, at high $De$ and low $c$, collapsed polymer layers—predicted by our simulations to contribute to negative particle-polymer interaction stresses—become extremely thin, and depletion effects may become comparable and non-negligible. These effects could be incorporated in future simulations by modeling slip boundary conditions at the particle surface or spatial non-uniformity in $c$, following \citet{mavrantzas1992theoretical}, within either the semi-analytical framework (section~\ref{sec:DiluteRheology}) for dilute systems or the DNS approach (section~\ref{sec:ConcentratedRheology}) for concentrated polymeric solutions.

The effects of particles on the extensional rheology of a viscoelastic liquid can be either beneficial or harmful in industrial applications. Reducing extensional strain hardening through particle addition could lower power consumption in industrial processes such as hydraulic fracturing and extrusion molding, where polymeric fluids experience strong extensional flow (section \ref{sec:Intro}). Conversely, in polymer processing operations like film blowing, strain hardening is crucial for achieving uniform film thickness \citep{munstedt2018extensional}. In this context, while the addition of particles can be beneficial, it must be carefully managed to ensure that the accumulated Hencky strain remains below the threshold where large positive interaction stress transitions to a negative value.

\section*{Acknowledgements}
This work was supported by NASA Grant 80NSSC23K0348 and NSF Grant 2206851. This work used the Bridges-2 supercomputer at the Pittsburgh Supercomputing Center through allocations CHM240066 and PHY210025 from the Advanced Cyberinfrastructure Coordination Ecosystem: Services \& Support (ACCESS) program \citep{boerner2023access}, supported by U.S. National Science Foundation Grants \#2138259, \#2138286, \#2138307, \#2137603, and \#2138296.\\
The first author (AS) is currently an employee of NTESS. This paper describes objective technical results and analysis. Any subjective views or opinions that might be expressed in the paper do not necessarily represent the views of the U.S. Department of Energy or the United States Government. Sandia National Laboratories is a multimission laboratory managed and operated by National Technology and Engineering Solutions of Sandia, LLC (NTESS), a wholly owned subsidiary of Honeywell International Inc., for the U.S. Department of Energy's National Nuclear Security Administration under contract DE-NA0003525. SAND-NO2024XXXJ.\\
Declaration of Interests. The authors report no conflict of interest.

\bibliographystyle{jfm}
\bibliography{MainDocument}
	
\appendix	
\section{More details on rheology constituents}\label{sec:Moredetails}
The particle-polymer interaction stress, given by ${c\phi}(\hat{\boldsymbol{\Pi}}^{PP} + \hat{\text{\textbf{S}}}^\text{PP})$ as defined in section \ref{sec:Formulation} and expressed in equation \eqref{eq:ConstitutiveRheology}, encompasses the influence of fluid traction (from both the solvent and the polymers) on the particle surface, as well as the alteration of polymer stress due to the presence of the particle. The first component, which represents the particle-polymer interaction stresslet, is denoted as $c\phi\hat{\text{\textbf{S}}}^\text{PP}$. The second component, referred to as the particle-induced polymer stress (PIPS), is represented by ${c\phi}\hat{\boldsymbol{\Pi}}^{PP}$. In this section, we provide a detailed mathematical analysis of these two stress components.

\subsection{Particle stresslet}\label{sec:Stresslet}
In a dilute particle suspension with well-separated particles, inter-particle interaction is rare. Thus, studying the imposed flow (uniaxial extension in our case) around an isolated particle in an unbounded fluid suffices to characterize the suspension rheology. The knowledge of the constitutive relation within the particle is not necessary to obtain the particle stresslet. Using the divergence theorem, \cite{batchelor1970stress} showed that it is the following area integral of a tensor product of stress on the particle surface, with position vector $\mathbf{r}_p$,
\begin{equation}
	\hat{\text{\textbf{S}}}(\boldsymbol{\sigma}) = \int_{\mathbf{r} = \mathbf{r}_p} \text{d}A \hspace{0.1in} \frac{1}{2} [\text{\textbf{rn}} \cdot \boldsymbol{\sigma} + \text{\textbf{n}} \cdot \boldsymbol{\sigma} \text{\textbf{r}}] - \frac{1}{3} \boldsymbol{\delta} \Big(\text{\textbf{n}} \cdot \boldsymbol{\sigma} \cdot \text{\textbf{r}}\Big).\label{eq:Stresslet1}
\end{equation}
The stresslet, being a linear function of stress, can be expressed as a sum of the solvent, $\hat{\text{\textbf{S}}}(\boldsymbol{\tau})$, and polymeric stresslet, $c\hat{\text{\textbf{S}}}(\boldsymbol{\Pi})$. However, the solvent stress is perturbed by the polymers, and thus the non-Newtonian nature of the underlying fluid also influences $\hat{\text{\textbf{S}}}(\boldsymbol{\tau})$. Denoting the polymer-induced solvent stress as $c\boldsymbol{\tau}^{\text{PI}}$, such that,
\begin{equation}
	\boldsymbol{\tau} = \boldsymbol{\tau}^{\text{Stokes}} + c\boldsymbol{\tau}^{\text{PI}},\label{eq:StressSplit}
\end{equation}
with $\boldsymbol{\tau}^{\text{Stokes}}$ being the stress on the particle if the fluid were Newtonian, the total solvent stresslet, $\hat{\text{\textbf{S}}}(\boldsymbol{\tau})$, can be written as the sum of the Stokes ($\hat{\text{\textbf{S}}}(\boldsymbol{\tau}^{\text{Stokes}})$) and polymer-induced solvent ($c\hat{\text{\textbf{S}}}(\boldsymbol{\tau}^{\text{PI}})$) stresslet. In the case of a sphere, $\hat{\text{\textbf{S}}}(\boldsymbol{\tau}^{\text{Stokes}}) = 5\langle\boldsymbol{e}\rangle V_p$, as initially shown by \cite{einstein2005neue}. The interaction stresslet, ${c\phi}\hat{\text{\textbf{S}}}^\text{PP}$, introduced in equation \eqref{eq:ConstitutiveRheology}, is related to the stresslets arising from $c\hat{\text{\textbf{S}}}(\boldsymbol{\Pi})$ and $c\boldsymbol{\tau}^{\text{PI}}$ via,
\begin{equation}
	V_p \hat{\text{\textbf{S}}}^\text{PP} = V_p \hat{\text{{S}}}^\text{PP} \langle\boldsymbol{e}\rangle = \hat{\text{\textbf{S}}}(\boldsymbol{\tau}^\text{PI}) + \hat{\text{\textbf{S}}}(\boldsymbol{\Pi}).
\end{equation}
It can therefore also be termed as the net non-Newtonian stresslet.

Irrespective of the polymer concentration, in an inertia-less fluid, the mass and momentum equations are linear in fluid pressure and velocity, and the polymer stress can be considered a forcing term. This realization, along with the fluid stress decomposition of equation \eqref{eq:StressSplit}, allows the governing fluid mass and momentum equation \eqref{eq:MassMomentum} to be written as the sum of a Stokes and a polymer-induced problem, as demonstrated in \cite{NumericalMethodPaper}. The Stokes problem is the same as the original problem described by equation \eqref{eq:MassMomentum}, but without the polymeric stress, $\boldsymbol{\Pi}$. These equations are,
\begin{equation}
	\nabla \cdot \mathbf{u}^\text{Stokes} = 0, \hspace{0.2in} \nabla \cdot \boldsymbol{\tau}^\text{Stokes} = 0,\label{eq:MassMomentumNewt}
\end{equation}
where $\boldsymbol{\tau}^\text{Stokes} = -p^\text{Stokes} \boldsymbol{\delta} + 2\boldsymbol{e}^\text{Stokes}$ is the Newtonian solvent stress, and $p^\text{Stokes}$, $\mathbf{u}^\text{Stokes}$, and $\boldsymbol{e}^\text{Stokes} = (\nabla \mathbf{u}^\text{Stokes} + (\nabla \mathbf{u}^\text{Stokes})^\text{T}) / 2$ are the Stokes pressure, velocity, and strain rate tensor fields. On these equations, the velocity boundary conditions of the original or complete problem described by equation \eqref{eq:MassMomentum} are imposed (i.e., replace $\mathbf{u}$ with $\mathbf{u}^\text{Stokes}$ in equation \eqref{eq:BCs}). The polymer-induced (labeled with a superscript PI) problem governs a polymer-induced solvent stress, $\boldsymbol{\tau}^\text{PI} = -p^\text{PI} \boldsymbol{\delta} + 2\boldsymbol{e}^\text{PI}$, that is driven by the polymer stress divergence,
\begin{equation}
	\nabla \cdot \mathbf{u}^\text{PI} = 0, \hspace{0.2in} \nabla \cdot \boldsymbol{\tau}^\text{PI} = -\nabla \cdot \boldsymbol{\Pi},\label{eq:NonNewtProblem}
\end{equation}
subject to zero boundary conditions for the velocity on the particle surface {and in the far-field}. Here,
\begin{align}\begin{split}
		&c\mathbf{u}^\text{PI} = \mathbf{u} - \mathbf{u}^\text{Stokes}, \hspace{0.2in} cp^\text{PI} = p - p^\text{Stokes}, \hspace{0.2in} 2\boldsymbol{e}^\text{PI} = \nabla \mathbf{u}^\text{PI} + (\nabla \mathbf{u}^\text{PI})^\text{T},\label{eq:NonNewtonianFields}
\end{split}\end{align}
$\mathbf{u}^\text{PI}$ and $p^\text{PI}$ are the polymer-induced velocity and pressure fields.

Since the stress $\boldsymbol{\tau}^\text{PI}$ originates due to $\boldsymbol{\Pi}$, the polymer-induced solvent stresslet, $\hat{\text{\textbf{S}}}(\boldsymbol{\tau}^\text{PI})$, must also originate due to $\boldsymbol{\Pi}$. In \cite{SteadyStatePaper}, using a generalized reciprocal theorem, we obtain an alternative decomposition of the total non-Newtonian stresslet arising from the particle-polymer interaction as a function of just the polymer stress field and its undisturbed value,
\begin{equation}
	V_p \hat{\text{\textbf{S}}}^\text{PP} = \hat{\text{\textbf{S}}}(\boldsymbol{\Pi}^U) + \hat{\text{\textbf{S}}}_\text{volume}(\boldsymbol{\Pi}; \boldsymbol{\Pi}^U),\label{eq:InteractionStresslet}
\end{equation}
where for a particle suspension with {each particle's volume $V_p$},
\begin{equation}
	\hat{\text{\textbf{S}}}(\boldsymbol{\Pi}^U) = V_p \boldsymbol{\Pi}^U, \text{ and, }
	\hat{\text{\textbf{S}}}_\text{volume}(\boldsymbol{\Pi}, \boldsymbol{\Pi}^U) = \int_{V_f} \text{d}V \hspace{0.1in} (\boldsymbol{\Pi}^U - \boldsymbol{\Pi}) : \nabla \mathbf{v}.\label{eq:NewStressletDecomp}
\end{equation}
For spheres, the particle shape-dependent 3-tensor, $\mathbf{v}$, is
\begin{equation}
	v_{ijk} = \left(\frac{5}{2r^5} - \frac{5}{2r^7}\right) r_i r_j r_k + \frac{1}{2r^5} \left(r_j \delta_{ik} + r_k \delta_{ij}\right) + \left(\frac{1}{2r^5} - \frac{5}{6r^3}\right) r_i \delta_{jk}.\label{eq:AuxillaryVelocity}
\end{equation}
Physically, $v_{ijk} \langle e \rangle_{jk}$ is the velocity disturbance generated by the presence of the particle under consideration in any imposed strain $\langle e \rangle_{jk}$. It can be obtained by Stokes flow solution around a sphere in different straining flows available in textbooks such as \cite{leal2007advanced}. Alternatively, as mentioned by \cite{koch2016stress}, $v_{ijk}$ is the decaying Newtonian fluid `velocity' field that extensionally deforms at the particle surface.

The alternative stresslet decomposition of equation \eqref{eq:InteractionStresslet}, valid for all values of polymer concentration, lends further insight into the suspension rheology as we discuss in section \ref{sec:RheologySplit}. The undisturbed part, $\hat{\text{\textbf{S}}}(\boldsymbol{\Pi}^U)$, is simply the stresslet on a fictitious particle placed in the far field such that the stress on its surface is the same as the undisturbed value. The change in the polymer stress in the fluid volume between the particle surface and the far field, due to the presence of the particle, leads to the stresslet $\hat{\text{\textbf{S}}}_\text{volume}(\boldsymbol{\Pi}, \boldsymbol{\Pi}^U)$. Viewing the fluid velocity and pressure through the decomposition into Stokes and polymer-induced parts allows us to obtain deeper mechanistic insights into particle-polymer interaction as considered in sections \ref{sec:Mech1} and \ref{sec:Mech2}. Additionally, while solving the fluid momentum equation for obtaining the results discussed in section \ref{sec:ConcentratedRheology}, the Stokes component is known analytically, and only the polymer-induced momentum equation \eqref{eq:NonNewtProblem} is evaluated numerically.

\subsection{Particle-induced polymer stress (PIPS)}\label{sec:PIPS}
Similar to the particle stresslet, the ensemble-averaged polymer stress, $c\langle \hat{\boldsymbol{\Pi}}\rangle$, can also be expressed as a function of relevant flow variables around an isolated particle in an unbounded fluid. However, this requires extra care, as simply replacing it with the volume average of the polymeric stress in the suspension \citep{jain2019extensional} leads to a logarithmic divergence when characterizing the rheology of a suspension of spheres in a dilute polymeric liquid. Numerically, this may manifest as absent or slow convergence of the required stress with the domain size, as we found through personal communication with the authors of \cite{jain2019extensional}. This divergence was first brought to attention by \cite{koch2006stress} for a second-order fluid and later for an Oldroyd-B fluid by \cite{koch2016stress}.

By identifying the source of this logarithmic divergence as the linearized polymeric stress, \cite{koch2016stress} obtained the correct form of $c\langle \hat{\boldsymbol{\Pi}}\rangle$ in terms of flow around an isolated particle. The linearization of the polymer stress is done about the undisturbed polymer stress field, $c\boldsymbol{\Pi}^U$, and velocity field, $\langle\mathbf{u}\rangle$ {with the particle-induced velocity disturbance driving the linear perturbations in the polymer stress.} We have recently extended the original formulation of \cite{koch2016stress} for Oldroyd-B fluid, implemented for steady-state shear rheology, to investigate the steady-state extensional rheology of the suspension of spheres in dilute FENE-P liquid \citep{SteadyStatePaper}. This formulation can be applied even to the transient case and to different polymer constitutive models. We briefly present the required formulation for the transient rheology of suspensions in FENE-P and Giesekus liquids and refer to the studies mentioned above for details.

The polymer stress and polymer configuration are a sum of the undisturbed (by the particle's presence), linear, and non-linear parts,
\begin{equation}
	{\boldsymbol{\Pi}} = {\boldsymbol{\Pi}}^U + {\boldsymbol{\Pi}}^L + {\boldsymbol{\Pi}}^{NL}, \hspace{0.2in} {\boldsymbol{\Lambda}} = {\boldsymbol{\Lambda}}^U + {\boldsymbol{\Lambda}}^L + {\boldsymbol{\Lambda}}^{NL},
\end{equation}
and the velocity field is the sum of the imposed (or undisturbed) velocity field and a perturbation about this,
\begin{equation}
	\mathbf{u} = \langle\mathbf{u}\rangle + \mathbf{u}'.\label{eq:VelPert}
\end{equation}
The undisturbed polymer stress, ${\boldsymbol{\Pi}}^U$, is obtained by using the undisturbed velocity, $\langle\mathbf{u}\rangle$ (equation \eqref{eq:UndisturbedStrainRate}), in the polymer constitutive model given by equations \eqref{eq:constitutive2} and \eqref{eq:Configuration}. The linearized polymer stress, ${\boldsymbol{\Pi}}^L$, is given by,
\begin{equation}
	\boldsymbol{\Pi}^{L} = \begin{cases}
		\frac{1}{De} \Big(f^{U} {\boldsymbol{\Lambda}}^{L} + \frac{\text{tr}({\boldsymbol{\Lambda}}^{L})(f^{U})^2}{L^2} \boldsymbol{\Lambda}^{U}\Big), & \text{FENE-P} \\
		\frac{1}{De} \boldsymbol{\Lambda}^{L}, & \text{Giesekus},
	\end{cases}\label{eq:LinPolyStress}
\end{equation}
where the linearized configuration, $\boldsymbol{\Lambda}^{L}$, is governed by,
\begin{eqnarray}
	\begin{split}
		\frac{\partial {\boldsymbol{\Lambda}}^{L}}{\partial t} +& \langle\mathbf{u}\rangle \cdot \nabla {\boldsymbol{\Lambda}}^{L} - {\langle\nabla\mathbf{u}\rangle}^\text{T} \cdot {\boldsymbol{\Lambda}}^{L} - {\boldsymbol{\Lambda}}^{L} \cdot \langle\nabla\mathbf{u}\rangle - \nabla {\mathbf{u}'}^\text{T} \cdot \boldsymbol{\Lambda}^{U} - \boldsymbol{\Lambda}^{U} \cdot \nabla \mathbf{u}' = \\
		&-\begin{cases}
			\frac{1}{De} \Big(f^{U} {\boldsymbol{\Lambda}}^{L} + \frac{\text{tr}({\boldsymbol{\Lambda}}^{L})(f^{U})^2}{L^2} \boldsymbol{\Lambda}^{U}\Big), & \text{FENE-P} \\
			\frac{1}{De} ((1 - 2\alpha) {\boldsymbol{\Lambda}}^{L} - \alpha ({\boldsymbol{\Lambda}}^{L} \cdot {\boldsymbol{\Lambda}}^{U} + {\boldsymbol{\Lambda}}^{U} \cdot {\boldsymbol{\Lambda}}^{L})), & \text{Giesekus}
		\end{cases}
	\end{split}\label{eq:LinPolyConfig}
\end{eqnarray}
Following the derivation of \cite{koch2016stress} and \cite{SteadyStatePaper}, one obtains that the ensemble average of $\boldsymbol{\Lambda}^{L}$ and hence $\boldsymbol{\Pi}^{L}$ is zero. This procedure is described as follows: after taking the ensemble average, the above equation can be recast as a linear operator acting on $\langle\boldsymbol{\Lambda}^{L}\rangle$ (the ensemble average of $\boldsymbol{\Lambda}^{L}$) with zero forcing. Zero forcing occurs because the ensemble average of $-\nabla {\mathbf{u}'}^\text{T} \cdot \boldsymbol{\Lambda}^{U} - \boldsymbol{\Lambda}^{U} \cdot \nabla \mathbf{u}'$ is zero. The initial condition is $\langle\boldsymbol{\Lambda}^{L}\rangle = 0$ because initially, the polymers are unstretched and hence have zero stress everywhere. Therefore, $\langle\boldsymbol{\Lambda}^{L}\rangle = 0$ and $\langle\boldsymbol{\Pi}^{L}\rangle = 0$ at all times (leading to $\langle\hat{\boldsymbol{\Pi}}\rangle=\hat{\boldsymbol{\Pi}}^U+\langle\hat{\boldsymbol{\Pi}}^{NL}\rangle$). {The volume average of $\boldsymbol{\Lambda}^{L}$ and hence $\boldsymbol{\Pi}^{L}$, however, diverges logarithmically in the steady state as discussed in previous studies. The non-linear polymer stress, $\boldsymbol{\Pi}^{NL}$ decays faster than $r^{-3}$ in the far-field. In case of transient flows, such as the start-up extensional flow considered here, the divergence is observed beyond a certain time proportional to $De$ (as well as the model parameters $L$ and $1/\alpha$).} Thus, removing the linearized stress before approximating the ensemble average with the volume average enables us to express $c\langle\hat{\boldsymbol{\Pi}}\rangle$ with a converging expression that can be obtained by the solution of an imposed flow around an isolated particle,
\begin{equation}
	c\langle\hat{\boldsymbol{\Pi}}\rangle = c\hat{\boldsymbol{\Pi}}^U + c{\phi}\hat{\boldsymbol{\Pi}}^{PP}.
\end{equation}
The first term is the polymer stress in the absence of the particles, and the second term is the ensemble average of the nonlinear polymer stress, $\hat{\boldsymbol{\Pi}}^{NL}$, or the particle-induced polymer stress in the suspension, expressed as,
\begin{equation}
	\hat{\boldsymbol{\Pi}}^{PP} = \frac{1}{V_p} \int_{V_f + V_p} \text{d}V \hspace{0.1in} \hat{\boldsymbol{\Pi}}^{NL} = \frac{1}{V_p} \int_{V_f + V_p} \text{d}V \hspace{0.1in} \hat{\boldsymbol{\Pi}} - \hat{\boldsymbol{\Pi}}^{L} - \hat{\boldsymbol{\Pi}}^{U}.\label{eq:PIPSTotal}
\end{equation}
It arises due to the change in the polymer configuration by the presence of the particles and has contributions from both the particle and the fluid phase. Here, $\phi$ is the particle volume fraction, and $V_f$ and $V_p$ are the fluid and the particle volume.

\subsection{Small time estimates of rheology}
Through a regular expansion in $t$ of flow variables in the mass, momentum, and constitutive equations, \cite{jain2019extensional} obtained small time estimates for an Oldroyd-B fluid. These estimates are also valid for FENE-P and Giesekus equations, as initially, the polymers are in equilibrium and the effect of finite $L (< \infty)$ and $\alpha (> 0)$, respectively, is not exhibited in these models at small times. At the linear order in time, $t$, as in \cite{jain2019extensional}, one finds the particle-induced polymer stress, $\hat{\Pi}^{PP}$, to be zero and the total interaction stresslet, $\hat{S}^{PP}$, to be that due to the sphere in a Newtonian fluid with viscosity $c t / De$,
\begin{equation}
	\hat{\Pi}^{PP} \langle\boldsymbol{e}\rangle = \mathcal{O}(t^2), \hspace{0.2in} \hat{S}^{PP} \langle\boldsymbol{e}\rangle = 5 \langle\boldsymbol{e}\rangle \frac{ct}{De} + \mathcal{O}(t^2).\label{eq:SmallTime3}
\end{equation}
With the definitions,
\begin{equation}
	\hat{\text{S}}^{U} = \hat{\Pi}^U, V_p \hat{\text{S}}^{\text{Vol}} \langle\boldsymbol{e}\rangle = \hat{\text{\textbf{S}}}_\text{volume}(\boldsymbol{\Pi}, \boldsymbol{\Pi}^U), V_p \hat{\text{S}}^{\boldsymbol{\tau}\text{PI}} \langle\boldsymbol{e}\rangle = \hat{\text{\textbf{S}}}(\boldsymbol{\tau}^\text{PI}), \text{ and } V_p \hat{\text{S}}^{\boldsymbol{\Pi}} \langle\boldsymbol{e}\rangle = \hat{\text{\textbf{S}}}(\boldsymbol{\Pi}),
	\label{eq:StressetSplit}\end{equation}
the components of the interaction stresslet from the two decompositions are
\begin{equation}
	\hat{\text{S}}^{U} = \hat{\text{S}}^{\boldsymbol{\tau}\text{PI}} = \frac{2}{De} t + \mathcal{O}(t^2), \hspace{0.2in} \hat{\text{S}}^{\text{Vol}} = \hat{\text{S}}^{\boldsymbol{\Pi}} = \frac{3}{De} t + \mathcal{O}(t^2).\label{eq:SmallTime4}
\end{equation}
The small time equivalence between individual components of the two stresslet decompositions is fortuitous, for, in general, the individual components are not identical.

\section{Regular perturbation and semi-analytical methodology for rheology of particles in viscoelastic fluids with small polymer concentration}\label{sec:MethodSemiAnalytical}
This section outlines the semi-analytical method employed to derive the results discussed in section \ref{sec:DiluteRheology}. The approach integrates a regular perturbation expansion in polymer concentration ($c$), the method of characteristics, and a generalized reciprocal theorem.

A regular perturbation of the relevant flow variables in $c$ is expressed as: $\boldsymbol{\sigma} = \boldsymbol{\sigma}^0 + c\boldsymbol{\sigma}^1 + \mathcal{O}(c^2)$, $\boldsymbol{\tau} = \boldsymbol{\tau}^0 + c\boldsymbol{\tau}^1 + \mathcal{O}(c^2)$, ${p} = {p}^0 + c{p}^1 + \mathcal{O}(c^2)$, $\mathbf{u} = \mathbf{u}^0 + c\mathbf{u}^1 + \mathcal{O}(c^2)$, $\boldsymbol{\Lambda} = \boldsymbol{\Lambda}^0 + c\boldsymbol{\Lambda}^1 + \mathcal{O}(c^2)$, $\boldsymbol{\Lambda}^L = \boldsymbol{\Lambda}^{0L} + c\boldsymbol{\Lambda}^{1L} + \mathcal{O}(c^2)$, $\boldsymbol{\Pi} = \boldsymbol{\Pi}^0 + c\boldsymbol{\Pi}^1 + \mathcal{O}(c^2)$, and $\boldsymbol{\Pi}^L = \boldsymbol{\Pi}^{0L} + c\boldsymbol{\Pi}^{1L} + \mathcal{O}(c^2)$. At the leading order in $c$, momentum and mass conservation are the Newtonian flow equations, as observed from equations \eqref{eq:MassMomentum}-\eqref{eq:constitutive1} or from equations \eqref{eq:MassMomentumNewt}-\eqref{eq:NonNewtonianFields}. Thus, at the leading order, the velocity is that around a force- and torque-free sphere suspended in a uniaxial extensional flow of a Newtonian fluid,
\begin{eqnarray}
	&u_i^0 = \begin{cases}
		\langle{e_{ij}}\rangle r_j + \frac{5}{2} \Big(\frac{1}{r^7} - \frac{1}{r^5}\Big) \langle{e_{jk}}\rangle r_j r_k r_i - \frac{1}{r^5} \langle{e_{ij}}\rangle r_j, & r \ge 1, \\
		0, & r < 1,
	\end{cases}\label{eq:u_field}
\end{eqnarray}
where $\langle{e_{ij}}\rangle$ is defined in equation \eqref{eq:UndisturbedStrainRate} \citep{leal2007advanced}. We obtain the transient extensional rheology of a dilute suspension in a dilute polymeric fluid up to $\mathcal{O}(c)$ by using just the velocity field of equation \eqref{eq:u_field}. In equations \eqref{eq:constitutive2} and \eqref{eq:Configuration}, $\mathbf{u}^0$ drives $\boldsymbol{\Lambda}^0$ and $\boldsymbol{\Pi}^0$. The undisturbed polymer stress $c\boldsymbol{\Pi}^U$ is obtained from equations \eqref{eq:constitutive2} and \eqref{eq:Configuration} for the FENE-P model with $\langle\nabla\mathbf{u}\rangle = \langle\boldsymbol{e}\rangle$ and truncates at the first order in $c$. We obtain the leading-order linearized polymer stress, $\boldsymbol{\Lambda}^{0L}$ and $\boldsymbol{\Pi}^{0L}$, from $\mathbf{u} = \mathbf{u}^0$ and $\langle\mathbf{u}\rangle = \mathbf{r} \cdot \langle\boldsymbol{e}\rangle$ by using equations \eqref{eq:VelPert} to \eqref{eq:LinPolyConfig}. Using $\boldsymbol{\Lambda}^0$ and $\boldsymbol{\Pi}^{0L}$, we obtain the leading-order components of the extensional viscosity required in equation \eqref{eq:ExtVisc},
\begin{equation}
	\mu_\text{ext} = 1 + 2.5\phi + 0.5c(\hat{\Pi}^U + {\phi}(\hat{\Pi}^{0PP} + \hat{\text{S}}^{0PP} + \mathcal{O}(c))),\label{eq:ExtVisc1}
\end{equation}
where,
\begin{equation}
	\hat{\Pi}^{0PP} \langle\boldsymbol{e}\rangle = \frac{1}{V_p} \int_{V_f + V_p} \text{d}V \hspace{0.1in} \hat{\boldsymbol{\Pi}}^0 - \hat{\boldsymbol{\Pi}}^{0L} - \hat{\boldsymbol{\Pi}}^{U}, \text{ and, }
	\hat{\text{S}}^{0PP} = \hat{\text{S}}^{U} + \hat{\text{S}}^{0\text{Vol}} = \hat{\text{S}}^{0\boldsymbol{\tau}\text{PI}} + \hat{\text{S}}^{0\boldsymbol{\Pi}},\label{eq:StressetSplit0}
\end{equation}
and,
\begin{equation}
	\hat{\text{S}}^{U} = \hat{\Pi}^U, V_p \hat{\text{S}}^{0\text{Vol}} \langle\boldsymbol{e}\rangle = \hat{\text{\textbf{S}}}_\text{volume}(\boldsymbol{\Pi}^0; \boldsymbol{\Pi}^U), V_p \hat{\text{S}}^{\boldsymbol{\tau}\text{0PI}} \langle\boldsymbol{e}\rangle = \hat{\text{\textbf{S}}}(\boldsymbol{\tau}^\text{PI}), \text{ and } V_p \hat{\text{S}}^{0\boldsymbol{\Pi}} \langle\boldsymbol{e}\rangle = \hat{\text{\textbf{S}}}(\boldsymbol{\Pi}^0).\label{eq:StressetSplit1}
\end{equation}
Obtaining $\boldsymbol{\tau}^{0\text{PI}}$ requires a solution of the $\mathcal{O}(c)$ momentum and mass conservation equation obtained from equation \eqref{eq:NonNewtProblem}. This is a partial differential equation and requires sophisticated numerical discretization such as finite difference, finite volume, finite element, or spectral methods. We can circumvent this numerical calculation for the $c \ll 1$ regime if only $\hat{\text{S}}^{0\boldsymbol{\tau}\text{PI}}$ and not the entire field $\boldsymbol{\tau}^{0\text{PI}}$ is required. This is because from $\hat{\boldsymbol{\Pi}}^0$ and $\hat{\boldsymbol{\Pi}}^{0U}$ (already known without evaluating $\boldsymbol{\tau}^{0\text{PI}}$), we can obtain $\hat{\text{S}}^{U}$, $\hat{\text{S}}^{0\text{Vol}}$, and $\hat{\text{S}}^{0\boldsymbol{\Pi}}$, thus evaluating $\hat{\text{S}}^{0\boldsymbol{\tau}\text{PI}}$ (from equation \eqref{eq:StressetSplit0}).

Similar to \cite{koch2016stress} and \cite{SteadyStatePaper}, we use the method of characteristics to obtain $\boldsymbol{\Lambda}^0$ and $\boldsymbol{\Lambda}^{0L}$ from $\mathbf{u}^0$ and $\langle\mathbf{u}\rangle$. The leading-order partial differential equations \eqref{eq:Configuration} and \eqref{eq:LinPolyConfig} are converted into ordinary differential equations along the streamlines of $\mathbf{u}^0$ and $\langle\mathbf{u}\rangle$, respectively. In our simulations, we generate these streamlines or the "mesh" through numerical integration. To capture regions with large gradients in $\boldsymbol{\Lambda}^0$ and $\boldsymbol{\Lambda}^{0L}$, we employ a higher streamline density or ``mesh" resolution near the surface of the particle and along the extensional axis. The streamlines are initiated at a distance of 5000 particle radii upstream and extend a distance of 50000 particle radii downstream from the center of the particle. Additionally, the streamlines extend up to about 500 particle radii at a 45$^\circ$ angle from the extensional axis, covering a significant region around the sphere.

Previous studies using this technique, such as \cite{koch2016stress} and \cite{SteadyStatePaper}, were concerned with the steady-state values of $\boldsymbol{\Lambda}^0$ and $\boldsymbol{\Lambda}^{0L}$. Here, we require the transient values, and the following augmentations to the numerical implementation are necessary. Initially, $\boldsymbol{\Lambda}^0 = \boldsymbol{\delta}$, and $\boldsymbol{\Lambda}^{0L} = 0$ everywhere in the domain. This implies polymers to be in a stress-free or equilibrium state, as expected at the beginning of the startup of a uniaxial extensional flow. In the following, we refer to the calculation of $\boldsymbol{\Lambda}^0$ and streamlines of $\mathbf{u}^0$, but the procedure is similar for $\boldsymbol{\Lambda}^{0L}$ and streamlines of $\langle\mathbf{u}\rangle$. We define the locations, $s(t_n)$, along each streamline/characteristic where the configuration at time step $t_n$, $\boldsymbol{\Lambda}^0(s(\mathbf{x}, t_n); t_n)$, is available. We perform integration along a particular streamline for a specified small time step $\Delta t = t_{n+1} - t_n$ to obtain the configuration at the next time step, $\boldsymbol{\Lambda}^0(s(\mathbf{x}, t_{n+1}); t_{n+1})$. The convected locations remain on the same streamline as the Stokes velocity field is temporally constant. Therefore, using the information only on the concerned streamline at time step $t_{n+1}$, we interpolate $\boldsymbol{\Lambda}^0(s(\mathbf{x}, t_{n+1}); t_{n+1})$ along each streamline to obtain the values at the starting locations, $\boldsymbol{\Lambda}^0(s(\mathbf{x}, 0); t_{n+1}) \leftarrow \boldsymbol{\Lambda}^0(s(\mathbf{x}, t_{n+1}); t_{n+1})$. We specify boundary conditions at the point on each streamline which is farthest upstream. At that location, the configuration at each time is the spatially constant undisturbed configuration, $\boldsymbol{\Lambda}^{0U}(t_{n+1})$, consistent with the current time, $t_{n+1}$. Since the calculation on each streamline is independent, the process described above to calculate $\boldsymbol{\Lambda}^0$ and $\boldsymbol{\Lambda}^{0L}$ is massively parallelizable. This parallelizability and the fact that only a simple ODE integration and interpolation are needed make this process computationally efficient.

\section{Methodology for direct numerical simulations of viscoelastic fluid with an arbitrary polymer concentration}\label{sec:DNSMethodology}
This section describes the numerical method utilized to obtain the results presented in section \ref{sec:ConcentratedRheology}. The method involves direct numerical simulations of the equations governing the fluid's mass and momentum, as well as the polymer constitutive equation \eqref{eq:Configuration}, which are discussed in detail in \cite{NumericalMethodPaper} along with validation in several flow scenarios. Since we know the Newtonian component of the velocity and pressure fields (solutions of equation \eqref{eq:MassMomentumNewt}) analytically, we solve the polymer induced mass and momentum equations \eqref{eq:NonNewtProblem} for the fluid. Here, we provide a summary of the method, outline the additional calculations required, and present the validation for the particle-induced polymer stress (PIPS), as defined in section \ref{sec:Formulation} and elaborated upon in appendix \ref{sec:PIPS}.

\begin{figure}
	\centering
	\subfloat{\includegraphics[width=0.33\textwidth]{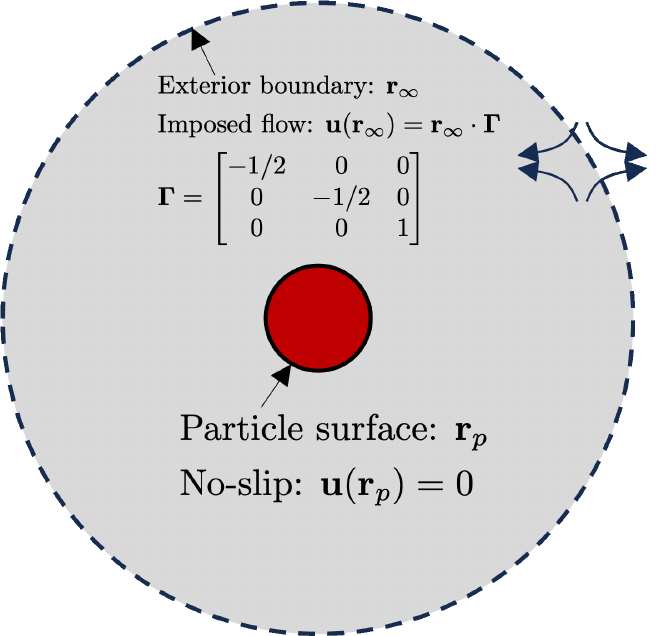}\label{fig:ComputationalDomain3D}}\hfill
	\subfloat{\includegraphics[width=0.33\textwidth]{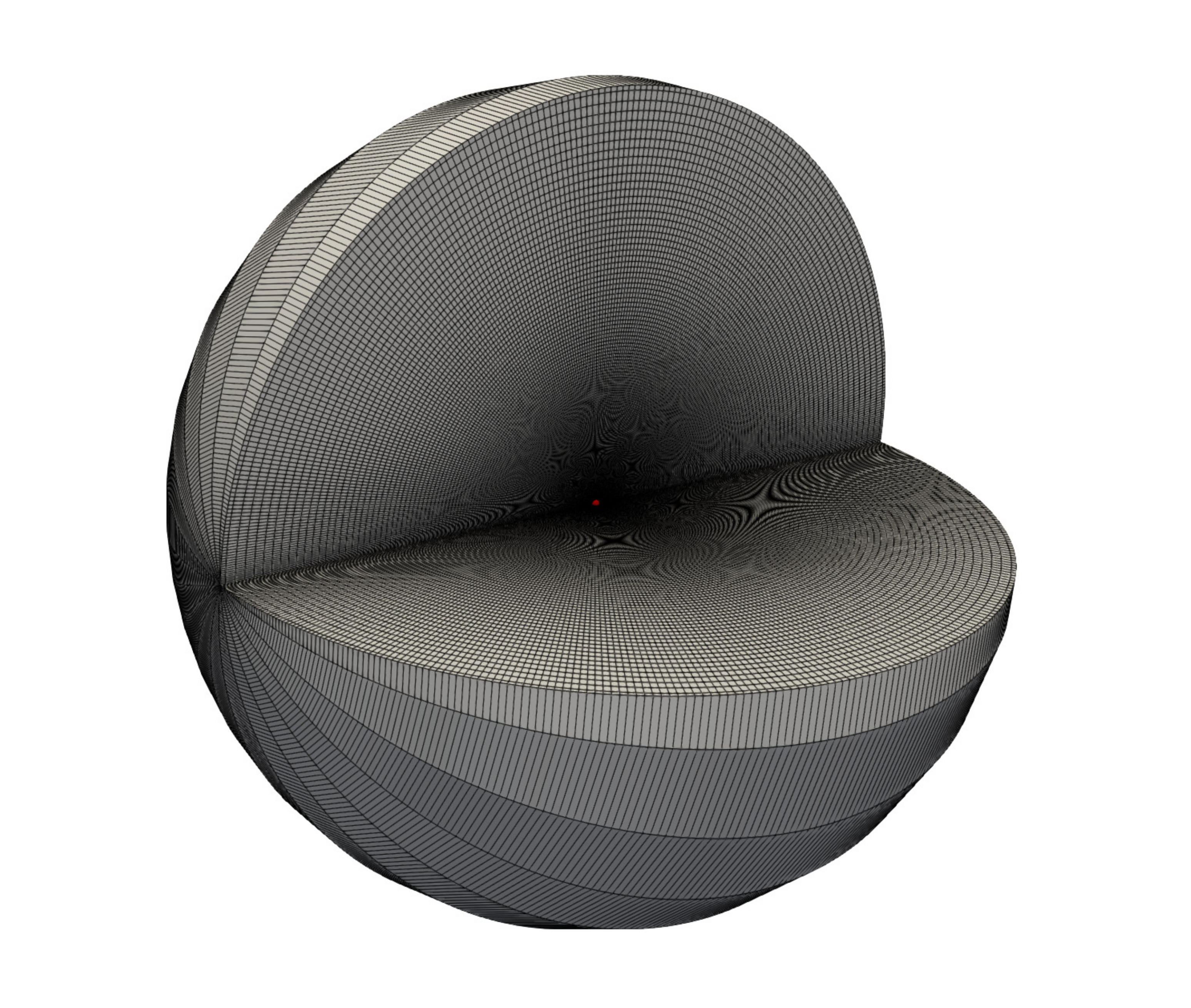}\label{fig:WholeDomain}}\hfill
	\subfloat{\includegraphics[width=0.33\textwidth]{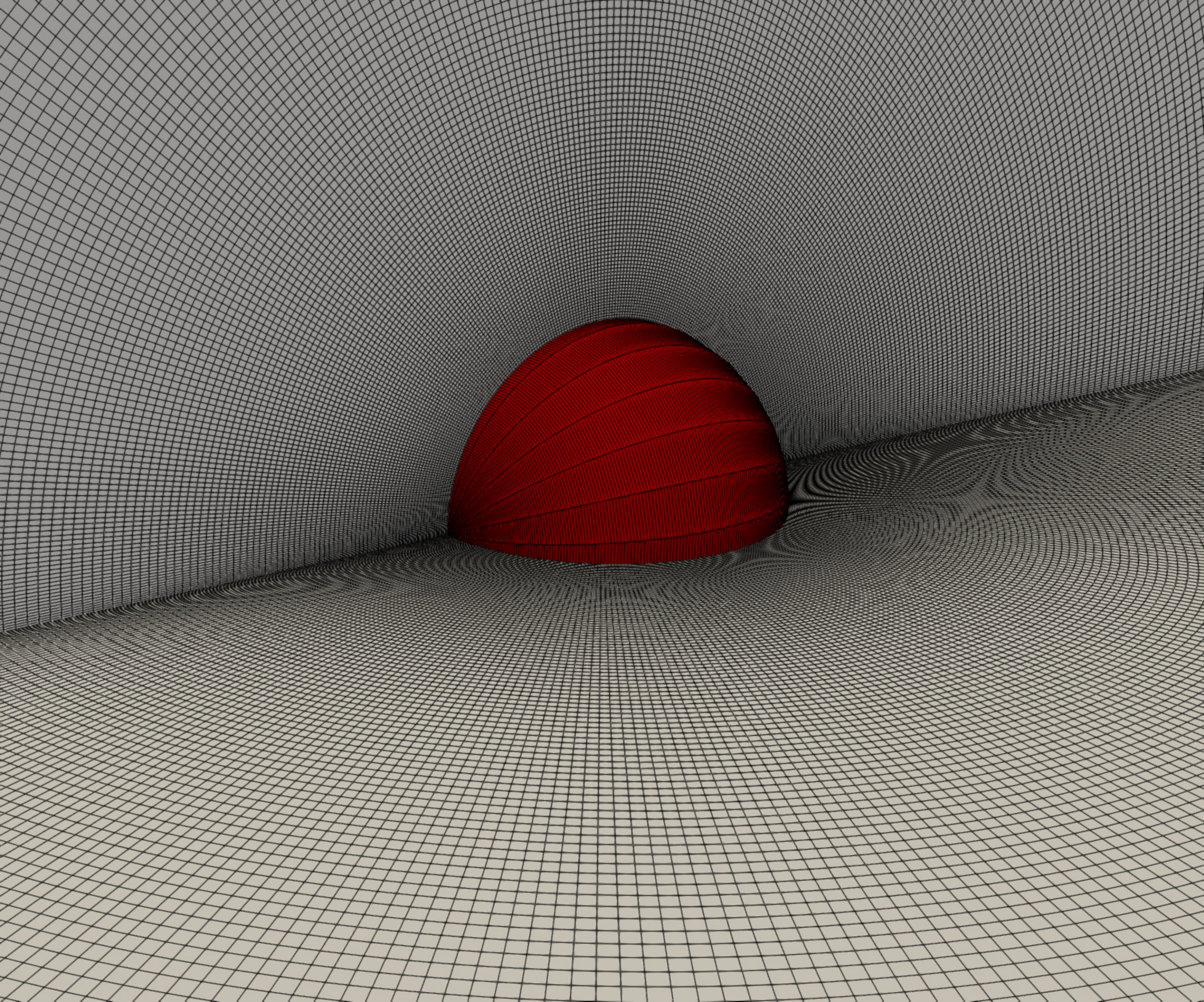}\label{fig:ZoomedDomain}}
	\caption{Left panel: A schematic representation of the computational domain (2D slice). The middle panel displays the discretized computational domain in its entirety, while the right panel provides a zoomed-in view of a region near the particle surface (red). A higher mesh density is utilized in proximity to the particle surface and along the extensional axis to enhance accuracy in these critical areas. \label{fig:CompDomain2}}
\end{figure}

Our numerical solver is written in prolate spheroidal coordinates, where the particle is one of the coordinate surfaces. Hence, the shape of any prolate spheroid is exactly modeled, and the no-slip velocity boundary conditions are imposed here. In the current work, a prolate spheroid with an aspect ratio of 1.001 represents a sphere. We varied the aspect ratios in the range 1.01 and 1.0001 and found the results insensitive to this change. The schematic of the computational domain is shown in the left panel, and a rendering of the discretized grid is displayed in the middle and right panels of figure \ref{fig:CompDomain2}. The computational domain is bounded by a no-slip particle surface on the inside and a far-field outer boundary where a uniaxial extensional flow is imposed as the velocity boundary condition. The polymer conformation tensor, $\boldsymbol{\Lambda}$, is governed by a first-order hyperbolic equation \eqref{eq:Configuration}. Therefore, we only need to apply one boundary condition at the locations where the characteristics of this equation (velocity streamlines) enter the computational domain. Thus, at the incoming streamlines on the outer boundary, the undisturbed configuration, $\boldsymbol{\Lambda}^U$, is used as the boundary condition. Finite difference spatial discretization is used to represent the fluid mass and momentum conservation as well as the polymer constitutive equations. Fourth, sixth, and eighth-order accurate schemes are used in the radial, polar, and azimuthal directions, respectively.  The Schur complement method \citep{furuichi2011development} is used to solve the quasi-steady mass and momentum equations induced by the polymers (equation \eqref{eq:NonNewtProblem}). The polymer constitutive equations are expressed in log-conformation form \citep{fattal2004constitutive, hulsen2005flow}. The benefit of evolving the logarithm of $\boldsymbol{\Lambda}$ (as first illustrated by \cite{fattal2004constitutive}) is that it preserves the positive definiteness of $\boldsymbol{\Lambda}$ and prevents numerical instabilities at large $De$.
	
Higher-order upwinding central schemes (HOUC) of \cite{nourgaliev2007high} are used to stabilize the convective derivative in the polymer equations. In the FENE-P model, to ensure that polymer stretch, $\text{tr}(\boldsymbol{\Lambda})$, at any location in the flow is upper bounded by the prescribed maximum extensibility, $L$, the numerical technique of \cite{richter2010simulations} is used to evolve a bespoke equation for $f = {L^2}/({L^2 - \text{tr}(\boldsymbol{\Lambda})})$ that ensures $f > 0$ at all locations and times. A second-order accurate Crank-Nicholson method is used to time integrate the polymer constitutive equations.

\subsection{Additions to the numerical method of \cite{NumericalMethodPaper} to calculate PIPS}
The method for the equations governing the fluid's mass and momentum (equation \eqref{eq:MassMomentum}) and the polymer constitutive relations (equations \eqref{eq:constitutive2} and \ref{eq:Configuration}) was considered in \cite{NumericalMethodPaper}. The results were validated for several examples considering different particle aspect ratios, viscoelastic fluid models, levels of fluid inertia, and imposed flows. The only missing piece required for rheological calculations is the evaluation of the particle-induced polymer stress, or PIPS. As discussed in appendix \ref{sec:PIPS}, to obtain the ensemble-averaged PIPS using a volume average, the linearized polymer stress, ${\boldsymbol{\Pi}}^L$, must be removed from the extra polymer stress (relative to the undisturbed). This requires solving equations \eqref{eq:LinPolyStress} and \eqref{eq:LinPolyConfig} governing the evolution of ${\boldsymbol{\Pi}}^L$.

The hyperbolic equation \eqref{eq:LinPolyConfig} governing $\boldsymbol{\Lambda}^{L}$ is evolved alongside and in a similar manner as the polymer constitutive equation \eqref{eq:Configuration} described in \cite{NumericalMethodPaper}. The linearized polymer configuration, $\boldsymbol{\Lambda}^{L}$, is defined everywhere and hence must be evaluated inside the particle in addition to the fluid region outside the particle. The only boundary conditions for this are in the far field, i.e., $\boldsymbol{\Lambda}^{L}_\text{far} = \boldsymbol{0}$. A log-conformation formulation is not required or possible for $\boldsymbol{\Lambda}^{L}$ as it is not a positive definite tensor. We do not experience any stability issues arising during the evolution of equation \eqref{eq:LinPolyConfig}. The computational domain for this equation extends to the region within the particle. As seen from its governing equation, $\boldsymbol{\Lambda}^{L}$ is driven by the undisturbed velocity field $\langle\mathbf{u}\rangle = \mathbf{r} \cdot \langle\boldsymbol{e}\rangle$ and the perturbation of the local velocity from the undisturbed, $\mathbf{u}'$. The spatial discretization required in the fluid domain for the $\boldsymbol{\Lambda}^{L}$ equation is similar to that for $\boldsymbol{\Lambda}$. Inside the particle, $\mathbf{u}' = -\langle\mathbf{u}\rangle$, and hence only the convective term in equation \eqref{eq:LinPolyConfig} needs to be discretized. This is done in a similar fashion as the discretization of the convective term in the fluid domain, i.e., using the HOUC schemes of \cite{nourgaliev2007high}. The additional implementation and computational time for evolving $\boldsymbol{\Lambda}^{L}$ are minimal. We obtain the linearized stress, $\boldsymbol{\Pi}^{L}$, from equation \eqref{eq:LinPolyStress}, and now we have all the ingredients necessary to evaluate the PIPS or $\hat{\boldsymbol{\Pi}}^{PP}$ from equation \eqref{eq:PIPSTotal}. The volume integration is performed separately in the fluid and particle domain using Gaussian quadratures to obtain $\hat{\boldsymbol{\Pi}}^{PP} = \hat{\Pi}^{PP} \langle\boldsymbol{e}\rangle$ defined in equation \eqref{eq:PIPSTotal}. From the discussion of the low $c$ regime in section \ref{sec:ResultsSemiAnalytical}, we find that the sphere's influence on the polymer configuration field extends to large distances downstream of the particle where the extra stretch of the polymers is confined to regions close to the extensional axis. This region becomes thinner in the far field. Therefore, to better resolve it, we cluster more points around the extensional axis.

Several parameters introduced by \cite{NumericalMethodPaper} dictate the shape of the computational domain and the mesh density. The radius of the sphere (or the minor radius of a non-spherical prolate particle) is fixed at 1, while the outer boundary radius is denoted as $r_\text{far}$. As mentioned earlier, the equations are solved in prolate spheroidal coordinates. Consequently, the structured grid within the computational domain is defined by $N_r$, $N_\theta$, and $N_\phi$, which represent the number of mesh points in the spheroidal, hyperboloidal, and azimuthal directions around a spheroid. For a sphere, these correspond to the number of mesh points in the radial, polar, and azimuthal directions. To account for the rapid changes in polymer configuration near the particle surface and along the extensional axis, a non-uniform grid is employed, clustering more mesh points in these critical regions (as shown in the middle panel of figure \ref{fig:CompDomain2}). This clustering is controlled by parameters $c_1$ (for the radial direction) and $c_2$ (for the polar direction), as defined in \cite{NumericalMethodPaper}, with values fixed at -1.0 and -1.5, respectively.
	
\subsection{Validation and grid convergence of PIPS}\label{sec:validation}
Figure 12 of \cite{NumericalMethodPaper} validates the interaction stresslet, $\hat{\text{S}}^{PP}$, for a sphere in uniaxial extensional flow. The stresslet, evaluated using direct numerical simulations (DNS), demonstrates strong agreement with results from our semi-analytical methodology in section \ref{sec:MethodSemiAnalytical} and the simulations conducted by \cite{jain2019extensional}. In this section, we focus on validating the particle-induced polymer stress (PIPS), $\hat{\Pi}^{PP}$, against these two sources.

\begin{figure}
	\centering
	\subfloat{\includegraphics[width=0.33\textwidth]{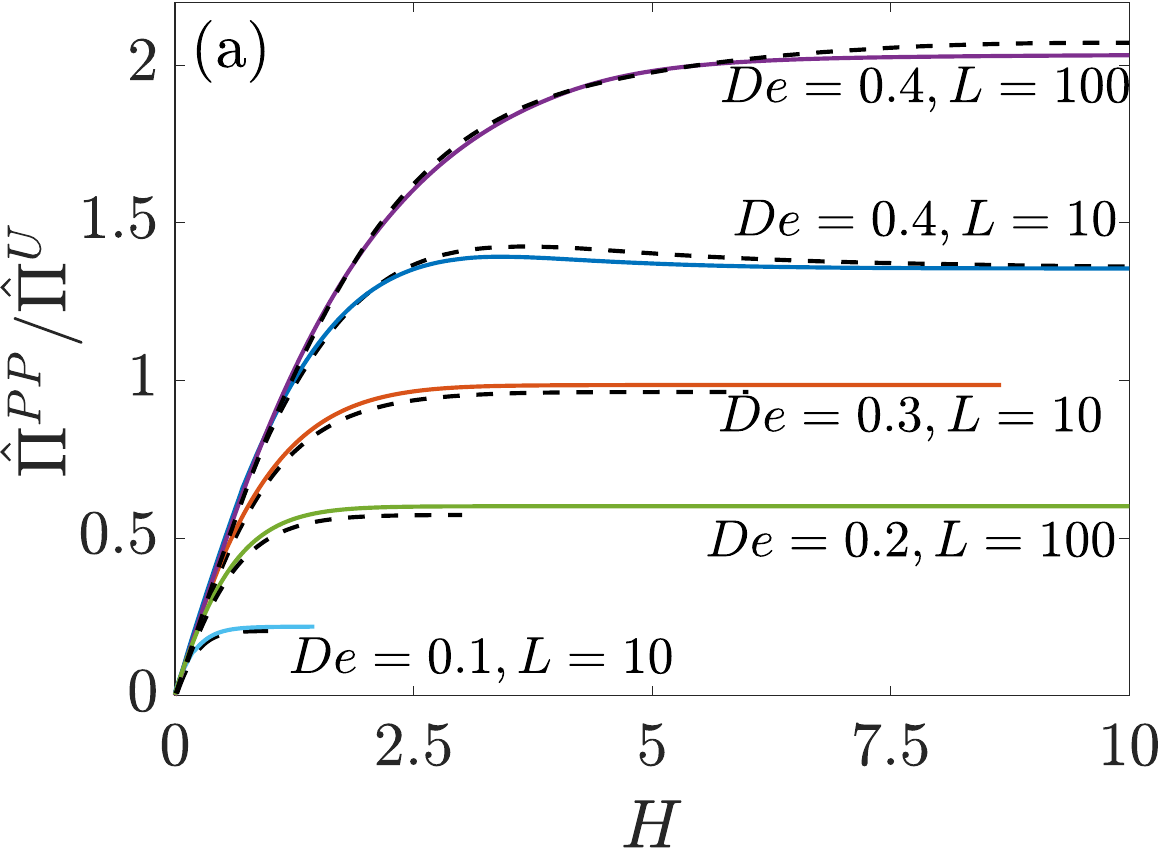}\label{fig:ValidationPIFSmallcSmallDe}} \hfill
	\subfloat{\includegraphics[width=0.33\textwidth]{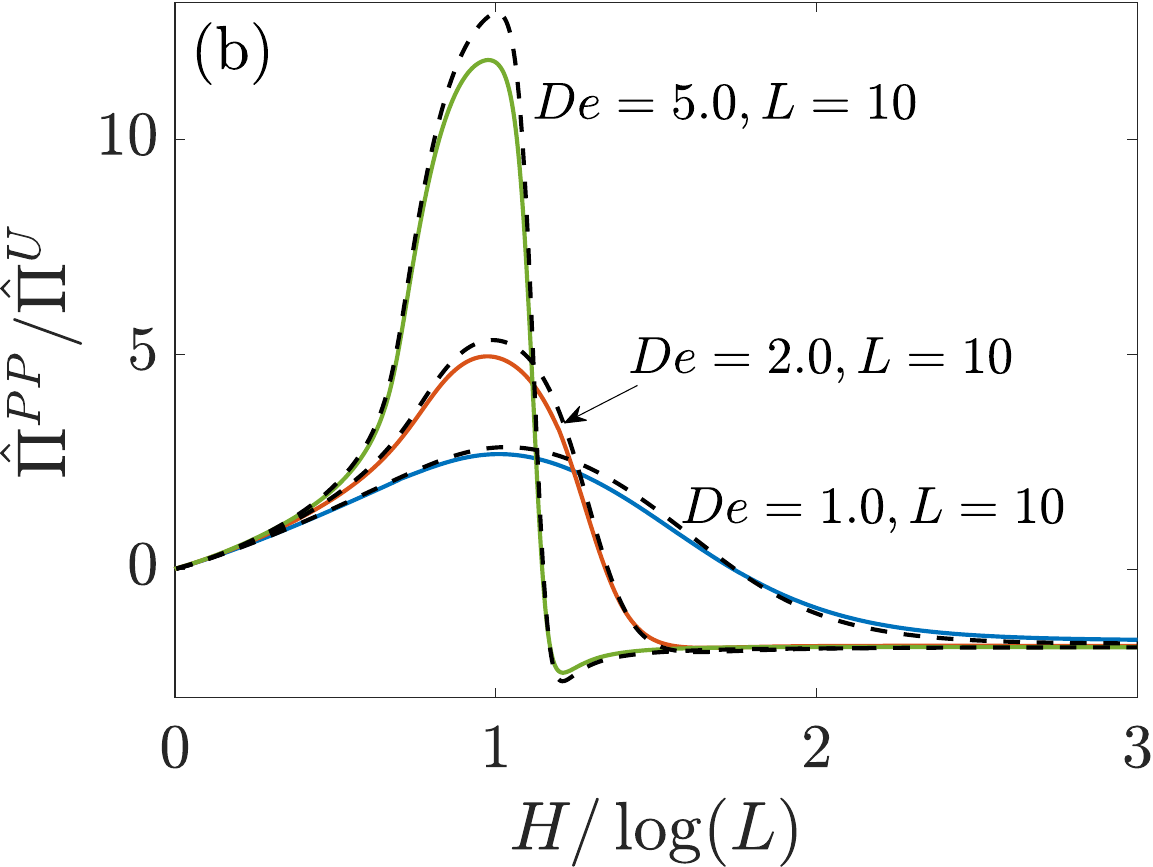}\label{fig:ValidationPIFSmallcLargeDeL10}} \hfill
	\subfloat{\includegraphics[width=0.33\textwidth]{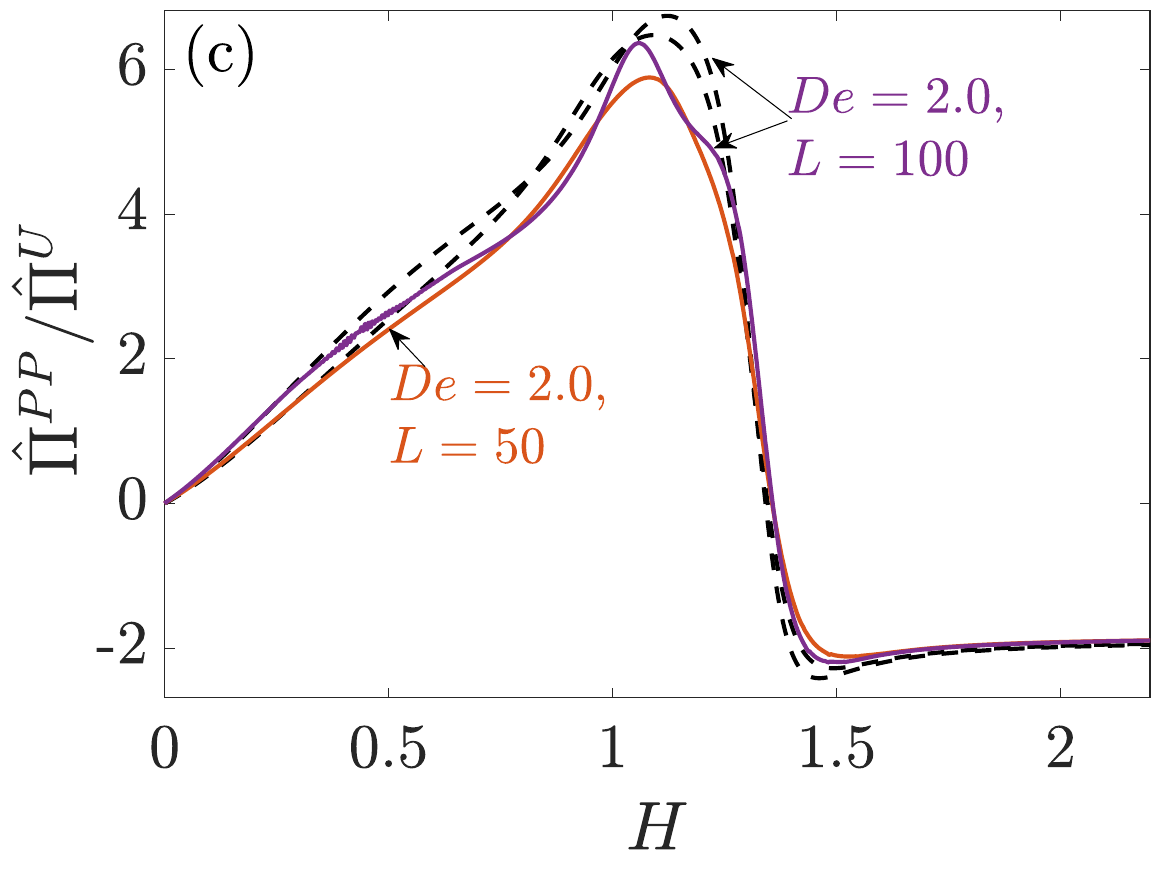}\label{fig:ValidationPIFSmallcLargeDeLargeL}}
	\caption{Comparison of the evolution of normalized particle-induced polymer stress (PIPS), $\hat{\Pi}^{PP}/\hat{\Pi}^U$, from the two methodologies described in sections \ref{sec:MethodSemiAnalytical} and \ref{sec:DNSMethodology}. The results from the direct numerical simulations (DNS, section \ref{sec:DNSMethodology}) at a polymer concentration of $c = 10^{-5}$ are shown with colored solid lines, while the results from the semi-analytical methodology (section \ref{sec:MethodSemiAnalytical}) are represented by dashed black lines.}
	\label{fig:Validationationc1em5}
\end{figure}

We start by validating the method using a polymer concentration of $c = 10^{-5}$ in the DNS. We then increased $c$ to $10^{-4}$ and observed similar results. The mesh sizes in these validation simulations varied, ranging from 200 to 350 in the radial direction, from 150 to 451 in the azimuthal direction, and from 21 to 25 in the polar direction. The domain size was set between $r_\text{far} = 100$ and 300. Convergence was achieved for mesh density, time step, and domain size in each case presented. In figure \ref{fig:Validationationc1em5}, we observe a good match between the semi-analytical (dashed curves) and DNS (solid curves) results across a wide range of $De$, $L$ and $H$. The small discrepancies are expected, as PIPS is calculated as a volume integral and we cannot achieve the extremely high resolution and large computational domain size of the semi-analytical methodology in our simulations.

As indicated by the results in section \ref{sec:U0Rheology}, the polymer stress increases with the Deborah number ($De$), polymer concentration ($c$), and polymer extensibility in FENE-P liquids ($L$, or equivalently the inverse of mobility in Giesekus fluids, $1/\alpha$). This translates to more rapid spatial variation in flow and polymer fields near the particle surface. Therefore, as these parameters are increased, the DNS require a finer mesh resolution to accurately resolve the polymer and fluid velocity fields. For some of the simulations reported in section \ref{sec:ResultsFinitec}, we increased $N_r$ to about 1500 and $N_\theta$ to about 1000 to ensure stable numerical simulations. Since the flow is axisymmetric, a relatively low value of $N_\phi = 25$ is sufficient. These equations are numerically stiff, and time step convergence is obtained for each numerically stable result, ensuring mesh size convergence.
		
\begin{figure}
	\centering
	\subfloat{\includegraphics[width=0.33\textwidth]{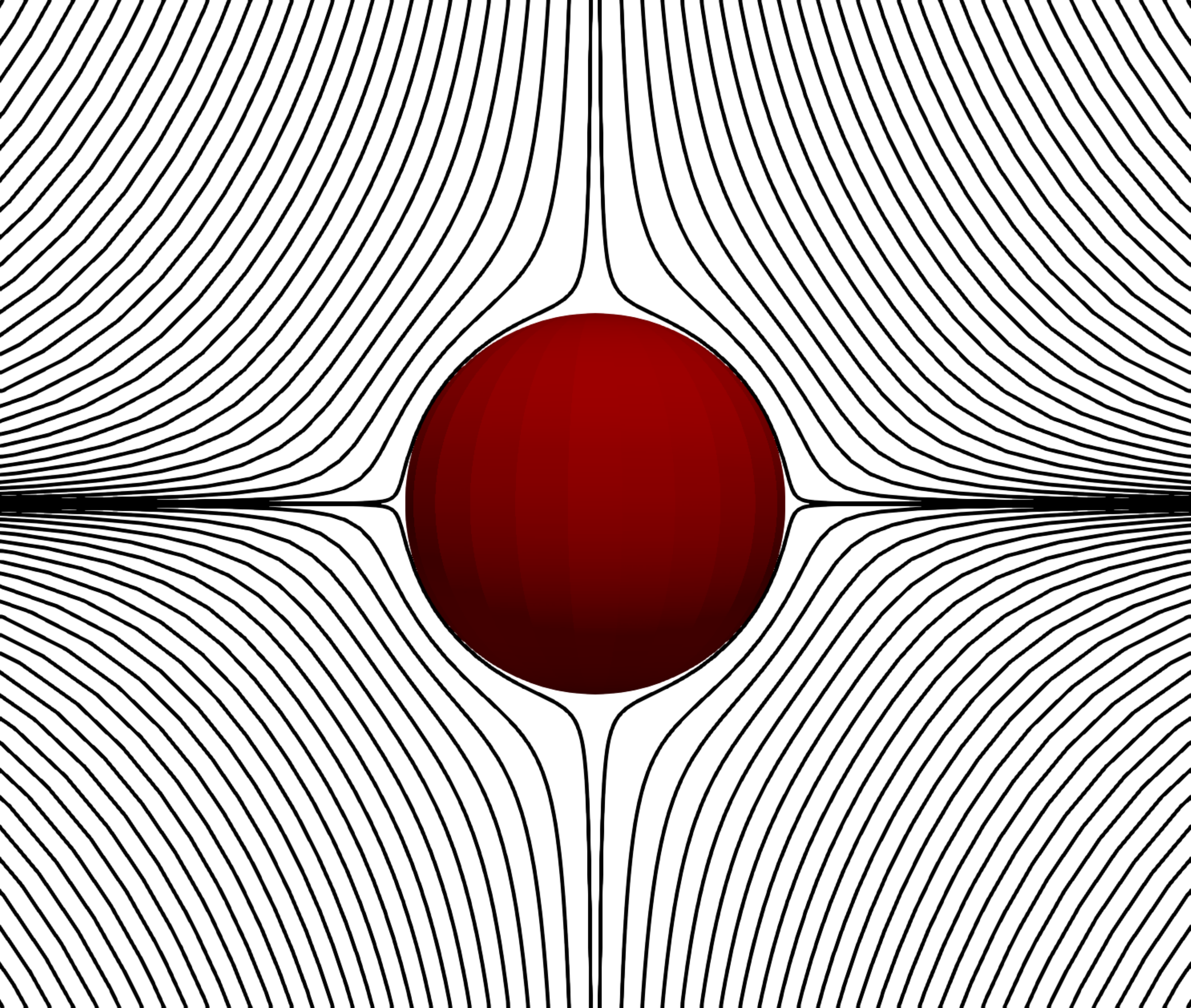}\label{fig:Streamlines}}\hspace{0.2in}
	\subfloat{\includegraphics[width=0.33\textwidth]{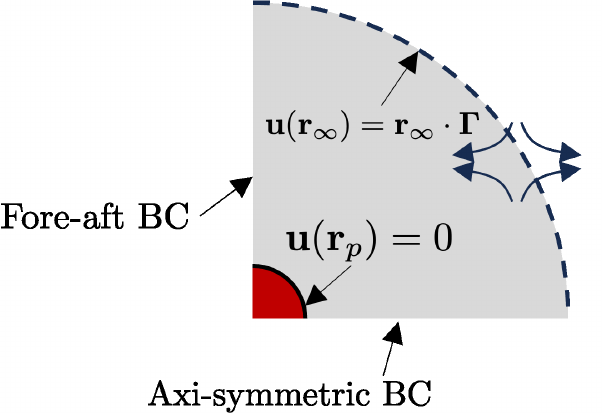}\label{fig:ComputationalDomainAxiSymmetric}}
	\caption{Left panel: Fore-aft symmetric flow in a compressional plane across the sphere (red). Right panel: A schematic illustrating the fore-aft and axisymmetric computational domain employed for some of the larger values of $c$, $L$, and $1/\alpha$ in section \ref{sec:ConcentratedRheology}. \label{fig:CompDomain}}
\end{figure}

The particle shape and imposed extensional flow are axisymmetric about the extensional axis. Furthermore, the flow is fore-aft symmetric, as shown by the left panel in figure \ref{fig:CompDomain}. Therefore, we can make the simulations $2N_\phi$ times faster by simulating a fore-aft and axisymmetric domain shown on the right panel of figure \ref{fig:CompDomain}. The numerical method of \cite{NumericalMethodPaper} is slightly altered by imposing a fore-aft symmetric and an axisymmetric boundary condition on the vertical and horizontal axes, respectively. As shown for a FENE-P liquid with $De = 1.0$, $c = 0.2$, and $L = 10$, the normalized PIPS, $\hat{\Pi}^{PP}/\hat{\Pi}^U$, from the axisymmetric method is indistinguishable from the original three-dimensional method.

\begin{figure}
	\centering	
	\includegraphics[width=0.4\textwidth]{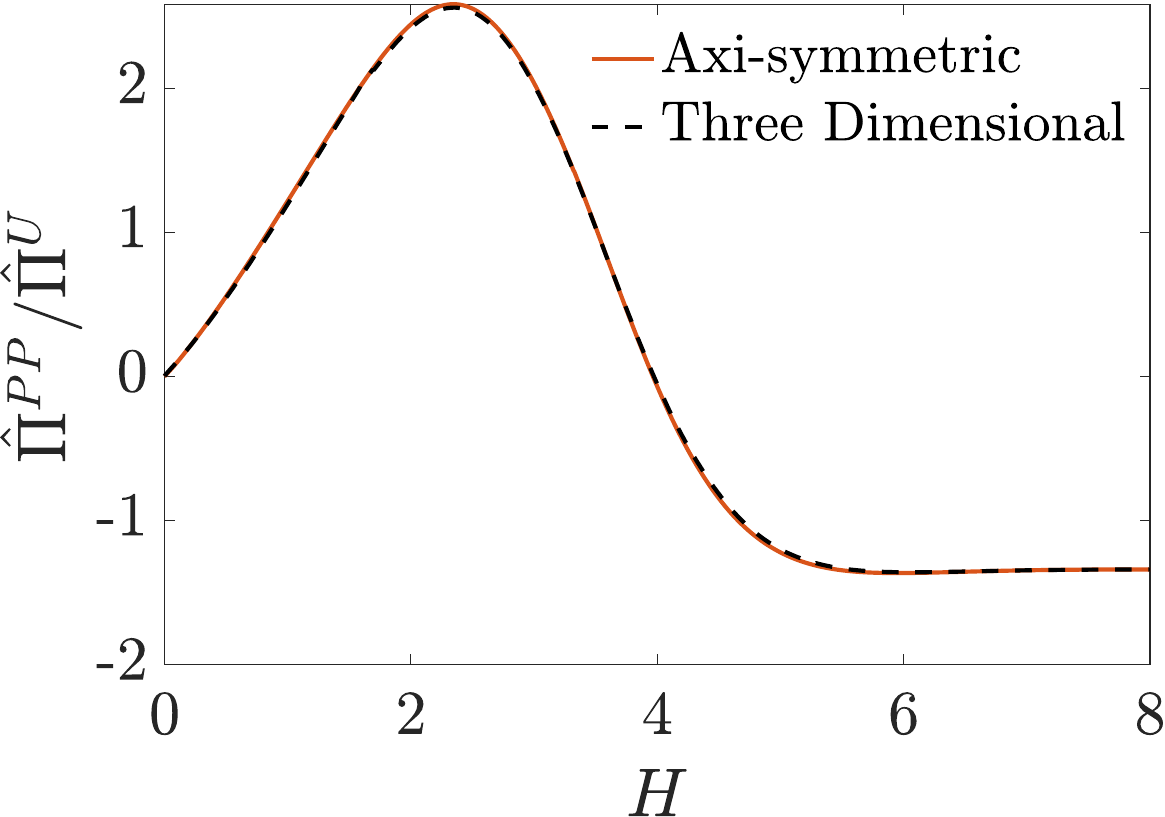}
	\caption{Three-dimensional versus axisymmetric calculation of normalized PIPS, $\hat{\Pi}^{PP}/\hat{\Pi}^U$, for $De = 1.0$, $c = 0.2$, $L = 10$ using $N_r = 500$ and $N_\theta = 351$.}
	\label{fig::ThreeDvsAxiSymmetric}
\end{figure}

\begin{figure}
	\centering	
	\subfloat{\includegraphics[width=0.48\textwidth]{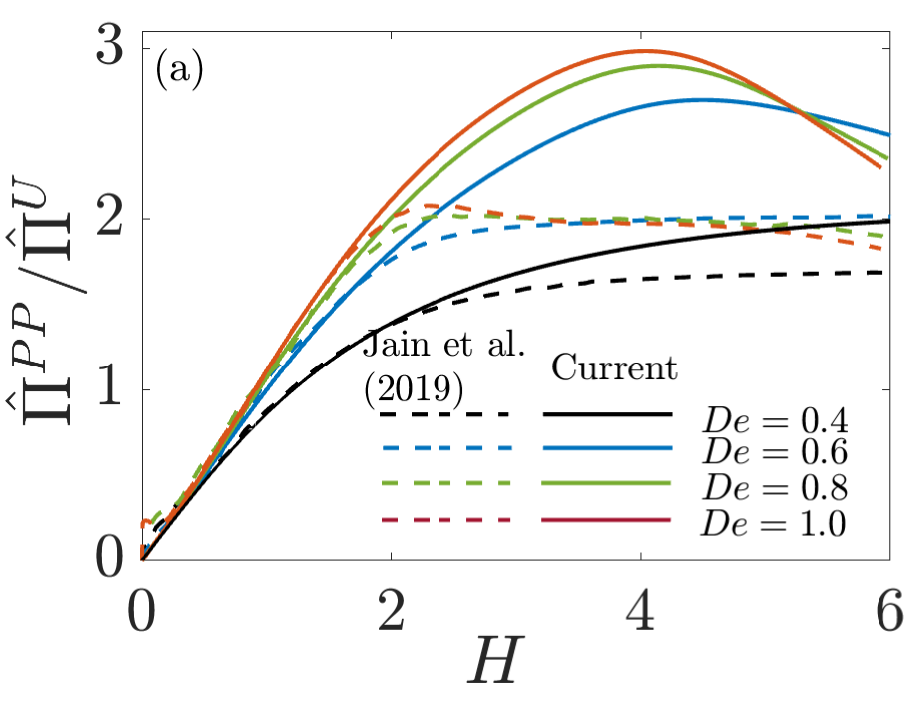}\label{fig:CompareShaqFENE}}\hfill
	\subfloat{\includegraphics[width=0.48\textwidth]{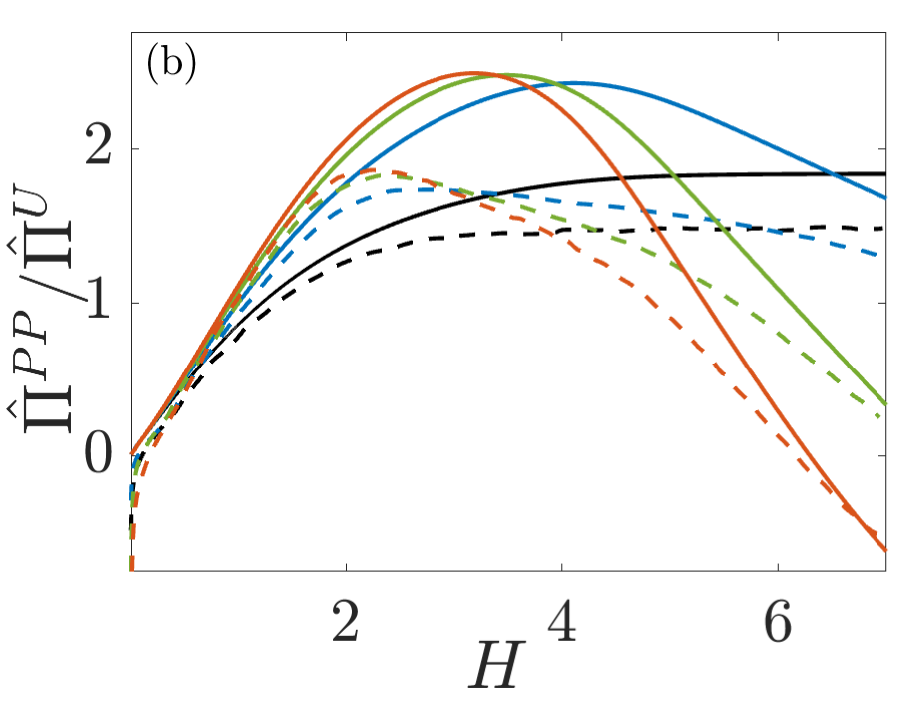}\label{fig:CompareShaqGiesekus}}
	\caption{Comparison of normalized PIPS, $\hat{\Pi}^{PP}/\hat{\Pi}^U$, from our calculations with that of \cite{jain2019extensional} across four different $De$ and $c = 0.471$ for (a) FENE-P liquids with $L = 100$ and (b) Giesekus liquids with $\alpha = 0.001$. Both figures share the same legend shown on the left figure.}
	\label{fig:Comparison with JainandShaqfeh}
\end{figure}

Comparison of $\hat{\Pi}^{PP}/\hat{\Pi}^U$, the PIPS normalized with the undisturbed stress, evaluated from our method described above with that from \cite{jain2019extensional} shown in figure \ref{fig:Comparison with JainandShaqfeh} shows a good agreement during the initial part of the transition. However, our simulations predict higher values thereafter for the FENE-P liquids (figure \ref{fig:CompareShaqFENE}) and during the intermediate period for Giesekus liquids (figure \ref{fig:CompareShaqGiesekus}). This discrepancy is likely due to the neglect of the linearized polymer stress, ${\boldsymbol{\Pi}}^L$, by \cite{jain2019extensional} as discussed in the remainder of this section.

As defined in equation \eqref{eq:PIPSTotal} and discussed in appendix \ref{sec:PIPS}, PIPS, $\hat{\Pi}^{PP}$, is the volume integral of $\hat{{\Pi}}^{NL} = \hat{{\Pi}} - \hat{{\Pi}}^{L} - \hat{{\Pi}}^{U}$ normalized with particle volume. {Due to the symmetry of uniaxial extensional flow and the sphere's shape, the PIPS integral has the same tensor symmetry as the imposed rate of strain tensor $\langle\boldsymbol{e}\rangle$ and thus we focus on the extensional (33) component of the various stress fields below. Here, $\hat{{\Pi}}$, $\hat{{\Pi}}^{U}$, $\hat{{\Pi}}^{L}$, and $\hat{{\Pi}}^{NL}$ refer to the 33 component of the stress tensors $\hat{\boldsymbol{\Pi}}$, $\hat{\boldsymbol{\Pi}}^{U}$, $\hat{\boldsymbol{\Pi}}^{L}$, and $\hat{\boldsymbol{\Pi}}^{NL}$ respectively.} The extra stress in the fluid domain created by the particle relative to the undisturbed or far-field from the particle is $\hat{{\Pi}} - \hat{{\Pi}}^{U}$, and \cite{jain2019extensional} use the particle volume normalized volume integral of this quantity to represent PIPS. In order to replace the ensemble-averaged suspension stress with a volume average, as previously discussed, the linearized stress, $\hat{\boldsymbol{\Pi}}^{L}$ (which has zero ensemble average) must be subtracted to ensure the convergence of the volume integral.

\begin{figure}
	\centering	
	\subfloat{\includegraphics[width=0.33\textwidth]{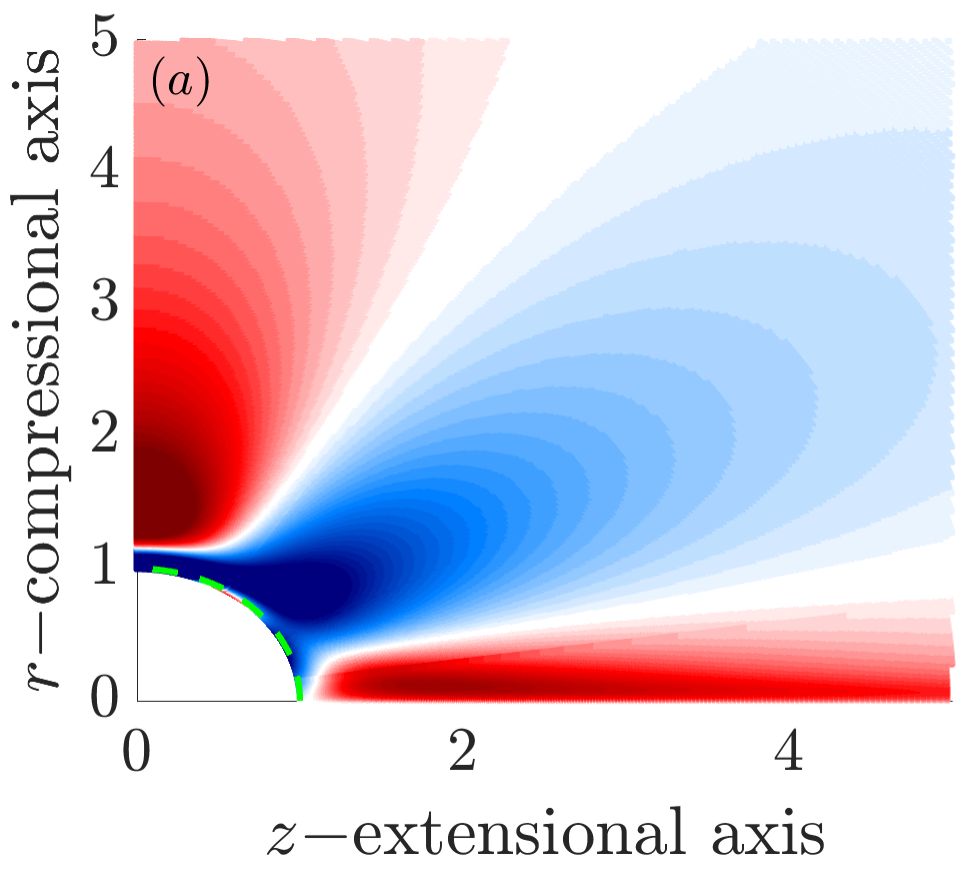}\label{fig:ActualStressDep4Grid14v7Zoomed}}
	\subfloat{\includegraphics[width=0.33\textwidth]{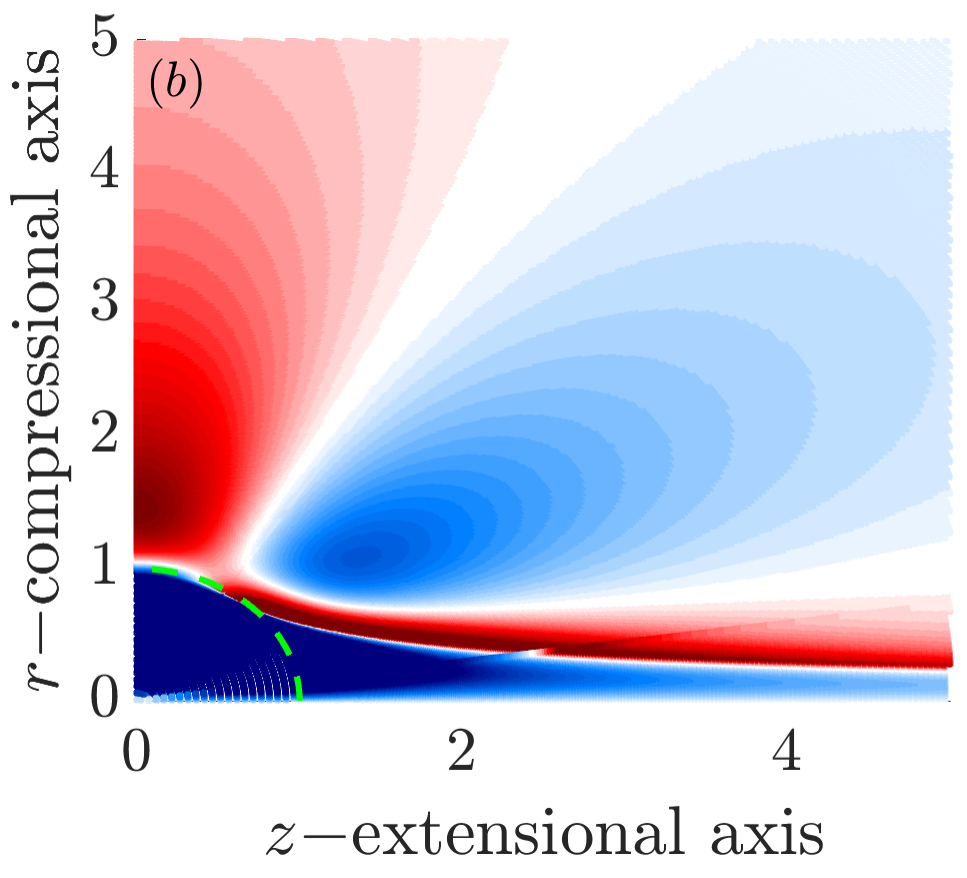}\label{fig:LinStressDep4Grid14v7Zoomed}}
	\subfloat{\includegraphics[width=0.334\textwidth]{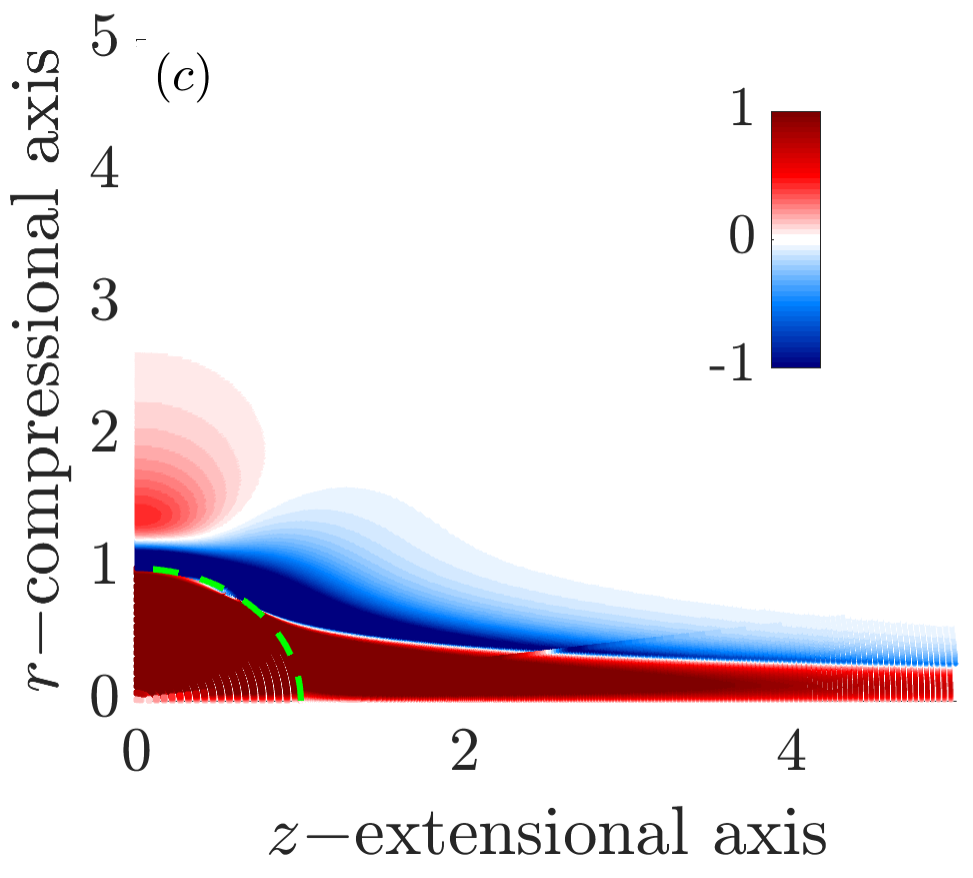}\label{fig:NLStressDep4Grid14v7Zoomed}}
	\caption{(a) Extra polymer stress, $\hat{{\Pi}} - \hat{{\Pi}}^{U}$, (b) linearized polymer stress, $\hat{{\Pi}}^{L}$, and (c) non-linear polymer stress, $\hat{{\Pi}}^{NL}$ around a sphere (boundary denoted by dashed green curve at $r^2 + z^2 = 1$) in a uniaxial extensional flow of a FENE-P liquid with $De = 0.4$, $L = 10$, and $c = 0.471$ at the steady state (Hencky strain, $H = 6$). All three figures share the same color scale labeled in (c).}\label{fig::ActualandNLZoomed}
\end{figure}

\begin{figure}
	\centering	
	\subfloat{\includegraphics[width=0.33\textwidth]{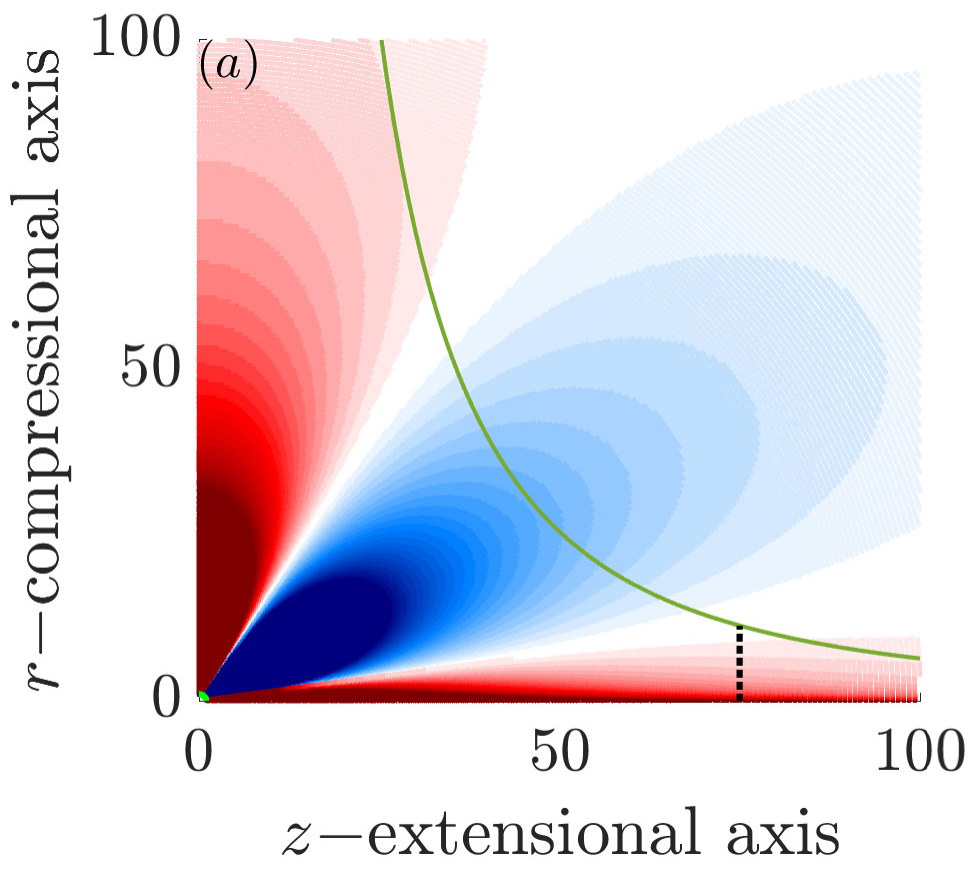}\label{fig:ActualStressDep4Grid14v7}}
	\subfloat{\includegraphics[width=0.33\textwidth]{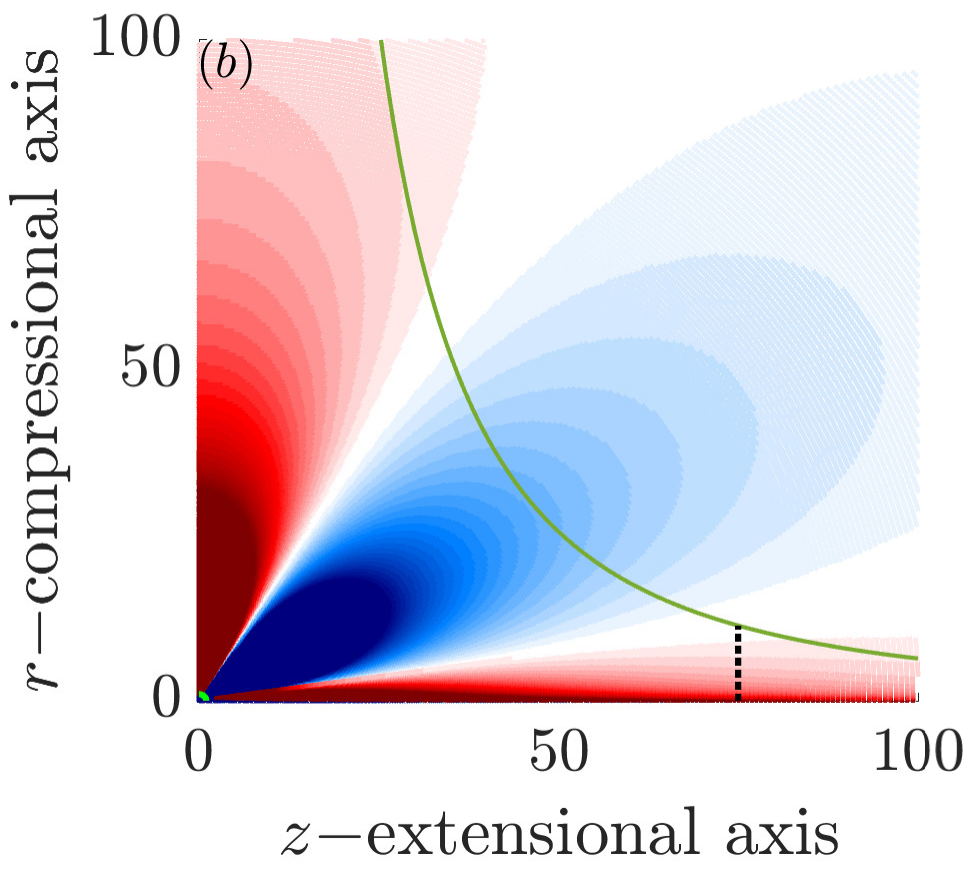}\label{fig:LinStressDep4Grid14v7}}
	\subfloat{\includegraphics[width=0.334\textwidth]{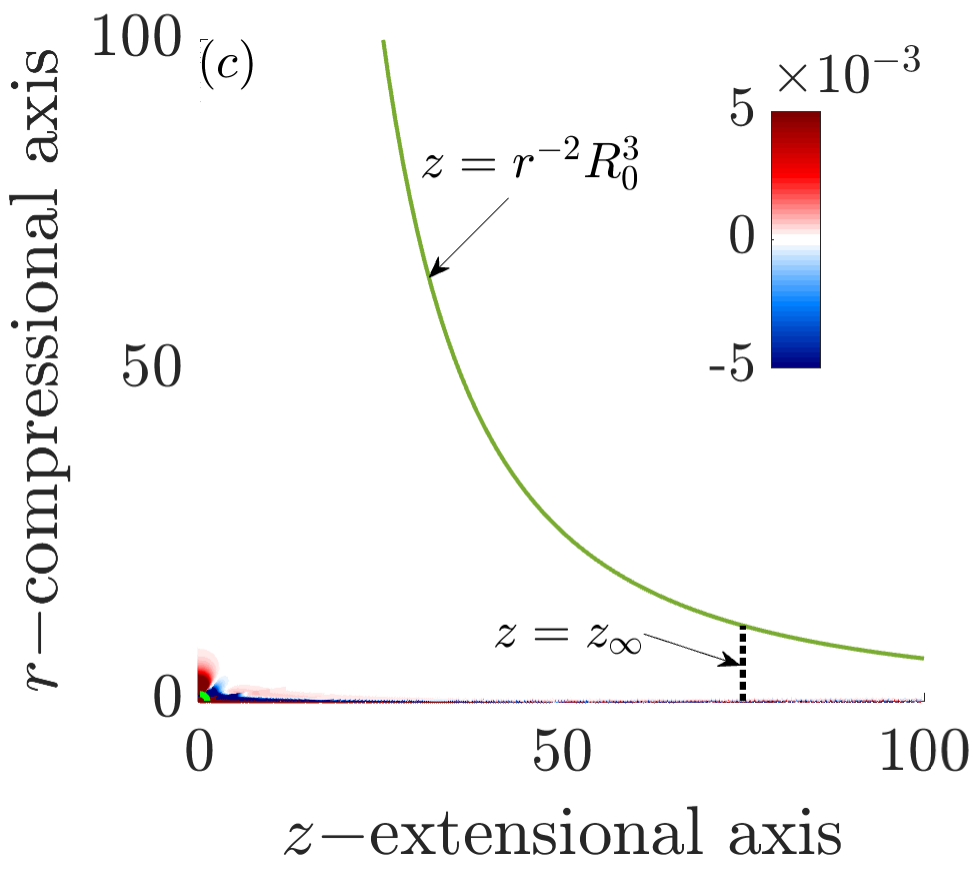}\label{fig:NLStressDep4Grid14v7}}
	\caption{Same parameters and caption as figure \ref{fig::ActualandNLZoomed}, but showing a larger region around the particle. The dashed black curve at $z = z_\infty = 75$ and a solid green curve representing $z = r^{-2} R_0^3$ ($R_0 = 40$) are relevant for the volume integral in PIPS.}
	\label{fig::ActualandNL}
\end{figure}

The three panels of figure \ref{fig::ActualandNLZoomed} show $\hat{{\Pi}} - \hat{{\Pi}}^{U}$, $\hat{{\Pi}}^{L}$, and $\hat{{\Pi}}^{NL} = \hat{{\Pi}} - \hat{{\Pi}}^{L} - \hat{{\Pi}}^{U}$ respectively for a FENE-P liquid with $L = 100$, $De = 0.4$, and $c = 0.471$, corresponding to the black curve in figure \ref{fig:CompareShaqFENE}. The computation domain used in these simulations extends to $r_\text{out} = 800$, and the number of mesh points is $N_r = 1000$ and $N_\theta = 451$. Figure \ref{fig::ActualandNL} shows the same quantities as figure \ref{fig::ActualandNLZoomed} but over a much larger region (up to 100 particle radii in each direction). From these figures, we can notice that most of the far-field variation in the extra polymer stress, $\hat{{\Pi}} - \hat{{\Pi}}^{U}$, is within $\hat{{\Pi}}^{L}$. Thus, the non-linear polymer stress $\hat{{\Pi}}^{NL}$ is concentrated much closer to the particle surface.

The simulations conducted by \cite{jain2019extensional} employed a rectangular box with extents of 40 and 40-80 particle radii in the compressional and extensional directions, respectively. This is perhaps an appropriate size if $\hat{{\Pi}}^{NL}$ is used in the PIPS integral. However, from the relatively unbounded simulation conducted here with a spherical domain of 800 particle radii, we can observe a substantial volumetric effect of $\hat{{\Pi}} - \hat{{\Pi}}^{U}$ from figure \ref{fig:ActualStressDep4Grid14v7}. Thus, at around 40-80 particle radii, the extra polymer stress leads to a significant contribution. In our simulations, increasing the computational domain affects $\hat{{\Pi}} - \hat{{\Pi}}^{U}$ near the boundaries of the smaller domain, but not $\hat{{\Pi}}^{NL}$.

\begin{figure}
	\centering	
	\subfloat{\includegraphics[width=0.4\textwidth]{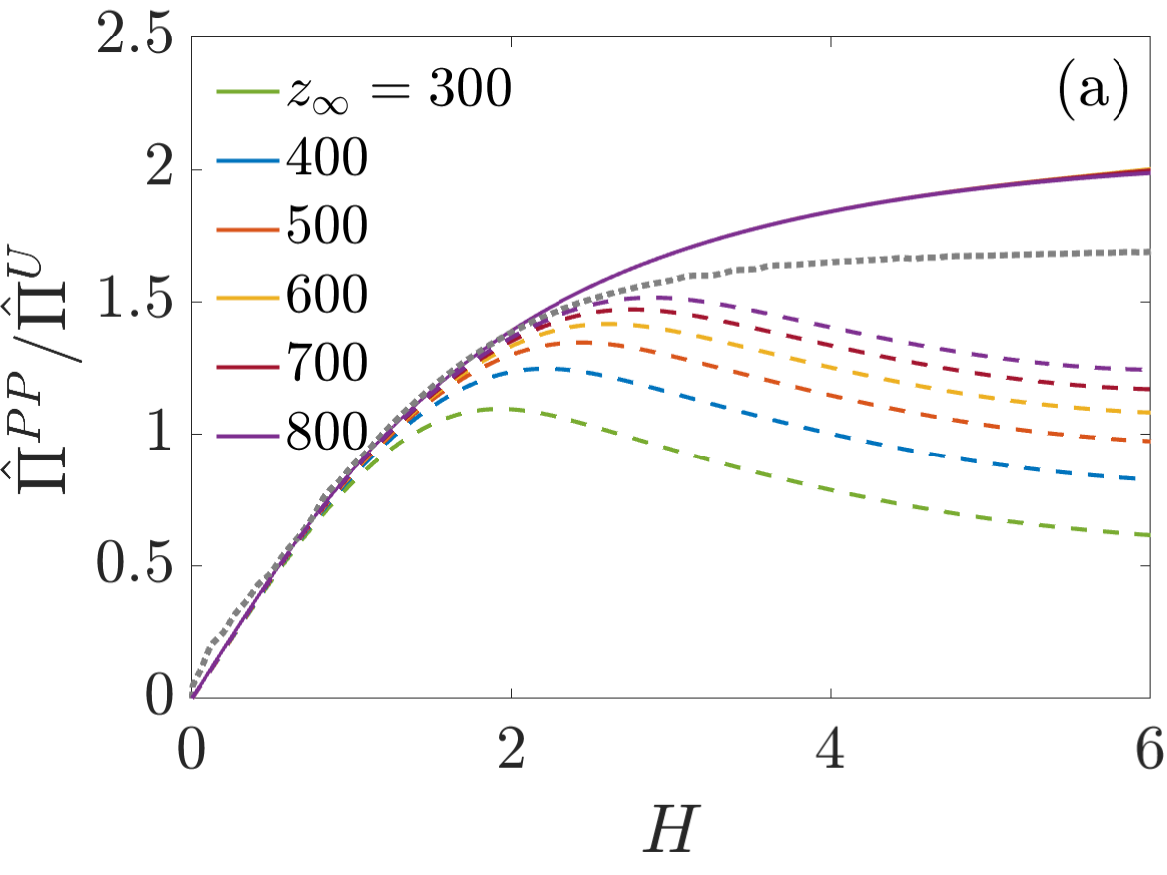}\label{fig:EffectofZcutoffNxNy1000451R040}}\hspace{0.2in}
	\subfloat{\includegraphics[width=0.4\textwidth]{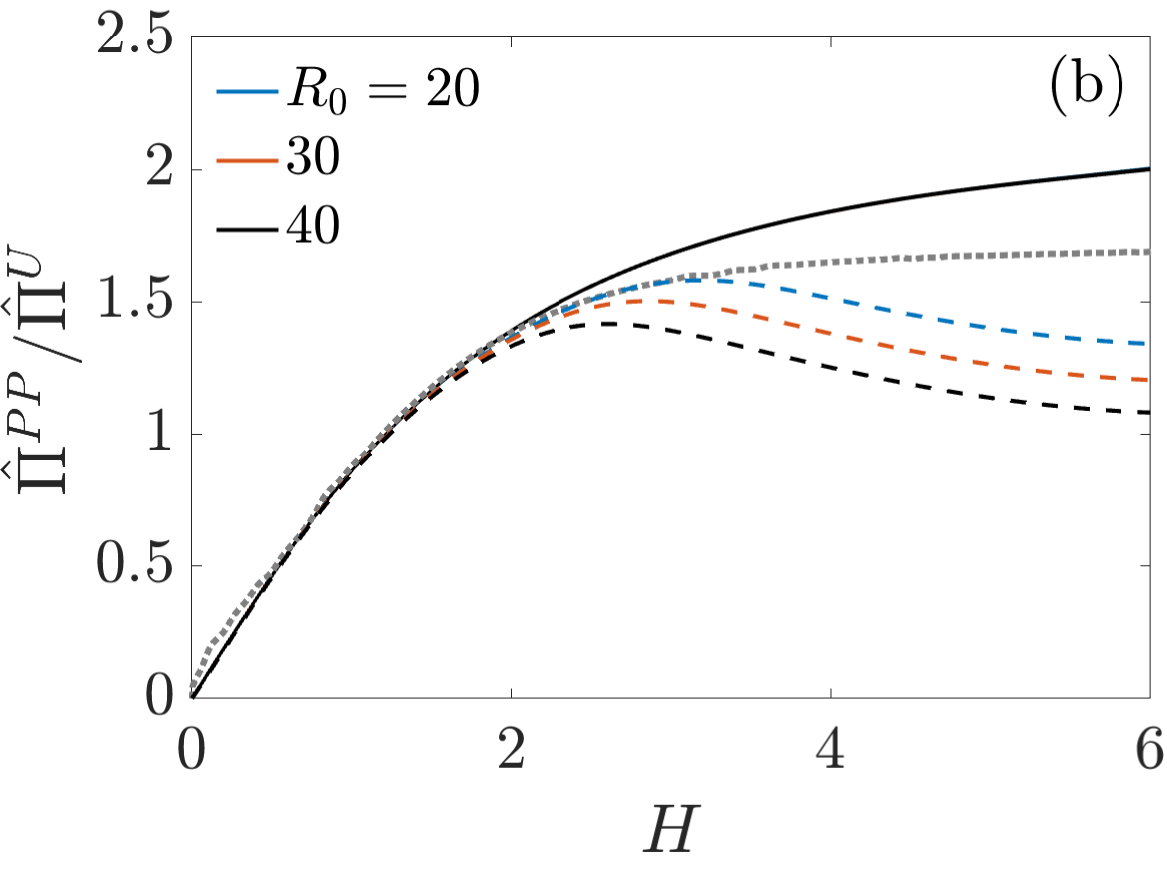}\label{fig:EffectofR0NxNy1000451Zcut600}}
	\caption{Effect of $z_\infty$ and $R_0$ (defined in figure \ref{fig:NLStressDep4Grid14v7}) on the evaluation of PIPS defined by the volume integral of non-linear polymer stress, $\hat{{\Pi}}^{NL}$, (solid lines) and extra polymer stress, $\hat{{\Pi}} - \hat{{\Pi}}^{U}$ (dashed lines) from our simulations for a FENE-P liquid with $L = 100$, $De = 0.4$, and $c = 0.471$ conducted for $r_\text{far} = 800$. The dotted line is from \cite{jain2019extensional}. Panels (a) and (b) represent $R_0 = 40$ and $z_\infty = 600$, respectively, with different curves representing different $z_\infty$ and $R_0$ labeled in the legend.}
	\label{fig::EffectofZcutoff}
\end{figure}

{Now we explicitly consider the effect of numerical parameters, such as domain size and mesh resolution, on the spatial distributions of $\hat{{\Pi}}$, $\hat{{\Pi}}^{L}$, and $\hat{{\Pi}}^{NL}$, and their subsequent effects on the volume integrals.} We first concentrate on the effect of the spatial extent of the volume integral of $\hat{{\Pi}} - \hat{{\Pi}}^{U}$ in the simulation conducted with a computational domain size $r_\text{far} = 800$ corresponding to the figures \ref{fig::ActualandNLZoomed} and \ref{fig::ActualandNL}, i.e., a FENE-P liquid with $L = 100$, $De = 0.4$, and $c = 0.471$. The volume integral is performed in a region within $z \le r^{-2} R_0^3$ and $z \le z_\infty$, and the parameters $R_0$ and $z_\infty$ are varied from 20 to 40 and 300 to 800, respectively. The curve $z \le r^{-2} R_0^3$ is a streamline of the undisturbed uniaxial extensional flow. Far from the particle, a polymer incoming into the computational domain at the upstream outer boundary ($\sqrt{z^2 + r^2} = r_\text{far}$) approximately travels along this curve. Figure \ref{fig::EffectofZcutoff} shows the evolution of PIPS as defined by the integral of $\hat{{\Pi}} - \hat{{\Pi}}^{U}$ in dashed lines and as defined by the integral of $\hat{{\Pi}}^{NL}$ (equation \eqref{eq:PIPSTotal}) in solid lines of different colors. In panel (a), $R_0$ is fixed to 40 and $z_\infty$ is varied, while in panel (b) $z_\infty$ is fixed to 600 and $R_0$ is varied. A dotted black line representing the result from \cite{jain2019extensional} is also included in each plot. Initially, all the curves coincide as the polymers are primarily affected by the region very close to the particle. However, at larger Hencky strains, $H$, the dashed curves for the integral of $\hat{{\Pi}} - \hat{{\Pi}}^{U}$ diverge, while the solid curves for the integral of $\hat{{\Pi}}^{NL}$ remain indistinguishable from one another. As $z_\infty$ or $R_0$ is increased, the integral of $\hat{{\Pi}} - \hat{{\Pi}}^{U}$ at large $H$ seemingly approaches that of $\hat{{\Pi}}^{NL}$, and it remains different from the calculations of \cite{jain2019extensional}.

\begin{figure}
	\centering	
	\subfloat{\includegraphics[width=0.4\textwidth]{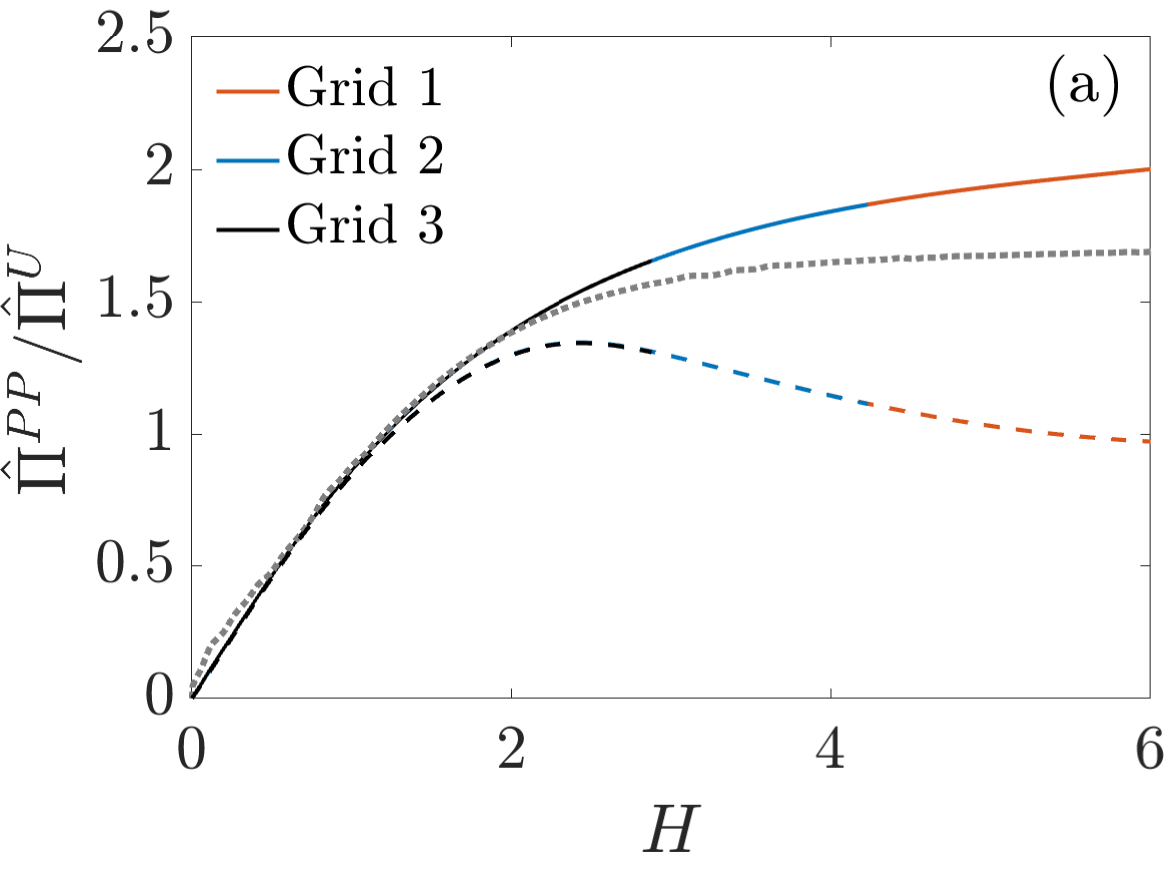}\label{fig:EffectofGridNxNy1000451Zinf500R040}}\hspace{0.2in}
	\subfloat{\includegraphics[width=0.4\textwidth]{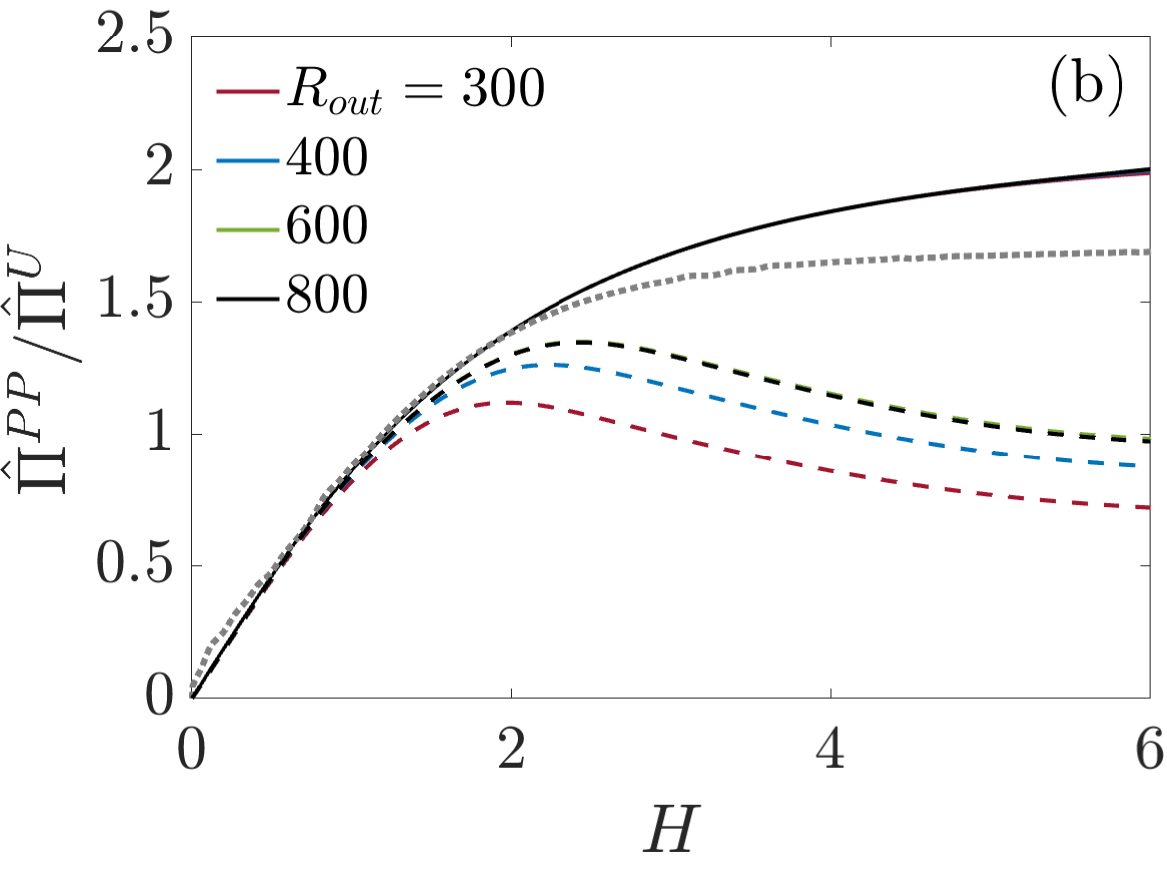}\label{fig:EffectofRoutNxNy1000451Zinf500R040}}
	\caption{(a) Effect of grid size on the $\hat{{\Pi}}^{NL}$ (solid lines) and extra polymer stress, $\hat{{\Pi}} - \hat{{\Pi}}^{U}$ (dashed lines) integral curves with Grid 1 corresponding to $N_r = 1000$, $N_\theta = 451$, Grid 2 to $N_r = 1000$, $N_\theta = 551$, and Grid 3 to $N_r = 1000$, $N_\theta = 551$. (b) Effect of computational domain size $R_\text{out}$. The simulations are for a FENE-P liquid with $L = 100$, $De = 0.4$, and $c = 0.471$. In these curves, $R_0 = 40$ and $z_\infty = 500$ is used, and a dotted curve representing results from \cite{jain2019extensional} is added.}
	\label{fig:EffectofRout}
\end{figure}

Each of the curves presented in figure \ref{fig::EffectofZcutoff} is converged with respect to the mesh size (as well as time step). This is indicated by figure \ref{fig:EffectofGridNxNy1000451Zinf500R040}, where each type of curve (solid or dashed) is indistinguishable across different mesh sizes used (different colors). The effect of the size of the computational domain, $r_\text{out}$, is shown in figure \ref{fig:EffectofRoutNxNy1000451Zinf500R040}. As expected, increasing the computational domain improves the $\hat{{\Pi}} - \hat{{\Pi}}^{U}$ integral (dashed lines) by bringing it closer to the integral of $\hat{{\Pi}}^{NL}$ (solid lines). However, solid curves representing the volume integral of $\hat{{\Pi}}^{NL}$ are unaffected by the change in mesh size (all other solid curves are hidden behind the black curve). Therefore, besides being mathematically more appropriate, removing the linearized stress, $\hat{\boldsymbol{\Pi}}^{L}$, from the extra polymer stress before evaluating the volume integral for PIPS is numerically beneficial in obtaining a faster convergence with mesh size.

	\section{An analogy: Polymers stretch like lines of dye released in the Newtonian flow at previous times}\label{sec:PolymerDye}
\begin{figure}
		\centering
		\subfloat{\includegraphics[width=0.33\textwidth]{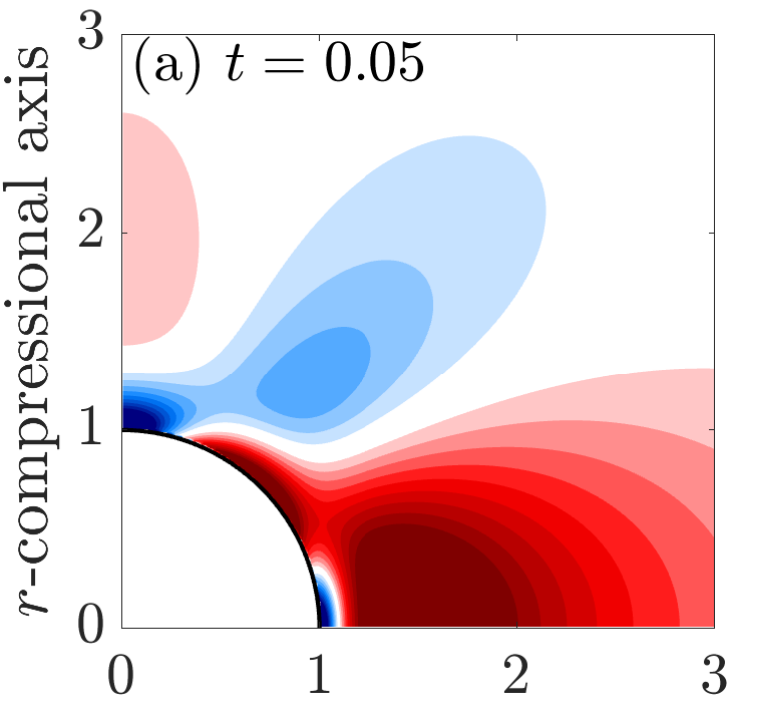}\label{fig:S_pt1}}\hfill
		\subfloat{\includegraphics[width=0.33\textwidth]{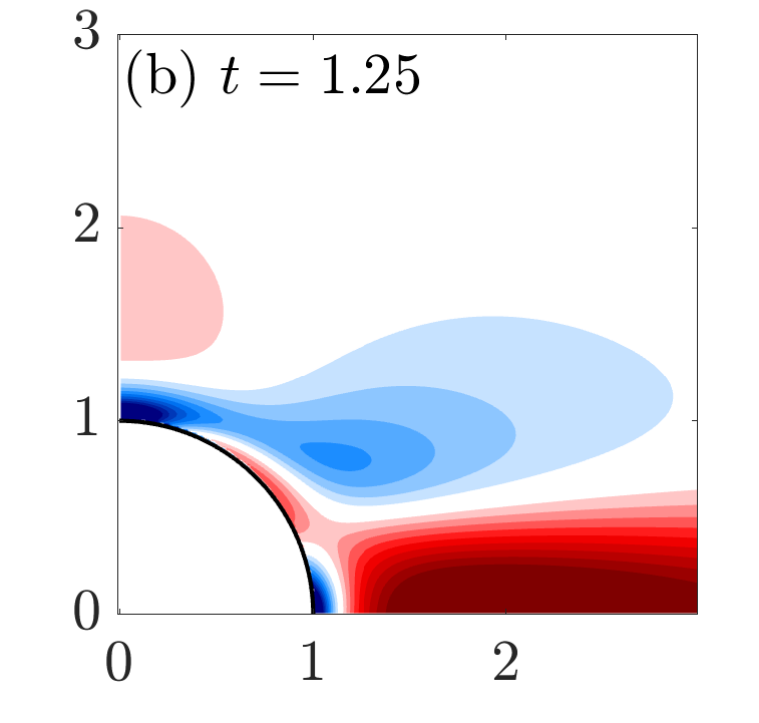}\label{fig:S_1}}\hfill
		\subfloat{\includegraphics[width=0.33\textwidth]{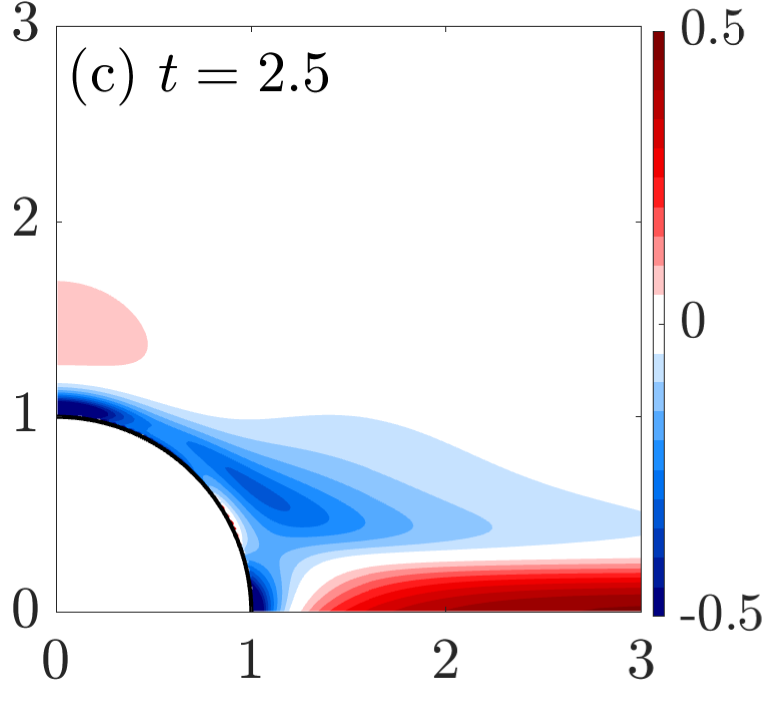}\label{fig:S_3}}\\
		\subfloat{\includegraphics[width=0.33\textwidth]{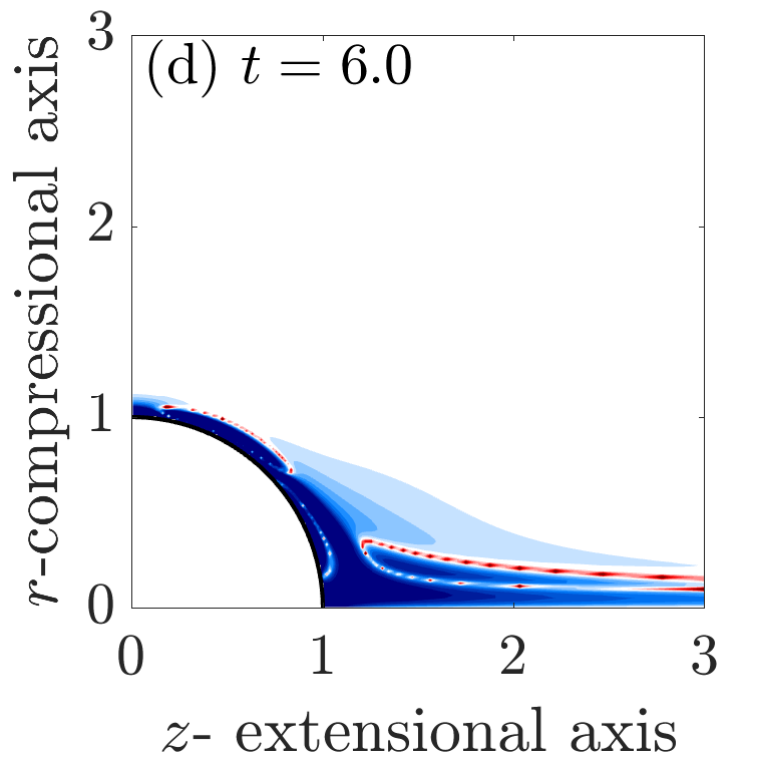}\label{fig:S_5}}\hfill
		\subfloat{\includegraphics[width=0.33\textwidth]{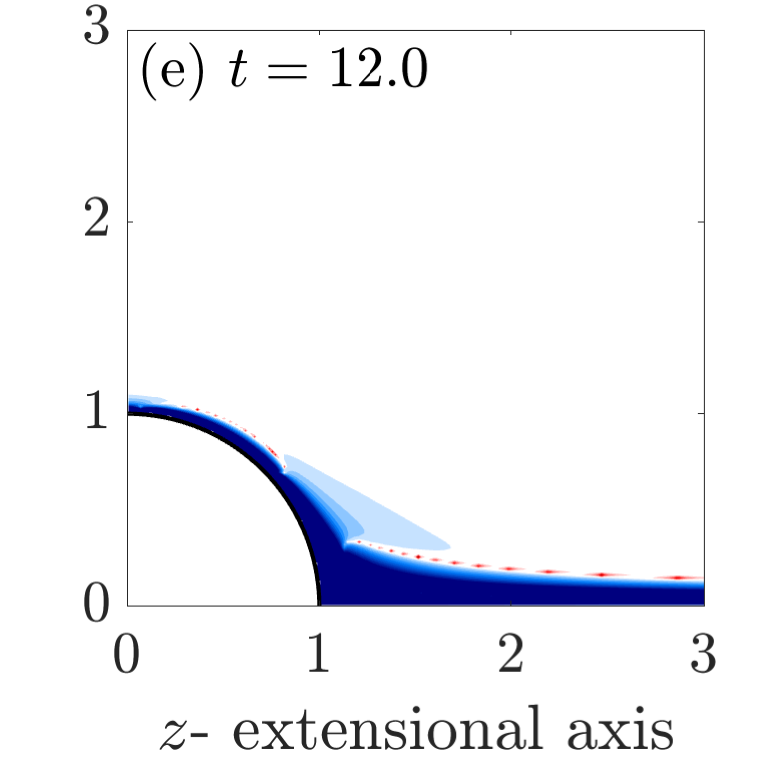}\label{fig:S_10}}\hfill
		\subfloat{\includegraphics[width=0.33\textwidth]{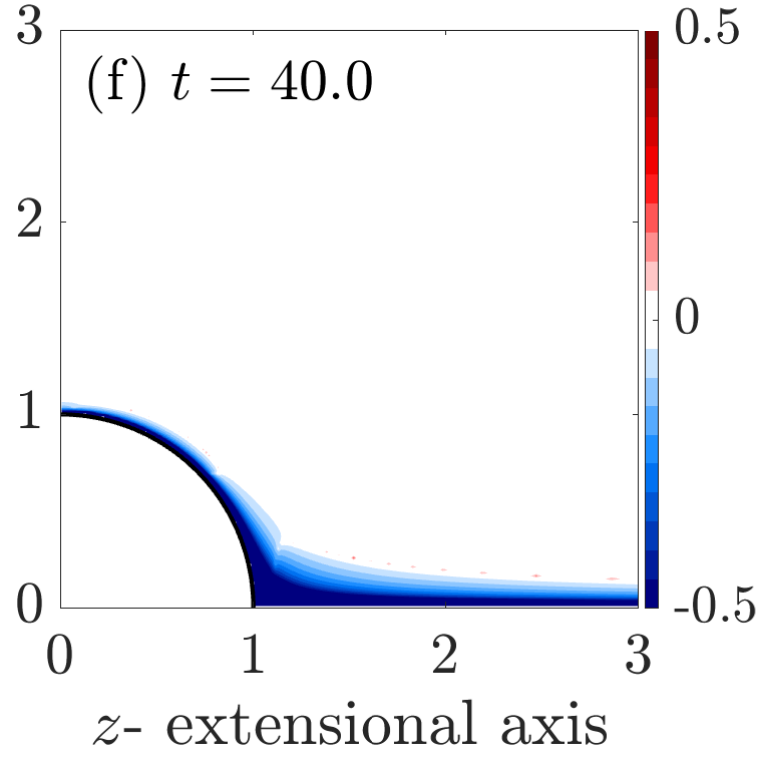}\label{fig:S_50}}
		\caption {The change in the finite time stretch field, $\Delta{\text{FTS}}(t; \boldsymbol{x}_0)$, due to the sphere in extensional flow for various $t$. \label{fig:S-field}}
	\end{figure}

	In our previous work \citep{SteadyStatePaper}, we noted an interesting analogy between the way the presence of a sphere in a uniaxial extensional flow changes the stretch of a line of dye in a Newtonian fluid and the way it changes the polymer stretch shown here through the $\Delta\mathcal{S}$ field. To quantify the former, we introduced a finite time stretch field, or FTS, defined as 
	\begin{equation}
		\text{FTS}(t; \boldsymbol{x}_0) = \frac{1}{2t} \ln\Big(\frac{1}{\tilde{\lambda}_1(t; \boldsymbol{x}_0)}\Big),
	\end{equation}
	where $\tilde{\lambda}_1(t; \boldsymbol{x}_0)$ is the minimum eigenvalue of the Cauchy-Green tensor,
	\begin{equation}
		\tilde{\boldsymbol{C}}_{0}^t = \nabla_{\boldsymbol{x}_0} \boldsymbol{\tilde{x}}(t; \boldsymbol{x}_0)^\text{T} \cdot \nabla_{\boldsymbol{x}_0} \boldsymbol{\tilde{x}}(t; \boldsymbol{x}_0),
	\end{equation}
	of the backward time flow map,
	\begin{equation}
		\boldsymbol{\tilde{x}}(t; \boldsymbol{x}_0) = \boldsymbol{x}_0 - \int_{0}^{t} \mathbf{u}(\boldsymbol{\tilde{x}}(\tau), \tau) \text{d} \tau
	\end{equation}
	generated by the velocity field $\mathbf{u}$. We refer to \cite{SteadyStatePaper} for a detailed derivation and discussion. In the undisturbed uniaxial extensional flow, $\text{FTS}(t; \boldsymbol{x}_0) = 1$ for all $\boldsymbol{x}_0$ and $t$. The change in the local $\text{FTS}(t; \boldsymbol{x}_0)$ at location $\boldsymbol{x}_0$ from this value for a chosen $t$,
	\begin{equation}
		\Delta\text{FTS}(t; \boldsymbol{x}_0) = \text{FTS}(t; \boldsymbol{x}_0) - 1,
	\end{equation}
	provides information about the way a fluid element or non-diffusive line of dye released time $t$ ago is currently stretched. The regions of $\Delta\text{FTS}(t; \boldsymbol{x}_0) > 0$ are the locations where the dye is more stretched, and the regions of $\Delta\text{FTS}(t; \boldsymbol{x}_0) < 0$ appear where the dye is less stretched in the presence of a sphere compared to the dye undergoing stretching in an undisturbed uniaxial extensional flow. We show the $\Delta\text{FTS}(t; \boldsymbol{x}_0)$ figures in figure \ref{fig:S-field} for various $t$.
	
	In section \ref{sec:InteractionMech}, we noted that at a particular $De$, $\Delta\mathcal{S}$ undergoes all the changes with $H$ during its transient that $\Delta\mathcal{S}$ at a lower $De$ undergoes until the latter's steady state. However, the changes at the lower $De$ occur in a shorter time at the larger $De$. From these $\Delta{\text{FTS}}(t; \boldsymbol{x}_0)$ figures, we observe that the effect of the sphere on the stretch of a polymer at large $De$ is the same as its effect on the stretching of a non-diffusive line of dye released in the flow at a certain time, $t$ ago. The analogy is valid up to a $t$ in proportion with the polymer's memory or its $De$.
	
	For small $De$ such as 0.2 with limited memory, the steady state $\Delta\mathcal{S}$ field at $H = 2.0$ in figure \ref{fig:DelSNormHencky2Dep2L100} is analogous to $\Delta{\text{FTS}}(t; \boldsymbol{x}_0)$ at $t = 0.05$. At $De = 0.4$, $\Delta\mathcal{S}$ first undergoes the same transition as $De = 0.2$ does up to its steady state. Then, the $\Delta\mathcal{S}$ field at $H = 2.85$ (close to steady state) and 10.0 (at steady state) for $De = 0.4$ shown in figures \ref{fig:DelSNormHenckyp2p85Dep4L100} and \ref{fig:DelSNormHenckyp10Dep4L100} is similar to the $\Delta{\text{FTS}}(t; \boldsymbol{x}_0)$ field at $t = 1.25$. At $De = 0.4$, $\Delta\mathcal{S}$ does not change much from $H = 2.85$ to its steady state. 
	
	The same qualitative changes as $\Delta\mathcal{S}$ at $De = 0.4$, but within a smaller range of $H$, are observed for $\Delta\mathcal{S}$ at $De = 0.6$ and $L = 100$. However, $\Delta\mathcal{S}$ at $De = 0.6$ evolves further, as represented through figures \ref{fig:DelSNormHenckyp5Dep6L100}, \ref{fig:DelSNormHencky1p5Dep6L100}, and \ref{fig:DelSNormHencky8Dep6L100} at $H / \log(L) = 0.5$, 1.5, and 8.0. This is similar to the changes in the stretch of the line of dye in a Newtonian fluid from $t = 1.25$ to 12.0, as shown by the $\Delta{\text{FTS}}(t; \boldsymbol{x}_0)$ plots at these times in figure \ref{fig:S-field}.
	
	At a very large $De = 5.0$ and $L = 100$, further qualitative changes occur in the $\Delta\mathcal{S}$ field in the later part of its evolution from $H / \log(L) = 1.15$ to 2.0, as shown in figures \ref{fig:DelSNormHencky1p15De5L100} and \ref{fig:DelSNormHencky2De5L100}. These are similar to the qualitative changes in the $\Delta{\text{FTS}}(t; \boldsymbol{x}_0)$ plots of figure \ref{fig:S-field} from $t = 12$ to 40. 
	
	At $De > 0.5$, lower $L$ stops the analogy at a smaller $t$. For example, at $De = 0.6$ with $L = 100$, the evolution of $\Delta\mathcal{S}$ up to its steady state is analogous to $\Delta{\text{FTS}}(t; \boldsymbol{x}_0)$ up to $t = 12$, as noted above. However, for $L = 10$, the steady state $\Delta\mathcal{S}$ shown in figure \ref{fig:DelSNormHencky8Dep6L10} is analogous to $\Delta{\text{FTS}}(t; \boldsymbol{x}_0)$ at $t = 2.5$. The loss in memory due to limited $L$ was also noted in \cite{SteadyStatePaper}. In that case, the analogy was made with $\Delta{\text{FTS}}(t; \boldsymbol{x}_0)$ at different $t$ across the steady state $\Delta\mathcal{S}$ at different $De$. Here, we have identified the analogy of $\Delta{\text{FTS}}(t; \boldsymbol{x}_0)$ at different $t$ with the transient evolution of $\Delta\mathcal{S}$ at a fixed $De$.
\end{document}